\documentclass[rmp,aps,twocolumn,nopreprintnumbers,eqsecnum]{revtex4-1}

\usepackage{amsmath}
\usepackage{amssymb}
\usepackage{amsfonts}
\usepackage{graphicx}
\usepackage{subfigure}
\usepackage{textcomp}
\usepackage{slashed}
\usepackage{color}
\usepackage{wrapfig}
\usepackage{dcolumn}
\definecolor{lred}{RGB}{180,10,10}
\definecolor{lblue}{RGB}{5,5,255}

\usepackage[colorlinks,,linktoc=section,linkcolor=lred,citecolor=lblue,filecolor=lblue,urlcolor=lblue]{hyperref}

\def\beq{\begin{equation}}
\def\eeq{\end{equation}}
\def\SM{$SU(3)\times SU(2)\times U(1)$}
\def\half{\frac12}
\definecolor{darkgreen}{RGB}{13,50,11}
\definecolor{darkgreen}{RGB}{13,70,11}
\definecolor{darkblue}{RGB}{13,13,101}
\definecolor{darkred}{RGB}{80,0,0}
\definecolor{red}{RGB}{255,15,0}

\begin{document}

 \rightline{NIKHEF/2013-010}


\newif\ifExtendedVersion 
\ExtendedVersionfalse

\title{Life at the Interface of Particle Physics and String Theory}
\ifExtendedVersion
\homepage{Extended version; see also www.nikhef.nl/$\sim$t58/Landscape}
\else
\homepage{Extended version available at www.nikhef.nl/$\sim$t58/Landscape}
\fi

\author{A N Schellekens}

\affiliation{Nikhef, 1098XG Amsterdam (The Netherlands)}
\affiliation{IMAPP, 6500 GL Nijmegen (The Netherlands)}
\affiliation{IFF-CSIC, 28006 Madrid (Spain)}

\begin{abstract}

If the results of the first LHC run are not betraying us, many decades of particle physics are culminating
in a complete and consistent theory for all non-gravitational physics: the Standard Model. 
But despite this monumental achievement  there is a clear sense of disappointment: many questions 
remain unanswered. Remarkably, most unanswered questions could just be environmental, and disturbingly
(to some) the existence of life may depend on that environment. Meanwhile there has been increasing
evidence that 
the seemingly ideal candidate
for answering these questions, String Theory, gives an answer few people initially expected: a huge ``landscape"
of possibilities, that can be realized in a multiverse and populated by eternal inflation. At the interface of ``bottom-up"
and ``top-down" physics, a discussion of anthropic arguments becomes unavoidable. We review developments
in this area, focusing especially on the last decade.

\end{abstract}

\maketitle

\tableofcontents

\section{\label{Introduction}Introduction}



In popular accounts, our universe is usually described as unimaginably huge. Indeed, during the
last centuries we have seen our horizon expand many orders of magnitude beyond any 
scale humans can relate to.

But the earliest light we can see has traveled a mere 13.8 billion years, just about three times the age of
our planet. We might be able to look a little bit further than that using intermediaries  other than light,
but soon we inevitably reach a horizon beyond which we cannot see. 

We cannot rule out the possibility that beyond that horizon there is 
just more of the same, or even nothing at all, but widely accepted theories suggest something else. In the theory of inflation,
our universe emerged from a piece of a larger ``space"  that expanded by at least sixty e-folds. 
Furthermore, in most theories of inflation our universe is not a ``one-off" event. It is much more plausible that the
mechanism that gave rise to our universe was repeated a huge, even infinite, number of times. 
Our universe could
just be an insignificant bubble in a gigantic cosmological ensemble, a ``multiverse".  
There are several classes of ideas that lead to such a picture, but there is no need 
to be specific here. The main point is that other universes than our  own may exist, at least in
a mathematical sense. The universe we see is really just {\it our} universe. Well, not just ours, presumably.

The existence of a multiverse may sound like speculation, but one may as well ask how we can possibly be certain that this is
{\it not} true. Opponents and advocates of the multiverse idea are both limited by the same horizon. On whom rests 
the burden of proof? What is the most extraordinary statement: that what we can see is precisely all
that is possible, or that other possibilities might exist? 

If we accept the logical possibility of a multiverse, the question arises in which respects other universes might be different.
This obviously includes quantities that vary even within our own universe, such as the distribution of matter and the
fluctuations in the cosmic microwave background. But the cosmological parameters themselves, and not just their fluctuations,
might vary as well. And there may be
more that varies: 
the ``laws of physics" could be different. 

Since we observe only one set of laws of physics it is a bit precarious to contemplate others. Could there exist alternatives to quantum
mechanics, or could  gravity ever be repulsive rather than attractive? None of that makes
 sense in any way we know, and hence it seems unlikely that anything useful can be learned by speculating about this. If we want to consider variations in the laws of physics, we
should focus on laws for which we have a solid underlying theoretical description. 

The most solid theoretical framework we know is that of quantum field theory, the language in which the Standard Model of particle
physics is written. Quantum field theory provides a huge number of theoretical possibilities, distinguished by some discrete and some continuous choices.
The discrete choices are a small
set of allowed Lorentz group representations, a choice of gauge symmetries (such as the strong and electroweak interactions), and a choice of gauge-invariant
couplings of the remaining matter. The continuous choices are the low-energy parameters 
that are not yet fixed by the aforementioned symmetries. In our universe we observe a certain choice among all of these options, called the Standard Model, sketched in section \ref{StandardModelSection}. 
But the quantum field theory we observe is just a single point in a discretely and continuously infinite space. Infinitely many  other choices are mathematically equally consistent.

Therefore the space of all quantum field theories provides the solid underlying description we need if we wish to consider alternatives
to the laws of physics in our own universe. This does not mean that nothing else could vary, just that we cannot 
discuss other variations with the same degree of confidence. 
But we can certainly theorize in a meaningful way about universes where the gauge group or the 
fermion masses are different, or where the matter does not even consist of quarks and leptons.

We have no experimental evidence about the existence of such universes, although there are speculations about possible 
observations in the Cosmic Microwave Background (see section \ref{CosmologySection}).
We may get lucky, but our working hypothesis will be the pessimistic  one that
all we can observe is our own universe. 
But even then, the
claim that the only quantum field theory we can observe in principle, the Standard Model of particle physics,  
is also the only one that
can exist mathematically, would be truly extraordinary.

Why should we even care about alternatives to our universe?  
One could adopt the point of view that the only reality is what we can observe, and that talking about anything else amounts to
leaving the realm of science. But even then there is an important consequence. 
If other sets of laws of physics are possible, even just mathematically, this implies that our laws of physics cannot be derived from first principles.
They would be -- at least partly -- environmental, and deducing them would require  some experimental or
observational input. Certainly this is not what many leading physicist have been hoping  for in the last decades. 
\ifExtendedVersion {\color{darkred}
Undoubtedly, many of them hoped for a  negative answer to
Einstein's famous question ``{\em I wonder if God had any choice in creating the world}". }\fi
Consider for example Feynman's question about the value of 
 the
fine-structure constant $\alpha$:
``{\em Immediately you would like to know where this number for a coupling comes from: is it related to pi or perhaps to the base of natural logarithms?"}.
Indeed, there exist several fairly successful attempts to express $\alpha$ in terms of pure numbers.
But if $\alpha$ varies  in the multiverse, such a computation would be impossible, and any  successes would be mere numerology. 

There is a  more common ``phenomenological" objection, stating that even if a multiverse exists, still the only universe of phenomenological interest is our own. 
The latter attitude denies the main theme of particle physics in the last three decades.
Most activity has focused on the ``why questions" and on the problem of ``naturalness". This concerns the discrete
structure of the Standard Model, its gauge group, the couplings of quarks and leptons,  the questions why they
come in three families and  
why certain parameters  have strangely small values. The least one can say is that
if these features could  be different in other universes, this might be part of the answer to those
questions.

But there is a more important aspect to the latter discussion that is difficult to ignore in a multiverse.
If other environments are possible, one cannot avoid questions about the existence of life. 
It is not hard to imagine entire universes where
nothing of  interest can exist, for example because the only stable elements are hydrogen and helium.
In those universes there would be no observers. Clearly, the only universes in the multiverse that can be observed are those
that allow the existence of observers. This introduces a bias: what we observe is  not a typical
sample out of the set of possible universes, unless all universes that (can) exist contain entities one might plausibly
call ``observers". 
If the Standard Model features we are trying to understand vary over the multiverse, this is already crucial information.
If there is furthermore a possibility that {\it our own existence} depends on the values of these parameters, it is downright
irresponsible to ignore this when trying to understand them.  Arguments  of this kind are called ``anthropic", and tend to stir up
strong emotions. These are the kind of emotions that always seem to arise when our own place in the cosmos and its history 
is at stake. One is reminded of the resistance against heliocentricity and evolution. But history is not a useful guide to the right answer,
it only serves as reminder that arguments should be based on facts, not on emotions. 
We will discuss some \ifExtendedVersion {\color{darkred}of the history and some }\fi  general objections in section \ref{AnthropicSection}.

The fact that at present the existence of other universes and laws of physics cannot be demonstrated experimentally does not mean that we will never know.
One may hope that one day we will find a complete theory of all interactions by logical deduction, starting from a principle of physics. For more than half a  century, it has been
completely acceptable to speculate about such theories provided the  aim was a unique answer. But it is equally reasonable to pursue such a theory even if it
leads to a huge number of possible realizations of quantum field theories. This is not about ``giving up" on the decade long quest for
a unique theory of all interactions. It is simply pointing out a glaring fallacy in that quest. 
Nothing we know, and nothing we will argue for here, excludes the possibility that the traditional path of particle physics
towards shorter distances or higher energies will lead to a unique theory. The fallacy is to expect that there should be a unique
way back: that starting with such a theory we might derive our universe uniquely using pure mathematics.

\ifExtendedVersion {\color{darkred}
Nowadays few physicist would describe their expectations in such a strong way. There is a variety of points of view,
spread between two extremes, 
the {\em uniqueness  paradigm} and the {\em landscape paradigm}. 
The former states that ultimately everything can be derived, whereas
the most extreme form of the latter holds that -- from now on -- nothing can be derived, because our universe is just a point in a huge distribution. 
Neither can be correct as stated. The first is wrong because some features
in our universe are clearly fluctuations, and hence not derivable. So we will have to decide which observables are fluctuations. The fact that we do no {\it see} them fluctuate is not sufficient to
conclude that they {\it do not}. We cannot decide this on the basis of the single event that is our universe.
The second paradigm necessarily involves a moment in time. In the past
many physical quantities
(such as molecules, atoms and nuclei) 
have been derived  from simpler input data. 
So if we want to argue that, in some sense, this will no longer be possible, we must argue that we live in a very special moment in the history of physics. The Standard Model has been pointing in that direction for decades already, and its current status strengthens
the case.
}\fi

\ifExtendedVersion {\color{darkred} On the other side of the energy scale,}
there \else { There }\fi exists a theoretical construction that may have a chance to fulfill the hope of finding the underlying theory: String Theory. 
It is the third main ingredient of the story, and will be introduced in section \ref{StringTheorySection}. It describes both gravitational and gauge interactions,
as well as matter. Initially 
it seemed to deliver the unique outcome many were hoping for, as 
the strong constraints it has to satisfy appeared to allow only very few solutions.

But within two years, this changed drastically.  The ``very few solutions" grew exponentially to astronomically large numbers. 
One sometimes hears claims that string theorists were promising a unique outcome. But this is simply incorrect. 
In several papers from around 1986 one can find  strong statements about large numbers of possibilities, 
starting with \textcite{Narain:1985jj}, shortly thereafter followed by
\textcite{Strominger:1986uh}; \textcite{Kawai:1986ah}; \textcite{Lerche:1986cx}; and \textcite{Antoniadis:1986rn}. Large numbers of
solutions had already been found earlier in the context of Kaluza-Klein supergravity, reviewed by \textcite{Duff:1986hr}, but the demise of uniqueness
of string theory had a much bigger impact.

The attitudes towards these results differed. Some blamed  the huge number of solutions on our limited knowledge of string theory, 
and speculated about a dynamical principle that would determine the true ground state, see for example \textcite{Strominger:1986uh}.
Others accepted it as a fact, and adopted the phenomenological point of view that the right vacuum would have to be selected by
confrontation with experiment, as stated by \textcite{Kawai:1986ah}. 
In a contribution to the EPS conference in 1987 
the hope
for a unique answer was  described as ``unreasonable and unnecessary wishful thinking" \cite{Schellekens:1987zy}. 

It began to become clear to some people that string theory was not providing evidence {\it against} anthropic reasoning, but 
{\it in favor} of it. But the only person to state this explicitly at that time was Andrei \textcite{Linde:1986fd}, who
simply remarked that {\em ``the emergent
plenitude of solutions should not be seen as a difficulty but as a virtue"}.
It took ten more years for a string theorist to put this point of view into writing \cite{Speech}, 
and fifteen years before the message was advertised  loud and clear by \textcite{Susskind:2003kw}, already in the title of his paper: ``The Anthropic Landscape of String Theory".

In the intervening fifteen years a lot had changed. An essential r\^ole in the story is played by {\it moduli}, continuous parameters of string theory. 
String theorists like to emphasize
that ``string theory has no free parameters", and indeed this is true, since the moduli can be understood in terms of vacuum expectation values  (vevs) of
scalar fields, and hence are not really parameters.  All parameters of quantum field theory, the masses and couplings of particles, depend on these scalar vevs. The number of moduli
is one or two orders of magnitude larger than the number of Standard Model parameters. 
\ifExtendedVersion {\color{darkgreen}
This makes those parameters  ``environmental" almost by definition, and 
the possibility that they could vary over an ensemble of universes  in a  multiverse is now wide open. 
}\else {
This makes those parameters  ``environmental"  by definition, and opens
the possibility that they could vary over an ensemble of universes.}\fi

The scalar potential governing the moduli is flat in the supersymmetric limit. Supersymmetry is a symmetry between boson and fermions, which is 
-- at best -- an approximate symmetry in our universe, but
also a nearly indispensable tool in the formulation of string theory. If supersymmetry is broken, there is no reason  why the potential should be flat.
But this potential could very well have a disastrous run-away behavior towards large scalar vevs  or have
computationally inaccessible local minima \cite{Dine:1985he}.
Indeed, this potential catastrophe was looming over string theory 
until the beginning of this century, when a new ingredient known as ``fluxes" was discovered by \textcite{Bousso:2000xa}.
This gave good reasons to believe that the potential can indeed have controllable
local minima, and that the number of minima  (often referred to as ``string vacua") is huge: 
an estimate of $10^{500}$ given by \textcite{Douglas:2004zg} is leading a life of its own in the literature. These minima  are not expected to be absolutely stable; a lifetime of about $14\times 10^9$ years is sufficient.

This ensemble  has been given the suggestive name ``the Landscape of String Theory".
Our universe would correspond to one of the minima of the potential. The minima are sampled by means of tunneling processes 
from an eternally inflating de Sitter (dS) space \cite{Linde:1986fc}.  If this process continues eternally,  if all vacua are sampled and if our universe
is one of them (three big IF's that require more discussion), then this
provides a concrete setting in which anthropic reasoning is not only meaningful, but inevitable.

This  marks a complete reversal of the initial expectations
of string theory, and is still far from being universally accepted or formally established. Perhaps it will just turn out to be 
a concept that forced us to rethink our expectations about the fundamental theory. But a more optimistic attitude is that we have 
in fact reached the initial phase of the discovery of that theory.

The landscape also provided a concrete realization of an old idea regarding
 the value of the cosmological constant $\Lambda$, which is smaller by more than 120 orders of magnitude than its naive size in Planckian units.    If $\Lambda$ varies over the
 multiverse, then its smallness is explained at least in part by the fact that for most of its values life would not exist. The latter statement is not debatable. What 
 can be debated is if $\Lambda$ does indeed vary, what the allowed values are and if anthropic arguments can be made sufficiently precise to determine its value.
  The anthropic argument, already noted by various authors,  was sharpened by  \textcite{Weinberg:1987dv}. It got little attention for more than a decade, because $\Lambda$ was 
 believed to be exactly zero and because  a physical mechanism allowing the required variation of $\Lambda$ was missing. In the string theory landscape the allowed values  of $\Lambda$ form a
 ``discretuum" that is sufficiently dense to accommodate the observed small value.

This
gave a huge boost to the Landscape hypothesis in the beginning of this millennium, 
and led to an explosion of papers in a remarkably broad range of scientific areas: string theory, particle physics, nuclear physics, astrophysics,
cosmology, chemistry, biology and geology, numerous areas in mathematics, even history and philosophy, not to mention theology.
It is impossible to cover all of this in this review.
It is not easy to draw a line, but on the rapidly inflating publication landscape we will use a measure that has its peak at the interface of the Standard Model and String Theory.

\ifExtendedVersion {\color{darkred}

An important topic which will not be covered are the various possible realizations of the multiverse. Especially in popular accounts,
notions like ``pocket universes", ``parallel universes", ``the many-world interpretation of quantum
mechanics", the string landscape and others are often uncritically jumbled together. They are not mutually exclusive, but do not all require
each other. For example, the
first three do not require variations in the laws of physics, and in particular the Standard Model.

To conclude this introduction we provide a brief list of popular books and reviews covering various points of view.
The anthropic string theory landscape is beautifully explained in \textcite{Susskind:2005bd}.  Another excellent popular book is \textcite{VilenkinBook}.
A very readable account of anthropic reasoning in cosmology is \textcite{Rees:1999cu}.
The classic book on the anthropic principle in cosmo\-logy is \textcite{BarrowTipler}, a mixture of historical, technical, philosophical and controversial material, that however can hardly be
called ``popular".

Precursors of the present review are \textcite{Hogan:1999wh} and  \textcite{Douglas:2006es}. 
The point of view of the author is presented more provocatively in \textcite{Schellekens:2008kg}. A very accessible review of the cosmological 
constant problem and the Bousso-Polchinski mechanism is presented in \textcite{Bousso:2007gp} and \textcite{Polchinski:2006gy}.  The book ``Universe or Multiverse" \cite{UniOrMulti} is
an interesting collection of various thoughts on this subject.

But there is also strong opposition to the landscape, the multiverse and the anthropic principle. One of the earliest works to recognize the emergent string theory landscape as well
as the fine-tunings in our universe is  \textcite{SmolinCosmos}, but
the author firmly rejects anthropic arguments.  The very existence of fine-tuning is denied in \textcite{StengerBook} 
(see however \textcite{Barnes:2011zh} for a detailed criticism, and an excellent review). The existence of the string theory landscape, as well as the validity of anthropic arguments
is called into question by \textcite{Banks:2012hx}, which is especially noteworthy because the author pioneered some of the underlying ideas.

}\fi

%
%
%

 \section{The Standard Model}\label{StandardModelSection}
 

Despite its modest name (which we will capitalize to compensate the modesty a little bit), the Standard Model is one of the greatest successes
in the history of science. It provides an amazingly accurate description of the three non-gravitational interactions we know: the strong, electromagnetic
and weak interactions. It successes range from the almost 10-digit accuracy of the anomalous magnetic moment of the electron to the  stunningly precise description of
a large number of high energy processes currently being measured at the LHC at CERN, and prior to that at the Tevatron at Fermilab, and many other
accelerators around the world. Its success was crowned on July 4, 2012, with the announcement of the discovery of the Higgs boson at CERN, the last particle that was still missing.
But this success has generated somewhat mixed reactions. In addition to the understandable euphoria, there are clear overtones of disappointment. Many particle physicists
hoped to see the first signs of failure of the Standard Model. A few would even have preferred {\it not} finding the Higgs boson.

This desire for failure on the brink of success can be explained in part by the hope of simply discovering something new and exciting, something that requires new theories and
justifies further experiments. But there is another reason. Most particle physicists are not satisfied with the Standard Model because 
it is based on a large number of seemingly {\it ad hoc} choices. Below we will enumerate them.


We start with the ``classic" Standard Model, the version without neutrino masses and right-handed neutrinos.
In its most basic form it fits on a T-shirt, a very popular item in the CERN gift shop these days. 
Its Lagrangian density is given by
\begin{equation}\label{StandardModel}
\begin{aligned}
{\cal L} &=  -\frac14  F_{\mu\nu} F^{\mu\nu} \\
&+ i \bar\psi \slashed{D}\psi + {\rm conjugate} \\
&+ \bar \psi_i Y_{ij} \psi_j \phi + {\rm conjugate} \\
& + | D_{\mu} \phi |^2 - V(\phi) \ .
\end{aligned}
\end{equation}
In this form it looks barely simple enough to be called ``elegant", and furthermore many details are hidden by the notation.

\ifExtendedVersion {\color{darkgreen}

\paragraph{Gauge group.}

The first line is a short-hand notation for the kinetic terms of the twelve gauge bosons, and their self-interactions. One recognizes the expression familiar 
from electrodynamics. There is an implicit sum over eleven additional gauge bosons, eight of which are the {\it gluons} that mediate the
strong interactions between the quarks, and three more that are responsible for the weak interactions. The twelve bosons are in one-to-one
correspondence with the generators of a Lie algebra, which is $SU(3)\times SU(2)\times U(1)$, usually referred to as the 
Standard Model ``gauge group", although strictly speaking we only know the algebra, not the global group realization. 
The generators of that Lie algebra
satisfy commutation relations $\left[T^a,T^b\right]=if^{abc} T^c$, and the real and fully anti-symmetric (in a suitable basis) constants $f^{abc}$ specify
the coupling of the gauge bosons labeled $a,b$ and $c$ to each other. Hence the $SU(3)$ vector bosons (the gluons) self-interact, as do the
three $SU(2)$ vector bosons. The field strength tensors $F^a_{\mu\nu}$ have the form $F_{\mu\nu}^a = \partial_{\mu} A_{\nu}^a -
\partial_{\nu} A_{\mu}^a + g f^{abc} A_{\mu}^b A_{\nu}^c$, where $g$ is the coupling constant. 
There are tree such constants in the
Standard Model, one for each factor in the gauge group. The will be denoted as $g_3, g_2$ $g_1$.
The coupling constant $g_1$ of the abelian factor does not appear yet, because
so far there is nothing the $U(1)$ couples to. Nothing in the formulation of the Standard Model fixes the choice of the gauge group
(any compact Lie algebra can be used) or the values of the coupling constants. All of that information is experimental input.

\paragraph{Fermions.}

The second line displays, in a short-hand notation, all kinetic terms of the fermions, the quarks and leptons, and their coupling to the twelve gauge bosons.
These couplings are obtained by  minimal substitution, and are encoded in terms of the covariant derivatives $D_{\mu}$
\begin{equation}
D_{\mu}=\partial_{\mu}-ig_i T^a A^a{_\mu}
\end{equation}
where $A^a_{\mu}$ is the vector field, and $T^a$ is a unitary $SU(3)\times SU(2)\times U(1)$ representation matrix, and $g_i$ is the
relevant coupling constant, depending on the label $a$. 
Representations of this Lie algebra are
combinations of representations of the factors, and hence  the choice can be parametrized as $(r,\ell,q)$, where $r$ is an irreducible representation of 
$SU(3)$, $\ell$ is a non-negative half-integer indicating an $SU(2)$ representation, and $q$ is a real number.   
If we write all fermions in terms of left-handed Weyl fermions, as is always possible, the fermion representation of the Standard Model is
\begin{equation*}
({\bf 3},{\bf 2},\frac16)+({\bf 3},{\bf 1},-\frac23)+({\bf 3},{\bf 1},\frac13)+({\bf 1},{\bf 2},-\frac12)+({\bf 1},{\bf 1},1)  \
\end{equation*}
This repeats three times for no known reason. These sets are called ``families". There is no theoretical reason why this particular combination of representations is the one we observe, but there is an important restriction on the
fermions from {\it anomaly cancellation}. This condition arises from triangle Feynman diagrams with three external gauge bosons or
two gravitons and a gauge boson, with a parity violating ($\gamma_5$) coupling  of at least one of the fermions.
These amplitudes violate gauge invariance, unless their group-theory factors cancel.
 This requires four cubic and one linear trace over the gauge group generators to vanish. This makes the structure of a single family a bit less arbitrary than it may seem
at first sight, but still leaves an infinity of other possibilities.

The first two lines are nearly completely fixed by
symmetries and depend only on the discrete choices of gauge group and representations, plus the numerical value of the three real coupling constants.
}
\else

\paragraph{The gauge sector.}
The first two lines are nearly completely fixed by
symmetries and depend only on the discrete choices of gauge group and representations, plus the numerical value of the three real coupling constants
of the gauge group $SU(3)\times SU(2)\times U(1)$. The left-handed fermions couple to this gauge group according to the following representations 
\begin{equation*}
({\bf 3},{\bf 2},\frac16)+(\overline{{\bf 3}},{\bf 1},-\frac23)+(\overline{{\bf 3}},{\bf 1},\frac13)+({\bf 1},{\bf 2},-\frac12)+({\bf 1},{\bf 1},1)  \
\end{equation*}
This repeats three times for no known reason. There is no theoretical reason why this particular combination of representations is the one we observe, although there is an important restriction on four cubic traces  and one linear trace of the representation matrices
from a condition called ``anomaly cancellation". \fi

\paragraph{Yukawa Couplings.}

The third line introduces a new field $\phi$, a complex Lorentz scalar coupled to the gauge group as $(1,2,\frac12)$, another choice dictated by
observation, and not by fundamental physics.
This line consists of all terms allowed by the gauge symmetry, with an arbitrary complex coefficient $Y_{ij}$, the Yukawa coupling,  for each term. 
The allowed couplings constitute three complex $3\times 3$ matrices, for a total of 54 parameters (not all of which are observable, see below). 

\paragraph{Scalar Bosons.}

The last line specifies the kinetic terms of the scalar boson,
with a minimal coupling to the gauge bosons. The last term is a potential, a function of $\phi$. This potential has the form
\begin{equation}
\label{ScalarPotential}
V(\phi)=\frac12 \mu^2 \phi^{*}\phi + \frac{1}{4}\lambda (\phi^{*}\phi)^2 .
\end{equation}
This introduces two more real parameters.  
\ifExtendedVersion {\color{darkgreen}
Despite the misleading notation, $\mu^2$ is just an arbitrary real number, which can have either sign.
In the Standard Model it is assumed to have a negative value, and once again this is a choice that is not dictated by any principle. Because of the
sign, the potential takes the shape of Mexican hat, and the minimum occurs for a non-zero value of $\phi$, and has the topology of a sphere in four dimensions.
}
\else
 { 
By means of the  Higgs mechanism this sector of the theory gives masses to the $W$ and $Z$ bosons and all quarks and leptons, and
to four weak mixing angles 
[the Cabibbo-Kobayashi-Maskawa (CKM) matrix].
}
\fi

\ifExtendedVersion {\color{darkgreen}
\paragraph{The Higgs Mechanism.}


The experimentally observed form of the Standard Model is obtained by picking an arbitrary point  (the choice does not affect the outcome) 
on the sphere and expanding $\phi$ around it. After this expansion, the Standard Model Lagrangian takes a considerably more complicated form, which occupies 
several pages, but everything on those pages is full determined by all the discrete and continuous choices mentioned above.
if we ignore the gauge couplings the three modes of the variation of $\phi$ along the sphere appear in the spectrum as massless Goldstone bosons.
But if we take the gauge couplings into account, three of the twelve gauge bosons acquire a mass by using the three Goldstone bosons as 
longitudinal components. These are the $W^{\pm}$ and $Z$ bosons with masses $80.4$ and $91.2$ GeV that mediate the weak interactions. 
The one remaining mode of the $\phi$ field appears in the spectrum as a massive real boson with mass $\sqrt{-2\mu^2}$, the famous
Higgs boson that has now finally been discovered, and has a mass of about 126 GeV. The eight gluons remain massless, as does a linear
combination of the original $U(1)$ vector boson (usually called ``$Y$") and a generator of $SU(2)$. This linear combination is the photon.
 The Yukawa
couplings, combined with the Higgs vev, produce mass matrices for the quarks and charged leptons. These can be diagonalized by unitary rotations of
the fermion fields. In the end, only 13 of the original 54 parameters are observable, 
6 quark masses, 3 charged lepton masses and 4 mixing angles [the Cabibbo-Kobayashi-Maskawa (CKM) matrix] which appear in the coupling of the W bosons to the charge $\frac23$ quarks   
and the charge $-\frac13$ quarks. 
}\fi

\paragraph{The CKM matrix.}\label{CKM}

The CKM matrix is obtained by diagonalizing two complex matrices, the up-quark mass
matrix $M_u$ and the down-quark mass matrix $M_d$, which are the product of the corresponding Yukawa coupling matrices and the Higgs 
vev $v$:
\begin{equation}
D_u = U^{\dagger}_L M_u U_R;\ \   D_d = V^{\dagger}_L M_d V_R;\ \  U_{\rm CKM} = U_L^{\dagger}V_L  
\end{equation}
where $D_u$ and $D_d$ are real, positive diagonal matrices. For three families, $U_{\rm CKM}$ can be parametrized by three angles and a phase.
It turns out to be nearly diagonal, which presumably is an important clue. An often used approximate parametrization is
\begin{equation*}
U_{\rm CKM} \approx \begin{pmatrix} 1-\lambda^2/2 & \lambda & A \lambda^3 (\rho-i\eta) \\
-\lambda & 1-\lambda^2/2 & A\lambda^2 \\
A\lambda^3 (1-\rho-i\eta) & -A\lambda^2 & 1
\end{pmatrix} 
\end{equation*}
where $\lambda=0.226$, and corrections of order $\lambda^4$ have been ignored. For values of the other parameters see \textcite{PDG}.  They will
not matter in the rest of this review, because the
current state of the art does not go beyond getting the leading terms up to factors of order 1, especially the hierarchy of the 
 three mixing angles,  $\theta_{12}=\lambda,  \theta_{23} \propto \lambda^2$
and $\theta_{13}\propto \lambda^3$.
The degree of non-reality of the matrix can be expressed in terms of the Jarlskog invariant $J$, which is defined as
\begin{equation}
{\rm Im}\left[V_{ij}V_{kl}V_{il}^*V_{kj}^*\right]=J \sum_{m,n} \epsilon_{ikm}\epsilon_{jln}\ .
\end{equation}
This is  a very small number: $J\approx 3 \times 10^{-5}$.

\paragraph{Quark and Lepton masses.}
The values of the quark and lepton masses, in GeV, are listed below. See \textcite{PDG} for errors and definitions.
\begin{align*}
\begin{tabular}{ddd}
\multicolumn{3}{c}{ $ u,c,t\ \ \ \ \ \ \ \  \ \    d,s,b \ \ \  \ \ \ \  \ \ \ \  e, \mu, \tau$} \\
\hline
 0.0023\quad & 0.0048 & 0.000511  \\
 1.275\quad & 0.095 &  0.105 \\
  173.5\quad & 4.5 &   1.777 
\end{tabular}
\end{align*}
The masses and hierarchies are not explained within the Standard Model; they are simply put in by means of the Yukawa coupling matrices.

\paragraph{The number of parameters.}
We now have
 a total of 18 observable parameters, which have now finally all been measured. From the measured values of the $W^{\pm}$ and $Z$ masses and 
the electromagnetic coupling constant $e$ we can compute $g_1=(M_Z/M_W) e$, $g_2=M_Z/(\sqrt{M_Z^2-M_W^2})$ and the vacuum expectation value $v$ of the
scalar $\phi$, using $M_W=\frac12 g_2 v$. This vacuum expectation value is related to the parameters in the potential as $v=2\sqrt{-\mu^2/\lambda}$, 
and has a value of about 246 GeV.
The Higgs mass determines $\mu^2$, and hence now we also know $\lambda$.

\paragraph{CP violating terms.}\label{StrongCPsection}

There is, however, one more dimensionless parameter that does not appear on the T-shirt. One can consistently add a term of the form
\begin{equation}
\label{StrongCP}
\theta \frac{g_3^2}{32 \pi^2} \sum_{a=1}^8 F_{\mu\nu}^a F_{\rho\sigma}^a \epsilon^{\mu\nu\rho\sigma} \ .
\end{equation}
where the sum is over the eight generators of $SU(3)$. 
\ifExtendedVersion {\color{darkgreen}
This term is allowed by all gauge symmetries, but forbidden by $P$ and $CP$. Neither is a symmetry of nature, however, and hence they cannot be invoked here.
The parameter $\theta$, an angle with values between $0$ and $2\pi$, is not an observable by itself. By making suitable phase rotations of the fermions its value can be changed, but then these phase rotations  
end up in the mass-matrices of the quarks.  In the end, this leads to one new physical parameter, $\bar\theta= \theta-{\rm arg}\ {\rm det}\ (M_u M_d)$, where $M_u$ and $M_d$ are the quark mass matrices.
A non-zero value for this parameter would produce a non-zero dipole moment for the neutron and certain nuclei, which so far has not been observed. This puts an upper limit on $\bar\theta$ of about
$10^{-10}$. 
}
\else
 {
This term is not forbidden by any symmetries. The parameter $\theta \in [0,2\pi)$ is shifted by the quark mass diagonalization. The physical combination,
$\bar\theta= \theta-{\rm arg}\ {\rm det}\ (M_u M_d)$, is observable in dipole moments of the neutron and nuclei. Nothing has
been seen so far, which implies that $\bar\theta < 10^{-10}$.
}\fi
Note that one could also introduce a similar term for the $SU(2)$ and $U(1)$ gauge groups, with parameters $\theta_2$ and $\theta_1$.   
\ifExtendedVersion {\color{darkgreen}
However $\theta_1$ is not observable,
because in an abelian theory  (\ref{StrongCP}) is a total derivative of a gauge-invariant operator. 
In non-abelian gauge theories such terms are total derivatives of
operators that are not gauge invariants, and that can be changed by non-perturbative effects (instantons). 
The  CP violating parameter $\theta_2$ of $SU(2)$ can be set to zero by means of baryon number phase rotations, using the anomaly of baryon number with respect to $SU(2)$. This
works provided baryon number is not broken by anything else than that anomaly. If there are explicit baryon number violating terms, $\theta_2$ might be observable in baryon number
violating processes, but no such processes have been seen so far, and -- by definition -- the Standard Model does not contain such terms. Hence it is unlikely that $\theta_2$ will ever
be observed, and in any case we would be beyond the Standard Model already.
}
\else 
{
However $\theta$ parameters of abelian theories  are not observable, and $\theta_2$ can be rotated to zero
using baryon number phase rotations. 
}\fi
 Therefore we get only one extra parameter, $\bar\theta$, bringing the total to 19. 
\ifExtendedVersion {\color{darkgreen}
Just as with all the other parameters, the Standard Model does not fix its value.
}\fi

\paragraph{Renormalizability.}

The 19 parameters were obtained by writing down all interactions allowed by the symmetry with a mass dimension less than or equal to 4.
Without this restriction,
infinitely many terms could be added to (\ref{StandardModel}), such as four-fermion interactions or polynomials in $(\phi^*\phi)$. Any such term defines a new mass scale, and
we can consistently ``decouple" these terms by sending these mass scales to infinity.
\ifExtendedVersion {\color{darkred}

 Such terms are sometimes called {\em irrelevant operators}.
Conversely, the presence of any such term requires, for quantum consistency, the presence of
infinitely many others. In this sense there is, for example, no arbitrariness in limiting the scalar potential to terms of at most order four in $\phi$; this is a consequence of the 
consistent assumption that there no negative dimension terms. The correctness of this assumption is under permanent experimental scrutiny. 
For example, compositeness of a Standard Model particle would manifest itself through operators with 
dimension larger than 4. For many such operators, the lower limit on the mass scale are now around 1 TeV.}\fi

\ifExtendedVersion {\color{darkred}

Virtual process in quantum field theory make all physical quantities depend, in principle, on all unknown physics. 
Loops of particles in Feynman diagrams depend on arbitrarily large momenta, and are therefore sensitive to
arbitrarily short distances. Furthermore, all particles, including those that have not been discovered yet, are pair-produced. It might
appear that this inhibits any possibility for making predictions. But in quantum field theories with only non-negative dimension 
operators, such as the Standard Model,  this problem is solved by lumping all unknowns together in just a finite
number of combinations, corresponding precisely to the parameters in the Lagrangian. Since they encapsulate unknown physics, 
the values of those parameters are
fundamentally unknown: they can only be measured. But a finite number of measurements produces an unlimited amount of
predictive power. Furthermore this is not just true for the precise values of the Standard Model parameters we measure, but also for 
other parameter values. A quantum field theory with twice the observed electron mass is equally consistent as the Standard Model.

This property is called ``renormalizability". 
}
\else 
{
In theories like the Standard Model, all unknown (and unknowable) virtual short-distance contributions are lumped together in a finite
number of parameters. This is known as ``renormalizability". This property does not depend on parameter values and discrete choices,
and remains just as valid if we make the electron mass twice as large. }\fi
\ifExtendedVersion {\color{darkred}
In the seventies of last century this was treated as a fundamental principle
of nature, but it has lost some status since then. It is now more common to say that the Standard Model is just an effective field
theory. }

\fi
As soon as evidence for a new term with dimension larger than four is found this will define a limiting mass scale 
$M_{\rm new}$ (where ``new" stands for new physics). 
All computations would be off by unknown
contributions of order $Q/M_{\rm new}$, where $Q$ is the mass scale of the process of interest. Since 
such new terms can be expected to exist on many grounds, including ultimately quantum gravity (with a scale 
$M_{\rm new}=M_{\rm Planck}$), the Standard Model is just an effective field theory valid up to some
energy scale.

\paragraph{Running couplings.}

As a direct consequence of the renormalization procedure, the values of the constants in the Lagrangian depend
on the energy scale at which they are measured. In the simplest case, the loop corrections to a gauge coupling
constant have the form
\begin{equation}
\label{Renorm}
g(Q)=g+ \beta_0 g^3 {\rm log}(Q/\Lambda) + \hbox{higher order} \ldots \ ,
\end{equation}
where $g$ is the coupling constant appearing in the Lagrangian, and $\Lambda$ is a manually introduced ultraviolet
cutoff of a momentum integral. We may use $g(Q)$ as the physical coupling constant to be compared to experimental
results at a scale $Q$. This then removes the dependence on $\Lambda$ in all physical quantities to this order. But if we had used
instead a different scale $Q'$ we would have measured a different value for the coupling constant, $g(Q')$. 
The value of $g(Q')$ can be expressed in terms of $g(Q)$ using Eq. (\ref{Renorm}), and involves a term $\beta_0 {\rm log}(Q/Q')$.
One can do better than this and sum up the leading contributions (``leading logs") of Feynman diagrams of any order in the
loop expansion. This leads to the {\it renormalization group equations}, with a generic form
\begin{equation}
\frac{dg_i(t)}{dt}= \beta(g_i(t)) \ ,
\end{equation}
where $\beta$ is a polynomial in all parameters in the Lagrangian. Here $t={\rm log}(Q/Q_0)$, 
where $Q_0$ is some reference scale.

\ifExtendedVersion {\color{darkred}
These equations can be solved numerically and sometimes exactly to determine how the parameters in the Lagrangian
evolve with energy. Of particular interest is the question how the parameters evolve if we increase $Q$ to energies beyond those
explored by current experiments. In many quantum field theories, this has disastrous consequences. Typically, these functions
have poles (``Landau poles") where couplings go to infinity and we loose perturbative control over the theory. A famous exception are 
non-abelian gauge theories, such as QCD, with not too much matter. In these theories the leading parameter, $\beta_0$, is negative
and the coupling approaches zero in the ultraviolet. In that case there is a Landau pole in the infrared, so that we loose
perturbative control there. In QCD, the energy scale where that happens is the QCD scale.}

{\color{darkblue}
The loss of perturbative control in the infrared limit can usually be remedied by means of a non-perturbative definition of the
action using discretised space-times (lattices), as is indeed the case for QCD. But the loss of perturbative control in the ultraviolet
limit cannot be handled by methods that can be deduced from known physics. This requires unknown, new physics. }

 {\color{darkred}
Note that not only the dimensionless parameters change logarithmically with $Q$, but
also the parameter $\mu^2$ in the Higgs potential, even though Eq. (\ref{Renorm} ) looks different in this case: there are
additional divergent contributions proportional to $\Lambda^2$. This implies that $\mu^2$ may get quantum contributions 
many orders of magnitude larger than its observed value, but this by itself does not invalidate the Standard Model, nor its 
extrapolation. The parameter $\mu^2$ is a renormalized input parameter, just as all others.
 }\fi

\paragraph{Range of validity.}
Now that we finally know all Standard Model couplings including the Higgs self-coupling $\lambda$ we can
see what happens to them if we assume that there is nothing but the Standard Model. It turns out that until
we reach the Planck scale they all remain finite; all Landau poles \ifExtendedVersion {\color{red} }\else { (points where the coupling constants diverge) }\fi are beyond the Planck scale. 

Note that not only the dimensionless parameters change logarithmically with $Q$, but
also the parameter $\mu^2$ in the Higgs potential, even though Eq. (\ref{Renorm}) looks different in this case: there are
additional divergent contributions proportional to $\Lambda^2$. This implies that $\mu^2$ may get quantum contributions that are 
many orders of magnitude larger than its observed value. But this by itself does not invalidate the Standard Model, nor its 
extrapolation: the parameter $\mu^2$ is a renormalized input parameter, just as all others.

\ifExtendedVersion {\color{darkred}
This is a remarkable fact.
If there would be a Landau pole, the Standard Model would predict its own downfall. Surely, new physics would then be
needed to regain computational control. In the present situation, the Standard Model is not only mathematically complete, but it
also remains valid until the Planck scale, leaving us rather clueless about new physics. Note that a randomly chosen quantum field theory would not necessarily have that range of validity, but that does not yet make it invalid as alternative laws of physics 
in different universes. All that is required is that new physics can be introduced that can remove the singular behavior 
and that this new physics is sufficiently decoupled from low energy physics.
}\fi

\ifExtendedVersion {\color{darkgreen}
\paragraph{The stability bound.}

The current value of the Higgs mass, and the corresponding value of $\lambda$ does have a slightly worrisome
consequence. The self-coupling $\lambda$ decreases and may become negative. 
If and where where that happens
depends rather sensitively on the top quark mass and the QCD coupling constant $\alpha_s=g_3^2/4\pi$, and can be
anywhere from about $10^{11}$ GeV to $M_{\rm Planck}$. A negative value for $\lambda$ in (\ref{ScalarPotential}) looks
catastrophic, since it would appear to make the potential unbounded from below, if probed at high energies. But that is 
too naive. First of all, in the real world, including quantum gravity, their will be higher order terms in the potential
of order $(\phi\phi^*)^n/(M_{\rm Planck})^{n-2}$, and secondly even in the absence of gravity 
one should consider the behavior of the complete potential at high energy, and not just evolve $\lambda$. This requires
the computation of the effective potential, and is discussed in detail in \textcite{Sher:1988mj}. It turns out that what really happens
is that the potential acquires a ``false vacuum", a new global minimum below the one of the Standard Model.  This is not a problem,
provided the tunneling amplitude towards that vacuum is sufficiently small to yield a lifetime of our vacuum larger than 
$13.8 \times 10^9$ years. Furthermore there must be a non-vanishing probability that we ended up and stayed in our vacuum, and not
in the false vacuum, during the early stages of the universe. Note that even if the probability is small this does not really matter,
because the false vacuum  has a very large Higgs vev and therefore is unlikely to allow life. The implications of the Higgs mass
on the stability of the vacuum are illustrated in 
 \cite{Ellis:2009tp,Degrassi:2012ry}. Especially figure 5 in the latter paper shows in a fascinating way where the Standard Model
 is located relative to the regions of (meta)stability. The stability bound can be avoided in  a very easy way by
 adding a weakly coupled singlet scalar \cite{Lebedev:2012zw}. Since we cannot distinguish this modification from the
 Standard Model at low energies, in this sense the Standard Model can be extrapolated to the Planck scale even without
 encountering stability problems. Furthermore, it has been argued that the current data also allow the interpretation that the Higgs
 coupling is running to the value zero at the Planck scale \cite{Bezrukov:2012sa}, with entirely different implications. 
Note that, contrary to some beliefs,  vacuum stability is {\it not} automatic in broken supersymmetric theories \cite{Abel:2008ve}.
}
\else 
{
\paragraph{The stability bound.}\label{StabBound}
The only potential problem in the extrapolation of the Standard Model couplings is that the Higgs self-coupling $\lambda$ may become
negative before the Planck scale, which may signal an instability. More precise determinations of the top quark mass and the
QCD coupling are needed to be certain if $\lambda$ does indeed go negative, and even if it does, it only implies a meta-stability of
our vacuum with a lifetime that exceeds the current age of the universe. Perhaps this is problematic for the evolution of the early universe,
but certainly not for its current state. Furthermore the problem can easily be avoided by adding 
a weakly coupled singlet scalar \cite{Lebedev:2012zw}, and hence it does not offer a clear hint at elaborate new structures beyond
the Standard Model. 
}\fi

\paragraph{Neutrino masses.}
 
The observation of neutrino oscillations implies that the ``classic" Standard Model needs to be modified, because at least
two neutrinos must have masses.
Only squares of mass differences can be determined from these experiments. They are
\begin{eqnarray*}
\Delta m^2_{21} &= (7.5 \pm 0.2) \times 10^{-5} \ {\rm  eV}^2 \\
|\Delta m^2_{23}| &= (2.3 \pm 0.1) \times 10^{-3}\  {\rm  eV}^2
\end{eqnarray*}
In principle, neutrinos could be nearly degenerate in mass with minute differences, but from various cosmological observations
we know that the sum of their masses must be less than about half an eV (see \textcite{dePutter:2012sh} for a recent update).
The masses can have a  normal hierarchy, $m_1 < m_2 \ll m_3$ or an inverted hierarchy, $m_3 \ll m_1 < m_2$.  They 
are labeled 1, 2, and 3 according to their $\nu_e$ fraction, in descending order.

The simplest way  of accommodating neutrino masses is to add  $N$ fermions  $\psi_S$ that are Standard Model singlets\footnote{One may give Majorana masses to the  left-handed neutrinos without introducing
extra degrees of freedom, but this requires adding  non-renormalizable operators or additional Higgses.}. 
The number $N$ is not limited by anomaly constraints, and in particular does not have to be three. 
To explain the data one needs $N \geq 2$, but $N=2$ looks inelegant. Better motivated options
are $N=3$, for right-handed neutrinos as part of families, as in $SO(10)$-related GUTs, or $N \gg 3$, in  string models with an abundance
of singlets.

As soon
as   singlets  are introduced, not only Dirac, but also Majorana  masses are allowed (and hence perhaps obligatory).
The most general expression for couplings and masses is  then (omitting spinor matrices)
\begin{equation}
{\cal L}_{\nu}=\sum_{i=1}^3 \sum_{a=1}^{N} \bar\psi^i_{\nu_L}Y_{ia} \psi^a_{S} + \sum_{ab}^N {\cal M}_{ab}\psi^a_{S}\psi^b_{S}\ .
\end{equation}
The first term combines the three left-handed neutrino component
with three (or two)   linear combinations of singlets into a Dirac mass $m$, and the second term provides a 
Majorana mass matrix $M$
for the singlets.   
This gives rise to a six-by-six neutrino mass matrix with three-by-three blocks,
of the form
\begin{equation}
M_{\nu}=\begin{pmatrix} 0 & m \\
m & M 
\end{pmatrix}
\end{equation}
The  mass scale of ${\cal M}$ is not related to any other Standard Model scale and is usually assumed to be large.
In the approximation $m \ll M$ one gets three light neutrinos with masses of order $m^2/M$ and $N$ heavy ones.
 This is called the see-saw mechanism. It gives a very natural explanation for the smallness of neutrino masses
 (which are more than eight orders of magnitude smaller than the muon mass) without unpalatable side-effects.
 The optimal value of the Majorana mass scale is debatable, and can range from $10^{11}$ to $10^{16}$ GeV depending
 on what one assumes about ``typical" lepton Dirac masses. 
  
 If we assume $N \geq 3$ and discard the parameters of the heavy sector,  which cannot be seen in low-energy neutrino physics,
 this adds nine parameters to the Standard Model: three light neutrino masses, four CKM-like mixing angles and
 two additional phases that cannot be rotated away because of the Majorana nature of the fermions. 
 This brings the total number of parameters to 28. However,
 as long as the only information about masses is from oscillations, the two extra phases and the absolute
 mass cannot be measured.
 
 The current values for the mixing angles are
\begin{eqnarray*}
{\rm sin}^2(2\theta_{12}) &= 0.857 \pm  0.024 \\ {\rm sin}^2(2\theta_{23})& > 0.95 \\ {\rm sin}^2(2\theta_{13})&=0.09 \pm 0.01
\end{eqnarray*}
Note that the lepton mixing angles, are not all small, unlike the CKM angles for quarks. 
The fact that $\theta_{13} \not=0$  is known only since 2012, and implies that  the CKM-like phase of the neutrino mixing matrix is measurable, in principle.  This also rules out the once popular idea of tri-bi maximal mixing 
\cite{Harrison:2002er}, removing a possible hint at an underlying symmetry.  
 
\ifExtendedVersion {\color{darkred}
It is also possible to obtain massive neutrinos without adding new degrees of freedom to the classic Standard Model, by adding
an irrelevant operator \cite{Weinberg:1980bf}
\begin{equation}
\label{WeinbergOperator}
\frac{1}{M} (\phi \psi_L)^T C (\phi \psi_L)
\end{equation}
where $\psi_L$ denotes the Standard Model lepton doublets $(1,2,-\frac12)$. This gives rise to neutrino masses of order
$v^2/M$, where $v$ is the Higgs vev of $246$ GeV, so that $M$ must be of order $10^{14}$ GeV to get a neutrino mass of order 1 eV. 
An operator of this form is generated if one integrates  out the massive neutrinos of the see saw mechanism, but it might also
have a different origin. Just as a direct Majorana mass, this operator violates lepton number.

One cannot detect the presence of any of  these lepton number violating terms with only neutrino oscillation experiments, 
not even using the two extra phases in the mixing matrix \cite{Langacker:1986jv,Doi:1980yb,Bilenky:1980cx}.
Experiments are underway to detect lepton number violation (via neutrinoless double beta decay) and to observe neutrino masses 
directly (by studying the endpoint of the $\beta$-decay spectrum of tritium), so that we would know more than just their differences.

In the rest of this paper the term ``Standard Model" refers to the classic Standard Model plus some mechanism to provide
the required neutrino mass differences. Since the {\it classic} Standard Model is experimentally ruled out, it is inconvenient 
to insist strictly on the old definition and reserve the name ``Standard Model" for it.   
 }
\fi

\section{Anthropic Landscapes}\label{AnthropicSection}

\setcounter{paragraph}{0}

The idea that our own existence might bias our observations has never been popular in modern science,
but especially during the last forty years a number of intriguing facts have led scientists from several areas of particle physics, astrophysics and cosmology
in that direction, often 
with
palpable reluctance. 
Examples are
Dirac's large number hypothesis in astrophysics \cite{Carter:1974zz,Carr:1979sg},  
chaotic inflation \cite{Linde:1986fd}, quantum cosmology \cite{Vilenkin:1986cy},  the cosmological constant
 \cite{Davies:1980ji,BarrowTipler,Weinberg:1987dv},
the weak scale in the Standard Model \cite{Agrawal:1997gf},  quark and lepton masses in the Standard Model \cite{Hogan:1999wh}, the Standard Model in string theory \cite{Speech} and  the cosmological constant in 
string theory \cite{Bousso:2000xa,Susskind:2003kw}.

This sort of reasoning goes by the generic name ``Anthropic Principle" \cite{Carter:1974zz}, 
which will be referred to as ``AP" henceforth. \ifExtendedVersion {\color{darkred}
Hints at anthropic reasoning can already be found much earlier in the history of science and philosophy.  An extensive 
historical overview   
can be found in \textcite{BarrowTipler}  and \textcite{Bettini:2004dj}.
In modern science the AP first started making its appearance
in astrophysics and cosmology, in the seventies of last century. 
At that time, particle physicist were just moving out of fog of nuclear and hadronic physics
into the bright new area of the Standard Model. In 1975 Grand Unified Theories were discovered, and it looked like a realization
of the ultimate dream of a unique theory of everything was just around the corner. 

In 1984 string theory re-entered the scene
(which it had occupied before as a theory of hadrons) as a promising theory of all interactions, including gravity. Within months,
everything seemed to fall into place. Grand Unified Theories emerged almost automatically, as a consequence of just a few
consistency conditions, which seemed to allow very few solutions. At that time, nobody in this field had any interest in 
anthropic ideas. They were diametrically opposite to what string theory seemed to be suggesting. It still took almost two decades 
before the ``A-word" made its appearance in the string theory literature, and even today mentioning it requires 
extensive apologies.

The name  ``anthropic principle" does not really help its popularity, and is doubly
unfortunate.
The word ``anthropic" suggests
that human beings are essential, whereas we should really consider any kind of observer. If observers exist in
any universe, then our existence is not a bias.  Furthermore the name suggests a principle of nature. Indeed in some forms of the AP  -- but not the one considered here --  it {\it is} an additional principle of nature. However, it is pointless to try and change the name. This is what it is called.

There exist many
different formulations of the AP, ranging from tautological to just plain ridiculous. See \textcite{BarrowTipler} for a discussion of many of these
ideas. 
We will avoid old terms like ``weak" and ``strong anthropic principle"  because historically they have been  used with different
meanings in different contexts, and tend to lead to confusion. In the present context, the AP is
merely a consequence of the true principles of nature, the ones we already know and the ones we still hope to discover, embodied in
some fundamental theory.  Discovering those underlying laws of physics is the real goal. 
We assume that those laws do not contain any statements regarding  ``life" and
``intelligence".
 This assumption is an important fork in the road, and making a different choice here leads to an entirely different
class of ideas. 
This assumption may be wrong. Some people point, for example, to the importance of the r\^ole of observers in the
formulation of quantum mechanics, or to the poorly understood notion of ``consciousness" as possible counter indications
(see {\it e.g.} \textcite{Linde:2002gj} for an -- inconclusive -- discussion).

}\fi In the rest of this review, the term AP is used in the following sense. We assume a multiverse, with some physical mechanism for
producing new universes. In this process, a (presumably large) number of options for the laws of physics is sampled. The possibilities
for these laws are described by some fundamental theory; they are ``solutions" to some ``equations". Furthermore we assume that we are able to conclude that some
other sets
of mathematically allowed laws of physics do not allow the existence of observers, by any reasonable definition of the latter (and one can indeed argue about that, see 
for example \textcite{Gleiser:2010ai}).

This would be a rather abstract discussion if we had no clue what such a fundamental theory might look like. But fortunately there
exists a rather concrete idea that, at the very least, can be used as a guiding principle: the String Theory Landscape described in the
introduction.
The rest of this section does not depend on the details of the string landscape, except that at one point we will  assume discreteness. However, 
the existence of some kind of landscape in some fundamental theory is a prerequisite. Without that, all anthropic arguments
lose there scientific credibility.

\subsection{What Can Be Varied?}\label{WhatVaries}

In the anthropic literature many variations of  our laws of physics are considered. 
\ifExtendedVersion {\color{darkred}
It has even been argued that life depends crucially on the special physical properties of water, which in its turn depend
on the bond  angle of the two hydrogen atoms. But this angle is determined completely by thee-dimensional geometry plus
computable small corrections. It cannot be changed. There is no version of chemistry where it is different. }\fi
Often it is realized years later that
a variation is invalid, because the parameter value is fixed for some previously unknown fundamental reason. 
 One 
also encounters statements like: we vary parameter X, but we assume parameter Y is kept fixed. But perhaps this is not
allowed in a fundamental theory. So what can we vary, and what should be kept fixed?

In one case we can give a clear answer to these questions: we can vary the Standard Model within the domain of quantum
field theory, provided we keep a range of validity up to an energy scale well above the scale of nuclear physics. Furthermore,
we can vary anything, and keep anything we want fixed. For any such variation we have a quantum field theory that is equally
good, theoretically, as the Standard Model. For any such variation we can try to  investigate the conditions for life. 
We cannot be equally confident about variations in the parameters of cosmology (see section \ref{CosmologySection}).

\ifExtendedVersion {\color{darkred}
Of course this does not mean that a more fundamental theory does not impose constraints on the Standard Model parameters. 
The Standard Model is  just an effective field theory that will break down somewhere, almost certainly at
the Planck scale and quite possibly well before that. The new physics at that scale may even fix all parameters completely,
as believers in the uniqueness paradigm are hoping. Even then, as long as the scale of new physics is sufficiently decoupled,
it is legitimate and meaningful to consider variations of the Standard Model parameters. }\fi
 
Even though it is just an effective field theory,
 it goes too far to say that the Standard Model is just the next nuclear physics.  In nuclear physics the limiting, new physics 
scale $M_{new}$ is within an order of magnitude of the scale of nuclear physics. Computations in nuclear physics depend on many
parameters, such as coupling constants, form factors and nucleon-nucleon potentials. These parameters are determined by
fitting to data, as are the Standard Model parameters. But unlike the Standard Model parameters, they cannot be varied outside
their observed values, in any way that makes sense. There is no theory of nuclear physics with twice the observed pion-nucleon coupling, and anything else unchanged.

This difference is important in many cases of anthropic reasoning. Some anthropic arguments start with unjustified variations
of parameters of nuclear physics. If life ceases to exist when we mutilate the laws of physics, nothing scientific can be concluded.
The only admissible variations in nuclear physics are those that can be derived from variations
in the relevant Standard Model parameters: the QCD scale $\Lambda_{\rm QCD}$, and the quark masses.


This raises an obvious question. If the Standard Model is
just an effective field theory, made obsolete one day by some more fundamental theory, then why can we
consider variations in its parameters? What if the fundamental theory fixes or constrains its parameters, just as QCD does with 
nuclear physics? The answer is that  the relevant scale $Q$  for  anthropic arguments is that of chemistry or nuclear physics.
This is far below the limiting scale $M_{new}$, which is more than a TeV or so. 
New physics at that scale  is  irrelevant
for chemistry or nuclear physics.

If we ever find a fundamental theory that fixes the quark and lepton masses, the anthropic argument will still be valid, but starts
playing a totally different r\^ole in the discussion. It changes from an argument for expectations about fundamental physics
to a profound and disturbing puzzle. In the words of \cite{Ellis:2006fy}: ``{\em in this case the Anthropic issue returns with a vengeance: (...) 
Uniqueness of fundamental physics resolves the parameter freedom only at the expense of creating an even deeper mystery, with no way of resolution apparent.}"

\subsection{The Anthropocentric Trap}

There is another serious fallacy one has to avoid: incorrectly assuming that something is essential for life, whereas it is only essential for {\it our} life. 
Any intelligent civilization (either within our own universe or in an entirely different one with different
laws of physics) might be puzzled about properties in their environment that seem essential for their existence. But that
does not imply that life cannot exist under  different circumstances.

\ifExtendedVersion {\color{darkred}
Let us take this from one extreme to another, from obvious fallacies to assumptions  that are generally made in anthropic arguments, but
should be considered critically. 

\subsubsection{Humans are irrelevant}

A tiny, instantaneous variation of the electron mass by one part in a million would be fatal for us, even if just one percent of the energy
difference were converted to heat. 
But it would clearly by nonsense
to claim that the electron mass is fine-tuned to one part in a million. We evolved in these conditions, and would have evolved
equally well with a slightly different value  of the electron mass (note that even the word ``evolved" already implies an anthropocentric assumption). 
Our health is believed to depend crucially on about twenty different
elements, but this not mean that all twenty  are really needed. The hormones produced by our thyroids contain iodine,
but it is easily imaginable that if no iodine were available in our environment, evolution would have solved the problem in a different way.
It is usually assumed that water is required, but that may also be too anthropocentric. 
This is even more true for the existence of DNA. It is
impossible for us to decide theoretically whether there are other ways of encoding life,
although this issue might one day be solved with real data in our own
universe, by discovering different forms of life.

\subsubsection{Overdesign and Exaggerated Claims}

Another potential fallacy is to overlook the fact that some features that are needed in principle are vastly
``overdesigned" in out universe: there is much more of it then is really required anthropically.
The formation of our solar system and evolution require a certain degree of smoothness in
our environment, but there is no reason why that should extend to the entire universe. The proton has to be sufficiently stable,
but it does not have to live $10^{31}$ years; the anthropic limit is about $10^{20}$ years (below that decaying protons would
produce too much radiation). Biological processes need energy, but that does not mean life requires stars employing nuclear fusion.
Only a fraction of about $10^{-9}$ of the sun's energy actually reaches the earth. Furthermore there is life in deep oceans getting
its energy from volcanic activity. 

Indeed, perhaps one can imagine life in universes where stars do not ignite, or where there are no nuclear fusion reactions at all
\cite{Adams:2008ad}.
With just gravity, fermionic matter, photons and quantum mechanics one can have radiating black holes, and various analogs of white
dwarfs and neutron stars, where the force of gravity is balanced by degeneracy pressure (the Pauli Principle). These ``stars" could
radiate energy extracted from infalling matter or electromagnetic annihilations.  

}{\color{darkgreen}

Another important set of anthropic constraints comes from abundances of some basic building blocks, like Carbon
in our universe. But these  do not have to be large over the entire universe either. In our universe, the relative Carbon abundance produced by 
Big Bang Nucleosynthesis is only about $10^{-15}$. The Carbon in our universe
 must therefore have been produced in early stars. Here one encounters the ``Beryllium Bottleneck": the fact that there is
 no bound state of two $\alpha$ particles ($^8{\rm Be}$) appeared to cripple carbon production in stars.
Famously, Hoyle   predicted the existence
 of a resonance in the Carbon nucleus that would enhance the process, and indeed this resonance was found. 
 
 This is often referred to as a successful anthropic prediction, because Carbon is essential for {\it our} kind of life. But it is in fact
 just a prediction based on the observed abundance of some element. 
 The puzzle would have been equally big if an element
 irrelevant for life had an anomalously high abundance. 
 Indeed, Hoyle himself apparently did not make the link between the
 abundance of Carbon and life until much later (see \textcite{HoyleHistory} for a detailed  account of the history as well as the physics).
 
The current status of the Hoyle state and its implications will be summarized in section \ref{TripleAlpha}. 
Based on what we know we cannot claim that life is impossible without this resonance. 
 We do not know which element abundances  are required for life, nor do we know how they vary over the Standard Model parameter space. 
 Perhaps there even exists a parameter region where $^8$Be is stable, and
 the beryllium bottleneck is absent \cite{Higa:2008dn}.  This would turn the entire anthropic argument on its head.

The abundance of Carbon in our
 own body, about $20\%$, is several orders of magnitude larger than in our environment, demonstrating the possibility of chemical
 processes to enhance abundances.   If we discover that we live near an optimum in parameter space, this would be a strong indication
  of multiverse scanning (a unique theory is not likely to land there), but as long as the maximum is broad or other regions exist there is no need to
 over-dramatize. Most observers will observe conditions that are most favorable to their existence.

  }{\color{darkred}
  
 \subsubsection{Necessary Ingredients}
  
 On the other end of the anthropocentric scale one finds requirements that are harder to argue with. 
Four dimensions (three space and one time) may be required
anthropically (see \cite{Tegmark:1997jg} and
references therein). 
The arguments include lack of stability of planetary orbits and atoms in more than three space dimensions, and
the topological triviality of less than three, which does not allow the biological plumbing 
needed for the kind of living organisms we know. These statements are obviously somewhat anthropocentric, 
but the differences are radical enough to focus on four dimensions henceforth. 

Fermionic matter and non-gravitational interactions are undoubtedly needed. 
If we assume the validity of quantum mechanics and special relativity and hence quantum field theory, there are only a
limit number of possibilities. Interactions can be mediated by scalar or vector bosons, and the latter can belong to
abelian or non-abelian gauge groups.
It is hard to avoid the conclusion that at least one abelian gauge interaction is needed,
like electrodynamics in our universe. Electrodynamic provides a carrier of energy and information, a balancing force
in stars, and chemistry. Photons play a crucial r\^ole during the early stages of cosmology. The existence of repulsive and attractive forces and opposite charges allows exact cancellation
of the force among macroscopic bodies, and all of this can work only if charges are conserved. 
Scalars interactions are an unlikely candidate, because they have none of these properties,
and scalars tend to be massive. Purely non-abelian interactions 
cannot be ruled out so easily. They can have
a complicated phase diagram, and the only non-abelian gauge theory we can study in our universe, QCD, may have
an infinite number of bound states (nuclei) if we switch off electromagnetism. Can there be life based on some purely 
non-abelian gauge theory? With current knowledge we cannot decide that.
}
\else 

Arguments based on water or DNA should be viewed with suspicion. Perhaps we do not even need fusion-fueled stars \cite{Adams:2008ad};
degenerate stars (white dwarfs or neutron stars) may provide sufficient energy.

 Arguments based on abundances are equally suspect. 
Fred Hoyle famously   predicted the existence
 of a resonance in the Carbon nucleus that would enhance Carbon production, and indeed this resonance was found. 
 This is often referred to as a successful anthropic prediction, because Carbon is essential for our kind of life. But it is in fact
 just a prediction based on the observed abundance of some element. 
 Indeed, Hoyle himself  did not make the link between the
 abundance of Carbon and life until much later \cite{HoyleHistory}.
 
The current status of the Hoyle state and its implications will be summarized in section \ref{TripleAlpha}. 
Based on what we know we cannot claim that life is impossible without this resonance. 
 We do not know which element abundances  are required for life, nor do we know how they vary over the Standard Model parameter space. 
 Perhaps there even exists a parameter region where $^8$Be is stable, and
 the beryllium bottleneck is absent \cite{Higa:2008dn}.  This would turn the entire anthropic argument on its head.

 If we discover that we live near an optimum in parameter space, this would be a strong indication
  of multiverse scanning (a unique theory is not likely to land there), but as long as the maximum is broad or other regions exist there is no need to
 over-dramatize. Most observers will observe conditions that are most favorable to their existence.\fi
 
 In view of the difficulties in defining anthropic constraints some authors have proposed other criteria that are
 under better control and still are a good ``proxy" for life. In particular, it seems plausible that the formation of complex structures will
 always be accompanied by entropy production in its environment, a criterion that would certainly work in our own universe. This ``entropic
 principle" has led to some successes for cosmological parameters \cite{Bousso:2010vi}, but seems less useful for the subtle details of the
 Standard Model parameter space.

\ifExtendedVersion {
\subsubsection{Other Potentially Habitable Universes}\label{OtherHabitable}
}\fi

\ifExtendedVersion {\color{darkred}
\paragraph{Purely electromagnetic universes?}

Going to extremes, let us ignore the problem of abundances and energy sources and focus only on
the building blocks of life. 
We need to assume some electromagnetic theory, simply because we know too little about anything else.  
So let us restrict attention to universes with
at least one massless photon species.
There must be charged particles, and there are many choices for their charges.
Presumably in any fundamental theory the choices are rational only.  A sensible theory is widely expected to
have both electric and magnetic charges, and then Dirac quantization implies rational charges; this is indeed expected to be true in string theory.
This still leaves us with
a large number of choices of electromagnetic theories. In the weak coupling limit
we could in principle work out the ``atomic" spectra
in many cases and even say something about molecules. 

We know one example which leads to observers: simply 
take a set of particles with masses and charges equal to those of the stable nuclei, and a world such as ours can be built,
with solid planets and living beings. We are treating the nuclei here as fundamental point particles, with spin 0 or $\frac12$.
Perhaps something in the chemistry of life is sensitive to fine details such as nuclear structure or magnetic moments, which
cannot be mocked up with fundamental particles, and perhaps there are bottlenecks in evolution that depend on such details,
but in the spirit of looking for the extremes we can not exclude this. 
We cannot exclude life in universes with only electromagnetism,
with fundamental nuclei and electrons whose abundances are due to some kind of baryogenesis, and with stars radiating 
energy without nuclear fusion, like white dwarfs or neutron stars \cite{Adams:2008ad}. Perhaps we do not need the strong and the weak interactions at all!
Furthermore, if this extreme possibility works for our kind of fundamental nuclei, there is going to be a huge number of variations that
work as well. We may vary nuclear masses freely, changes charges, allow additional photons. 

}{\color{darkgreen}

\paragraph{The weakless universe.}

A much more convincing case can be made if only the weak interactions are eliminated \cite{Harnik:2006vj}. 
These authors made some clever changes in the theory to mimic physics in our universe as
closely as possible. Then one can rely on our experience with conventional physics. In particular, the strong interactions
are still present to power stars in the usual way. In our universe, the weak interactions provide chiral gauge symmetries,
that protect quark and lepton masses and reduce the mass hierarchy to just one scale, the weak scale. In the weakless universe
only the $u, d$ and $s$ quarks and the electron are kept, and are given small Dirac masses (of order $10^{-23}$ in Planckian units;
alternatively, one may choose extremely small Yukawa couplings and move the weak scale to the Planck scale). 

In our universe,
before electroweak freeze-out proton and neutrons are in equilibrium because of weak interactions. This leads to a computable Boltzmann
suppressed neutron to proton ratio $n/p$ at freeze-out, that does not depend on primordial quark abundances, and is the main source of  the observed Helium/Hydrogen ratio.
Without the weak interactions,
the initial neutron to proton does depend on the quark abundances produced in baryogenesis. If the number of up quarks and down quarks is the same, then $n/p=1$,
and conventional BBN will burn all baryons into $^4He$ by strong interactions only. In the weakless universe, 
BBN can be made to produce the same hydrogen to helium ratio as in our universe by adjusting either the primordial quark abundances or the baryon-to-photon ratio. In the
later case one can get
a substantially larger deuterium abundance, and a surviving stable neutron background contributing a fraction of about $10^{-4}$ 
to the critical density. 

The larger deuterium abundance comes in handy for hydrogen burning in stars, because they cannot use the weak process $pp\rightarrow De^+\nu$,
but can instead can work via $pD \rightarrow ^3{\rm He}\gamma$. Despite very different stability properties of nuclei, stellar nucleosynthesis to any relevant nuclei appears possible,
and stars can burn long enough tot match the billions of years needed for evolution in our universe (although those stars have a significantly reduced luminosity in comparison to the sun). 

Another obvious worry is the r\^ole of supernova explosions. In our universe, stars can collapse to form neutrons stars. The neutrinos released in this weak interaction process can blast
the heavy nuclei formed in the star into space. This process is not available in the weakless universe. What is possible  is a type-Ia supernova that originates from accumulation by a white dwarf
of material from a companion. In this case the shock wave is generated by nuclear fusion, and does not require the weak interactions.

While this is a compelling scenario, there are still many differences with our universe: no known mechanism for baryogenesis, different stellar dynamics and stars with lower luminosity, a presumably
far less efficient process for spreading heavy elements, and the absence of plate tectonics and volcanism (driven by the core of the earth, which is powered mostly by weak decays), which the authors regard 
as ``just a curiosity".  In the weakless universe, after BBN one is left with a potentially harmful  \cite{Hogan:2006xa,Cahn:1996ag} stable neutron background.
\textcite{Clavelli:2006di}  pointed out that material ejected from type-Ia supernova has a low oxygen abundance. Since oxygen is the most abundant element  (by mass) in the earth's crust and
oceans and in  the human body, this would seriously undermine the claim that the weakless universe exactly mimics our own. However, it certainly seems plausible that such a universe might support some
kind of life, although perhaps far less vigorously. 

It is noteworthy that all differences would seem to diminish the chances of life.  However, that may just be due to a too anthropocentric way of looking at our own universe. 
Unfortunately our universe is the only one where the required computations have been done completely, by nature itself. In our own universe we may not know the mechanism for baryogenesis, but at least we know that
such a mechanism must exist.

\paragraph{Other cosmologies}

Instead of changing the quantum field theory parameters underlying our own universe, one can also try to change cosmological
parameters, such as the baryon-to-photon ratio, the primordial density perturbations, the cosmological constant and the curvature density parameter $\Omega$. 
This was done by \textcite{Aguirre:2001zx}, and also in this case regions in parameter space could be identified where certain parameters differ by
many orders of magnitude, and yet some basic requirements of life are unaffected. These cosmologies are based on the cold big bang model.

\paragraph{Supersymmetric Universes.}\label{SusyAnth}

Of all the quantum field theories we can write down, there is a special class that is the hard to dismiss on purely
theoretical grounds: supersymmetric field theories. Any problem in quantum field theory, and especially fine-tuning and
stability problems, are usually far less severe in supersymmetric theories. In the string theory landscape, supersymmetric
vacua are the easiest ones to describe, and getting rid of supersymmetry is a notoriously difficult problem. 
If we cannot rule out supersymmetric theories on fundamental grounds, we should find anthropic arguments against them.

Fortunately, that is easy. In supersymmetric theories electrons are degenerate with scalars called selec\-trons.  These scalars are not
constrained by the Pauli principle and would all fill up the s-wave of any atom \cite{Cahn:1996ag}. Chemistry and stability of matter
\cite{Dyson1967}\cite{Lieb}
would be lost. This looks sufficiently devastating, although
complexity is not entirely absent in such a world, and some authors have speculated about the possibility of life under
these conditions, for entirely different reasons, see {\it e.g.} \cite{Clavelli:2005vm,Banks:2012hx}. 

Even if supersymmetric worlds are ruled out anthropically, there is still a measure problem to worry about.  
Supersymmetric landscapes often have flat directions, so that they form continuous ground states regions. If we are aiming
for  a discrete landscape, as is the case in the string theory landscape, the question naturally arises why the continuous
regions do not dominate the measure-zero points by an infinite factor. In the string landscape,  the discreteness is related
to local minima of a potential, and if one ends up anywhere in such a potential  one reaches one of the minima. Claiming that
only the surface area of the minima matters is therefore clearly too naive. 
This should be
discussed in its proper context, the problem of defining a measure for eternal inflation.
}
\else 
{
\subsubsection{Other Habitable Universes.}\label{OtherHabitable}
Going to extremes, one can imagine habitable universes with only electromagnetic  and gravitational interactions, with fundamental
nuclei and electrons created by some kind of generalized baryogenesis and with only dim stars stabilized by degeneracy pressure
of fermions, radiating gravitational energy built up during their collapse. These universes would still have solid matter, chemistry and biology like ours.  

A less extreme possibility is a universe without weak interactions  \cite{Harnik:2006vj}.
These authors made some clever changes in the theory to mimic physics in our universe as
closely as possible, so that one can rely on our experience with conventional physics. Quarks and leptons have small masses (in Planck units)
not because of a light Higgs boson, but by having extremely small Yukawa couplings. 
Type-II supernovae are not available, but type-Ia supernovae,
whose explosions are driven by the strong interactions, can take over their r\^ole in spreading heavy elements.
However, there are some serious worries: there is no
known mechanism for baryogenesis\footnote{In our own universe we are not certain about the mechanism either, but at least we are sure that one exists.}, 
stars are less bright, there may be no plate tectonics and volcanism (which are fueled to a large extent by weak decays), 
type-I supernovae may not produce enough oxygen \cite{Clavelli:2006di}, and there is a potentially harmful  \cite{Hogan:2006xa,Cahn:1996ag} stable neutron background.

Instead of changing the quantum field theory parameters underlying our own universe, one can also try to change cosmological
parameters, such as the baryon-to-photon ratio, the primordial density perturbations, the cosmological constant and the curvature density parameter $\Omega$. 
This was done by \textcite{Aguirre:2001zx}, and also in this case regions in parameter space could be identified where certain parameters differ by
many orders of magnitude, and yet some basic requirements of life are unaffected. 
}\fi
 
\ifExtendedVersion 
\else 
{
Alternative universes that {\it must} probably be ruled out anthropically are the exact supersymmetric ones, because supersymmetric theories are the hardest to dismiss
on fundamental grounds.
Fortunately, ruling them out is easy. In supersymmetric theories electrons are degenerate with scalars called selectrons.  These scalars are not
constrained by the Pauli principle and would all fill up the s-wave of any atom \cite{Cahn:1996ag}. Chemistry and stability of matter
\cite{Dyson1967,Lieb}
would be lost. Although this may look sufficiently devastating, it has not stopped speculation about the possibility of life under
these conditions,  see {\it e.g.} \textcite{Clavelli:2005vm,Banks:2012hx}. }\fi

\subsection{Is Life Generic in QFT?}\label{Gedanken}

It may seem that  we are heading towards the conclusion that any quantum field theory (QFT) allows the existence of life and intelligence.
Perhaps any complex system will eventually develop self-awareness \cite{Banks:2012hx}. Even if that
is true, it still requires sufficient complexity in the underlying physics. But that is still not enough to
argue that all imaginable universes are on equal footing. We can easily imagine a universe with just
electromagnetic interactions, and only particles of charge $0, \pm1, \pm2$.
Even if the clouds of Hydrogen and Helium in such a universe somehow develop 
self-awareness and even intelligence, they will have little to be puzzled about in their QFT 
environment. Their universe remains unchanged over vast ranges of its parameters. There are no
``anthropic" tunings to be amazed about. Perhaps, as argued by \textcite{Bradford:2011zz}, fine tuning is an 
inevitable consequence of complexity and hence any complexity-based life will observe a fine-tuned environment. But this
just strengthens the argument that we live in a special place in the space of all quantum field theories, unless one drops the
link between complexity and life. But  if life can exist without  complexity, that just begs the question why the problem was solved in
such  a complicated way in our universe.

\begin{figure}
\includegraphics[width=2.9in]{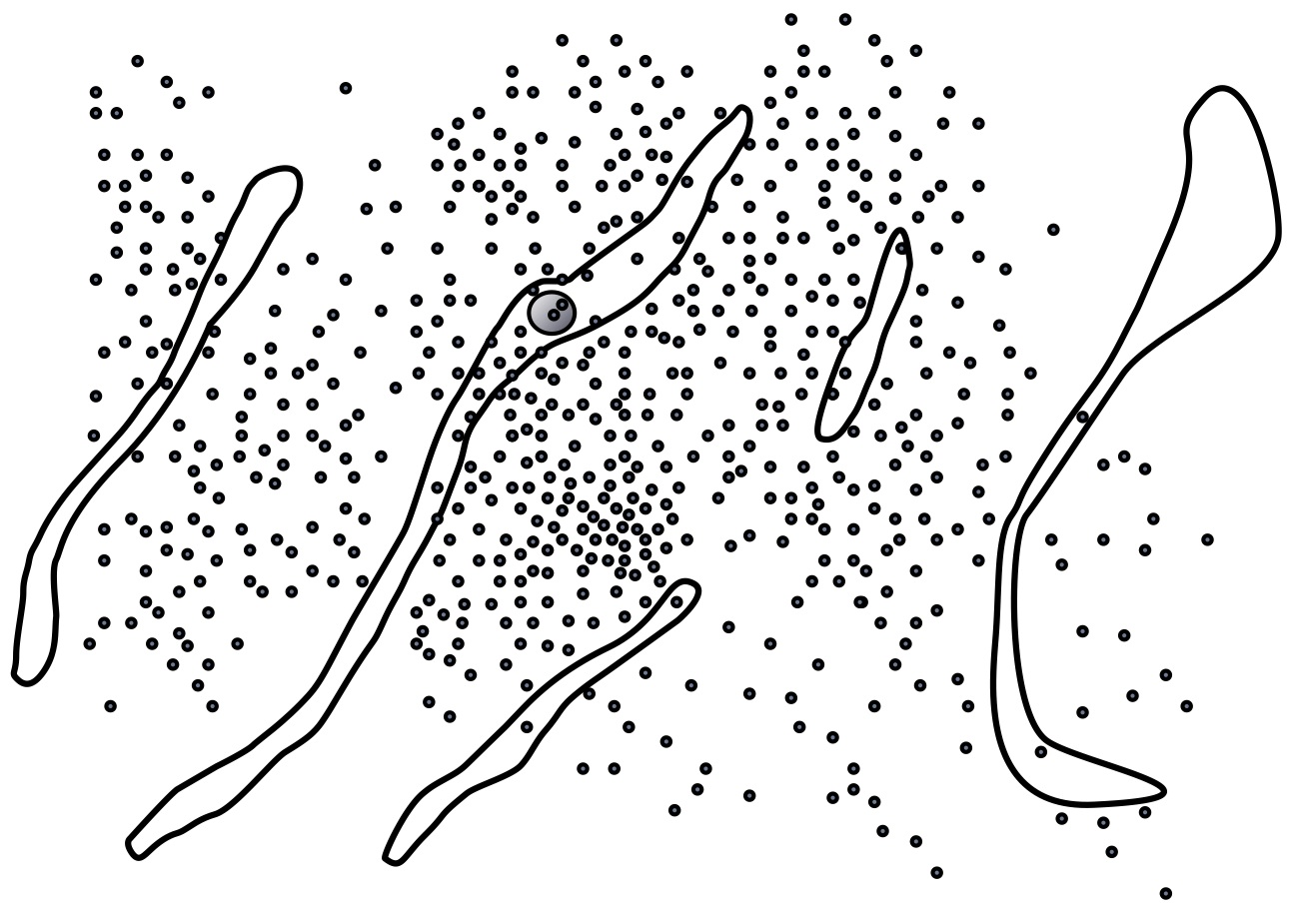}
\caption{Habitable regions in QFT space. The gray circle represents the experimental bounds on the Standard Model. The dots show the distribution of QFT points
in a hypothetical landscape.}
\label{LandscapeDistributions}
\end{figure}

If we put everything we know and everything we do not know together, the picture that emerges is one of many domains where life might exist, and
many more where it definitely does not. Presumably the habitable regions are narrow in certain directions, and very elongated in others. A cartoon version
of such regions in part of QFT space is shown in Fig. \ref{LandscapeDistributions}, with the gray circle 
showing our own location
and the experimental uncertainties.


This diagram  represents two unrelated {\it gedanken} computations \cite{Schellekens:2008kg}.
The contours are the result of the anthropic gedanken computation explained above.  The dots show the results of a very different one. They represent
points in QFT space obtained from some fundamental theory, such as string theory. Here the implicit assumption is made that 
such a theory will lead to a discrete set of points.  
In this concrete setting, it is clear that the
two gedanken computations are completely unrelated. The first one involves low-energy physics: nuclear and atomic physics and chemistry. The second
one involves geometry and topology of manifolds with membranes and fluxes wrapped around them, and determining minima of potentials 
generated by all this structure. We can actually do both kinds of computations only in simple cases, but we know enough to conclude that it would take
a miracle for them to match each other, if the second computation were to produce a unique answer. The obvious way out is precisely what string theory suggests: 
that there is not a single point, but a cloud of points, covering a substantial part of the QFT parameter space. 
\ifExtendedVersion {\color{red}
}\else {
Note that no such cloud is required for a point to land precisely in the gray, experimental circle, because unlike the anthropic contours this circle cannot
be determined by a computation.
}\fi


These contours are sharp lines in the case of particle physics thresholds, such as  reactions that stop being exothermic 
or stability of essential 
building blocks (although there is usually a small transition region where
a particle is just stable enough). In other cases they are more like contour lines of  distributions. Most papers make
different assumptions about the definitions of these lines ({\it i.e.} the necessary conditions for life), and consider
different slices through the parameter space. 

Moving out of our own location, the first line we encounter is the end of our region. There our kind of life ends, and we have to rely on speculation
to know if other kinds of life are possible. This happens for example if one of the crucial processes in the functioning of stars is shut off. Other processes
may take over, but stellar  lifetimes and/or heavy element abundances may differ by orders of magnitude, and we cannot rely on experimental data to  be certain
that such a universe will ``work". Beyond this {\it terra incognita} (perhaps more appropriately called ``{\it no man's land}")
there is usually another boundary where the conditions  become so adverse that any kind
of complexity can be ruled out. For a discussion along similar lines see \textcite{Hall:2007ja}. 
In the rest of this review we shall not make this distinction over and over again, and
use the adjective ``anthropic" rather loosely for any parameter change that is likely to affect life, whether it is our life or life in general.

Real plots of this kind can be found in many papers, {\it e.g.} \textcite{Agrawal:1997gf}; \textcite{Tegmark:1997qn}; \textcite{Hogan:1999wh}; \textcite{Tegmark:2005dy}; \textcite{Hellerman:2005yi}; \textcite{Graesser:2006ft}; \textcite{Hall:2007ja}; \textcite{Barr:2007rd}; \textcite{Jaffe:2008gd}; \textcite{Elor:2009jp}; and \textcite{Barnes:2011zh}.

Even without drawing further conclusions, 
it is simply incredibly exciting to see where we are located on the parameter space map, and to
see the lines of minor and major catastrophes surrounding us. It is a bit like seeing our fragile planet in the vastness of space, on the first Apollo 8 pictures.
It is also a great way of appreciating how our universe really works. If we do indeed understand that, we should be able to change something and work out the consequences. 

Fig. \ref{LandscapeDistributions} was deliberately drawn in this way to illustrate a few fallacies that are perhaps blatantly obvious, but that are nevertheless
repeated incessantly in the literature.

\begin{itemize}
\item{Anthropic reasoning will never completely determine the Standard Model. It is quite clear that even in our own environment there are variations that have no
conceivable impact on life, such as the $\tau$ mass. \ifExtendedVersion {\color{darkred} Furthermore, everything we know suggests that there might very well exist other, disconnected habitable regions, and
even far in the foreseeable future our technical abilities will be insufficient to rule them out.}\fi}
\item{Anthropic reasoning combined with a fundamental theory  is not likely to determine the Standard Model either. This would require the density of the cloud to match
the size of the anthropic region, in such a way that precisely one point lands inside it. That would be another miracle.}
\item{There is no reason to expect the maximum of the density distribution, even when folded with sampling probabilities, to select our vacuum. Computing these maxima is
another {\it gedanken} computation that cannot be sensitive to the location of the domains\ifExtendedVersion\else\footnote{Unless life in a universe somehow affects the sampling probability of its offspring. 
This includes science fiction ideas like scientists making copies of their own universe in experiments. A related idea was proposed by \textcite{Smolin:1994vb},
who argued that collapsing black holes create new universe with slightly changed parameters. This would make the maximum of black hole production a point of attraction in a multiverse. However, black holes are hardly
the optimal environment for life, nor a suitable device for transferring information. For further discussion see \textcite{RothmanEllis}; \textcite{BarrowCLAP}; \textcite{Vilenkin:2006hq}; and \textcite{Smolin:2006gt}. Note that the {\it existence} of a landscape is in any case a prerequisite for such a proposal.}\fi, the other gedanken computation.}
\ifExtendedVersion {\color{darkred}
\item{The gray region decreases monotonically with time. 
Some people have tried to reduce  arguments of this kind to absurdity, by claiming that it leads to the conclusion that
the density of points would have to increase as well, in order to have a chance of getting one point inside the gray circle.
But no cloud of points is needed to explain the presence of one point in the experimental domain. There is no miracle here that requires such an explanation, because the experimental domain
is not determined by a calculation.  }}\fi
\item{Bounds on parameters may disappear as others are allowed to vary. Obviously the projection
of the regions on the axes cover essentially everything, but if we intersect them with horizontal or vertical lines, we get narrow bounds. 
\ifExtendedVersion {\color{darkred} But that this was going to be true was
obvious from the start. In the $(m_e,m_\tau)$ plane (keeping everything else fixed), the anthropic domain is a extremely elongated along the $m_\tau$ axis, and narrow along the 
$m_e$ axis. If for some reason we decide that the fundamental parameters are $(m_e-m_\tau)$ and $(m_e+m_\tau)$ we would conclude that if the latter linear combination is
kept fixed, the former is tightly constrained. Then someone might point out that if $(m_e+m_\tau)$ was also allowed to vary, the bound goes away. People committing this obvious fallacy
are apparently either assuming that anthropic arguments should constrain all parameters to  a small circle, or do not understand the notion of small regions in more than one dimension.}\fi
}
\end{itemize}

If  one can show that a  parameter is anthropically constrained, keeping all others fixed, that is a tremendous success. If one can do it
while allowing others to vary, that is an even bigger success. Only in cases where strong claims are made about the actual value of a parameter
(especially that it must be small), it becomes really important to ask if the smallness is a consequence of fixing other parameters. 

\ifExtendedVersion {\color{darkgreen}
There is one interesting exception to the third point, namely if life in a universe somehow affects the sampling probability of its offspring. 
This includes  science fiction ideas such as  scientists making copies of their own universe in experiments. Another variant was proposed by \cite{Smolin:1994vb},
who argued that collapsing black holes create new universes with slightly changed parameters, and that universes that are optimized for black hole
production are also optimal for life. This would make the maximum of black hole production a point of attraction in a multiverse. But black holes are not
a friendly environment for life, nor a suitable device for transferring information. For further discussion see \textcite{RothmanEllis}; \textcite{BarrowCLAP}; \textcite{Vilenkin:2006hq}; and \textcite{Smolin:2006gt}. 
}\fi

\ifExtendedVersion {\color{darkred}
Note that any argument based on sampling probabilities, such as the one mentioned above, needs a landscape of possibilities in the first place. Then the anthropic genie is already out of the bottle.
In any situation where the dots in Fig. \ref{LandscapeDistributions} are sampled eternally, nothing else is needed. Sampling probabilities, either determined
by fundamental physics or by life itself, may help explain our location within the anthropic domain, but cannot help in explaining why we are inside a domain. The
anthropic argument already explains that. Any second explanation just creates a new mystery.

There exist numerous proposals for avoiding the anthropic principle even given the existence of a huge landscape, see for example 
\cite{Firouzjahi:2004mx,Brustein:2005yn,BouhmadiLopez:2006pf,Kobakhidze:2004gm,Linde:2010nt,Kawai:2011qb}, but they tend to focus on one 
anthropic issue (in particular the cosmological constant), and ignore all others. For the cosmological constant, the value zero is a natural point of attraction,
and hence it is on quite general grounds not too surprising that it can be singled out by some mechanism. But this will not work for the more subtle, though
far less extreme tunings of the Standard Model, unless a miracle happens. 
}\fi

\subsection{Levels of Anthropic Reasoning}

Even in the interpretation used in this review, one may distinguish several versions of the AP:
\begin{enumerate}
\item{AP0: A mere tautology. }
\item{AP1: An explanation for certain fine tunings. } 
\item{AP2: A predictive method.}
\end{enumerate}

\vskip .3cm\noindent
{\em AP0:} 
If the fundamental theory allows many universes that do not allow observers, we should not be puzzled to find ourselves
in one that does. This is true, but not very useful.

\noindent
{\em AP1:}  Suppose we conclude that some variable $x$, {\it a priori} defined on an interval $[0,1]$ has to lie in
an extremely narrow band of size $\epsilon$ for observers to exist.
If the fundamental theory contains $N$ values of $x$ evenly scattered over the interval, the chance that
none of them is in the observer range is $(1-\epsilon)^N$. For $N=M/\epsilon$ and small $\epsilon$ this goes like $e^{-M}$. For sufficiently large $M$, we would
agree that there is nothing surprising about the existence of a point in the observer band.  
For concreteness, one may think
of numbers like $10^{-120}$ for $\epsilon$ and $10^{500}$ for $N$, so that $M=10^{380}$. 
\ifExtendedVersion {\color{darkred}
These are the kind of numbers that appear in the discussion of
one  fine-tuned parameter, the cosmological constant. }\fi
The chance that a flat distribution contains no points in the observer range would then be
the absurdly small number ${\rm exp}(-10^{380})$. Obviously, the fine-tuning is then explained.
Note that we are talking about landscape density distributions here, not about sampling probabilities in eternal inflation (see section \ref{EternalInflation} for 
various approaches towards defining the latter).

\vskip .3cm \noindent
{\em AP2:} 
It may be possible to go one step further, and determine the most probable point where we should expect to find ourselves 
within the anthropic window. This requires additional information compared to AP1. 
We should be able to assign a probability to each point,  work out
the probability distribution, and determine its maximum. 
\ifExtendedVersion {\color{darkred}
We need to know how often a certain
solution is sampled with respect to others, and we will have to compare  the ``fertility" of distinct universes that allow observers. Universes with more observers 
are more often observed.}\fi This brings some very serious measure problems into the discussion. What counts as an observer, and
what counts as an observation? Should we sum over the entire history of the universe, and how do we include parts of the universe that
are currently behind the horizon? How do we even define probabilities in the context of eternal inflation, where anything that
can happen happens an infinite number of times? Furthermore there is the issue of ``typicality" \cite{Vilenkin:1994ua}.  If we can define and compute a probability distribution, 
should we expect to find ourselves at its maximum? Are we ``typical"? Does statistics even make sense if we can observe just a 
single event? 

Many criticisms of anthropic reasoning are aimed at the measure and typicality problems in  AP2, and
especially its use for predicting the cosmological constant. See for example  \textcite{Muller:2001ym,Smolin:2004yv,Starkman:2006at,Neal:2006py,Bostrom:2007zz,Maor:2008zza,Armstrong:2011qa} for a variety
of thoughts on this issue. We will return to the measure problem in section \ref{EternalInflation}.

\ifExtendedVersion {\color{darkred}
Sampling probabilities are not relevant for AP1. Suppose $x_a$ is a value of $x$ for which observers can exists, whereas for $x_b$ they cannot exist.
If $N$ is much smaller than $1/\epsilon$ the existence of a habitable value $x_a$  in the fundamental
theory is mysterious. This becomes no less mysterious if  $x_a$ has a vastly larger sampling probability. Would we conclude that we find ourselves in
$x_a$ because that is far more probable, or because nothing can live in $x_b$? Similarly, if $x_b$ had a much larger sampling probability, would that affect
the conclusion in any way? }\fi

Perhaps AP1 is as far as we can ever get. We may determine the boundaries of our domain, and find out how a fundamental theory spreads its ``vacua" over
that domain. There is a lot of interesting physics and mathematics associated with all of these questions. In the end we may just be satisfied that we roughly
understand where we are, just as we are not especially obsessed with deriving the orbit and size of our planet in the landscape of astrophysical objects.
Establishing the fundamental theory will have to be done by other means, perhaps purely theoretically, and by ruling out alternatives. 

\ifExtendedVersion {\color{darkred}
Sharp predictions, rare even in the golden age of particle physics, are not likely to come from this. An AP1-based prediction would say nothing more than
that we live within an anthropic domain, and almost by definition that can never be more than a post-diction. AP2-based predictions may be possible in
cases where some contribution to the probabilities is strongly skewed in one direction. In that case we may predict that we should live close to the edge
of an anthropic domain, or the intersection point of several edges.

 In the context of a multiverse sampling the string landscape
 the ``anthropic principle" is nothing more than a bias we have because we exist as observers.
Therefore it makes no sense to ask if
the anthropic principle is falsifiable or require that it makes any predictions. It is the underlying theory that has to be 
falsifiable. 

}\fi

\ifExtendedVersion {\color{darkred}

\paragraph{Measures of fine-tuning}

It is impossible to make a mathematically precise statement about the amount of (fine)-tuning. To compute the
surface areas we need a measure on the space of quantum field theories, but this space does not come with
a measure. By ``measure" we mean here a prescription to compute a surface area in Fig. \ref{LandscapeDistributions} in 
order to decide that an anthropic region is large or small. This should not be confused with the measure problem of eternal
inflation (see section  (\ref{MeasureProblem}), which has to do with the relative sampling rates of vacua. 

The most common situation is that of a dimensionful parameter that is small in Planck units, such
as the strong scale (or the proton mass), the weak scale or the cosmological constant. If one assumes a flat
distribution one may say that the parameter is fine-tuned. 
But if instead we use a logarithmic distribution the conclusion would be different. If the proton mass must be
as small as it is for anthropic reasons, we need to tune it with 19 digit precision if the distribution is flat, but with
only two digit precision if we only need to tune the exponent. 

Therefore it is rather pointless to ask if our universe is ``fine-tuned" as long as we only know one vacuum.
In a sufficiently dense discrete landscape this question becomes well-defined: the discrete points provide a 
measure. But if the existence of a discrete landscape has already been established, the question looses much of its
relevance.

}\fi

\subsection{First Signs of a Landscape?}


\ifExtendedVersion {\color{darkred}
Historically, we are in a remarkable situation. The Standard Model is consistent, and can be extrapolated all the way to the Planck scale. 
Our vacuum may be metastable, but this is not an inconsistency nor a disagreement with any experimental result. 
We have a complete theory of the strong electromagnetic and weak interaction that seems to be perfectly adequate.
All remaining problems of the Standard Model are matters of taste. There are many unmotivated choices, and many dimensionless
parameters are very small. 
It is worth emphasizing  this point
because this is a rare moment in the history of physics that may already have passed when this review appears in print.  
}\fi

The current situation in particle physics invites an appeal to Occam's razor. 
\ifExtendedVersion {\color{darkred}
Although there are shelves
full of scientific work proposing new physics beyond the Standard Model, it is strange that all of that new physics
has managed to hide itself so well.  Perhaps nature is playing a really evil trick with us, presenting us with a 
complete Standard Model just to deceive us. But we cannot avoid asking the obvious question: }\else
We cannot avoid asking the obvious question: \fi
Could it be that the Standard Model, including a minor extension to accommodate neutrino
oscillations, is really all there is? 
Indeed, suggestions in that direction  have already been
made some time ago by \textcite{Shaposhnikov:2006xi}, albeit not in the context of a landscape.

It is undeniable that this state of affairs has contributed to the interest in ``anthropic" and ``landscape" thinking in particle physics.
Could it be true that the Standard Model is like a dart that was thrown repeatedly at the space of all quantum field theories, until
one of them landed in one of the anthropic domains of Fig.  \ref{LandscapeDistributions}? 
This is the central question of this review.  

\ifExtendedVersion {\color{darkred}
It is clear that the discovery of major new  structures, especially new strong interactions, would 
indicate that this sort of thinking is, at best, premature. But this would also raise the alarming possibility of an indefinite series of 
matryoska's of substructure, which we might never be able to disentangle completely.
Would  that be a more attractive option than a landscape?
The absence of major new physics in any accelerator experiment in the last decade may be an indication that we are
living at a special time in the history of particle physics, that we might have our first view of the foothills of a landscape. 
Perhaps we have reached the interface of bottom-up 
and top-down thinking, and our next task is to locate the Standard Model of particle physics within the landscape of some
fundamental theory, perhaps string theory.

Low energy supersymmetry, by far the most discussed option for new physics,  is an attractive compromise. It might be the final step towards
a theory of all interactions, allowing a smooth interpolation towards the Planck scale, and could give us important clues about
that theory. Nothing we know about low energy supersymmetry gives evidence {\it against} the idea of a landscape. It does not
reduce the number of parameters, quite the contrary, and supersymmetry itself plays a crucial r\^ole in understanding the only 
(part of) a landscape we can even begin to discuss. 
But if nature wanted to provide us with a soft, supersymmetric, landing on the landscape, it could have given some clues
a little earlier. 
}\fi

But even in the most extreme landscape scenario, there are plenty of problems left that require a solution. It is just
that the nature of the remaining problems has shifted in a remarkable way in a certain direction: most problems are now
``environmental", and many have anthropic implications.


One can roughly order the open problems according to their urgency, in the following way.
\begin{itemize}
\item{No consistent  theory.}
\item{Disagreement between theory and experiment.}
\item{Environmental, but not anthropic problems.}
\item{Potentially anthropic problems.}
\end{itemize}

We will make an -- admittedly rather artificial -- separation between particle physics and cosmology.

 \subsubsection{Particle Physics}

The main item in the first category is quantum gravity.
The Standard Model does not contain gravity, and adding it using standard QFT methods leads to 
inconsistencies. \ifExtendedVersion {\color{darkred} Some people would argue that  a violation of the stability
bound on the Higgs self-coupling belongs in this category as well.}\fi

In the second category
there is a long list of deviations of low statistical significance that may one day develop into
real problems, astrophysical phenomena for which there is no good theoretical model, but which may point to
new particle physics, a hint of a gamma-ray line in cosmic rays
at 130 GeV \cite{Weniger:2012tx}
and a 4$\sigma$ indication for spatial variations of the fine structure constant \cite{Webb:2010hc}. 

\ifExtendedVersion {\color{darkred}
The last two categories refer to the so-called ``why" problems: why are certain parameters or discrete choices the way they are.
This kind of problem may never be solved. Nothing goes wrong if we cannot find a solution, and there is no
experiment we can do to force nature to give us a clue. If in addition different values of a parameter have a negative impact on
the prospects for life, then this fact already provides a clue.
Then it becomes even less obvious that a ``solution", in the traditional sense, is required. But we should not jump to conclusions.
Simply saying ``parameter X has value $y$, because otherwise we would not exist" is too simplistic.}\fi

In the third category are all Standard Model parameters that have peculiar values, without any reason to hope
that anthropic arguments are going to be of any help. The most important one is the CP-violating angle $\bar\theta$ of the strong
interactions, arguably the most important Standard Model problem in the context of a landscape \cite{Banks:2003es,Donoghue:2003vs}.
Another example of non-anthropic  parameters
with small values are the CKM angles, and some of the quark mass ratios.
\ifExtendedVersion {\color{darkred}
The famous questions ``why are there three families of quarks and leptons" 
is probably in this category as well, although the number ``3" is not very peculiar from a landscape perspective, so this does not even
belong on a list of problems.}\fi

The last category consists of all problems related to parameters whose values {\it do} potentially have an impact on the
existence of life. This includes the group structure and representations of the Standard Model, the scales of the strong and the weak interactions (the ``gauge hierarchy problem", see subsection \ref{GaugeHierarchy}), the light quark masses and
the electron mass (assuming the heavier fermions stay heavy), neutrino masses and
perhaps even the mass of the top quark. The environmental impact of the fermion masses will be discussed in section \ref{QuarkMasses}.

\subsubsection{Cosmology}\label{CosmologySection}

\ifExtendedVersion {\color{darkred}
For the sake of the argument we will treat cosmology as a theory with input parameters, although its theoretical
underpinnings are far less robust. There are certainly plenty of ``category 1" problems here, especially in inflation and
pre-big bang cosmology. For Standard Model parameters
 it is easier to accept the extreme form of the landscape paradigm, namely that the Standard Model might be merely a point in a huge landscape about
which only few details can be derived from fundamental physics, just as only a few general features of Mount Everest can be derived 
from geology. In cosmology we are simply not in such a situation, or at least not yet.
}\fi

The main cosmological parameters are
the cosmological constant $\Lambda$, the density parameter $\Omega$, the matter density fluctuations  $Q=\delta\rho/\rho$, the 
dark/baryonic matter ratio $\zeta$, the baryon-to-photon ratio $\eta$
and the parameters of inflation (see \textcite{Tegmark:2005dy} for a systematic survey of all parameters).
\ifExtendedVersion {\color{red}
}
\else 
{
The theoretical foundations of cosmology belong to the first category defined above.
}\fi
There is no effective theory of cosmology where all of these parameters can manifestly be varied independently and without worrying about the impact of changes in our understanding of gravity. 
For example, the cosmological constant only has an observable meaning in a theory of gravity. 
The notion of decoupling it from gravity, as one can do for Standard Model parameters, does not even
make sense. \ifExtendedVersion {\color{darkred}
Cosmological variations are highlighted in the book ``Just Six Numbers" \cite{Rees:1999cu}; just one of his six numbers, the
fraction of energy released in nuclear reactions, is directly related to Standard Model physics. }\fi

Anthropic issues in cosmology will not be discussed in detail in this review, except for the cosmological constant, the focal
point of a lot of attention. Here we will just briefly mention some interesting observations.

The main item in the second category is ``dark matter", or more precisely the complete set of problems that 
is elegantly solved if we postulate the existence of dark matter: galaxy rotation curves, the bullet cluster, structure formation, the
features of the Cosmic Microwave Background (CMB), the amount of deuterium produced in Big Bang Nucleosynthesis and
the matter density of the Universe.  
There is a minority point of view that holds  that these problems belong in the first category,
and require a modification of 
gravity. But should we really be
so surprised if dark matter exists? Is it not a typical example of anthropocentric  hubris to assume that anything that exists
in the universe must be observable by {\it us}, or made out of the same stuff that {\it we} are made of?
Postulating dark matter moves this problem largely to category four, although there are still  serious problems 
in computer simulations of galaxy formation which may point to  a more fundamental problem (see \textcite{Famaey:2013ty} for a list of open problems). 

The  dark-to-baryonic matter ratio $\zeta$, which is
$\approx 5$ in our universe, may have anthropic
implications, since dark matter plays an important r\^ole in structure formation. This was first discussed for
axion dark matter \cite{Linde:1987bx}, because
the most popular solution to the strong CP problem, the Peccei-Quinn mechanism,
predicts an additional particle, the axion, that contributes to dark matter.  In contrast to the more popular WIMP dark matter\footnote{WIMPs are
``weakly interacting massive particles", which are present, for example,  in certain supersymmetric extensions of the Standard Model.},
whose abundance is predicted by its interactions, axionic dark matter must satisfy constraints which are in part anthropic in nature (for more on
axions see section 
\ref{Axions}). The constraints were made more precise by
\textcite{Hellerman:2005yi}, who found $\zeta < 10^{5}$ and \textcite{Tegmark:2005dy} who
concluded that $2.5 < \zeta < 10^2$, using some additional anthropic requirements. These papers also discuss the effect of 
other parameter variations (in particular $Q$ and $\Lambda$) on these bounds.
Using assumptions about a multiverse measure and the number of observers per baryon, \textcite{Freivogel:2008qc} gave an
anthropic statistical prediction for $\zeta$ roughly in agreement with the observed value. Although the emphasis on all  these
papers is on axionic dark matter, some of the conclusions on $\zeta$ do not really depend on that. 
 
Most other cosmological parameters are also in the fourth category.
Changing any of these substantially has an impact on some 
feature in the history and/or current status of the universe that would appear to be catastrophic at least for {\it our}
kind of life, and hence it is at least possible that this is part of the reason we observe the values we do.

\ifExtendedVersion {\color{darkred}
But once again, we should not jump to conclusions. 
}\else {
But we should not jump to conclusions. }\fi
An extreme example
is the smoothness and isotropy of the cosmic microwave background. This fact may be regarded as environmental, and if it
were a wildly fluctuating distribution this could have a very negative impact on the prospects for life \cite{Tegmark:1997in}.
But surely one cannot
assume that the entire density perturbation function is tuned this way just for life to exist in one galaxy. The most popular solution
to this ``horizon problem" is inflation, which  solves another problem with anthropic relevance, the flatness problem, but also
introduces some new fine-tunings. \ifExtendedVersion {\color{darkred} 

Inflation is an especially rich area for anthropic and landscape ideas. An early example is
\textcite{Vilenkin:1994ua}, giving arguments that typical civilizations will see an extremely flat inflaton potential. In \textcite{Freivogel:2005vv}
anthropic landscape arguments are given suggesting that the number of e-folds of inflation will not be much larger than the observed
lower bound of about 60. According to these authors, $59.5$ e-folds are required for structure formation, and we should not expect
to see much more than that.

 Many features 
of string theory have the potential to be relevant here such as  moduli, axions \cite{Dimopoulos:2005ac} and D-branes \cite{Kachru:2003sx}. In  \textcite{Liddle:2006qz} the possibility of a common origin of inflation, dark matter and dark energy in the
string landscape is discussed. 
See \textcite{Quevedo:2002xw,Kallosh:2007ig,Burgess:2007pz,Cicoli:2011zz} for reviews of inflation in string theory. 
But inflation in the string landscape also introduces new problems, see {\it e.g.} 
\textcite{Hall:2006ff,Allahverdi:2007wh,Huang:2008jr}.
}\fi

\ifExtendedVersion 
Inflationary cosmology also offers interesting opportunities for predictions  
of features of the CMB, see {\it e.g.}
\textcite{Tegmark:2004qd,Holman:2006ny,Ashoorioon:2010vw,Frazer:2011tg,Yamauchi:2011qq}. 
\else
Inflationary cosmology offers interesting opportunities for predictions based on landscape and/or anthropic ideas, especially
for observations of the CMB, see {\it e.g.}
\textcite{Tegmark:2004qd,Holman:2006ny,Ashoorioon:2010vw,Frazer:2011tg,Yamauchi:2011qq}. 
\fi
Furthermore, the CMB may even
give direct hints at the existence of a multiverse.
There is a chance of observing collisions with other bubbles in the multiverse, see for example 
\textcite{Aguirre:2007an} and WMAP results  presented by  \textcite{Feeney:2010dd}. \textcite{GonzalezDiaz:2011ke} 
consider an even more exotic possibility involving non-orientable tunneling.
In principle there might be information about other universes in the detailed structure of the cosmic microwave background, but
at best only in the extreme future \cite{Ellis:2006mi}.

Anthropic predictions
for the density parameter $\Omega$ were already made a long time ago by \textcite{Garriga:1998px}. This work, as well as 
\textcite{Freivogel:2005vv}, points out the plausibility of observing negative spatial curvature, ({\it i.e.} $\Omega_k  >  0$,  
where $\Omega_k \equiv 1-\Omega$)
 in  a multiverse picture. They argue that sixty e-folds of inflation are anthropically needed, and having
 a larger number of e-folds is statistically  challenged.
 The current observational constraint is $|\Omega_k| < 10^{-2}$.
Furthermore, \textcite{Guth:2012ww,Kleban:2012ph}
point out that observation of even a small positive curvature ($\Omega_k < -10^{-4})$ would falsify most ideas of eternal inflation,
because tunneling in a landscape gives rise to open Friedmann-Robertson-Walker (FRW) universes. 

That the  baryon to photon ratio $\eta \approx 6\times 10^{-10}$ may have anthropic implications was already observed a long time ago
(see \textcite{Carr:1979sg}; \textcite{Nanopoulos:1979hw}; \textcite{Linde:1985gh} but also 
\textcite{Aguirre:2001zx} for critical comments), but it is not simply a tunable free parameter. Inflation would dilute any such initial condition, as would any baryon number violating process that gets into
equilibrium in the early stages of the universe. See \textcite{Shaposhnikov:2009zzb}
for a list of 44 proposed solutions to the baryogenesis problem. Most of these solutions generate new anthropic issues themselves.


This brief summary does not do justice to the vast body of work on string and landscape cosmology. Further references can be
found in reviews of string cosmology, {\it e.g.}
\textcite{Burgess:2011fa}.

\subsubsection{\label{Cosm}The Cosmological Constant}

The cosmological constant  $\Lambda$  is a parameter of classical general relativity that is allowed by general coordinate invariance.
It has dimension $[{\rm length}]^{-2}$ and appears in the Einstein equations as
(the metric signs are $(-,+,+,+)$)
\beq \label{LambdaDef}
 R_{\mu\nu}-\half g_{\mu\nu} R +\Lambda g_{\mu\nu} = 8\pi G_N T_{\mu\nu}\ .
 \eeq
Without a good argument for its absence one should therefore consider it as a free parameter
that must be fitted to the data. It contributes to the equations of motion with an equation of state $P=w\rho$, where $P$ is
pressure and $\rho$ is density, with $w=-1$ (matter has $w=0$ and radiation $w=\frac13$). As the universe expands, densities 
are diluted as (the initial values are hatted)
\begin{equation}
\rho_{w}=\hat\rho_{w} \left(\frac{a}{\hat a}\right)^{-3(1+w)}\ .
\end{equation}
As a result, 
if $\Lambda \not= 0$ it will eventually dominate if the universe lasts long enough
\ifExtendedVersion {\color{darkred}
 (and if there is no ``phantom matter" with $w < -1$)}\fi.
 \ifExtendedVersion
 \else 
 {
 The natural length scale associated with $\Lambda$ is the size of the universe.
 }\fi
 \ifExtendedVersion {\color{darkgreen}
 
However, $\Lambda$ itself affects the expansion. For $\Lambda < 0$ the universe collapses in a time 
$ct=\pi\sqrt{3/\Lambda}$   whereas for $\Lambda > 0$ the universe goes into exponential expansion as ${\rm exp}(\sqrt{\Lambda/3} ct)$.
These two cases correspond to exact maximally symmetric solutions to the Einstein with $\Lambda\not=0$ and
without matter, and are called Anti-de Sitter (AdS) and de Sitter (dS) spaces respectively. The latter has a horizon at a distance $(\sqrt{\Lambda/3}$
from the observer. Light emitted by matter beyond that horizon can never reach the observer because of the expansion. The fact
that our universe has existed billions of years and that we observe galaxies at distances of billions of light years gives 
immediately an upper limit on $|\Lambda|$ (see Eq. (\ref{ObservationalBound}) below) which is already known for decades \cite{BarrowTipler}).
}\fi

\ifExtendedVersion {\color{darkgreen}
The fact that the length associated with $\Lambda$ is of cosmological size is not surprising in itself, but there is second
interpretation of $\Lambda$ that puts this in an entirely different perspective. }\fi
The parameter $\Lambda$ contributes   to the equations of motion in the same way as vacuum energy density $\rho_{\rm vac}$, 
which has an energy momentum tensor $T_{\mu\nu}=-\rho_{\rm vac}g_{\mu\nu}$. Vacuum energy is a constant contribution to
any (quantum) field theory Lagrangian. It receives contributions from classical effects,
for example different minima of a scalar potential and quantum corrections ({\it e.g.} zero-point  energies of
oscillators). However, it plays no r\^ole in field theory as long as gravity is ignored. It can simply be set to zero. Since vacuum energy
and the parameter $\Lambda$ are indistinguishable it is customary to identify  $\rho_{\rm vac}$ and $\Lambda$.  The precise
relation is
\begin{equation}\label{RhoLambdaDef}
 \frac{\Lambda}{8\pi}=\frac{G_N \rho_{\rm vac}}{c^2} := \rho_{\Lambda} \ .
\end{equation}
This immediately relates the value of $\Lambda$ with all other length scales of physics, entering in $\rho_{\Lambda}$, which of course
are very much smaller than the size of the universe. The extreme version of this 
comparison is to express  $\rho_{\Lambda}$ in Planck mass per (Planck length)$^3$, which gives a value smaller than $10^{-120}$. This
was clear long before $\rho_{\Lambda}$ was actually measured. 

\ifExtendedVersion {\color{red}}
\else
{
  More recently, observations of redshifts of distant type-Ia supernovae gave evidence for accelerated expansion \cite{Riess:1998cb,Perlmutter:1998np}, which can be fitted with the $\Lambda$-parameter. Combined
with more recent data on the cosmic microwave background, this indicates that the contribution of $\Lambda$ to the
density of the universe is about $70\%$ of the critical density $\rho_c \approx
9.9 \times 10^{-27} {\rm kg}/{\rm m}^3$, assuming the standard $\Lambda$CDM model of cosmology.
This then leads to an ``observed" value 
\begin{equation}
\rho_{\Lambda}\approx + 1.3 \times 10^{-123} \ .
 \end{equation}
}\fi

\ifExtendedVersion {\color{darkred}

This huge difference in length scales implies a huge fine-tuning problem.
It was noted a long time 
ago by \cite{Linde:1974at,Veltman:1974au} that the Standard Model Higgs mechanism induces a huge change in vacuum energy.
Other  contributions are expected to come from dynamical symmetry breaking in QCD and inflation. The latter is especially hard to 
avoid, because in most models the exponential is driven by vacuum energy, which must therefore have been vastly larger in the
inflationary period than it is now. Quantum corrections to vacuum energy are due to vacuum bubble diagrams 
(coupling to gravitons to generate the $\sqrt{-g}$ factor).  There are 
contributions from all particles, with opposite sign for bosons and fermions. These diagrams are quartically ultra-violet divergent: they
are infinite if we naively integrate over arbitrarily large momenta, and they are proportional to $M_{\rm cutoff}^4$ if we assume
that nature cuts off the divergence at some scale  $M_{\rm cutoff}^4$ (note that 
that quantum corrections contribute to the density  $\rho_{\rm vac}$, and hence $\Lambda$ gets quartic corrections, 
not quadratic ones as its dimension might suggest). It is likely that the divergent integral are cut off by a consistent theory of quantum 
gravity (and indeed, string theory does that), and in that case the cut off scale would be the Planck scale. In that case, the naive
order of magnitude for $\rho_{\Lambda}$ is the Planck density, one Planck mass per Planck volume ($5.15 \times 10^{96}\ {\rm kg}/{\rm m}^3$). 
In these units
 the aforementioned old observational limits, using 
$y \times 10^{9}$ (light)years for the assumed cosmic time (length) scale, are\footnote{In the rest of this section we use $\hbar=c=G_N=1$.}
\begin{equation}
\label{ObservationalBound}
|\rho_{\Lambda}| < 3.4 y^{-2} \times 10^{-121}
 \end{equation}
The fact that this number is so absurdly small is called ``the cosmological constant problem". The problem can be mitigated
by assuming a smaller cutoff scale for the quantum corrections, but even if we choose the TeV scale  there are 
still sixty orders of
magnitude to be explained. It seems unlikely that the cut-off can be less than that, because then we are in known quantum
field theory territory, and furthermore we then have the classical contributions to worry about as well. One may consider 
radical changes to the theory, so that gravity somehow does not couple to vacuum energy at all, but so far no working
proposals exist. See for example \cite{Weinberg:1988cp,Polchinski:2006gy,Bousso:2007gp} for a summary of some of these
ideas and why they do not work. 

A recent very detailed  review is  \cite{Martin:2012bt}. This paper, just as \cite{Koksma:2011cq} makes
the remarkable claim that vacuum energy diagrams are not divergent in quantum field theory, but finite and proportional
to $m^4 {\rm ln}\  m$, with opposite signs for bosons and fermions, and vanishing contributions for photons and gravitons. 
This would still get us only halfway (there are still vacuum contributions, a bare contribution, and the loop contributions do not magically cancel).
The trouble with this claim is that, even if correct, it requires knowing the full particle spectrum at all mass scales. 
If QED is embedded in a GUT, there
would be extra GUT  particles contributing, and the claim that photons do not contribute to vacuum energy looses its meaning. The whole
point of renormalization is that everything unknown beyond a certain mass scale is combined into a few constants, such as 
$\Lambda$. This proposal violates that principle. It can only be useful if we know the complete theory, but in that case we would
also know $\Lambda$. 
This just illustrates how hard it is to get rid of contributions from vacuum diagrams.

The small observational limit led most people to believe that $\Lambda$ had to be identically zero, for reasons that
were still to be discovered. But then from observations of redshifts of distant type-Ia supernovae gave evidence for accelerated expansion \cite{Riess:1998cb,Perlmutter:1998np}. This expansion can be fitted with the $\Lambda$-parameter. Combined
with more recent data on the cosmic microwave background this indicates that the contribution of $\Lambda$ to the
density of the universe is about $70\%$ of the critical density $\rho_c \approx
9.9 \times 10^{-27} {\rm kg}/{\rm m}^3$, assuming the standard $\Lambda$CDM model of cosmology.
Then
\begin{equation}
\rho_{\Lambda}\approx + 1.3 \times 10^{-123} 
 \end{equation}
}\fi

\paragraph{Anthropic arguments.}
The foregoing discussion already implies that there will be an anthropic range for $\Lambda$, assuming everything else is kept fixed. 
Although this may have been clear to some much earlier, it appears that
the first paper stating this is  \textcite{Davies:1980ji}.  They did not make it quantitative, though. In subsequent years 
\textcite{Linde:1984ir,Banks:1984cw,Sakharov:1984ir} also discussed anthropic implications of $\Lambda\not=0$. Sakharov's paper contains the remarkable 
 statement: {\it ``If the small value of the cosmological constant is determined by ``anthropic selection", then it is due to the discrete parameters. 
This obviously requires a large value of the number of dimensions of the compactified space or (and) the presence in some topological factors of a complicated topological structure."}

\ifExtendedVersion {\color{darkred}

Since $\Lambda$ must be small to get a large cosmological length or time scale, it is obvious that $\Lambda$ is anthropically
bounded in {\it any} universe that has life based on some kind of nuclear or chemical complexity, for any underlying field theory.
If one requires that a compact object consisting of $N$ constituents of mass $\mu$ (in Planck units) fits inside a dS horizon, one gets 
$\Lambda \lesssim \mu^2 N^{-2/3} \lesssim N^{-4/3}$, where the second inequality holds for 
pressure-balanced objects (stars, planets or humans, but not galaxies).
In AdS space the minimal time for biological evolution,  some large number $M$ times 
the constituent time scale $\mu^{-1}$, must be smaller than the collapse time,  $\approx \sqrt{3/ \Lambda}$. This gives a 
limit $-\Lambda \lesssim M^{-2} \mu^{2}$. Clearly, both $N$ and $M$ must be large but are
not computable from first principles; in our universe $N\approx 10^{27}$ for a human brain and $M \gg 10^{31}$ (assuming full biological evolution in a single year). 
In dS space a better limit can be obtained by weakening the assumptions underlying Weinberg's argument or by determining the maximal size
of an object that can break away from the exponential expansion. Using results from \textcite{Hellerman:2005yi}, \textcite{Harnik:2006vj} derived a limit
\begin{equation}
\rho_{\Lambda} \lesssim {\rm min}\left[\mu^4, \mu^{-2} N^{-2}\right]
\end{equation}
No matter which limit one uses, the conclusion is in any case that in universes with a value for $\Lambda$ that is not extremely
small life is impossible. Obviously these are very crude limits, and there is no guarantee that other universes exist where $|\rho_{\Lambda}|$
is as large as one of the bounds.
But, crude as it may be,  this arguments implies that if $\Lambda$ can be treated as a free parameter on a Planckian interval, it is
clearly true that the reason for its smallness is anthropic. }

If all other parameters are kept fixed at their observed values, much 
tighter bounds can be obtained.
\else
{
Crude bounds on $\rho_{\Lambda}$ in {\it any} habitable universe can already be obtained by requiring that complex objects with a large
number of constituents (for example brains) can form and fit inside the horizon in dS (see the last section of  \textcite{Harnik:2006vj}), 
or that non-gravitational interaction time scales are much smaller than the collapse time in AdS. This implies that if $\rho_{\Lambda}$ can vary
on Planckian scales, its observed value is in any case at least partly anthropic. 

Much tighter bounds can be obtained if we
fix the other parameters at their observed value. }\fi
\textcite{BarrowTipler} pointed out that if $\Lambda$ is too large and negative, the universe would collapse before
life has evolved. 
\ifExtendedVersion {\color{darkred} The precise limit depends on the time needed for stars to produce heavy elements, and the time needed
for biological evolution, for which we have no theoretical estimate, and just one data point. }\fi The authors used the average
life-time of a main-sequence star to get a limit. 
This quantity can be entirely expressed in terms of Standard Model parameters and the Planck mass, and leads to 
a limit 
\begin{equation}\label{LowerBound}
| \rho_{\Lambda} | \ \lessapprox\ \alpha^{-4}\left(\frac{m_e}{m_p}\right)^4  \left(\frac{m_p}{M_{\rm Planck}}\right)^6 = 6.4 \times 10^{-120}.
\end{equation}
\ifExtendedVersion {\color{darkred}
For negative $\Lambda$ the collapse of the entire universe is unquestionably a limit, but there are still some important uncertainties in this bound. 
Because we have observational evidence for a {\it positive} $\Lambda$, the bound for negative $\Lambda$ may seem irrelevant, but it is interesting for several
reasons, so it is worthwhile to see if it can be made more precise. The formula on which it is based is actually the Eddington limit, a
lower limit on the lifetime for objects radiating at maximum luminosity. This limit is 
 \begin{equation*}
 T_{\rm min}= \frac{2}{3}  \epsilon \left[ \left(\frac{M_{\rm Planck}}{m_p}\right)^3 \right]
 \frac{\alpha^2  m_p^2}{m_e^2 } t_{\rm Planck}  \approx 3\ \hbox{million years}
 \end{equation*}
where $\epsilon$ is the energy efficiency, the fraction of mass that can be converted to energy,  $\epsilon \approx .007$. 
The actual limit is equal to $t_{\rm tot}=t_*+t_{\rm ev}$, where $t_*$ is the time  of the first supernova explosion after the big bang
(which produced the required heavy elements) and $t_{\rm ev}$ the shortest time
needed for geological and biological evolution. The latter cannot be determined from first principles. It depends
on biological time scales, the evolution of the sun, plate tectonics, volcanism, the movement of the solar system through the spiral arms of the galaxy, subtle
variations in the earth's axis, the moon, Jupiter, meteor impacts and other catastrophic events, effect we do not even know about, and most importantly, chance. From the
one data point we have we know that $13.8$ billion years is sufficient. If we assume that we are typical, faster evolution must be far less probable, but as long as it is
possible it affects the limit. The most conservative assumption is that $t_{\rm ev}$ under ideal circumstances could be less than a million years, so that the limit is determined by
the minimal stellar lifetime.
So $t_{\rm tot}$, the minimal time
between the big bang and evolution until intelligent life,  must lie between 3 million years and
13.8 billion years. 
}
\else
{
Rather than theoretical lifetimes of stars, one may consider observational extremes: the minimal stellar life-time of about 3 million years, and
the current age of the universe. The fastest time in which intelligent life can form must lie between these extremes.
}\fi
Requiring that this is less than the time of collapse, $\pi \sqrt{3/\Lambda}$, gives $\rho_{\Lambda} > - \rho_{\rm min}$, with
\begin{equation}
1.8 \times 10^{-122}  <  \rho_{\rm min} < 3.8 \times 10^{-115} \ .
\end{equation}

 The limit (\ref{LowerBound}) was argued to be valid 
 for positive $\Lambda$  as well.  However, \textcite{Weinberg:1987dv}
 pointed out that structure that has already formed will not
be ripped apart by an expanding universe.  Once galaxies have formed, it makes no
difference how much time is needed to make stars or evolve life, because the expansion will not inhibit that from happening. 
He then derived a limit based on the assumption
that life would not form if the universe expands too fast to inhibit galaxy formation.
The exact form of Weinberg's bound is
\begin{equation}
\rho_{\Lambda} < \frac{500}{729}\ \Delta^3 \rho_0\ ,
\end{equation}
and was derived by studying
the collapse of a spherical overdensity $\Delta$ using a Robertson-Walker metric. 
The overdensity starts expanding at $t=0$ when the universe has a matter  density $\rho_0$. For  $\rho_{\Lambda}=0$ it recollapses and forms structure, but as $\rho_{\Lambda}$ 
is increased a point is reached beyond which
the recollapse does not occur anymore. This gives the maximum value of $\rho_{\Lambda}$ for the overdensity $\Delta$.  The absolute upper limit
in  a given universe is given by determining the maximal overdensity that can occur. Since density fluctuations are distributions, there will not be
a strict upper limit, but the number of galaxies that can be formed will drop off rapidly beyond a certain $\rho_{\Lambda}$. 

\ifExtendedVersion {\color{darkred}
A ``lower bound on
the anthropic upper bound" can be obtained
by observing quasars (centers of young galaxies) at high redshift $z$, and then extrapolating
the observed matter density back to the corresponding age by 
multiplying the present $\rho_{\rm matter}$ by a factor $(1+z)^3$.  
If we know, empirically, that galaxies can form at such densities,
then vacuum density can not play a r\^ole if is smaller than the density of matter. With the information available in
1987 this argument gave an upper limit of at least $550 \rho_{\rm matter}$ based on quasars observed at $z=4.4$ (the exact pre-factor is $\frac13 \pi^2 (1+z)^3$).
However, meanwhile dwarf galaxies 
have been observed at $z=10$, increasing the bound by almost an order of magnitude \cite{Loeb:2006en}. The observed value of 
$\rho_{\Lambda}/\rho_{\rm matter}$is about 2.5, more than three orders of magnitude below this upper bound. 
There is also an upper bound to this upper bound: at a redshift of $z \approx 1000$ we reach the time of decoupling, and we are certainly not going to observe
galaxies that formed that early. This implies an absolute upper limit on $\rho_{\Lambda}$ of order $10^{-113}$.
}
\else 
{
In 1987 precision cosmology did not exist yet, and no theoretical estimate of the upper limit was possible. Hence an
empirical estimate was made. If proto-galaxies can be observed at high redshift $z$, when the matter density was larger by a factor $(1+z)^3$, 
a cosmological constant density of the same size would not obstruct galaxy formation either. In 1987 this led to an upper limit $\rho_{\Lambda} < 550 \rho_{\rm matter}$
from quasars at $z=4.4$. However, meanwhile dwarf galaxies 
have been observed at $z=10$, increasing the bound by almost an order of magnitude \cite{Loeb:2006en}.
}\fi  

\paragraph{Estimates of the Value of $\Lambda$.}
Nowadays we can determine the density fluctuations using COBE and WMAP (and recently PLANCK) results. 
It is instructive to make a rough estimate using the time of matter-radiation equality as the starting point
of structure formation.  An order of magnitude estimate for the matter density at equality is \cite{Hellerman:2005yi}:
$\rho_{ \rm eq} \approx T_{\rm eq}^4$, $T_{eq} \approx  m_p \eta (\zeta+1)$, where $\eta=6.3 \times 10^{-10}$ is the baryon-to-photon ratio and 
$\zeta$ the cold dark matter to baryon ratio. Using for $\Delta$
the {\it average}  for the fluctuations,
$Q \approx 2 \times 10^{-5}$ yields $\rho_{\Lambda} < 7.3 \times 10^{-125}$ (with parameter values
 from \textcite{Tegmark:2005dy}). Putting in the correct factors of order 1, and taking into account the contribution of neutrinos to matter-radiation equality,
lowers this number substantially. 
Clearly 
 a more careful treatment of galactic-size density perturbations (which contribute with a third power) is needed. 

Furthermore the ``bound" is  not a step function. One expects a mean density of galaxies that falls of with increasing $\rho_{\Lambda}$.
Such a  function was computed by \textcite{Efstathiou:1995ne} based on the results of COBE (but prior to  the observation of accelerated expansion). 
Although the observation of a positive $\Lambda$ in 1998 came as a shock to many, there were already several indications in that direction because
of the density contribution needed for spatial flatness (as predicted by inflation) and the age of the universe. This had already been pointed out by 
\textcite{Weinberg:1987dv}. The results  of \textcite{Efstathiou:1995ne} predicted a value for $\rho_{\Lambda}$ in agreement with that expectation, although
with large uncertainties, and subject to some criticisms \cite{Vilenkin:1995nb,Weinberg:1996xe}. This computation was improved and done analytically rather than numerically by \textcite{Martel:1997vi}, with similar results. Distributions for $\rho_{\Lambda}$ based on more recent cosmological data can be found in  \textcite{Pogosian:2006fx,Tegmark:2005dy}.

Computations of this kind rely on several assumptions. The distribution of theoretically allowed values of $\rho_{\Lambda}$ must be essentially flat
near $\Lambda=0$. Since $\Lambda=0$ is not a special point from the perspective of quantum gravity, and since the relevant range is extremely small
in Planck units, this seems plausible. Furthermore, the vacuum selection mechanism -- for example eternal inflation -- must not prefer special values either. This is less obvious, see section \ref{CosmoConstTWO}. 
It is assumed that observers are correlated with galaxies, and sometimes with stars, planets and baryons, and that we are typical
observers (the ``principle of mediocrity" of  \textcite{Vilenkin:1994ua}). 

The computations mentioned above assumed that only $\rho_{\Lambda}$ varies. The possibility that $Q$ also varies was considered by 
\textcite{Tegmark:1997in}, who computed the anthropic bounds  $10^{-6} < Q < 10^{-4}$ assuming $\Lambda=0$. They also pointed out that
without anthropic bounds on $Q$, the bound on $\Lambda$ is invalid. A potentially serious problem was raised in
\textcite{Banks:2003es}; \textcite{Graesser:2004ng}; \textcite{Garriga:2005ee};  and \textcite{Feldstein:2005bm}. Depending on models of inflation, 
the probability distribution may vary so steeply as a function of $Q$ that extreme values are strongly preferred, so
that the observed value $Q \approx 10^{-5}$, roughly in the middle of the anthropic range, has a very low probability of being observed (the ``Q-catastrophe").
But even when both $\rho_{\Lambda}$ and $Q$ vary, there is a robust bound on $\rho_{\Lambda}/Q^3$ \cite{Garriga:2005ee}. See \textcite{Vilenkin:2004fj} for a brief review 
of anthropic predictions for the cosmological constant.

We return briefly to the cosmological constant problem in section \ref{CosmoConstTWO}, after the string theory landscape and the measure problem have been explained.

\subsection{Possible Landscapes}

\subsubsection{Fundamental Theories}

The ``Anthropic Principle" discussed here is not a principle of nature, and not our ultimate goal. That goal is a fundamental theory in which
different quantum field theories are realized, and can be sampled. 
\ifExtendedVersion {\color{darkgreen}

We  are not interested in an anthropic principle added to the Standard Model and
General Relativity as an additional principle of physics. The ultimate goal remains finding a ``theory of everything", 
a theory that agrees with all current experiments and can be extrapolated to extreme limits without encountering inconsistencies. }{ This fundamental theory provides the input distributions for anthropic 
arguments, and may in principle be falsified with the help of such arguments. 
But it is the fundamental theory we should try to falsify, and not the anthropic principle, which is only a tool that may help us finding the theory.   
Once that has been achieved, the anthropic principle will only be a footnote. }{\color{darkgreen}

In the rest of this review we will avoid the term ``theory of everything", since it sounds a bit too pretentious, and just call it
 a ``fundamental theory". But whatever it is called, one can raise a number of objections. Perhaps such
a theory does not even exist, or perhaps our mental abilities are too limited
 to find it, or perhaps we simply do not have enough information to figure out the answer. Since, by assumption, we can only 
 extract all our information from an infinitesimal bubble in a multiverse, the latter possibility is a very serious one. One can also
 speculate about fundamental obstacles, along the lines of Goedel's theorem or Heisenberg's uncertainty principle that would
 make it impossible, {\it in principle} to ever achieve such a goal. Or one may believe in the onion shell paradigm, which states
 that as we explore nature at ever shorted distances, we will continue to discover new physics forever, like peeling layers
 from an onion with an infinity of shells. But most physicists believe that there would be a last shell at the Planck length, so that
 the number of unknown shells is presumably finite.  
 
 The marvelous success of the Standard Model gives some reason for optimism. It might even turn out to be
 the last shell. It brings three of the four known interactions under control, and hence it seems not too unreasonable
 to hope that the fourth one can also be mastered. The first attempts to apply the traditional methods of perturbative quantum
 field theory to quantum gravity were made decades ago \cite{tHooft:1974bx} but ran into problems. String theory
 can be considered the most conservative attempt to overcome these problems. It does indeed overcome them in special cases
 (ten space-time dimensions with supersymmetry). There are good reasons to believe that in that case
 it does indeed lead to the desired finite perturbative expansion. But that is not all we expect from a fundamental theory.

If one accepts the foregoing  anthropic arguments,  it is not plausible that the Standard Model is 
mathematically unique. Nothing said so far about a possible fundamental theory requires that. 
Non-uniqueness of the Standard Model may look like the end of last century's dreams, but it is only saying that those
dreams were to naive.
Then we need
a fundamental theory which allows several, presumably a large number, of alternatives. In general, we will refer to the list of options as the
``landscape" of that theory. This list consists of all universes
like our own (which of course must be on the list itself), but with a different point in the space of all quantum field theories governing matter and the non-gravitational interactions. 
In string theory we can -- in some cases -- give this notion a more precise meaning, where the different points correspond to minima of a potential.
The name ``landscape" was, presumably, partly inspired by the hills and valleys of that potential, as well as by the common usage in
other fields, such as the ``publishing landscape". 

But we will need more than just a list of options. There has to be a way to reach different points in that landscape through
some kind of physical process, such that they can all be sampled in a process like eternal inflation. This disqualifies the quantum
field theory landscape, where any QFT is allowed, but there is no mechanism to go from one to another.

\subsubsection{The r\^ole of gravity}\label{Gravity}

During three-quarters of last century, particle physics and gravity where mostly treated as separate subjects. 
All Lorentz covariant theories of particle physics can be coupled straightforwardly to classical general relativity. Furthermore,
gravity can be consistently decoupled from particle physics by sending Newton's  constant to zero. The fact that quantum gravity
did not enjoy the same consistency as quantum field theory (in particular the existence
of perturbative infinities that cannot be removed by renormalization) was often stated, but seen mostly as a problem of gravity.
On the other hand, 
the  remaining problems of particle physics -- understanding the discrete and continuous choices of groups,
representations and parameters -- have often been thought to be completely solvable without coupling to gravity. This might indeed be true. However, there are good grounds to
believe that these two theoretical areas must meet.

Indeed, there are a few concrete reasons why one might expect (quantum) gravity to play a r\^ole in limiting the number of options of quantum field theory.}
\else 
{
The fundamental theory provides the input distributions for anthropic 
arguments, and may in principle be falsified with the help of such arguments. 
But it is the fundamental theory we should try to falsify, and not the anthropic principle, which is only a tool that may help us finding the theory.   
Once that has been achieved, the anthropic principle will only be a footnote. 

We can try
to decide which properties such a fundamental theory should have, and which current ideas qualify. Indeed, there are a few concrete reasons to believe
quantum gravity should play an essential r\^ole. In particular, one cannot discuss parameter values without dealing with the problem
that they are fundamentally undetermined in a renormalizable quantum field theory. Furthermore
there are infinitely many of them in a non-renormalizable
theory like naively quantized gravity. One cannot consider changing parameters without discussing changes in vacuum energy, which
can only be done in the context of gravity. So we need a fundamental theory of quantum gravity with dynamics and connectivity
in the space of
couplings.}\fi

\ifExtendedVersion {\color{darkgreen}
\vskip .5cm \noindent{\em UV completion.} 
The renormalization procedure that works so well in quantum field theory requires an infinite number of fundamentally
undetermined parameters in quantum gravity. Although in all low-energy observables their contribution is suppressed by
powers of the Planck mass, this situation is untenable in a theory that aims at determining the Standard Model parameters.
Furthermore one cannot solve this problem in gravity alone. Gravity can be decoupled from the Standard Model, but not the
other way around. Loop corrections due to matter are equally important as loop corrections due to gravity itself.      
One has to find a theory of matter and gravity that does not have
these problems. This is the requirement of ``UV completeness". One may expect this to restrict the possibilities for matter.  }\fi 

\ifExtendedVersion {\color{darkgreen}
\vskip .5cm \noindent{\em Vacuum energy.}
The energy of the ground state is not an observable in quantum field theory, and can be set to zero
by subtracting a constant. This is fine as long as a single theory with fixed parameter values is considered, but as soon as 
different parameter values are allowed one might expect that constant to differ. Gravity does couple to vacuum energy, and
therefore it is plausible that in order to consider different  quantum field theories, we have to embed them in a theory of gravity.
\vskip .5cm \noindent{\em Holography.}
Black hole entropy is proportional to the area of the horizon. This has led to the idea of holography, stating that
fundamental degrees of freedom should be associated with surface area rather than volume. This 
suggests that theories which consistently describe black hole physics must place some restrictions on degrees of freedom.

}\fi
\vskip .5cm

\ifExtendedVersion {\color{darkgreen}
Using these guidelines, 
we can try
to decide which properties a fundamental theory should have, and which current ideas qualify.}\fi

\subsubsection{Other Landscapes?}

The String Theory Landscape seems to fit the bill, although there is a lot of work still to be done, and a lot that can go wrong. There are
many ideas that are presented as competitors, and here we list a few of them, to see if they qualify. 
We will not enter here in a discussion 
about the relative merits of some of these ideas as theories of quantum gravity.\ifExtendedVersion {\color{darkgreen}
\begin{itemize}
\item{Loop quantum gravity \cite{Ashtekar:1986yd} is a method for quantizing of gravity, but it does not appear to restrict matter.}
\item{Dynamical triangulations \cite{Ambjorn:2004qm} is method for defining sums over geometries in quantum gravity, 
with the interesting feature that four dimensions emerge dynamically. But it has nothing to say, so far, about matter coupled to gravity.}
\item{Asymptotically Safe Gravity  \cite{Weinberg:1976xy}, \cite{Reuter:1996cp} attempts to deal with the perturbative problems of 
quantum gravity by demonstrating the existence of a non-perturbative ultra-violet fixed point. The other interactions may share that
feature. However, this does not imply that they can be derived from such a principle. 
For example, non-abelian gauge theories have a perturbative ultra-violet fixed point, 
but this does not determine the value of the coupling constant at lower energies, nor the  choice
of gauge group. Furthermore there is no mechanism for change within this landscape. 
Interestingly, in this context (under some additional assumptions) the Higgs mass was predicted two year prior to its
observation \cite{Shaposhnikov:2009pv}. This prediction is that the Higgs self-coupling  vanishes at the Planck scale, and hence
the Higgs mass would be exactly at the stability bound.
With more up-to-date values of the top quark mass the prediction is now too high by about two
standard deviations.}
\item{Noncommutative geometry  \cite{Chamseddine:2007hz} is claimed to provide an essentially unique derivation
of the Standard Model, although some features such as the number of families and the masslessness of the photon do not
(yet) come out, and the appearance of the Standard Model gauge group involves additional assumptions. 
One can also obtain supersymmetric QCD with these methods  \cite{Broek:2010jw}, suggesting a larger 
landscape, but it is not clear if distinct theories are physically connected. In this context the Higgs mass was also predicted, at a value that is now ruled out, but it is claimed that can be fixed by means of an additional scalar field that was previously ignored \cite{Chamseddine:2012sw}. }
\item{Finite Unified theories \cite{Heinemeyer:2007tz} are $N=1$ supersymmetric GUTs that can be made 
all-loop finite. This has lead to a correct prediction for the Higgs bosons mass range, more than three years before its discovery. This approach does not have the ambition 
to provide a theory of all interactions, but only to find reductions in the parameter space. As in the two foregoing cases, there
is no mechanism for connecting different points in this landscape.}
\item{Spontaneously broken local conformal invariance was argued \cite{tHooft:2011aa} to be a physically motivated condition that
fixes all parameters, leaving only a (denumerably infinite) number of discrete choices of gauge groups and representations. 
}
\end{itemize}
All of these ideas may be correct, and they are not necessarily mutually exclusive. All we are saying is they do not
seem to give rise to an anthropic landscape, and observation that most of these authors will enthusiastically agree with.
When it comes to that, string really is the ``only game in town". 

Note that three of these approaches now claim to predict or postdict the Higgs mass correctly. This was also done
in the context of string theory \cite{Kane:2011kj}, with additional assumptions, although it is just barely a {\it pre}diction.} 
\else  
{

Some alternative approaches to quantum gravity, for example loop quantum gravity \cite{Ashtekar:1986yd} or dynamical triangulations \cite{Ambjorn:2004qm} have nothing to say about matter. Asymptotically safe gravity  \cite{Weinberg:1976xy}, \cite{Reuter:1996cp} strongly restricts matter if
quantum field theory is also required to be asymptotically safe, but cannot fix the couplings of asymptotically free gauge theories. There is no
known way of physically connecting different theories. The same is true for noncommutative geometry \cite{Chamseddine:2007hz}. In contrast
to earlier claims it does  not yield the Standard Model uniquely;  for example, one can also obtain supersymmetric  QCD \cite{Broek:2010jw}. 
But it is still far from providing a useful landscape. Finite unified theories \cite{Heinemeyer:2007tz} also limit the possible quantum field theories, but do not yield a connected landscape.
Spontaneously broken local conformal invariance was argued \cite{tHooft:2011aa} to be a physically motivated condition that
fixes all parameters, leaving only a (perhaps denumerably infinite) number of discrete choices of gauge groups and representations. 

Since all these authors will enthusiastically agree that they do {\it not} propose an anthropic landscape, it is fair to say that
in this respect String Theory really is the only game in town. }\fi


\subsubsection{Predictive Landscapes}\label{PredLandsubsection}

The existence of a landscape does not necessarily 
imply that all predictive power is lost. We just list some options here to counter some common
philosophical objections.

\vskip .5cm \noindent{\em Universal Predictions.}
A large ensemble of possibilities may still have one or more universal predictions. In the case of the
string landscape, what comes closest to that is a {\it negative} prediction, namely the {\it absence} of variations
in Standard Model parameters (see section \ref{VarCons}). There may be other opportunities for universal predictions because of the universal existence of moduli
and axions
in string theory. 

\vskip .5cm \noindent{\em Sparse Landscapes.} 
If a landscape is small enough, current data may already be sufficient to find the solution that corresponds to our universe.
Having determined that, all parameters would be known exactly. The Standard Model data has been estimated to provide
about 80 digits worth of information \cite{Douglas:2006es} so that a landscape of, say, $10^{30}$ points would realize this possibility, with
a lot of predictions left.
But this is not likely to be true in the string theory landscape,  if current ideas about the cosmological constant are correct. This
would already require more than $10^{120}$ solutions, and a computation of the cosmological constant with 120 digit
precision in each of them, if we want to pin down the solution exactly. See \textcite{deAlwis:2006nm} and \textcite{Denef:2006ad} for 
an exposition of some of the problems involved.

\vskip .5cm \noindent{\em Friendly Landscapes.}
It might happen 
that some parameters vary over a wide range, while others are sharply peaked at definite values.  
Toy examples of such landscapes have been constructed using scalar field potentials \cite{Distler:2005hi,ArkaniHamed:2005yv}.
For a large number $N$ of scalars, some parameters may be distributed
widely, whereas others vary by a fraction ${1\over \sqrt{N}}$. The widely distributed ones were argued to be
the dimensionful ones, {\it i.e.} the weak scale and the cosmological constant.
This would allow anthropic arguments for
the dimensionful parameters to be valid without eliminating the possibility for fairly sharp predictions for Yukawa couplings and hence
quark and lepton masses. 
There might be enough variability left to allow even the anthropic constraints on those masses to be met.
They might not be at the peak of their distribution, but anthropically pushed towards the tail. 
\ifExtendedVersion {\color{darkred} The required large number of scalars does indeed exist in string theory: the moduli.}\fi

\vskip .5cm \noindent{\em Overwhelming Statistics.}
The following example shows that the dream of an {\it ab initio} determination of the Standard Model and all its parameter values
is not even necessarily inconsistent with anthropic arguments. It requires a large hierarchy of sampling probabilities, the probability
for a vacuum to be selected during eternal inflation. Let us assume that the treacherous problem of defining these probabilities (see section \ref{EternalInflation}) has been solved, 
and let us order the vacua according to this probability.  Suppose that the $m^{\rm th}$ vacuum has probability $\epsilon^m$, where
$\epsilon$ is a small number. 
Furthermore, assume that, on average, only one out of $M$ vacua lands in the anthropic domain. For definiteness, let us take
$\epsilon=0.1$ and $M=1000$. The first anthropic vacuum is not likely to be the one with $m=0$, and hence it will
have a very small sampling probability, but that does not matter. The point is
that the second anthropic vacuum would typically have a probability of $10^{-1000}$ with respect to the first. 
\ifExtendedVersion {\color{darkred}
This prediction would have at least the level of confidence of any ``traditional" prediction in particle physics. }\fi
Such a scenario might be realized if one ``master" vacuum dominates the population of vacua by a huge statistical
factor, and all other vacua are obtained from it by a sequence of tunneling events (see section \ref{EternalInflation}). 
To actually compute the dominant
anthropic vacuum would require determining the master vacuum, the tunneling rates and the anthropic domains, all of which
are in principle computable without experimental input. In practice this seems utterly implausible, but in this example all 
observed anthropic miracles would  be explained, provided the complete set of vacua is large enough and distributed in the right way, and still
there would be a  nearly unquestionable prediction of all parameters.

\subsubsection{Catastrophic Landscapes}\label{Catastrophic}

The last scenario implicitly assumes that anthropic regions in QFT space are described by step functions, so that a given QFT either allows or does not allow life.
In reality there will be smooth distributions at the boundaries, and depending on how fast they fall off  
there is an important potential problem: outliers in distributions may be strongly selected. 
To illustrate that, consider an extreme version of overwhelming statistics, suggested by \textcite{Linde:2010nt}. They consider 
the possibility that landscape probabilities depend on the cosmological constant $\Lambda$ as ${\rm exp}(24\pi^2/\Lambda)$, and
that  $\Lambda$ can take only a discrete set of positive values, $\Lambda=n/N$, $n=1,\ldots N$. Here $\Lambda$ is expressed in Planck units, and $N$ is a large integer. In this situation, $n=1$ is strongly favored statistically. If we define $P(n)$ as the probability for vacuum $n$,
then we find
\begin{equation}
\frac{P(n)}{P(1)}=e^{-24\pi^2 N\left(\frac{n-1}{n}\right)} \ .
\end{equation}
If the most probable vacuum, $n=1$, is ours, then $N\approx 10^{120}$, and anything else is suppressed by behemothic factors.  
The authors conclude ``{\em This means that by finding the 
vacuum with the smallest 
$\Lambda$ we fix all other parameters; 
no additional anthropic reasoning is required}". 

But this is not likely to be true. If one can define strict anthropic boundaries in field theory space, as in fig (\ref{LandscapeDistributions}),
the vacuum with smallest $\Lambda$ has only a small chance of ending up within the anthropic contours. If any boundary line is in reality
a contour of a gaussian distribution, with a tail stretching over the entire parameter space, then the $n=1$ vacuum is vastly more likely
to lie somewhere in the tail. Suppose for example a variable $x$ has an anthropic distribution $\propto {\rm exp}[-(x-x_0)^2/(2\sigma^2)]$,
and suppose vacuum 2 happens, against all odds,  to lie near the peak. Then vacuum 1 can lie $\approx \sqrt{N}$ or about $10^{60}$ standard deviations
away from the peak, and still beat vacuum 2 in overall probability.  

This would be the worst possible outcome. It resembles uniqueness, but is catastrophically  inferior.
There would be a huge landscape that does not solve any problem. It would not explain any fine tunings, not even those
of the cosmological constant itself. It is very unlikely that we would ever be able to compute the  lowest $\Lambda$ vacuum, because
$\Lambda$ would depend on all intricacies of particle physics, cosmology and of a fundamental theory, which would have to be computed with
120 digits of precision.

\ifExtendedVersion {\color{darkred}
\subsubsection{Expectations and implications}

What kind of landscape can we expect? The special properties of the  Standard Model suggest a fairly large number of possible QFT choices. If we assume the correctness of quantum field theory with regard to vacuum energy, these choices will have to be
discrete, because any small changes in any parameter will generate huge contributions.
If we furthermore assume that
the cosmological constant  does indeed take arbitrary values 
on the interval $[-1,1]$ in Planck units, we need at least $10^{123}$ vacua more or less evenly distributed on the interval. It is also important
that such a dense distribution exists for arbitrary choices of the matter sector.  

Any demonstration that a given  ``vacuum"
might describe our universe has to rely on
the fact that the cosmological constant can always be adjusted to the required value by adding a sufficiently densely distributed constant.

It seems inevitable that this fact also ultimately ruins the hope of finding ``the" vacuum that precisely corresponds to our universe.
In principle, a discrete set of values keeps that hope alive. The right vacuum may not be derivable from scratch, but at least by closing in on it
with experimental data we might finally determine it, and reach a stage where all Standard Model parameters can be predicted with unlimited precision.
Unfortunately, the previous paragraph shows that for every vacuum that fits the Standard Model
data, there must be at least $10^{120}$ more that scan the cosmological constant axis, and we are unable to use that piece of information. It would be nice
if the two problems were completely decoupled, so that all  $10^{120}$ alternatives have {\it exactly} the same values for the Standard Model parameters.
But this is not plausible. The standard situation in physics is that all changes in one variable generate changes in any other variable, unless there is a 
symmetry forbidding that. So we should expect the Standard Model parameters of the $10^{120}$ variants to form a dense cloud, and the best we can hope for
is that the width of that cloud is extremely small, a friendly landscape.
 See \textcite{deAlwis:2006nm}) for a similar discussion.  In \textcite{Denef:2006ad} computational complexity is discussed in
 regions in the string theory landscape.
}\fi

 \section{String Theory}\label{StringTheorySection}

 Just as ``Standard Model" and ``Anthropic Principle", ``String Theory" is poorly named. It owes its name to
 its original formulation: strings propagating through space-time and interacting by splitting and joining. But nowadays this
 is merely a general name for an interconnected web of theories, including some that do not have a string interpretation at all.
 \ifExtendedVersion {\color{darkred}
 
 It was always clear that this formulation was perturbative as an expansion in the number of times strings split and join.
 But despite four decades of efforts, we still do not know of what theory it is a perturbative expansion. Furthermore, those
 efforts have made it clear that whatever that theory is, some of its regions are not described by perturbative strings at all,
 but perhaps by interacting higher dimensional membranes. The term ``string" only refers to some easily accessible 
 entry gates into this theoretical wonderland. Because that is indeed what it is: an amazing generator of theoretical 
 and mathematical insights, which any intelligent civilization in the landscape will encounter during their development,
 and which no intelligent civilization can afford to ignore. We cannot be sure what its ultimate fate will be in the history
 of science: merely an awe-inspiring tool, or a first hint at  what  we can expect from a fundamental theory of all interactions.
 Here we will speculate that the latter is true, and that what we are learning by studying the perturbative expansion, and 
 various non-perturbative effects, are indeed fundamental properties of a complete  theory. For lack of a better name, we will
 continue to refer to his whole interconnected web of computational methods and mathematical features as ``string theory".}\fi
 
  We will only introduce a few basic concepts of string theory here. There are many excellent books on this subject, such as the classic 
 \textcite{Green:1987sp}, the introductory course by \textcite{Zwiebach:2004tj}, the books by
 \textcite{Polchinski:1998rq} and \textcite{Kiritsis:2007zza} and the very recent one by \textcite{Blumenhagen:2013fgp}. These books also
provide extensive references to classic string theory papers, which we will omit here unless they have direct relevance to the landscape.
  
 \subsection{Generalities}

 In its most basic form, a string amplitude is derived from the following two-dimensional \ifExtendedVersion {\color{darkred} conformally invariant }\fi action
 \begin{equation}\label{FullAction}
 \begin{split}
 &S[X,\gamma]=\hfill \\ &-{1\over 4 \pi \alpha'}\int d\sigma d\tau
\sqrt{\!-\!\det\gamma}\sum_{\alpha\beta}\gamma^{\alpha\beta} \partial_{\alpha} X^{\mu} \partial_{\beta} X^{\mu}  g_{\mu\nu}.
\end{split}
\end{equation}   
Here $X^{\mu}(\sigma,\tau)$ is a map from the two-dimensional surface swept out by the string (the world-sheet, with coordinates $\sigma$ and $\tau$)
into  space time, $\gamma_{\alpha\beta}$ is the metric on that surface, and $g_{\mu\nu}$ is the space-time metric. 
\ifExtendedVersion {\color{darkgreen}
 The 
{\it Regge slope parameter} $\alpha'$ 
(the notation is an awkward
historical relic) has the dimension $[{\rm length}]^2$, and is related to the tension of the string as $T=1/2\pi \alpha'$.
}\else {
The parameter $\alpha'$ 
has the dimension $[{\rm length}]^2$, and is related to the tension of the string as $T=1/2\pi \alpha'$. The two-dimensional metric $\gamma$ can be
integrated out, so that the action takes the form of a surface area. Amplitudes are computed by performing a path-integral over surfaces
weighted by a factor $\exp(-iS/\hbar)$. 
}\fi
\ifExtendedVersion {\color{darkgreen}

The two-dimensional metric $\gamma$ can be
integrated out, so that the action takes the form of a surface area. Amplitudes are computed by performing a path-integral over surfaces
weighted by a factor $\exp(-iS/\hbar)$. For closed strings without external lines these surfaces are spheres with multiple handles, $n$-tori, which
can be described in complex coordinates by the theory of Riemann surfaces. Scattering amplitudes are obtained by considering surfaces with
external lines, obtained by gluing tubes to the surface. 
An important feature of this two-dimensional field theory is that it is not only Poincar\'e invariant
(which ensures that the amplitudes do not depend on the way the surfaces are parametrized) but also conformally invariant. This means that, as a first step,  the  Poincar\'e group is enlarged by adding scale transformations.
transformations. In higher dimensions with $p$ space and $q$ time dimensions this extends the Poincar\'e group to $SO(p+1,q+1)$, but in two dimensions
the conformal group has an infinite number of generators. This infinite symmetry is the reason why string theory is so much more powerful than its analogs with world sheets of one dimension (particles) and three or more dimensions (membranes).
Conformal invariance is important for the consistency of string theory (absence of unphysical states  (``ghosts") in the space-time spectrum, and it is therefore essential to maintain it when the theory is quantized.}\fi

\ifExtendedVersion {\color{darkgreen}
Upon quantization by standard methods, one finds a spectrum that describes an infinite tower of massive particles (the vibrational modes of the string)
plus a few massless ones and a tachyon. The massless ones are a symmetric tensor $G_{\mu\nu}$, an anti-symmetric tensor $B_{\mu\nu}$ and a scalar $\phi$, the dilaton.
The tree-level scattering amplitudes can be matched to field theory Feynman diagrams obtained by decomposing the string into its modes. In particular
one finds contributions that correspond to exchange of the symmetric tensor, which precisely match those obtained from quantized Einstein gravity.
This implies a relation between Newton's constant $G_N$ (in $D$ space-time dimensions) and $\alpha'$ }\else {
The modes of vibration of the propagating string are observed as particles. The particle spectrum consist of a tachyon, a massless symmetric tensor $G_{\mu\nu}$, an anti-symmetric tensor $B_{\mu\nu}$ and a scalar $\phi$, the dilaton, 
plus an infinite tower of excitations. 
The interpretation of $G_{\mu\nu}$ as the graviton field implies a relation between Newton's constant and $\alpha'$
}\fi
\begin{equation}
G_N \propto g_s^2 (\alpha')^{\frac{1}{2}(D-2) },
\end{equation}
where $g_s$ is the string coupling 
constant defined below. 
The parameter $\alpha'$ also sets the mass scale for the string excitations. Consequently, their spacing is in multiples of the Planck scale.
The space-time metric $g_{\mu\nu}$ in (\ref{FullAction}) should be viewed as a space-time background in which the string propagates. 
The background can be curved \ifExtendedVersion {\color{darkred} (then $g_{\mu\nu}$ depends on $X$)}\fi, but it is subject to consistency 
conditions that follow from the quantization. They imply Einstein's equations plus higher order corrections, but also restrict the number of space-time
dimensions. For a flat metric, this yields the requirement $D=26$. 
The other two massless fields, $B_{\mu\nu}$ and a scalar $\phi$, can  be included in a generalization of (\ref{FullAction}) as background fields.
The dilaton couples as
\begin{equation} S(X,\gamma,\phi) \propto \int d\sigma d\tau \sqrt{\gamma} R(\gamma) \phi\ .\label{dileul}
\end{equation}  
This
introduces a dependence of amplitudes on the Euler index  $\chi$ of the surface as $e^{-\chi\phi}$. 
\ifExtendedVersion {\color{darkred}
($\chi=2$ for a sphere, $\chi=0$ for a torus, and
$\chi=2(1-g)$ for a genus-$g$ surface, a sphere with $g$ handles; each external line adds a factor $g_s$). }\fi
Hence the constant mode $\phi_0$  of $\phi$  provides a weight factor for surfaces of different topology. This 
defines a loop expansion parameter: the string coupling constant $g_s=e^{\phi_0}$. It is not a constant set by hand in the action, but it is
the vacuum expectation value of a scalar field. Therefore its value can be set dynamically. The only  genuine parameter is
$\alpha'$, but this is a dimensionful quantity that sets the scale for everything else.


The bosonic string action can be generalized by  adding two-dimensional fermions $\psi^{\mu}$ to the two-dimensional bosons $X^{\mu}$, both with $\mu=0,\ldots, D-1$. 
Quantization consistency then requires the existence of a two-dimensional supersymmetry called {\it world-sheet supersymmetry} 
relating the bosons and the fermions. These are called  fermionic strings. In flat space, they can only be consistently quantized if $D=10$. 

Another
generalization is  to consider two-dimensional surfaces that are not oriented, such as the Klein bottle, and surfaces with boundaries, such as
the annulus. This leads to theories of open and closed strings, that can exist in 26 and 10  dimensions for bosonic and fermionic strings respectively.

Furthermore one can make use of the fact that in free two-dimensional theories left- and right-moving
modes can be treated independently. In closed string theories one can even use bosonic string modes for the left-movers and fermionic ones
for the right-movers. These are called heterotic strings,  and their flat space-time dimension is limited by the smaller of the two, namely $D=10$.

\subsection{Modular invariance}

\def\Im{{\rm Im~}}

Although the string theory spectrum consists of an infinite set of particles, string theory is not simple a quantum
field theory with an infinite number of fields.
The difference  becomes manifest in the simplest 
closed string one-loop graph, the torus.
At lowest order,  the relevant  integral takes the form\begin{equation*}
\int \frac {d^2\tau} {( \Im \tau)^2 } ( \Im \tau)^{(2-D)/2}\   \hbox{{\rm Tr}} \  e^{2 i\pi \tau (L_0-\frac{c}{24})}e^{-2 i\pi \bar\tau (\bar L_0-\frac{c}{24})} \ .
\end{equation*}
The operators $L_0-\frac{c}{24}$ and $\bar L_0-\frac{c}{24}$ are the two-dimensional Hamiltonians of the left- and
right-moving modes, and the trace is over the tensor product of the two Hilbert spaces.
The integral in QFT would be over the entire complex 
upper half plane, and is clearly divergent near $\tau=0$. But in string theory
the contributions to this integral consists of infinitely many 
identical copies
of each other, and
they would be over-counted
if we were to integrate over the entire upper
half plane. These  identical copies are related by the following transformation
\beq
\label{ModTrans}
\tau \rightarrow \frac{a\tau+b}{c\tau+d},  \  \  \  a,b,c,d \in {\bf Z},\  \ \  ad-bc=1.
\eeq
The restriction to a single copy is allowed provided that the
integrand is invariant under this  transformation, which implies strong
constraints on the spectrum of eigenvalues of $L_0$ and $\bar L_0$. These  are known as modular invariance constraints.
\ifExtendedVersion {\color{darkred}
To avoid the over-counting we can then
limit ourselves to one region, and in particular we may choose one
 that excludes $\tau=0$, thereby explicitly avoiding the field theory divergence. The latter is essentially factored out as an
 infinite over-counting multiplicity.}\fi
 
 \subsubsection{Finiteness and Space-time Supersymmetry}
 
 Modular invariance is the real reason why closed string theory is UV finite. This holds for any closed string theory,
 including  the bosonic string. There is a wide-spread belief that in order to deal with UV 
 divergences in quantum gravity and/or quantum field theory nature must be 
 supersymmetric at its deepest level.  However, the UV finiteness of closed strings 
 has nothing to do  with space-time supersymmetry.
  
 The $\tau$-integral may still diverge for
 another reason: the presence of tachyons in the spectrum. Furthermore, if the one-loop
 integral is non-zero, there is a dilaton tadpole, which leads to divergences at two loops
 and beyond because the dilaton propagator is infinite at zero momentum. But both of these
 problems are related to an inappropriate choice of the background, and are IR rather than UV. 
 The tachyon signals an instability, an expansion around a saddle point of the action. 
 They are absent in certain fermionic string theories. Their absence
 requires fermions in the spectrum, but does not require supersymmetry.

\ifExtendedVersion {\color{darkred}
The one-loop integral gives a contribution $\Lambda$ to the vacuum energy (the cosmological constant), and implies that
the one-loop diagram with one external dilaton line is nonzero: a tadpole. In general, tadpoles indicate that the equations of motion
are not satisfied. In this case
the dilaton
tadpole  signals that the flat background space-time that was used is not a solution to the
equations of motion; instead one must use de Sitter (dS) or Anti de Sitter (AdS) space with
precisely the value $\Lambda$ as its cosmological constant  \cite{Fischler:1986ci,Fischler:1986tb}. 
Unfortunately this argument only provides an explanation for the presence of the tadpole, but it does not provide
an exact (A)dS. For AdS spaces that problem has been solved in certain cases (using the AdS/CFT 
correspondence), but for dS space this is considerably more complicated. This is especially disturbing because the observation
of accelerated expansion of the universe suggests that we live in a dS space with an extremely small cosmological
constant. Even disregarding the smallness, the very existence of a positive cosmological constant is a problem in string theory
(and also in quantum field theory). 
Some see this as merely a technical problem, while
others regard it as an obstacle that is so fundamental that any attempts at making approximations must fail (see \cite{Banks:2012hx}).}\fi

Space-time supersymmetry automatically implies  
absence of tachyons and the dilaton tadpole, but it is not an exact symmetry of nature, and  therefore cannot be used to argue
for their absence.

\subsubsection{Ten-dimensional Strings}

The condition of modular invariance is automatically satisfied for the bosonic string, but imposes
relations among the boundary conditions of the world-sheet fermions.
These conditions have several solutions: supersymmetric ones and non-supersymmetric ones, 
 with
and without tachyons\ifExtendedVersion {\color{darkred}, and even fermionic  string theories with only bosons (including tachyons) in the spectrum (``type 0 strings").}\else.\fi

The best-known solutions are the supersymmetric ones. There are two closed fermionic superstrings, called type-IIA and
type-IIB, and two heterotic superstrings, distinguished by having a gauge algebra $E_8\times E_8$ or $SO(32)$. Open string theories
have to satisfy an additional constraint: cancellation of tadpoles for the $\chi=1$ surfaces,   the disk and the crosscap.
This  leads to
just one theory, called  type-I, with gauge group $SO(32)$. Apart from the type-IIA theory, all of these theories have chiral fermions
in their spectrum. \ifExtendedVersion {\color{darkred} In ten-dimensional field theories, chiral fermions were known to lead to disastrous inconsistencies (chiral anomalies)
which could only be avoided by contrived-looking combinations of fields \cite{AlvarezGaume:1983ig}. 
Precisely these fields came out of string theory. In heterotic strings, additional interactions are required to achieve this \cite{Green:1984sg}, and those
interactions are indeed found in the string effective field theory. These ``miracles" ignited the string revolution of 1984.
After 1984, closed strings (especially the $E_8\times E_8$ heterotic
strings) dominated the field, and there was a tendency to view all others as unwelcome artifacts of the construction methods that
would disappear on closer scrutiny. But precisely the opposite happened.}\fi

\subsection{D-branes, p-forms and Fluxes}

\ifExtendedVersion {\color{darkred} 
A second revolution took place around 1995, and originated in part from a new insight in open string boundary conditions.
It was always clear that one may consider
two possible boundaries: the Neumann boundary condition, which respects space-time Poincar\'e invariance, and the
Dirichlet boundary condition, that explicitly  violates it by fixing the endpoint of the open string to a definite space-time point. }\else {
Open strings can have two kinds of boundaries conditions: the Neumann boundary condition, that respects space-time Poincar\'e invariance, and the
Dirichlet boundary condition, that explicitly  violates it by fixing the endpoint of the open string to a definite space-time point. }\fi
\ifExtendedVersion {\color{darkred}
The breakthrough was understanding that this could have a perfectly consistent interpretation by assuming that the open string ends on a physical
object localized in space-time, and spanning a subspace of it, a membrane \cite{Polchinski:1995mt}. 
}\else {
However, they can have a perfectly consistent interpretation by assuming that the open strings end on a physical
object, localized in space-time and spanning a subspace of it, called a D-brane \cite{Polchinski:1995mt}.}\fi 
In $d$ space-time dimensions, the endpoints
of open strings with $d-k$ Neumann boundary conditions and $k$ Dirichlet boundary conditions  sweep out a 
$m$-dimensional surface called a ${\rm D}_{m}$-brane (where the ``D" stands for Dirichlet and $m=d-k-1$). 

These D-branes are part of string theory as non-perturbative solutions, like solitons in field theory (see \textcite{Duff:1994an} for a review).
Since they are non-perturbative, they
cannot be read off directly from the low energy effective action of string theory,  
but they do betray their existence because they are sources of massless fields 
which {\it do} appear in the spectrum. 
These fields are anti-symmetric tensors of rank $p$, called $p$-forms. The source for such $p$-form fields are membranes with
$p-1$ dimensional space-like surfaces ($\rm{M}_{p-1}$ branes) that sweep out a $p$ dimensional world volume $V_p$ as they propagate.
A $p$-form field $A_p$ has a field strength tensor $F_{p+1}$, which is an anti-symmetric tensor with $p+1$ indices. All of these
statements are fairly straightforward generalizations of Maxwell's theory of electrodynamics in four dimensions, which correspond to the case $p=1$. 
In this case the sources are ${\rm M}_0$ branes (particles) that sweep out a one-dimensional world line. The relation between fields, field strengths, source branes and
their world volumes can be summarized as follows:
\begin{equation}
A_p \rightarrow F_{p+1} \rightarrow {\rm M}_{p-1} \rightarrow  V_p\ .
\end{equation}
One can define a magnetic dual of these fields, again in analogy with electric-magnetic duality in electromagnetism. 
In general, this relates the field
strength $F_n$ to a field strength $F_{d-n}$ in the following way
\begin{equation}
F_{\mu_1\ldots \mu_n}=\epsilon_{\mu_1,\ldots \mu_d}F^{\mu_{n+1}\ldots \mu_d}\ .
\end{equation}
In this way the field $A_p$ is related to a field $A_{d-p-2}$, and  the source   ${\rm M}_{p-1}$ branes are dual to 
${\rm M}_{d-p-3}$ branes. For electromagnetism in $d=4$ dimensions  ($p=1$) this yields point-like electric charges, dual to point-like
magnetic charges. 

The analogy with electrodynamics extends to a quantization condition for the dual brane charges, analogous to
the Dirac quantization condition for electric and magnetic charges, $eg=2\pi k, k \in \mathbb{Z}$. This will play an important
r\^ole in the following. On compact manifolds, these $p$-form fields can wrap around  suitable topological 
cycles of the correct dimension to support them. These wrapped fields are called ``fluxes".  A very instructive toy model,  using the
monopole analogy, can be found in \textcite{Denef:2007pq}.

In the closed string spectrum of type-II strings, $p$-form fields originate
from the left-right combination of space-time spinors, which in their turn originate from world-sheet
fermions with periodic boundary conditions along the closed string, called Ramond fermions. For this reason the part of the
spectrum containing these fermions is referred to as the ``RR-sector".  In type-IIA string theories, the RR tensor fields have
odd rank $p$, and they are sources of $D_{p-1}$ branes, starting with the  $D_{0}$ branes that correspond to
particles. In type-IIB strings the p-form tensor fields have even rank, and the branes odd rank.

In string theory one always has 2-forms $B_{\mu\nu}$ which are sourced by 1-dimensional objects, the strings themselves. 
In ten dimensions, these are dual to five-branes. In type-II strings this gives rise to ``NS5-branes", called this way because the 
$B_{\mu\nu}$ field originates from the combination of left- and right moving Neveu-Schwarz fermions with anti-periodic boundary conditions
along the closed string. In heterotic strings they are called
heterotic five-branes.

\subsection{Dualities, M-theory and F-theory}

The discovery of branes led
to a plethora of  proven and conjectured relations between {\it a priori} different string constructions. 
The ten-dimensional
$E_8\times E_8$ and $SO(32)$ heterotic strings can be related to each other after compactifying each of them on a circle,
inverting its radius  ($R \rightarrow \alpha'/R$; this is called target space duality or T-duality), and giving vevs to suitable background fields \cite{Ginsparg:1986bx}.
The same is true for type-IIA and type-IIB strings \cite{Dine:1989vu,Dai:1989ua}. The $SO(32)$ heterotic string was shown to be related to the type-I $SO(32)$ string
under inversion of the string coupling constant,  $g \rightarrow 1/g$ (strong coupling duality or S-duality;  \textcite{Polchinski:1995df}).

S-duality, foreseen several years earlier by  \textcite{Font:1990gx},
produces a remarkable result for the remaining ten-dimensional theories. 
Type-IIA is mapped to an 11-dimensional
theory compactified on a circle \cite{Witten:1995ex,Townsend:1995kk}.
The radius of the circle is proportional to the string coupling constant and is inverted as in T-duality. 
For infinitely large radius 
one obtains an uncompactified 11-dimensional theory; in the limit of small radius this compactification describes the weakly coupled
type-IIA theory.
The 11-dimensional theory is not a string theory. It is called ``M-theory". Its field theory limit turned out to be the crown jewel of supergravity:  $D=11$ supergravity,
which until then had escaped the new developments in string theory. Because of the existence of a three-form field in its spectrum
it is believed that it is described by interacting
two-dimensional and/or five dimensional membranes.

A similar relation holds for the $E_8\times E_8$ heterotic string. Its strong coupling limit can be formulated in terms of 11-dimensional
M-theory  compactified on a line-segment \cite{Horava:1995qa},
the circle with two halfs identified. This is sometimes called ``heterotic M-theory".

Strong coupling duality maps type-IIB strings to themselves \cite{Hull:1994ys}. Furthermore the self-duality can be extended from an action just
on the string coupling, and hence the dilaton, to an action on the entire dilaton-axion multiplet. This action is mathematically
identical to the action of modular transformations on the two moduli of the torus, Eq. (\ref{ModTrans}), and corresponds to the
group $SL(2,\mathbb{Z})$. This isomorphism suggests a geo\-metric understanding of the self-duality in terms of a compactification torus $T_2$, whose
degrees of freedom correspond to the dilaton and axion field. An obvious guess would be
that the type-IIB string may be viewed as 
a torus compactification of some twelve-dimensional theory \cite{Vafa:1996xn}. But there is no such theory. The first attempts to develop this
idea led instead to a new piece of the landscape called ``F-theory", consisting only of compactifications and related to $E_8\times E_8$ heterotic strings
and M-theory by chains of dualities.

\subsection{The Bousso-Polchinski Mechanism}\label{BoussoPolchinski}

\ifExtendedVersion {\color{darkgreen}
It was realized decades ago \cite{Linde:1984ir} that
anti-symmetric tensor fields might play an important r\^ole in solving the cosmological constant problem.
Such tensors are natural generalizations of vector fields $A_{\mu}$ to tensors with an arbitrary number of anti-symmetrized
indices. The one of interest here is the field strength with the maximum number of indices
in four space-time dimensions, $F_{\mu\nu\rho\sigma}$. They appear as field strengths of three index anti-symmetric tensor fields $A_{\mu\nu\rho}$, which can exist in four space-time dimensions, but they have trivial dynamics, no propagating modes, and hence there are no ``photons" associated with them.  }\else {
It was realized decades ago \cite{Linde:1984ir} that rank-4 field strengths of rank-3 anti-symmetric tensors
might play an important r\^ole in solving the cosmological constant problem.
}\fi
Such four-index field strengths can get constant values without breaking Lorentz invariance, namely
$F_{\mu\nu\rho\sigma} = c \epsilon_{\mu\nu\rho\sigma}$, where 
$\epsilon_{\mu\nu\rho\sigma}$ is the Lorentz-invariant completely anti-symmetric four-index tensor\ifExtendedVersion {\color{darkred}
;  it is unique up to normalization, which is fixed in the standard way as $\epsilon_{\mu\nu\rho\sigma} \epsilon^{\mu\nu\rho\sigma} = -24$}\fi.
The presence of such a classical field strength in our universe is unobservable unless
we couple the theory to gravity. If we do, it gives a contribution similar to the
cosmological constant $\Lambda$, in such a way  that the latter is replaced by 
\begin{equation}
\Lambda_{\rm phys} = \Lambda 
 - \frac{1 }{48} F_{\mu\nu\rho\sigma} F^{\mu\nu\rho\sigma} = \Lambda + \frac12 c^2.
 \end{equation}

In string theory $c$ is not an arbitrary real number: it is quantized \cite{Bousso:2000xa}. This is due 
to a combination of the well-known Dirac quantization argument for electric charges in theories with magnetic
monopoles, and string theory dualities. The formula for 
the cosmological constant now looks something like this
 \beq
 \Lambda_{\rm phys} = \Lambda + \frac{1}{2} n^2 f^2 \ ,
 \eeq
where $f$ is some number derived from the string theory under consideration. If instead
of $F_{\mu\nu\rho\sigma}$ we were to consider an electromagnetic field, $f$ would be
something like  the strength of the electromagnetic coupling $e$: some number of order 1.
For generic negative values of $\Lambda$ we would
be able to tune $\Lambda_{\rm phys}$ to an extremely small value only if 
$f$ is ridiculously small.

However, it turns out that string theory typically contains hundreds of fields
 $F_{\mu\nu\rho\sigma}$.
Taking $N$ such fields into account, the result now
becomes 
 \beq
 \label{BPlambda}
 \Lambda_{\rm phys} = \Lambda + \frac{1}{2} \sum_{i=1}^N n_i^2 f_i^2 . \ 
 \eeq
 \ifExtendedVersion {\color{darkred}
One would expect the values for the real numbers $f_i$ to be different. 
Again an analogy with electromagnetic fields is useful. Although for the Standard Model we need
just one such vector field, string theory may contain more than one. Generically, these will
all have different fine-structure constants, or in other words different values for the
electromagnetic couplings $e_i$.
}\fi

If indeed the values of  $f_i$ are distinct and  incommensurate, then Eq.
(\ref{BPlambda}) defines a dense discrete set of values. Bousso and Polchinski called
it a ``discretuum". It is an easy
exercise to show that with $N$ equal to a few hundred, and values for $f_i$ of the order of
electromagnetic couplings and small integers $n_i$,  one can indeed obtain the required small value of $\Lambda_{\rm phys}$, given some negative $\Lambda$.

This realizes a dynamical neutralization  of $\Lambda$ 
first proposed by \textcite{Brown:1987dd,Brown:1988kg} (see  \textcite{Feng:2000if} for a related string realisation).
This makes any field strength $F_{\mu\nu\rho\sigma}$ (and hence $\Lambda$)
decay in discrete
steps by  bubble nucleation. This process stops as $\Lambda$ approaches zero. This is analogous to the decay of an electric field between
capacitor plates by pair creation of electron-positron pairs.  However, Brown and Teitelboim (as well as \textcite{Abbott:1984qf} in an
analogous model) already 
pointed out an important problem 
in the single field strength case they considered. First of all, as noted above, one has to assume  an absurdly small value for $f$. But even if one
does, the last transition from an expanding dS universe to ours would take so long to complete that 
all matter would have been diluted (the ``empty universe problem"). With multiple four-form field strengths, both problems are avoided; see \textcite{Bousso:2007gp}
for details.

All the ingredients used in the foregoing
discussion are already present in string theory; nothing was added by hand.
In particular large numbers of fields $F_{\mu\nu\rho\sigma}$ are
present, and the quantization of the field strengths follows using standard arguments. 

\subsection{Four-Dimensional Strings and Compactifications}\label{Compact}

There are essentially
two ways of building string theories in four dimensions.
One is to choose  another background space-time geometry,
and the other is  to change the world-sheet theory. The geometry can be chosen as a flat four-dimensional space combined with a compact six-dimensional
space. This is called ``compactification". This is not simply a matter of hand-picking a manifold: it must satisfy the equations of motion 
of string theory, and must be stable. Indeed, an obvious danger is that a given manifold simply ``decompactifies"  to six flat  dimensions.  The world-sheet
theory can be modified by choosing a different two-dimensional conformal field theory. In the action (\ref{FullAction}) and its supersymmetric analog only free bosons
$X$ or free fermions $\psi$ are used. One can choose another two-dimensional field theory that satisfies  the conditions of conformal invariance.  This is called a
conformal field theory (CFT). In particular one may use interacting two-dimensional theories. Only $X^{\mu}$ and $\psi^{\mu}$, 
$\mu=0,\ldots 3$, must remain free fields. 

\ifExtendedVersion {\color{darkred}
The simplest compactification manifold is a six-dimensional torus. This can be described both in terms of space-time
geometry, or by means of a modified world-sheet CFT (bosons with periodic boundaries). However, the resulting theories 
only have non-chiral fermions in their spectrum.

The next hurdle was taken very rapidly thereafter:  to construct {\it chiral} string theories in four dimensions. The first examples were obtained by assuming that
the extra six dimensions were not flat space, but a compact manifold. The equations of motion of string theory, plus some simplifying assumptions that
were relaxed in subsequent papers, required this space to be a Calabi-Yau manifold, a Ricci-flat manifold with three complex dimensions. 
Soon many other methods were found, and the bewildering choice of possibilities
led to
a massive effort that is still continuing today.}
\fi

As in ten dimensions, all four-dimensional string theories are related to others by strong-weak dualities, target space dualities and combinations thereof. This suggests
a connected ``landscape" of four-dimensional strings.   

We will present here
just a brief sketch of the string compactification landscape. For further details we recommend the very complete book by
\textcite{Ibanez:2012zz} and references therein.

\subsubsection{Landscape Studies versus Model Building}

The amount of work on string compactifications or four-dimensional string constructions is too vast to review here.
Most of this work is focused on finding examples that match the Standard Model as closely as possible. This is important, at
the very least as an existence proof, but it is not what we will focus on in this review. Our main interest is not  in finding a ``model" where
property X is realized, but the question if  we can understand {\it why} we observe property X in our universe, given anthropic and landscape
constraints. The relative importance of these two points of view depends on how optimistic one is about the chances of finding the exact Standard Model
as a point in the landscape.

\subsubsection{General Features}

\ifExtendedVersion {\color{darkred}
Even if the massless spectrum matches that of the Standard Model, 
such a theory contains infinitely many additional particles: massive string excitations, Kaluza-Klein modes as in field theory
compactifications, 
and winding modes due to strings wrapping the compact spaces.

Their masses are respectively proportional to the string scale, the inverse of the compactification
radius or the compactification radius itself. In world-sheet constructions the different kinds of modes are on equal footing,
and have Planckian masses. In geometric constructions one can consider large volume limits, where other
mass distributions are possible. But in any case, of all the modes of the string only the massless ones are relevant for providing
the Standard Model particles, which will acquire their masses from the Higgs mechanism and QCD, as usual.

All Standard Model fermions (by definition this does not include right-handed neutrinos) are chiral. Their left and right handed components couple in a different way to the weak interactions,
and this implies that they can only acquire a mass after weak symmetry breaking. This implies that their mass is proportional to the Higgs vev. 
Therefore one can say that the weak interactions protect them from being very massive. It is very well possible that for this reason all we have seen so far at
low energy is chiral matter. In attempts at getting the Standard Model from string theory, it is therefore reasonable to require that the chiral spectra match. If one does that,
one finds that in general large quantities of additional vector-like matter, whose mass is not protected by the weak interactions. Typically, if one requires three chiral families,
one gets $N+3$ families and $N$ mirror families. If the $N$ families ``pair off" with the $N$ mirror families to acquire a sufficiently large mass, the low energy spectrum agrees with the data.
}\fi


For phenomenological, but more importantly practical reasons most efforts have not focused on getting  the SM, but the
MSSM, the Minimal Supersymmetric Standard Model. 
 But it turns out that ``minimal" is not exactly what one typically finds. Usually there are many additional fields that have not (yet) been 
 observed. In addition to the superpartners of all the Standard Model particles and the additional Higgs field of the MSSM, they include
 moduli, axions, additional vector bosons, additional ``vector-like" matter and additional exotic matter.

Moduli are massless scalar singlets whose presence can be understood in terms of continuous deformations
of the compactification manifold or other features of the classical background fields.
The vacuum expectation values of these fields generate the deformations. Typically, there are tens or hundreds of them.
 In the more general setting of M-theory, the dilaton is part of this set as well.

Axions may be thought of as the imaginary part of the moduli, which are complex scalars in supersymmetric theories. It is useful to make
the distinction, because mechanisms that give masses to moduli, as is required for phenomenological reasons, sometimes leave
the imaginary part untouched. Axions may provide essential clues about the landscape, see section \ref{Axions}.
\ifExtendedVersion {\color{darkred}
These remain then as massless Goldstone bosons of a shift symmetry. On general grounds one expects that
theories of gravity have no exact global symmetries, so that the axions ultimately acquire a mass from non-perturbative effects. This mass can
be exponentially suppressed with respect to the string scale. One of the axions is the  four-dimensional $B_{\mu\nu}$ field.
}\fi

\ifExtendedVersion {\color{darkred}
Most string spectra have considerably  more vector bosons than the twelve we have seen so far in nature. Even if
the presence of $SU(3)$, $SU(2)$ and $U(1)$ as factors in the gauge group is imposed as a condition, one rarely finds just the Standard Model gauge group.
In heterotic strings one is usually left with one of the $E_8$ factors. Furthermore in nearly all string constructions additional $U(1)$ factors are found.  A very
common one is a gauged $B-L$ symmetry. 
}\fi

\ifExtendedVersion {\color{darkred}
}
\else 
{
Essentially all ``raw" string spectra contain, in addition to the chiral Standard Model particles, large numbers of scalars and vector-like ({\it i.e.} non-chiral) fermions.
Unlike chiral fermions, they can acquire a mass if the string spectrum is perturbed, for example by giving vevs to moduli. If this is not generically
what happens, string theory makes an incorrect prediction.}\fi
 
Furthermore one often finds particles that do not match any of the observed matter representations\ifExtendedVersion {\color{darkred}, nor their mirrors}\fi.
Notorious examples are particles with fractional electric charge or higher rank tensors.  \ifExtendedVersion {\color{darkred}These are generically called ``exotics". If there are exotics that are chiral with respect to \SM, these spectra should be rejected, because any
attempt to make sense of such theories is too far-fetched to be credible. }\fi These particles may be acceptable if they are vector-like, 
because one may hope that they become massive under generic perturbations. 
 \ifExtendedVersion {\color{darkred}
All particles that have not been seen in experiments must somehow
acquire a large enough mass. This is one of the main challenges of string phenomenology, the effort to connect string theory
with particle physics phenomenology. }\fi

Although superfluous particles may appear to be a curse, some of them may turn out to be a blessing. All quantum field theory parameters depend
on the moduli, and hence the existence of moduli is a first step towards a landscape of possibilities. 
\ifExtendedVersion {\color{darkred}
This should be viewed as a {\it good} feature of string theory. 
}\fi
\ifExtendedVersion {\color{darkred} Furthermore
the large number of moduli opens the way  to a solution of the cosmological constant problem, by allowing a large number of vacua densely covering the range of possibilities. A popular
estimate for the number of vacua is $10^{500}$, and the exponent is determined by the number of moduli. }\fi

Axions can play a r\^ole in  solving the strong CP problem, and may also provide a 
significant part of dark matter. Additional gauge groups are often needed as ``hidden sectors" in model building, especially for supersymmetry breaking. Extra $U(1)$'s may be observable
trough kinetic mixing \cite{Goodsell:2010ie} \ifExtendedVersion {\color{darkred}
with the Standard Model $U(1)$, via contributions to the action proportional to $F_{\mu\nu}V^{\mu\nu}$, where $F$ is the $Y$ field strength, and $V$ the one of the extra $U(1)$'s}\fi. 
Vector-like particles and exotics might  be observed and provide evidence for string theory, though this is wishful thinking.

\ifExtendedVersion {\color{darkred}
Some of the undesirable features may not be as generic as they seem. They may just be an artefact of the necessarily primitive methods at our disposal. Our intuition from many
years of four-dimensional string model building may well be heavily distorted by being too close to the supersymmetric limit, and by algebraically simple constructions. 
Perhaps a generic, non-supersymmetric, moduli-stabilized string ground state has no gauge group at all, so that the gauge group we observe is a rare, but anthropically
required deviation from normality. It may well be that a in a generic string ground state only chiral matter is light, as expected on the basis of standard QFT lore (any mass term
that is allowed is indeed present). If that turns out {\it not} to be the case, these features must be viewed as evidence against string theory.
}\fi

\subsubsection{Calabi-Yau Compactifications}

\ifExtendedVersion {\color{darkgreen}
A torus compactification preserves all space super symmetries, and hence on ends up with $N=4$ supersymmetry in
four dimension. The maximal number of super symmetries that allows chiral fermions in four dimensions is $N=1$
This problem can be overcome by choosing a different background geometry. In general, this means that one chooses a six-dimensional
compact manifold, with classical field configurations for all the massless fields: gravity, the dilaton, the $B_{\mu\nu}$ field and the gauge fields.
This was
first applied to the $E_8\times E_8$ heterotic string \cite{Candelas:1985en}.
These authors found consistent compactifications by using six-dimensional, Ricci-flat, K\"ahler manifolds with $SU(3)$ holonomy, called Calabi-Yau manifolds.
}
\else 
The first examples of compactifications with chiral spectra and $N\!\!=\!\!1$ supersymmetry were found for the $E_8\times E_8$ heterotic string by \textcite{Candelas:1985en}.
These authors used six-dimensional, Ricci-flat, K\"ahler manifolds with $SU(3)$ holonomy, called Calabi-Yau manifolds.\fi
They assumed that the $B_{\mu\nu}$ field strength $H_{\mu\nu\rho}$ vanishes, which leads to the consistency condition
\begin{equation}\label{Bianchi}
dH  = {\rm Tr}\  R \wedge R - \frac{1}{30}  {\rm Tr}\  F \wedge F  = 0.
\end{equation}
This implies in particular a relation between the gravitational and gauge field backgrounds. This condition can 
be solved by using a background gauge field that is equal to the spin connection of the manifold, embedded in an $SU(3)$ subgroup of one of the $E_8$ factors.
In compactifications
of this kind one obtains a spectrum with  a gauge group $E_6 \times E_8$. The group $E_6$ contains the Standard Model gauge group $SU(3)\times SU(2)\times U(1)$ plus two additional
$U(1)$'s. The group $E_8$ is superfluous but hidden  (Standard Model particles do not couple to it), and may play a r\^ole in  supersymmetry breaking. 
\ifExtendedVersion {\color{darkgreen}

If some dimensions of space are compactified, ten-dimensional fermion fields are split as
\begin{equation}
\Psi_{+}(x,y)=\Psi_L(x)\Psi_{+}(y)+\Psi_R(x)\Psi_{-}(y)
\end{equation} 
where $x$ denote four-dimensional and $y$ six-dimensional coordinates, $+$ denotes one chirality in ten six dimensions, and $L,R$ denote chirality  in four dimensions. The number 
of massless fermions of each chirality observed in four dimensions is determined by the number of zero-mode solutions of the six-dimensional Dirac equation in the background
of interest. These numbers are equal to two topological invariants of the Calabi-Yau manifold, the Hodge numbers, $h_{11}$ and $h_{12}$. As a result one obtains
$h_{11}$ chiral fermions  in the
representation $(27)$ and $h_{12}$ in the  $(\overline 27)$ of $E_6$. The group $E_6$ is a known extension of the Standard Model, an example of a Grand Unified Theory, in which
all three factors of the Standard Model are embedded in one simple Lie algebra. It is not the most preferred extension; a Standard Model family contains 15 or 16 (if we assume the existence
of a right-handed neutrino) chiral fermions, not 27. However, since the 11 superfluous fermions are not chiral with respect to \SM, they can acquire a mass without  the help of the Higgs mechanism,
in the unbroken Standard Model. Therefore these masses may be well above current experimental limits. }
\else 
{
In these compactifications one obtains
$h_{11}$ chiral fermions  in the
representation $(27)$ and $h_{12}$ in the  $(\overline{27})$ of $E_6$, where $h_{11}$ and $h_{12}$ are the topological Hodge numbers of the
Calabi-Yau manifold. }\fi

The number of Calabi-Yau manifolds is huge. \textcite{Kreuzer:2000xy}  enumerated a subset associated with four-dimensional reflexive polyhedra.
This list contains more than 470 million topological classes with 31,108 distinct Hodge number pairs.  The total number of topological classes of Calabi-Yau manifolds has been conjectured 
to be finite. 

\textcite{Strominger:1986uh} considered more  general geometric background geometries with torsion, leading to so many possibilities that the author concluded
``{\it all predictive power seems to have been lost}".

\subsubsection{Orbifold Compactifications}

One can also compactify on a six-dimensional torus, but this does not yield chiral fermions; the same is true for the
more general asymmetric torus compactifications \ifExtendedVersion {\color{darkred} with 6 left-moving and 22 right-moving ``chiral" bosons}\fi found by \textcite{Narain:1985jj}.
But string theory can also be compactified on tori with discrete identifications.  The simplest example is the
circle with the upper half identified with  the lower half, resulting in a line segment. These are called orbifold compactifications \cite{Dixon:1985jw}, and do
yield chiral fermions. These methods opened many new directions, such as orbifolds with gauge background fields (``Wilson lines")
\cite{Ibanez:1986tp}, and were soon generalized to {\it asymmetric orbifolds} \cite{Narain:1986qm},
where ``asymmetric" refers to the way left- and right-moving modes were treated. 

\ifExtendedVersion {\color{darkred}
\subsubsection{The Beginning of the End of Uniqueness}

Several other methods were developed around the same time. Narain's  generalized torus compactifications
lead to a continuous infinity of possibilities, but all without chiral fermions. Although this infinity of possibilities is not
really a surprising feature for a torus compactification, Narain's paper was an eye-opener because, unlike standard six-dimensional
torus compactifications, this approach allowed a complete modification of the gauge group. 

{\color{darkblue} Free field theory constructions, discussed in more detail below, allowed a more systematic exploration of certain classes of string theories.
It became clear very quickly that there was a plethora of possibilities. Unlike Narain's constructions, these theories can have chiral fermions, and
furthermore they did not provide a continuum of options, but only discrete choices.  With the benefit of hindsight, one can now say that all these 
theories {\it do} have continuous deformations, which can be realized by giving vacuum expectation values to certain massless scalars in the spectrum.
Since these deformed theories do not have a free field theory descriptions, these deformations are not manifest in the construction. They are the world sheet
construction counterparts of the geometric moduli. This does however not imply that the plethora of solutions can simply be viewed as different points in one continuous
moduli space. Since many spectra are chirally distinct, it is more appropriate to view this as the discovery of a huge number of distinct moduli spaces, all leading
to different physics. Fifteen years later, work on moduli stabilisation provided hints at the existence of non-trivial potentials on these moduli spaces, with a huge
number of metastable local minima. This explosive growth of possibilities comes on top of the one discovered in 1986.}


Already as early as 1986 it became customary to think of the different four-dimensional string theories  
or compactifications as ``vacua" or ``ground states" of a fundamental theory (see for example the discussion at the end of
\textcite{Kawai:1986ah}). Here one also finds the remark that perhaps our universe is merely a sufficiently long-lived metastable
state. All this work from 1986 gave the first glimpse of what much later became known as the ``string theory landscape". 
}\fi

\subsubsection{Free Field Theory Constructions}

World-sheet methods started being explored in 1986. The first idea was to exploit boson-fermion equivalence in two dimensions. In this way the artificial
distinction between the two can be removed, and one can describe  the  heterotic string entirely in terms of free fermions (\textcite{Kawai:1986va} and
 \textcite{Antoniadis:1986rn})
or free bosons \cite{Lerche:1986cx}. 
\ifExtendedVersion {\color{red}}\else {
These constructions are closely related. The free boson constructions have an elegant description
in terms of even self-dual lattices, for which remarkable counting formulas exist. Using such formulas and assuming a definite structure for the
(bosonized) fermionic string sector, the latter authors arrived at a rigorous (but far from saturated) upper limit of the total number of string theories
in this class: $10^{1500}$.
}\fi
\ifExtendedVersion {\color{darkred}
These constructions are closely related, and there is a huge area of overlap: 
constructions based on complex free fermions pairs
can be written in terms of free bosons.  However,  one may also consider  real fermions or free bosons on lattices that do not allow a
straightforward realization in terms of free fermions.

\paragraph{Free fermions}

Both methods have to face the problem of finding solutions to the conditions of modular invariance. In the fermionic constructions this is done
by allowing  periodic or anti-periodic boundaries  on closed cycles on the manifold for all fermions independently. Modular {\it transformations} change those boundary conditions, and hence 
they are constrained by the requirements of modular {\it invariance}. These constraints can be solved systematically (although in practice usually not exhaustively).  Very roughly (ignoring some of the constraints), the number of modular invariant combinations
is of order $2^{\frac12 n (n-1)}$ for $n$ fermions. There are 44 right-moving and 18 left-moving fermions, so that there are potentially
huge numbers of string theories. In reality there are however many degeneracies.

\paragraph{Free Bosons: Covariant Lattices}\label{FreeBosons}

In bosonic
constructions the modular invariance constraints are solved by requiring that the momenta of the bosons lie on a Lorentzian 
even self-dual lattice. This means that the lattice of quantized momenta is identical to the lattice defining the compactified space,
and that all vectors have even norm. Both conditions are defined in terms of a metric, which is $+1$ for left-moving bosons and
$-1$ for right-moving ones. These bosons include the ones of Narain's torus, plus eight right-moving ones representing the
fermionic degrees of freedom, $\psi^{\mu}$ and the ghosts of superconformal invariance. These eight bosons 
originate from the {\em bosonic string map} (originally developed 
for ten-dimensional strings by \textcite{Englert:1986na}) used by 
\textcite{Lerche:1986cx} to map the entire 
fermionic sector of the heterotic string was to a bosonic string sector. 
Then the Lorentzian
metric has signature $((+)^{22},(-)^{14}$)), and the even self-dual lattice is denoted $\Gamma_{22,14}$ called a {\it covariant lattice}
because it incorporates space-time Lorentz invariance for the fermionic string (starting from free fermions,
\textcite{Kawai:1986ah} found a related construction in terms of odd self-dual lattices).
Since the conditions for modular invariance are invariant under $SO(22,14)$ Lorentz transformations, and since the spectrum 
of $L_0$ and $\bar L_0$ is changed under such transformations, their would appear to be a continuous infinity
of solutions. But the right-moving modes of the lattice are strongly constrained by the requirement of two-dimensional 
supersymmetry, which is imposed using a non-linear realization discovered by \textcite{Antoniadis:1985az} (other realizations
exist, see for example \textcite{Waterson:1986ru,Schellekens:1987ij}). This leads to the ``triplet constraint" first formulated in 
\textcite{Kawai:1986va}.
This makes the right-moving part of the lattice rigid. The canonical
linear realization of supersymmetry, relating $X^{\mu}$ to $\psi^{\mu}$, on the other hand leads to lattices 
$\Gamma_{22,6} \times E_8$ with complete Lorentz rotation freedom in the first factor, which is just a Narain lattice.


\subsubsection{Early attempts at vacuum counting.}

The rigidity of the right-moving part of the lattice discretizes the number of solutions, which is in fact finite for a given
world-sheet supersymmetry realization. A very crude attempt to estimate the number of solutions was
made by \textcite{Lerche:1986ae}, and works as follows. One can map the right-moving bosons to a definite set of 66 left-moving
bosons, while preserving modular invariance. This brings us into the realm of even self-dual {\it Euclidean} lattices, 
for which powerful classification theorems exist.  

Such lattices exist only
in dimensions that are a multiple of eight, and have  been enumerated for dimensions $8,16$ and $24$, 
with respectively 1,2 and 24 solutions (in 8 dimensions the solution is the root lattice of $E_8$, in 16 dimensions they are
$E_8 \oplus E_8$ and the root lattice of $D_{16}$ plus a spinor weight lattice, and in 24 dimensions  the solutions were
enumerated by
  \textcite{Niemeier}).
There exists
a remarkable formula (the ``Siegel mass formula") which gives information about the total number
of distinct lattices $\Lambda$ of dimension $8k$ in terms of :
\begin{equation} \label{Siegel}
\sum_{\Lambda} g(\Lambda)^{-1}= \frac{1}{8k} B_{4k} \prod_{j=1}^{4k-1} \frac{B_{2j}}{4j}
\end{equation} 
Here $g(\Lambda)$ is the order of the automorphism group of the lattice $\Lambda$ 
and $B_{2j}$ are the Bernouilli numbers. Since the automorphisms include the reflection symmetry, $g(\Lambda) \geq 2$.  
If we assume that the lattice of maximal symmetry is $D_{8k}$ (the root lattice plus a spinor, which is a 
canonical way to get an even self-dual lattice)) we have a plausible guess
for the upper limit of $g(\Lambda)$ as well, namely the size of the Weyl group of $D_{8k}$, $2^{8k-1}(8k)!$.
 This assumption
is incorrect for $k=1$, where the only lattice is $E_8$, and $k=2$, where the lattice $E_8 \times E_8$ wins against $D_{16}$, but
for $k=3$ and larger the Weyl group of $D_{8k}$ is larger than the automorphism group of the lattice $(E_8)^k$. For $k=3$ the
assumption has been checked in \textcite{ConwaySloane} for all 24 Niemeier lattices. Making this assumption we get
\begin{equation}
  \frac{1}{4k} B_{4k} \prod_{j=1}^{4k-1} \frac{B_{2j}}{4j}   < N_{8k}  < 2^{8k-1}(8k-1)!\  B_{4k} \prod_{j=1}^{4k-1} \frac{B_{2j}}{4j}
\end{equation}
which for $k=11$ gives $10^{930} < N_{88} < 10^{1090}$ (in \textcite{Lerche:1986ae} this number was estimated rather inaccurately as
$10^{1500}$; all numbers quoted here are based on an exact computation). 

From a list of all $N_{88}$ lattices one could read off all the free bosonic CFTs with the world-sheet supersymmetry realization
discussed above.  In particular, this shows that the total number is finite. However, there is a very restrictive subsidiary constraint
due to the fact that 66 of the 88 bosons were obtained from the right moving sector. Those bosons must have their momenta
on a $D_3 \times (D_7)^9$ lattice and satisfy an additional constraint inherited from world sheet supersymmetry, 
the triplet constraint. Perhaps a more
reasonable estimate is to view this as a lattice with 32 orthogonal building blocks, $D_3 \times (D_7)^9 \times (D_1)^{22}$, which
should be combinatorially similar to $(D_1)^{32}$ then the relevant number would be  
$N_{32}$, which lies between $8 \times 10^7$ and $ 2.4 \times 10^{51}$. But unlike $N_{88}$, $N_{32}$ is not a strict limit, 
and furthermore is still subject to the triplet constraint.

All of this can be done explicitly for 10 dimensional strings. 
Then one needs the lattices of dimension 24, and eight of the 24 lattices satisfy the subsidiary constraints for ten-dimensional strings \cite{Lerche:1986ae}, namely the presence of a $D_8$ factor.

\subsubsection{Meromorphic CFTs.}

The concept of chiral conformal field theories and even self-dual lattices can be generalized to interacting theories, 
the so-called {\it meromorphic} conformal field theories \cite{Goddard:1989dp}. These can only exist if the central charge $c$ (the generalization
of the lattice dimension to CFT)  is a multiple of 8. For $c=8$ and $c=16$ these meromorphic CFTs are just chiral bosons
on even self-dual lattices, but for $c=24$  there  $71$ CFT's are conjectured \cite{Schellekens:1992db} to exist including the 24 Niemeier lattices (most of them have indeed been constructed). Gauge symmetries in the vast majority of the heterotic strings
in the literature
(for exceptions see for example \textcite{Candelas:1997eh})
are mathematically described in terms of affine Lie algebras, a kind of string generalization of simple Lie-algebras, 
whose representations are characterized by a Lie-algebra highest weight
and an additional integer parameter $k$ called the {\it level}. In the free boson theories the only representations one
encounters have $k=1$, and the total rank equals the number of compactified bosons in the left-moving sector, 22 for
four-dimensional strings, and 24 for Niemeier lattices.  All even self-dual lattices are direct sums of level 1 affine algebras plus a number of abelian 
factors (U(1)'s), which we will call the gauge group of the theory. 
In meromorphic CFT's the restriction to level one is removed. The list of $71$ meromorphic CFTs contains 70 cases
with a gauge group whose total central charge is 24, plus one that has no gauge group at all, the ``monster module". 
Just one
of these yields an additional ten-dimensional string theory with tachyons and an $E_8$ realized as an affine
Lie algebra at level 2. This solution was already known \cite{Kawai:1986vd}, and was obtained using free fermions. 

The importance of the meromorphic CFT approach is
that it gives a complete classification of all solutions without assuming a particular construction method. 
In four dimensions the same method can be used. For example,
from a list of meromorphic CFTs  with $c=88$ all four-dimensional string theories with a given realization of world-sheet supersymmetry (namely the same one used above) can be obtained, independent of the construction method. 
Unfortunately next to nothing is known about meromorphic CFTs for $c\geq 32$. It is not known if, like lattices, they are
finite in number. Their gauge groups can have central charges that are not necessarily 0 or the total central charge of the
meromorphic CFT. It is not known if the gauge groups are typically large or small. There is an entire landscape
here that is totally unexplored, but hard to access. 

So far this method of mapping a heterotic theory to a meromorphic CFT has only been applied to 
a world-sheet supersymmetry realization using the triplet constraint. But this can be generalized to other realizations
of world-sheet supersymmetry, including perhaps the ones discussed in the next section. 

The point we are trying to make here is that despite many decades of work, we  are probably still only able
to see the tip of a huge iceberg. 

}\fi

\subsubsection{Gepner Models.}

In 1987 world-sheet constructions were extended further by the use of interacting 
rather than free two-dimensional conformal field theories \cite{Gepner:1987qi}. The  ``building
blocks" of this construction are two-dimensional conformal field theories with  $N=2$ world-sheet supersymmetry. 
These building blocks are combined (``tensored") in such a way that they contribute in the same way to the energy
momentum tensor as six free bosons and fermions. This is measured in terms of the central charge of the Virasoro
algebra, which must have a value $c=9$. In principle the number of such building blocks is huge, but in practice only a very limited
set is available, namely an infinite series of ``minimal models" with central charge $c=3k/(k+2)$, for $k=1\ldots\infty$. There are
168 distinct ways of adding these numbers to 9. 
\ifExtendedVersion {\color{darkgreen}}\else {
For each of the 168 tensor combinations a  number of distinct modular invariant partition functions can be
constructed, for a grand total of about five thousand \cite{Fuchs:1989yv,Schellekens:1989wx}

There is a close relationship between these ``Gepner models" and
geometric compactifications on Calabi-Yau manifolds. Exact correspondences between their spectra were found, including
the number of singlets. This led to the conjecture that Gepner Models
are Calabi-Yau compactifications in  special points of moduli space. Evidence was provided by a 
conjectured relation between
$N\!\!=\!\!2$ minimal models and critical points of Landau-Ginzburg models \cite{Vafa:1988uu,
Lerche:1989uy}. 

Modular invariance requires the left- and right-moving sectors of Gepner
algebras to be the same.  There is no such limitation in free CFT constructions, but these are limited by being
non-interacting in two dimensions. But asymmetric {\it and} interacting CFT constructions also exist.
Examples in this class were obtained using a method called ``heterotic weight lifting" \cite{GatoRivera:2010xn}. 
In the left-moving
sector one of the  superconformal building blocks (combined with one of the $E_8$ factors) is replaced by 
another CFT that has no superconformal symmetry, but is isomorphic to the original building block as a modular
group representation. But this is just a small step into a part of the landscape that is hard to access. }\fi

\ifExtendedVersion {\color{darkgreen}
With the constraints of superconformal invariance solved, one now has to deal with modular invariance. In exact CFT
constructions the partition function takes the form
\begin{equation}
\label{PartFunc}
P(\tau,\bar\tau)=\sum_{ij} \chi_i(\tau) M_{ij} \bar\chi_j(\bar\tau) 
\end{equation}
where $\chi_i$ are characters of the Virasoro algebra, traces over the entire Hilbert space built on the ground state
labeled $i$ by the action of the Virasoro generators $L_n$:
\begin{equation}
\chi_i(\tau)= {\rm Tr} e^{2\pi i \tau (L_0-c/24})
\end{equation}
The multiplicity matrix $M$ indicates how often the ground states $| i\rangle | j \rangle$   occurs in the spectrum. Its entries
are non-negative integers, and it is severely constrained by modular invariance.
Note that in (\ref{PartFunc}) we allowed for the possibility that the left- and right-moving modes have a different symmetry
(a different extension of superconformal symmetry) with different sets of characters $\chi$ and $\bar\chi$.  But then 
the conditions for modular invariance are very hard to solved. They can be trivially solved if
the left and right algebras are the same. Then
 modular invariance demands that $M$ must commute with the matrices $S$ and $T$ that 
represent the action of the modular transformations $\tau \rightarrow -1/\tau$ and $\tau\rightarrow \tau+1$ on the characters. 
This has always at least one solution, $M_{ij}=\delta_{ij}$. 

However, assuming identical left and right algebras is contrary to the basic idea of the heterotic string. Instead 
Gepner model building focuses on a subset, namely those spectra that can be obtained from a symmetric type-II spectrum 
by mapping one of the fermionic sectors to a bosonic sector. For this purpose we can use the same bosonic string map
discussed above. This results in a very special and very limited subset of the possible bosonic sectors.

Using the discrete symmetries of the building blocks,
for each of the 168 tensor combinations a  number of distinct modular invariant partition functions can be
constructed, for a grand total of about five thousand \cite{Schellekens:1989wx}. 
Each of them gives a string spectrum with a gauge group 
$E_6\times E_8$ (or occasionally an extension of $E_6$ to $E_7$ or $E_8$)
with massless chiral matter in the representations $(27)$ and $(\overline{27})$ of $E_6$, exactly like the
Calabi-Yau compactifications discussed above.  

Indeed, it was understood not long thereafter that there is a close relationship between these ``Gepner models" and
geometric compactifications on Calabi-Yau manifolds. Exact correspondences between their spectra were found, including
the number of singlets. This led to the conjecture that Gepner Models
are Calabi-Yau compactifications in a special point of moduli space. Evidence was provided by a 
conjectured relation between
of $N=2$ minimal models to critical points of Landau-Ginzburg models \cite{Vafa:1988uu,
Lerche:1989uy}.  
}\fi

\ifExtendedVersion 
{\color{darkblue} Getting the right number of families in this class of models has been challenging, since this number 
turns out to be quantized in units of six or four in nearly all cases that were studied initially. The only exception is a class
studied by \textcite{Gepner:1987hi}. We will return to this problem later.}
\fi

\subsubsection{New Directions in Heterotic strings}

\ifExtendedVersion \paragraph{New embeddings.}\fi
   The discovery of heterotic M-theory opened many new directions. Instead of the canonical 
 embedding of the $SU(3)$ valued spin-connection of a Calabi-Yau manifold, 
some of these manifolds admit other bundles that can be embedded in the gauge group.
In general, condition (\ref{Bianchi}) is then not automatically satisfied, but in heterotic M-theory one may get extra contributions
from heterotic five branes \cite{Lukas:1998hk,Lalak:1998jg}.
 
In this way one can avoid getting the Standard Model via the complicated route of $E_6$ Grand Unification.
Some 
examples that have been studied are $SU(4)$ bundles  \cite{Braun:2005nv}, $U(1)^4$ bundles \cite{Anderson:2012yf} and
$SU(N)\times U(1)$ bundles \cite{Blumenhagen:2006ux}
which break $E_8$ to the more appealing $SO(10)$ GUTs, to
$SU(5)$ GUTs, or even directly to the Standard Model. Extensive and systematic searches are underway that have resulted in hundreds  of distinct examples \cite{Anderson:2011ns}
with the exact supersymmetric Standard Model spectrum, 
without even any vector-like matter (but with extra gauge groups and the usual large numbers of singlets). 
\ifExtendedVersion {\color{darkred}
However, the gauge group contains extra $U(1)$'s and an $E_8$ factor, and large numbers
of gauge singlets, including unstabilized moduli. There can be several Higgs multiplets. To break the GUT groups down to the
Standard Model background gauge fields on suitable Wilson lines are used. For this purpose one needs a manifold with
a freely acting ({\it i.e.} no point on the manifold are fixed by the action) discrete symmetry. One then identifies points on the manifold
related by this symmetry and adds a background gauge field on a closed cycle on the quotient manifold (a Wilson line). 
}\fi

\ifExtendedVersion {\color{darkred}
\paragraph{Free fermionic construction.} 
In-depth explorations \cite{Assel:2010wj} have been done of a subclass of fermionic
constructions using a special set of free fermion boundary conditions that allows spectra with three families
to come out. This work focuses on Pati-Salam model. Other work \cite{Renner:2011gs,Renner:2011yh} explores the variations of the
``NAHE" set of free fermion boundary conditions. This is a set of fermion boundary vectors proposed by 
\textcite{Antoniadis:1989zy} that are a useful starting point for finding
``realistic" spectra.
}\fi

\ifExtendedVersion {\color{darkred}
\paragraph{The Heterotic Mini-landscape.} 
This is a class of orbifold compactifications on a torus $T^6/{\mathbf{Z}_6}$
cleverly constructed so that the heterotic gauge group $E_8\times E_8$ is broken down to different subgroups
in different fixed points, such as $SO(10)$, $SU(4)^2$ and $SU(6)\times SU(2)$. This leads to the notion of 
{\em local unification} \cite{Forste:2004ie,Buchmuller:2004hv,Buchmuller:2005jr}. 
The Standard Model gauge group is the intersection of the various ``local" gauge realized at
the fixed points. Fields that are localized near the fixed points must respect its symmetry, and hence be in complete
multiplets of that group. Unlike field theory GUTs, these models
have no limit where $SO(10)$ is an exact global symmetry. 
In this way one can make sure that matter families are in complete spinor representations of $SO(10)$,
while Higgs bosons need not be in complete representations of $SO(10)$, avoiding the notorious 
doublet splitting problem of GUTs.  The number of 3-family models in this part of the landscape
is of  order a few hundred, and there is an extensive body of work on their phenomenological
successes and problems, see  for example \textcite{Lebedev:2006kn,Nilles:2008gq} and references therein.
}\else 
{
A more traditional orbifold approach is the ``heterotic mini-landscape". 
This is based on a class of orbifold compactifications on a torus $T^6/{\mathbf{Z}_6}$
cleverly constructed so that the heterotic gauge group $E_8\times E_8$ is broken down to different subgroups
at different fixed points, such as $SO(10)$, $SU(4)^2$ and $SU(6)\times SU(2)$. This leads to the notion of 
{\em local unification} \cite{Forste:2004ie,Buchmuller:2004hv,Buchmuller:2005jr}. The Standard Model gauge group is the intersection of the various ``local" gauge groups realized at
the fixed points. The number of 3-family models in this part of the landscape
is of  order a few hundred, and there is an extensive body of work on their phenomenological
successes and problems, see for example \textcite{Lebedev:2006kn,Nilles:2008gq} and references therein. But despite the name, work in this area is not
really aimed at landscape distributions, but at getting the Standard Model. 
}\fi

\ifExtendedVersion {\color{darkred}
\paragraph{Heterotic Gepner Models} 
As explained above, the original Gepner models are limited in scope by the requirement that the left and right
algebras should be the same.  There is no such limitation in free CFT constructions, but they are limited in being
non-interacting in two dimensions. What we would like to have is asymmetric, interacting CFT construction. 
Examples in this class have been obtained using a method called ``heterotic weight lifting" \cite{GatoRivera:2010xn}. 
In the left-moving
sector one of the  superconformal building blocks (combined with one of the $E_8$ factors) is replaced by 
another CFT that has no superconformal symmetry, but is isomorphic to the original building block as a modular
group representation. This opens up an entirely new area of the heterotic string landscape. It turns out that
the difficulty in getting three families now disappears.
}\fi

\subsubsection{Orientifolds and Intersecting Branes}\label{OrientifoldsAndIntersectingBranes}

\ifExtendedVersion {\color{darkred} The Standard Model comes out remarkably easily from the simplest heterotic strings.  But that is by no means
the only way. }\fi Another way to get gauge groups in string theory is from stacks of membranes. 
If open strings end on a D-brane that does not fill all of space-time, a distinction must be made between their fluctuations
away from the branes, and the fluctuations of their endpoints on the branes. The former are standard string vibrations leading
to gravity (as well as a dilaton, and other vibrational modes of closed strings), whereas fluctuations of the endpoints are
only observable on the brane, and give rise to fermions and gauge interactions. 

\ifExtendedVersion {\color{darkred}
\paragraph{Brane Worlds}

This then leads to the ``brane-world" scenario, where our universe is assumed to be a $D_{p-1}$ brane embedded in a higher
dimensional space-time. Then all observed matter and gauge interactions ({\it i.e.} the Standard Model)
would be localized on the brane, whereas
gravity propagates in the bulk. The additional dimensions must be assumed to be compact in order to make them unobservable,
but since they can only be observed using gravitation, the limits are completely unrelated to the distance scales probed at the LHC.
From tests of the $1/r^2$ dependence of Newton's law of gravity one get a limit of about $.1$ mm; anything smaller is currently
unobservable. The brane world idea was proposed as a possible solution to the gauge hierarchy problem 
\cite{ArkaniHamed:1998rs}. By allowing gravity to expand into extra dimensions below distances of $.1$ mm one can explain
the observed weakness of gravity in terms of a dilution of the field lines into extra dimensions. Our large observed Planck scale would
just be an illusion, related to the fundamental scale of gravity $M_{ \rm grav}$ as
\begin{equation}
M_{Planck}^2 \propto (M_{\rm grav})^{2+n} (R)^n
\end{equation} 
where $n$ is the number of extra dimensions and $R$ their compactification radius. For moderately large $R$ one can obtain a 
large 4-dimensional Planck scale from a moderately small (for example in the TeV range) fundamental scale of gravity. This inspired followed other constructions, where the extra dimensions were assumed to be not flat, but warped \cite{Randall:1999ee}. 

However, these solutions to the hierarchy problem are not our main interest here. Furthermore, they are put under severe stress by the recent
LHC results.} 
\fi

\paragraph{Chan-Paton groups.}
\ifExtendedVersion {\color{darkred} The possibility of getting gauge theories and matter from branes sparked another direction of research with the goal of
getting the Standard Model from open string theories. }\fi
To get towards the Standard Model, one starts with
type-II string theory, and compactifies six dimensions on a manifold\ifExtendedVersion {\color{darkred}. This compactified manifold 
may have a large radius, as in the brane world scenario, but this is optional}\fi. In these theories one 
finds suitable D-branes coinciding with four-dimensional Minkowski space, and intersecting each other in the
compactified directions.  These can be D5, D7 or D9 branes in type-IIB and  D6  branes in type-IIA (some
other  options can be considered, but require more discussion; see for example \textcite{Ibanez:2012zz}). Each such brane
can give rise to a gauge group, called a Chan-Paton gauge group, which can be $U(N)$, $Sp(N)$ or $O(N)$ \cite{Marcus:1986cm}. 
By having several different branes one can obtain a gauge group consisting of several factors, like the one of the
Standard Model. The brane intersections can give rise to massless string excitations of open strings with their ends on the two
intersecting branes. These excitations can be fermions, and they can be chiral. Each open string end endows the fermion with
a fundamental representation of one of the two Chan-Paton groups, so that the matter is in a bi-fundamental
representation of those gauge groups.

\begin{center}
\begin{figure}
\includegraphics[width=3.3in]{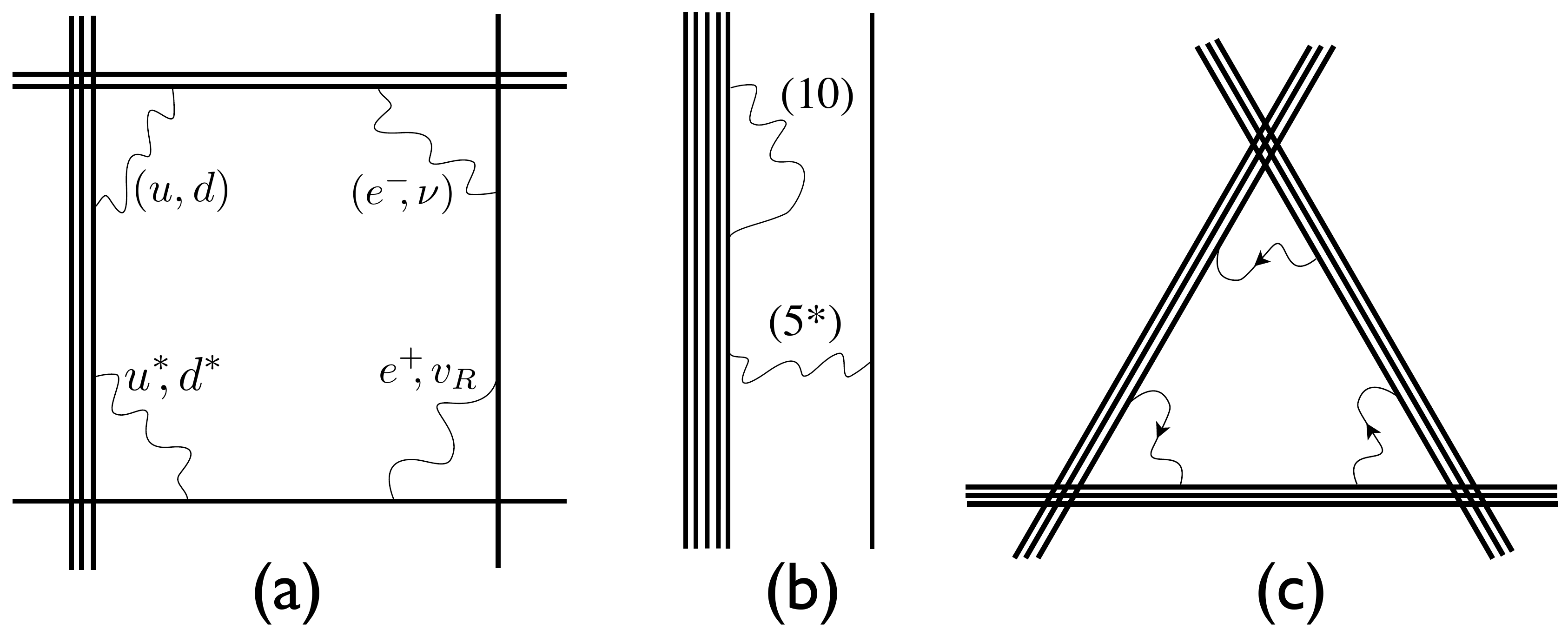}
\caption{Brane configurations: (a) the Madrid model, (b) SU(5) GUTs and (c) Trinification.\label{GUTTrinification}}
\end{figure}
\end{center}

Remarkably, a Standard Model family has precisely the right structure to be realized in this manner. The first example was 
constructed by  \textcite{Ibanez:2001nd} and is called the ``Madrid model". 
It consists of four stacks of branes, a $U(3)$ stack giving the strong interactions, a $U(2)$ or $Sp(2)$ stack for the weak
interactions, plus two  $U(1)$ stacks. The Standard Model $Y$ charge is a linear combination of the unitary phase
factors of the first, third and fourth stack (the stacks are labeled {\bf a} $\ldots$ {\bf d}) 
\begin{equation*}
Y=\frac16 Q_{\bf a} + \frac12 Q_{\bf c} - \frac12 Q_{\bf d}.
\end{equation*}
This configuration is depicted in Fig. \ref{GUTTrinification}(a).

 \ifExtendedVersion\else
 To build a  complete model requires another topological feature,  an orientifold plane, needed
 to cancel the tadpoles of the disk diagram. 
 This also cancels the leading contributions to chiral anomalies. 
 Anomalous
 $U(1)$ gauge bosons  acquire a mass by absorbing an axion field participating in a generalized Green-Schwarz mechanism.
 But this can also give a mass to anomaly-free $U(1)$ gauge bosons, and care must be taken that this does not happen to the
 Standard Model $U(1)$, $Y$. There are hundreds of papers where these conditions are solved, resulting in Standard Model spectra. 
 These are called orientifold models.
 An extensive review of the first five
 years of this subject can be found in 
 \textcite{Blumenhagen:2005mu}.
\fi

\paragraph{The three main classes.}
There are other ways of getting the Standard Model. If there are at most four brane stacks involved,
they fall into three broad classes, labeled by a real number $x$.
The Standard Model generator is in general some linear combination of all four brane charges (assuming stack {\bf b} is
$U(2)$ and not $Sp(2)$), and takes the form \cite{Anastasopoulos:2006da}
\begin{equation}\label{OrientifoldOptions}
Y=(x-\frac13) Q_{\bf a} + (x-\frac12) Q_{\bf b}  + x Q_{\bf c} + (x-1)  Q_{\bf d}.
\end{equation}
Two values of $x$ are special. The case $x=\frac12$ leads to a large class containing among others
the Madrid model, Pati-Salam models \cite{Pati:1974yy}  and flipped $SU(5)$ \cite{Barr:1981qv} models.
The value $x=0$ gives rise to classic $SU(5)$ GUTs \cite{Georgi:1974sy}. To get Standard Model families in this case one 
needs chiral anti-symmetric rank-2 tensors, which originate from
open strings with both their endpoints on the same brane. The simplest example is shown in  Fig. \ref{GUTTrinification}(b). It has one $U(5)$ stack giving rise to the GUT gauge group, but needs at least
one other brane in order to get matter in the $(5^*)$ representation of $SU(5)$.

Other values of $x$ can only occur for
oriented strings, which means that there is a definite orientation distinguishing one end of the string from the other end.
An interesting possibility in this class is the trinification model,  depicted in Fig. \ref{GUTTrinification}(c). 
\ifExtendedVersion {\color{darkred}
This configuration is naturally embedded in $E_6$ GUTs,
just as $SU(5)$ GUTs and the Pati-Salam model are naturally
embedded in $SO(10)$. However, these GUT groups cannot be obtained with these open string constructions. The reason is
 $SO(10)$ GUTs cannot be obtained although $SO(10)$ itself is a possible Chan-Paton group is that all matter in
 open string models must be in bi-fundamentals, and therefore it is impossible to get a spinor representation. 
 }\fi
 
 \ifExtendedVersion {\color{darkred}
 Note that it was assumed here that there are at most four branes participating in the Standard Model. If one
 relaxes that condition, the number of possibilities is unlimited.} \fi \ifExtendedVersion {\color{darkred}
 There exist other realizations of the Standard Model using branes, see for example
 \textcite{Antoniadis:2000ena} and \textcite{Berenstein:2006pk}.}\fi

 \ifExtendedVersion {\color{darkred}
 \paragraph{Orientifolds.}
 
 An important issue in open string model building is the cancellation of tadpoles of the disk diagram. These lead to
 divergences and can lead to chiral anomalies. These tadpoles can sometimes be canceled by adding another
object to the theory, called an orientifold plane. In fact, the usual procedure is to start with an oriented type-II string, and
consider an involution of the world-sheet that reverses its orientation. Then one allows strings to close up to that
involution. In terms of world-sheet topology, this amounts to adding surfaces with the topology of a Klein bottle. 
The combination of torus and Klein-bottle diagram acts like a projection on the closed string theory, removing
some of its states. In most cases, removing states from string theory comes at a price: other states must be added
to compensate what was removed. This is rigorously true in heterotic strings, and is evident in orbifold constructions,
where part of the spectrum is projected out (``modded out"), but then new states (``twisted states") must be added,
corresponding to strings that stretch between identified point.
Using a (somewhat misleading) analogy with orbifolds \cite{Horava:1989vt} one adds open strings to the
orientifold-projected closed strings which in some sense can be thought of as twisted sectors. The
analogy is not perfect: there exist orientifold-projected closed strings that are consistent all by themselves. But the
procedure (called the orientifold construction) is well-defined and does not require the analogy to work.  

\paragraph{Anomalies}

Canceling all tadpoles between the disk and crosscap diagram removes most anomalies, but some factorized
anomalies remain.
This also happens for the original heterotic strings, where modular invariance removes the non-factorizable
anomalies, so that the full anomaly polynomial factorizes into two factors \cite{Schellekens:1986xh},
\begin{equation}
A(F,R)=({\rm Tr} F^2 - {\rm Tr} R^2) A'(F,R)
\end{equation}
which can then be canceled by the Green-Schwarz mechanism \cite{Green:1984sg} involving tree-level diagrams
with exchange of the $B_{\mu\nu}$ axion. In open strings (and also in more general heterotic strings)
the anomaly factorizes also, but in terms of several factors. These anomalies are then canceled by a Green-Schwarz mechanism
involving multiple axions, which are available in the Ramond-Ramond sector of the closed theory. 

In four dimensions, a factorized anomaly always involves a $U(1)$. The corresponding $U(1)$ vector bosons acquire a mass
by ``eating" the axion , which provides the missing longitudinal mode. String theory will always remove anomalous symmetries
in this manner, but it turns out that this can happen for non-anomalous $U(1)'s$ as well. This can be traced back to anomalies
in six dimensions (see \textcite{Kiritsis:2003mc} and references therein). In the Madrid model shown above, the Chan-Paton gauge group is at least $U(3) \times
Sp(2) \times U(1)\times U(1)$. This contains the Standard Model $Y$-charge plus two additional $U(1)$'s. One of these
is anomalous, and a third linear combination corresponds to $B-L$, which is not anomalous (if there are three right-handed
neutrinos, as is the case here).   A massless $B-L$ gauge boson is one of the most common generically wrong
predictions of most string models. However, there is a way out: it can acquire a mass from axion mixing despite 
being anomaly-free. If that does not happen one has to resort to the standard procedure in string phenomenology: assume
that one of the many massless fields in the spectrum gets a vacuum expectation value that breaks the symmetry.

}\else {
}\fi

\paragraph{Boundary RCFT constructions.}

Just as in the heterotic string, one can construct spectra using purely geometric methods, orbifold methods or 
world-sheet constructions. \ifExtendedVersion {\color{darkred} Most work in the literature uses the second approach. }\fi
 
 World-sheet approaches use boundary CFT: conformal field theory on surfaces with boundaries and crosscaps. This
requires an extension of the closed string Hilbert space with ``states" \ifExtendedVersion {\color{darkred}(in fact not normalizable, and hence not in the closed string
Hilbert space) }\fi
that describe closed strings near a boundary, or in 
the presence of orientation reversal. An extensive formalism for computing boundary and crosscap states in (rational) CFT
was developed in the last decade of last century, starting with work by \textcite{Cardy:1989ir}, developed further  by several groups, including
\textcite{Bianchi:1990yu}; \textcite{Pradisi:1996yd}; \textcite{Fuchs:1997fu}; \textcite{Behrend:1999bn}; and  \textcite{Huiszoon:1999xq},  culminating in a simple and general formula \cite{Fuchs:2000cm}. 
For an extensive review of this field see \textcite{Angelantonj:2002ct}.
This was applied by \textcite{Dijkstra:2004cc} to orientifolds
of Gepner models, and led to a huge (of order $200.000$) number of distinct  string spectra that match the chiral Standard Model. 
This set provides an extensive scan over the orientifold landscape. \ifExtendedVersion 

{\color{darkred}
These spectra are exact in perturbative string theory and not only the massless but also all massive states are
known explicitly. There are no chiral exotics, but in general there are large numbers of the ubiquitous vector-like states
that plague almost all exact string spectra. All tadpoles are canceled, but in most cases this requires hidden sectors.
However, there are a few cases where all tadpoles cancel entirely among the Standard Model branes (hence no hidden 
sector is present) and furthermore the superfluous $B-L$ vector bosons acquires a mass from axion mixing. These
spectra have a gauge group which is exactly $SU(3)\times SU(2)\times U(1)$ (there are a few additional vector bosons from
the closed sector, but the perturbative spectrum contains no matter that is charged under these bosons; this is
the same as in the type IIA string, which contains a vector boson that only couples to non-perturbative states, D0-branes).}\fi
\ifExtendedVersion

{\color{darkgreen}
\paragraph{Results}Orientifold model building has been very actively pursued during the first decade of this century. It is impossible to
review all the different approaches and their successes and failures here, but fortunately an extensive review is available
\cite{Blumenhagen:2005mu}. The original work proposing the Madrid model \cite{Ibanez:2001nd} also found 
non-supersymmetric examples, but since only RR tadpoles were canceled and not NS-NS tadpoles, these were not stable.
The search for stable, supersymmetric examples took some time but was finally successful \cite{Cvetic:2001tj}
although initially the
spectra were plagued by the presence of chiral (but non-anomalous) exotic matter. 
}
\fi

\subsubsection{Decoupling Limits}\label{LocalModels}

Brane model building led to an interesting change in strategy. Whereas string theory constructions were originally
``top-down" (one constructs a string theory and then compares with the Standard Model), using branes one can
to some extent work in the opposite direction, ``bottom-up". The idea is to 
start with the Standard Model and construct a brane configuration to match it, using branes localized at (orbifold) singularities.
Then this brane configuration may be
embedded in string theory at a later stage. 
This point of view was pioneered by \textcite{Aldazabal:2000sa}.
This is a useful approach in open string models because the
gauge fields are localized on D-branes.  This makes it possible to decouple gravity by sending the compactification
radius to infinity. By contrast,
in heterotic string models both gravity and gauge 
interactions originate from closed string exchange, and such a decoupling limit would not make sense. 
Examples with  $\mathbb{Z}_3$ singularities were given by the aforementioned authors. 
\textcite{Berenstein:2001nk} considered the discrete group $\Delta_{27}$, and
\textcite{Verlinde:2005jr}  used D3-branes on a del Pezzo 8 singularity.

\ifExtendedVersion {\color{darkred}

The other extreme  is to take the details of the 
Standard Model for granted and focus on issues like moduli, supersymmetry breaking
and hierarchies. In this case one has to assume that once the latter are solved, the Standard Model can be added. 
Both points of view are to some extent a return to the ``old days" of quantum field theory. On the one hand, the
techniques of branes and higher dimensions are used to enrich old ideas in GUT model building; on the other hand,
string theory is treated as a ``framework", analogous to quantum field theory, where gauge groups, representations and
couplings are input rather than output. 
}\fi

Decoupling of gravity is an important element in recent work on
F-theory GUTs \cite{Beasley:2008dc,*Beasley:2008kw,*Donagi:2008ca} obtained
by compactifying F-theory on  elliptically fibered Calabi-Yau fourfolds. This allows the construction of models that may be
thought of as  non-perturbative realizations of the orientifold $SU(5)$ GUT models depicted in Fig. \ref{GUTTrinification}(b), 
solving some of their problems, especially absence of the top-Yukawa coupling, which is perturbatively forbidden. 
This has led to a revival of Grand Unified Theories, invigorated with features of higher dimensional theories. We will 
return to this in sections \ref{ConstUnif} and \ref{LanSym}.
See reviews by \textcite{Weigand:2010wm,Heckman:2010bq,Leontaris:2012mh,Maharana:2012tu} for further details.

\ifExtendedVersion {\color{red}}\else {
The other extreme  is to take the details of the 
Standard Model for granted and focus on issues like moduli, supersymmetry breaking
and hierarchies. In this case one has to assume that once the latter are solved, the Standard Model can be added. }\fi
\ifExtendedVersion {\color{darkred}

An example in the second category is recent work in the area of M-theory compactifications \cite{Acharya:2012tw}. }\else {
This is what is done in recent work on M-theory compactifications \cite{Acharya:2012tw}. }\fi
Getting chiral $N\!\!=\!\!1$ supersymmetric spectra in M-theory requires compactification on a
seven dimensional manifold with $G_2$ holonomy \cite{Acharya:2001gy}, also known as a Joyce manifold. 
Much less is known about M-theory than about string theory, and much less is known about Joyce manifolds 
than about Calabi-Yau manifolds, since the powerful tool of  complex geometry is not available. For this reason the Standard Model
is treated as input rather than output, in the spirit of QFT.

Another  kind of compactification that allows splitting the
problem into decoupled parts is the LARGE Volume Scenario \cite{Balasubramanian:2005zx}, originally invented for the purpose of
 moduli stabilization (see section \ref{ModStab}).  Here both kinds of decoupling limits have been discussed, and there have also been steps towards putting 
both parts together \cite{Conlon:2008wa}. This illustrates that focusing on decoupling limits does not mean that the original goal of a complete theory is
forgotten. Indeed, there also  exist {\it global} F-theory constructions \cite{Blumenhagen:2009yv,Marsano:2012yc}.

\subsection{Non-supersymmetric strings}

Although the vast majority of the literature on  string constructions concerns space-time supersymmetric spectra, in world-sheet
based methods -- free bosons and fermions, Gepner models, and certain orbifolds -- it is as easy to construct
non-supersymmetric ones. 
\ifExtendedVersion {\color{darkred}
In fact, it is easier, because space-time
supersymmetry is an additional constraint. }\fi These spectra are generally plagued by tachyons, but by systematic
searches one can find examples where no tachyons occur.  This was first done in ten dimensions by  
\textcite{Dixon:1986iz,AlvarezGaume:1986jb}. These authors found a heterotic string theory with a $SO(16)\times SO(16)$ gauge group, 
the only tachyon-free non-supersymmetric theory in ten dimensions, out of a total of  seven. 
Four-dimensional non-supersymmetric strings were already constructed shortly thereafter \cite{Lerche:1986cx,Kawai:1986vd}.
\ifExtendedVersion {\color{darkred}
All of these examples employ off-diagonal left-right pairings of partition functions. In the absence of space-time supersymmetry,
tachyonic character exist, but they may be paired with a non-tachyonic one so that there is no physical tachyon. 
The result can be interpreted by means of ``mis-aligned supersymmetry" \cite{Dienes:1994np}. Finiteness of the vacuum energy
at one-loop is due to alternating boson and fermion surpluses at subsequent levels.}

{\color{darkgreen}
In orientifold models there are two additional ways to remove the closed string tachyons. They may also be projected out
by the Klein-bottle diagram \cite{Sagnotti:1995ga},  or it is possible to consider supersymmetric closed string theories with supersymmetry broken only
in the open sector. An ten-dimensional example of the latter kind was described by \textcite{Sugimoto:1999tx}, and this is known
in general as ``Brane Supersymmetry Breaking" \cite{Antoniadis:1999xk}.

After adding open strings one also has to worry also about tachyons in the open sector. These may be avoided by judicious choices
of Chan-Paton multiplicities. In addition, one has to make sure that the crosscap and disk tadpoles cancel, which implies an
additional constraint on these choices. Examples satisfying all these constraints were found using 
orbifold methods by \textcite{Angelantonj:1998gj} (see also  \textcite{Angelantonj:2002ct} for further references)  and using Gepner orientifolds
by \textcite{GatoRivera:2008zn}. 
The latter authors even tried to obtain the chiral Standard Model spectrum in this manner, but without
success, presumably just because the sample size was too small. }\else {

Non-supersymmetric strings can also be constructed using orientifold methods,  see for example
 \textcite{Sagnotti:1995ga}; \textcite{Angelantonj:1998gj}; \textcite{Sugimoto:1999tx};  and \textcite{GatoRivera:2008zn}. This includes the interesting possibility of having broken
 supersymmetry only in the open sector (``Brane Supersymmetry Breaking" \cite{Antoniadis:1999xk}).}\fi

Non-supersymmetric strings can have a vacuum energy $\Lambda$ of either sign. See for example \textcite{Dienes:2006ut}  for  a distribution of values of the vacuum energy for a class
of heterotic strings.  
There also exist examples where $\Lambda$
vanishes exactly to all orders in perturbation theory \cite{Kachru:1998hd} but probably this feature does not hold beyond
perturbation theory  \cite{Harvey:1998rc}. 
\ifExtendedVersion {\color{darkred}

One might think that in the absence of any evidence for low energy supersymmetry, and because of the evidence in favor
of an accelerated expansion of the universe, non-supersymmetric strings with a positive 
cosmological constant are a better candidate  for describing our universe than the much more frequently studied supersymmetric ones. But the absence of supersymmetry is a serious threat for the stability of these theories, even in the absence of tachyons
in the perturbative spectrum. All of these theories have massless particles, which include at least the dilaton, and usually
many others. Absence of tachyons only says something about the second order terms in scalar potentials. Higher order terms
can still destabilize these theories. In many cases there are tachyons in spectra continuously connected to them. In
\textcite{Ginsparg:1986wr,Nair:1986zn} this was analyzed for the $O(16)\times O(16)$ string, where the continuous parameters were
obtained by compactification of one dimension. The (meta)-stability of brane supersymmetry breaking is discussed
by \textcite{Angelantonj:2007ts}.}
\else {

Because of the lack of evidence for low energy supersymmetry one might think that non-supersymmetric strings are to be preferred.
Unfortunately they tend to have  instabilities. They all have massless scalars (at least a dilaton) that can run off towards tachyonic regions
and have tadpoles that cause divergences in two-loop diagrams.}\fi 

There is always a  dilaton
tadpole. This signals that the flat background space-time that was used is not a solution to the
equations of motion; instead one must use de Sitter (dS) or Anti-de Sitter (AdS) space with
precisely the value $\Lambda$ as its cosmological constant  \cite{Fischler:1986ci,Fischler:1986tb}. 
Unfortunately this argument only provides an explanation for the presence of the tadpole, but  does not provide
an exact (A)dS solution.

\subsection{The String Theory Landscape}

\ifExtendedVersion {\color{darkred}
From the huge amount of work described in the previous chapter we have learned that string theory 
can describe all gross
features of  
a supersymmetrized version of the Standard Model. 
But there are still some major (and many minor) obstacles: supersymmetry must
be broken, all moduli must be stabilized,  a cosmological constant must be generated, and that constant must be absurdly small and positive.
 
All of these requirements are needed for phenomenological reasons, but they also have anthropic implications. 
If that were not the case, we
should already ask ourselves why string theory, to first approximation, tends to predict all of these features  
so catastrophically wrong.

It is possible that the supersymmetric vacua we are able to discuss are merely an irrelevant corner in a huge, predominantly
non-supersymmetric landscape. 
This would mean that our current view is completely distorted by our technical limitations. This
may be true, but it is not a useful working hypothesis. The exact non-supersymmetric string theories discussed above illustrate this
point: it has been very hard to to make progress in this area of string theory.

Fortunately there does exist a plausible anthropic argument why the -- apparently -- ubiquitous supersymmetric vacua are not
observed, see sec. \ref{OtherHabitable}. With supersymmetry out of the way, the other two major problems change character. It not
very plausible that in a non-supersymmetric theory there would be exactly flat moduli potentials or an exactly vanishing
cosmological constant. One could contemplate the possibility that we live in a universe where one or more moduli
are slowly running away to infinity. They would then drag various constants of nature along with them, 
and changing constants of nature
evidently create anthropic problems. This will be discussed in section \ref{VarCons}. It seems however more plausible that we
need to find stable points in the potential. 

If such stable point exist, all moduli will get masses. There are important restrictions on their masses. If they are extremely
small, moduli can mediate long-range forces that would be observed as violations of the equivalence principle. But since 
nature abhors light scalars this is not a major worry. Any mass below the scale of supersymmetry breaking would be unnatural,
so the expected mass of the moduli is the supersymmetry breaking scale.

}\fi

A crucial test for the string landscape is the existence of (meta)stable dS vacua. They are needed for three reasons: there is evidence
that our own universe approaches such a space at late times, eternal inflation requires the existence of at least one dS
vacuum, and cosmic inflation in our own universe may need, at least approximately, a dS space as well. Furthermore, for
explanations of apparent anthropic tunings we need a large number of such spaces, and they have to be distributed in the right way.

%
%

\subsubsection{Existence of de Sitter Vacua}\label{ModStab}

%
%
The art of constructing dS vacua is based on assembling the many ingredients of the string toolbox in a controlled way:
branes, fluxes, orientifold planes, non-perturbative effects (usually in the concrete forms of ``brane instantons" or gaugino condensation), world-sheet
perturbative corrections and string perturbative corrections. Fortunately, several fairly recent review articles
are available, {\it e.g.} \textcite{Grana:2005jc}; \textcite{Douglas:2006es}; \textcite{Blumenhagen:2006ci}; \textcite{Denef:2008wq} and the slightly more accessible one by \textcite{Denef:2007pq}.  Here we will  just  give a brief summary, and mention some recent developments. 


The most explicit results have been obtained in type-IIB (and related F-theory) compactifications.
One starts with a Calabi-Yau compactification. The continuous deformations of such manifolds are described by
moduli of two different kinds: $h_{21}$ complex structure (``shape") moduli and $h_{11}$ K\"ahler (``size") moduli, where $h_{21}$ and
$h_{11}$ are the Hodge numbers of the CY manifold. One can add 3-form RR and NS fluxes, 5-form fluxes, 
denoted $F3, H3$ and $F5$ respectively, and D3 and D7 branes.

In type-IIB theories the 3-form fluxes can stabilize all complex structure moduli. This stabilization is due to a tree-level 
term in the superpotential that takes the form \cite{Gukov:1999ya}
\begin{equation}
W_{\rm flux}=\int (F_3 -\tau H_3) \wedge \Omega\ ,
\end{equation}
where  $\tau=a+ie^{-\phi}$, and $a$ is the axion and $\phi$ the dilaton. The dependence on the complex structure moduli 
is through $\Omega$, the holomorphic three-form of the Calabi-Yau manifold. 
This term also fixes the dilaton and axion. However, $W_{\rm flux}$ does not depend on the K\"ahler moduli and hence cannot fix them. This leaves therefore at least one modulus unfixed, since
every CY manifold has at least one K\"ahler modulus.

The next step is to try and fix the size moduli with non-perturbative terms in the superpotential. These take
the form $W \propto {\rm exp}(i \lambda s)$, where $s$ is the size modulus and $\lambda$ a parameter. Such terms can be
generated by instantons associated with Euclidean D3-branes \cite{Witten:1996bn} or from gaugino condensation
in gauge groups on wrapped D7 branes. Assuming at least one of these effects to be present, \textcite{Kachru:2003aw} (usually referred to as KKLT)
obtained string vacua with all moduli stabilized. This work builds on several earlier results, such as
\textcite{Dasgupta:1999ss,Klebanov:2000hb,Giddings:2001yu} and other references cited.
KKLT considered the special case $h_{11}=1$, so
that only one size modulus needs to be stabilized. They argued that by suitable choices of fluxes one can obtain solutions 
where supersymmetry is unbroken, and all world-sheet  and string perturbative corrections ({\it i.e} the $\alpha'$ and $g_s$ expansion)
are small. The solution obtained in this way has a negative vacuum energy, and is a fully stabilized supersymmetric
AdS vacuum. This is achieved by choosing fluxes so that $W_{\rm flux}$ is small, the volume is large and the dilaton
(which determines the string coupling) is stabilized at a point where the coupling is small. Here ``small" and ``large" refer to tunings
by just a few orders of magnitude.

This is however just a ``scenario", since the existence of the non-perturbative effects still needs to be demonstrated. Many
would-be instantons do not contribute because of superfluous zero-modes. 
It turns out that
models with just one K\"ahler modulus do not work, and that instanton contributions are ``not generic" \cite{Robbins:2004hx,Denef:2004dm} but still occur sufficiently often to allow a huge number of solutions.

The next step is more problematic and more controversial. One must break supersymmetry and obtain a dS vacuum
(this is called ``up-lifting").
In KKLT
this is done by adding an anti-D3 brane in a suitable location on the Calabi-Yau manifold, such that the validity of the
approximations is not affected. Anti-D3 branes explicitly violate supersymmetry, and hence after introducing them
one loses the control offered by supergravity.  Of course, supersymmetry must be broken anyway, but it would be preferable
to break it spontaneously rather than explicitly. Attempts to realize the KKLT uplifting in supergravity or string theory have failed so far 
\cite{Bena:2012bk,Bena:2012vz}, but opinions differ on the implications of that result. There exist several alternatives
to D3-brane uplifting (see {\it e.g.} \textcite{Burgess:2003ic}; \textcite{Saltman:2004sn}; \textcite{Lebedev:2006qq}; and also \textcite{Westphal:2007xd,Covi:2008ea} for
further references.)

The result of a fully realized KKLT construction is a string  vacuum that is free of tachyons, but one still has to worry about non-perturbative
instability. The uplift contribution vanishes in the limit of large moduli, so there is always a supersymmetric vacuum in that
limit, separated from the dS vacuum by the uplifted barrier that stabilized the AdS vacuum. One can work out
the tunneling amplitude, and KKLT showed that it is generically much larger than the observed lifetime of our universe, yet
well below the theoretical upper limit in dS space, the Poincar\'e recurrence time. See also \textcite{Westphal:2007xd} for a systematic
analysis of several kinds of minima.

An alternative scenario was described by \textcite{Balasubramanian:2005zx}.
The starting point is  the same: type-IIB fluxes stabilizing the 
complex structure moduli and the dilaton and axion. But these authors use $\alpha'$ corrections to their advantage rather than
tuning parameters to minimize them. By means of suitable $(\alpha')^3$ corrections they were able to find minima where
all moduli are stabilized at exponentially large volumes in {\it non}-supersymmetric AdS vacua. The fact that  $\alpha'$ corrections
can be important at large volumes may be counter-intuitive, but can be understood in terms of the no-scale
structure of the underlying supergravity. For other work discussing the importance of perturbative corrections see 
\textcite{Becker:2002nn,vonGersdorff:2005bf,Berg:2005yu,Bobkov:2004cy}.
Additional mechanisms are then needed to lift the vacuum to dS.Aan explicit example
was presented recently by \textcite{Louis:2012nb}.
This scenario requires
special Calabi-Yau manifolds with $h_{21} > h_{11} > 1$ and a structure consisting of one large topological cycle and one or more small ones. This has been given the suggestive name ``Swiss Cheese manifold". Not every Calabi-Yau manifold has this property,
but several hundreds are  known \cite{Gray:2012jy,Cicoli:2011it}. A natural hierarchy can be obtained by associating Standard Model branes with the small cycles.
This is called the LARGE
volume scenario (LVS). 

Although type-IIA and type-IIB  string theories in ten dimensions only differ by a single sign flip, the discussion of moduli
stabilization for the compactified theories is vastly different. This is because  in type-IIA theories the available RR-fluxes are even-forms, and
the available D-branes are D-even branes. Since there still are three form NS-fluxes one now gets flux potentials that
depend on the complex structure moduli and others that depend on the K\"ahler moduli. As a result, {\it all} moduli can   
now be stabilized classically by flux potentials \cite{DeWolfe:2005uu} (see however \textcite{McOrist:2012yc}). Unfortunately, it can also be shown 
\cite{Hertzberg:2007wc} that none of the aforementioned ingredients can be used to lift these theories to dS.
There are more ingredients available, but so far no explicit examples are known (see \textcite{Danielsson:2011au} for a recent attempt).

Moduli stabilization for heterotic M-theory
was discussed by \textcite{Braun:2006th}. Supersymmetry is broken and a lift to dS achieved using heterotic five-branes and anti-five-branes.
For the perturbative heterotic strings in the ``mini-landscape" a scenario for moduli stabilization was presented by 
\textcite{Dundee:2010sb}. 
 \textcite{Acharya:2006ia}  discussed this for M-theory compactifications on manifolds with $G_2$ holonomy.
 These authors do not use fluxes, because in this class of models they would destroy the hierarchy. Instead,
all  moduli are stabilized by non-perturbative contributions generated by strong gauge dynamics. To this end they
introduce two ``hidden sector" gauge groups.  A similar mechanism was applied to type-IIB theories by \textcite{Bobkov:2010rf}. 
These arguments often rely on plausible but unproven assumptions about  terms in potentials and non-perturbative effects.
In explicit models the required terms may be absent, even though generically allowed. 

\subsubsection{Counting and Distributions}

Fluxes are characterized by integers specifying how often they wrap the topological cycles on the manifold.
However, the total number of possibilities is limited by conditions for cancellation of tadpoles.
 For a large class of F-theory constructions this condition takes the form 
\begin{equation}
N_{\rm D3}-N_{\overline {\rm{D3} }}+ \frac{1}{2\pi^4{{\alpha'}^2}}\int H_3\wedge F_3 = \frac{\chi(X)}{24}\ ,
\end{equation}
where the first two terms denote the net contribution from D3-branes, the third one the contribution due to fluxes
and the right hand side is a contribution  (``tadpole charge") from orientifold planes  \cite{Sethi:1996es}; $\chi(X)$  is the Euler number of a Calabi-Yau fourfold defining the F-theory under consideration.
 Since the flux contribution is always positive this makes the number
of possibilities finite.  

This has been the starting point for estimates of the total number of flux vacua. \textcite{Douglas:2004zg}
gave the following estimate  (based on \textcite{Ashok:2003gk,Denef:2004ze})
\begin{equation}\label{DouglasCounting}
N_{\rm vac} \approx \frac{(2\pi L)^{K/2}}{(K/2)!},
\end{equation}
where $L$ is the aforementioned tadpole charge and $K$ the number of distinct fluxes. 
For typical manifolds this
gives numbers of order $10^{N}$, where $N$ is of order a few hundred. This is the origin of the (in)famous estimate
$10^{500}$. Note  that Eq. (\ref{DouglasCounting}) should still be summed over distinct manifolds, that it only counts fluxes and no other 
gadgets from the string theory toolbox, and that none of these $10^{500}$ vacua includes the Standard Model, because
no structure (like intersecting D-branes or singularities) is taken into account to produce chiral matter. Indeed,
the presence of chiral matter may influence moduli stabilization in a negative way \cite{Blumenhagen:2007sm}.

It is noteworthy that this formula turns a nuisance (a large number of moduli) into a virtue: the large number of moduli
gives rise to the exponent of Eq. (\ref{DouglasCounting}), and it is this large exponent that makes neutralization of the
cosmological constant possible. This is not automatically true for all string compactifications and moduli 
stabilization mechanisms; the existence of a sufficiently large set of vacua has to be demonstrated in each case. 
\textcite{Bobkov:2009za} has shown that  fluxless $G_2$  compactifications of M-theory
also yield a large discretuum of  vacua.

In type-IIA constructions there are also tadpole conditions to satisfy, but in this case they do not 
reduce the vacuum count to a finite number. Instead it was found that supersymmetric AdS vacua exist at arbitrarily large volume,
in combination with an arbitrarily small cosmological constant. This implies that the total number of vacua is infinite, but
it can be made finite by making a phenomenologically inspired cut on the volume of the compactification. 
\textcite{Acharya:2006zw} presented general arguments suggesting that the number of
string vacua must be finite, if one puts upper bounds on the cosmological constant and the compactification volume.


The most important contribution not taken into account in  Eq. (\ref{DouglasCounting}) is the
effect of supersymmetry breaking. 
\ifExtendedVersion {\color{darkred}
These computations count supersymmetric AdS vacua. They must still be lifted to dS
by a supersymmetry breaking contribution. }\fi
Already in \textcite{Douglas:2004zg} the possibility was mentioned that most of the AdS vacua might become tachyonic if such a lift is 
applied. Recent work seems to indicate that this is indeed what happens. In \textcite{Chen:2011ac} this was investigated 
for type-IIA vacua and in
\textcite{Marsh:2011aa} for supergravity.  These authors analyze general scalar potentials using random matrices to determine
the likelihood that the full mass matrix is positive definite.
They find that this is exponentially suppressed  by a factor $\approx {\rm exp}(-cN^p)$, where $N$ is the number of complex
scalar fields and $p$ is estimated to lie in the range $1.3$ to 2. This suppression can be reduced if a large subset of the scalars
is decoupled by giving them large supersymmetric masses. Then only the number of light scalars contributes
to the suppression. Even more worrisome results were reported recently by \textcite{Greene2013}. In a study of  landscapes modeled with scalar fields, they
found a doubly exponential decrease of the number of meta-stable vacua as a function of the number of moduli, due to  dramatic increases in tunneling rates.

\subsubsection{Is there a String Theory Landscape?}

It is generally accepted that there exists a large landscape of fully stabilized supersymmetric AdS solutions. But these do not describe
our universe. Not in the first place because of the observation of accelerated expansion of the universe, but
because of the much more established fact that our vacuum is not supersymmetric. Supersymmetric vacua have a vacuum energy
that is bounded from above at zero. Supersymmetry breaking makes positive contributions to vacuum energy. Hence if stable
non-supersymmetric vacua exist (which few people doubt), it would be highly surprising if their vacuum energy could not surpass
the value zero. Most arguments for or against the existence of dS vacua do not really depend on the sign of the cosmological
constant; $+10^{-120}$ is nearly indistinguishable from $-10^{-120}$. Hence one would expect distributions to behave smoothly near zero,
although they may drop off rapidly. 

By  now there are many constructions of dS vacua, although there are always some assumptions,
and it is often not possible to check the effect of higher order world-sheet or string loop corrections. But given the large number of
possibilities, it would require a miracle for {\it all} of them to fail. If that is the case there should exist some general no-go theorem that was
overlooked so far.  

But the  mere existence of  vacua with positive $\Lambda$ is not enough.  To make use
of the Bousso-Polchinski neutralization of $\Lambda$
a sufficiently dense discretuum of such vacua is needed. This mechanism relies on the fact that whatever the
contribution of particle physics, cosmology and fundamental theory is, it can always be canceled to 120 significant
digits by flux contributions, {\it without making actual computations with that precision}.
If in reality these distributions are severely depleted in part of the range, or have a highly complicated
non-flat structure, this argument would fail. There might still exist  a huge landscape, but it would be useless. 

The mighty landscape of a decade ago has been eroding at an alarming rate. The actual number of vacua is the product of huge numbers divided by huge
suppression factors. Perhaps this will re-ignite dreams of a unique theory. Could it be that the product is exactly one, with the Standard Model and
 the observed cosmological constant  as the only survivor? That would be an absurd example of the second
gedanken computation of section \ref{Gedanken}. Any hopes that landscape erosion
will reduce the number of de Sitter vacua to {\it one} are unfounded, but there is a risk that it will be reduced to {\it zero}.

More fundamental objections against the use of effective potentials in quantum 
gravity or the formulation of QFT and string theory in de Sitter space have been raised by  \textcite{Banks:2012hx}. 
If these objections are
valid, we may not have {\it any} theoretical  methods at our disposal to deal with the apparent accelerated expansion of the universe.

\section{The Standard Model in the  Landscape}

In this chapter we will discuss how the main features of the Standard Model fit in the
String Theory Landscape, taking into account anthropic restrictions and analytical and numerical work on landscape distributions.

\ifExtendedVersion {\color{darkred}
It would be interesting to know if there are any quantum field theories that do not have a place somewhere
in the string landscape. For a discrete, non-supersymmetric landscape, obviously a continuous infinity of quantum
field theories is not realized in any vacuum, but one can phrase this question in a meaningful way for supersymmetric theories.
In \cite{Vafa:2005ui} it was argued that indeed such a ``swampland" of non-realizable quantum field theories indeed exists, one
of the examples being the ten-dimensional anomaly free $U(1)^{496}$ $N=1$ gauge theory. Further evidence is provided in 
\cite{Fiol:2008gn}. Perhaps  there is no swampland for supersymmetric theories in six dimensions, but this could be due to
the much more powerful chiral anomaly constraints \cite{Kumar:2009us}. In four dimensions this issue remains unsettled. 
}\fi

 \subsection{The Gauge Sector}

It is by now abundantly clear that string theory can reproduce the discrete structure of the Standard Model: the gauge group and
chiral fermion representations.  We cannot even begin to enumerate all the papers that succeeded in doing this.  

 \subsubsection{Gauge Group and Family Structure}

\ifExtendedVersion \paragraph{Why $SU(3)\times SU(2) \times U(1)$?} \fi
From the landscape perspective, one might hope that the  gauge group can be understood
using string theory plus anthropic constraints. The anthropic constraints are hard to determine,
 but all three factors of the gauge group are needed
for {\it our} kind of life. 
Electromagnetism is so essential that it is impossible to imagine life without it. One can imagine life 
without $SU(3)_{\rm color}$ and only electromagnetism, but 
it is by no means obvious that such universes will really come to life. 
The weak interactions also play a crucial r\^ole in our universe, but perhaps not in every habitable 
one (see section \ref{OtherHabitable}). 

The choice of fermion representation is also essential, but it is even harder to determine 
what happens if we change it. It is possible that it is chiral in order to keep the fermions light (a plausible reason why  $SU(2)_{\rm weak}$ 
might be needed). Chiral fermions have chiral anomalies that must be canceled. This fixes to some extent the particle
content of a single quark and lepton family, if one insists on simplicity. See \textcite{Shrock:2008sb} for some {\it gedanken} variations
of the representations in a family.

If life requires electromagnetism, a non-abelian strong interaction group, and a chiral spectrum that becomes non-chiral
after symmetry breaking at energies far below the Planck scale, perhaps the one-family Standard Model is
the simplest option one can write down.
More complicated possibilities are easy to find. For example, changing the number
of colors from 3 to some odd integer $N$ and the quark charges to $p/N$ for suitable $p$, one can find an infinite series of cousins of the Standard Model \cite{Shrock:1995bp} that,
for all we know, are anthropically  equally valid. It is likely that in the landscape small groups are statistically favored: then $N=3$ would be the
first acceptable value.  
If furthermore small numbers of gauge group factors are also favored, our Standard Model might be the 
statistically dominant anthropic choice. 

It has also been suggested that the choice $N=3$ for the number of colors (with everything else kept fixed) is a consequence of  the fact
that only for $N=3$ there is a simple GUT embedding \cite{Shrock:2007ai}. This explanation would require the 
landscape to be dominated by GUT gauge groups.  

\ifExtendedVersion \paragraph{Landscape scans of groups and representations.}\fi There have been several studies of distributions of groups and representations in sub-landscapes, but because of lack of a sufficiently
well-defined question there is no good answer either. 
\ifExtendedVersion
{\color{darkgreen}
 For free fermion
constructions of heterotic strings see {\it e.g} \textcite{Dienes:2006ut,Dienes:2007ms,Renner:2011gs,Renner:2011yh}. 

In \textcite{Blumenhagen:2004xx} this was done for orientifold models. \textcite{Kumar:2004pv} derived a formula of the
average gauge group rank (for $D3$-brane gauge groups of type-IIB flux vacua). 
\textcite{Anastasopoulos:2006da} gave a classification   of all brane models with at most four brane stacks that contain the 
Standard Model, and a scan for realizations of these options in Gepner orientifolds was presented. 
For other work on distributions of gauge group features see \textcite{Kumar:2006tn,Balasubramanian:2008tz}.
}\fi 
\ifExtendedVersion\else
See {\it e.g.} \textcite{Dienes:2006ut,Dienes:2007ms,Renner:2011gs,Renner:2011yh} for
free fermion heterotic strings and \textcite{Blumenhagen:2004xx,Kumar:2004pv,Anastasopoulos:2006da,Kumar:2006tn,Balasubramanian:2008tz}
for orientifold models. 
Note that all these studies, as well as others mentioned below, are for unstabilized points in supersymmetric moduli spaces.
Furthermore, drawing conclusions about correlations is  made difficult
because of limited sampling \cite{Dienes:2006ca,Dienes:2008rm}.\fi
\ifExtendedVersion
{\color{darkgreen}
 An important caveat is that all of these studies are done on exact realizations in special points in
moduli space. This is therefore not true landscape ``statistics". One would really like to see distributions
of physical quantities for fully stabilized dS vacua, but this is not technically possible at present. These studies are
also plagued by over-counting problems.  Only some features of a spectrum are sampled, and one cannot be certain
if two identical spectra still differ in some other respect. For this reason, comparing total numbers of ``vacua" between
different approaches is meaningless.
Drawing conclusions is also made difficult
because of limited sampling \cite{Dienes:2006ca,Dienes:2008rm}.}
\fi

 \subsubsection{The Number of Families}

%
\ifExtendedVersion \paragraph{Why three families?}\fi We are made out of just one family of fermions. 
There are no good arguments why three families should be anthropically required, 
although some unconvincing arguments can be pondered, based on the r\^ole of the $s$ quark in QCD, of the muon in biological mutations, the top
quark in weak symmetry breaking, or the CP-violating CKM angle in baryogenesis. See also \textcite{Schellekens:2008kg} and \textcite{Gould:2010kr} for arguments and counter-arguments.

Perhaps one day we {\it will}  discover a good anthropic reason
for three families.
If not, the number of families was just picked out of a distribution. Multiple families are a generic feature in string theory,
due to to topological quantities like Hodge numbers of compactification manifolds or intersection numbers of branes (although
often this notion is muddled by attempts to distinguish families in order to explain mass hierarchies).

\ifExtendedVersion
 \paragraph{Landscape scans of the number of families.}\fi

Landscape studies of the number of 
families
tend to suffer from lamppost artifacts: initial studies of simple models favor multiples of four or six families and disfavor three, but as more general
models are studied the number three becomes less and less challenged.
\ifExtendedVersion {\color{darkgreen}

The number of families was studied first in the context of heterotic
 Calabi-Yau compactifications (with $SU(3)$ spin connection embedded in $E_8$)
 and their CFT realizations, 
especially Gepner models and orbifolds, where the number of families is half the Euler number of the manifold.
Systematic studies of Gepner compactifications with $E_6$ and $SO(10)$ gauge groups were presented by 
\textcite{Schellekens:1989wx,Fuchs:1989yv}. In all but one exceptional case\footnote{This Gepner model is closely related
the first three-family Calabi-Yau manifold, see \cite{Schimmrigk:1987ke}.} \cite{Gepner:1987hi}, the number of families is a multiple of six or -- less often-- four. 
\textcite{GatoRivera:2010gv} enlarged the scope  by allowing
broken GUT groups and asymmetric realizations of space-time and world-sheet 
supersymmetry  (which are not required in the bosonic sector of the heterotic string, but automatically imposed in symmetric constructions). This did not lead to additional cases of
three-family models. The reason why  the number three was so hard to get has never been understood, but
the problem disappears in more general constructions. For example, a much larger scan of Euler numbers of classes of Calabi-Yau
manifolds \cite{Kreuzer:2000xy} does not show a  strong suppression of Euler number six. Furthermore, in Gepner 
constructions the problem 
disappears if one uses  different building blocks for the left- and the right-moving sector. Modular invariance makes this hard
to do, but matching building blocks which are isomorphic (in the sense of the modular group) but not identical can be constructed, 
and lead to family distributions that are peaked at zero (which is anthropically excluded) and fall of slowly
\cite{GatoRivera:2010xn,GatoRivera:2010fi}
In these distributions, 
three families are about as common as one, two and four, but numbers larger than six occur only rarely. This 
behavior persists in the more general class of orbifold permutations of Gepner models \cite{Maio:2011qn}.

A similar conclusion can be drawn for free-fermionic constructions of heterotic strings. Here modular invariance can be
solved in general, and hence this approach does not suffer from a bias towards symmetric constructions, as do the Gepner models.
However, it should be kept in mind that most scan done in this context are biased towards three families because a special
choice of fermion boundary conditions, the so-called NAHE set  \cite{Antoniadis:1989zy} is used, {\it a priori} designed to produce three families.  
It has been suggested that in this context 
the number of families can be understood as ``(10-4)/2", {\it i.e.} half the number of compactified 
dimensions \cite{Faraggi:2004rq}, but in a systematic scan of this class \cite{Faraggi:2006bc}
a distribution was found that is peaked around zero, and
with three families occurring in about $15\%$ of all cases.

In most other constructions only three family spectra have been studied, so that we cannot be certain how special they are.
In the ``heterotic mini-landscape", see {\it e.g.} \textcite{Lebedev:2006kn,Lebedev:2008un}, the requirement of having thee families
reduces the sample size by more than an order of magnitude. This is in rough agreement with the foregoing results.

\hyphenation{Dijk-stra}
In orientifold models the family distribution also peaks at zero, but falls off more rapidly. In a study of Gepner orientifolds
 with Standard Model gauge groups \cite{Dijkstra:2004cc} the number of three family spectra was about two orders of magnitude
 less than those with two families. Qualitatively similar results were obtained for $T^6/Z_2 \times Z_2$ orientifolds by \textcite{Gmeiner:2005vz}. These authors  found a large dip in
 the distribution precisely for three families. However, a more detailed analysis of the
 same class \cite{Rosenhaus:2009cs} does not show such a dip. These authors found most spectra in the tail of the distribution at large
 moduli, not considered in \cite{Gmeiner:2005vz} (however, they also found that in this tail there is a huge suppression due to K-theory 
 constraints, which was estimated, but not fully taken into account). A general analysis of brane intersection numbers (giving rise
 to chiral fermions) for this
 class, using both analytical and numerical methods, was presented by \textcite{Douglas:2006xy}. 
 These conclusions depend strongly on the kind of orbifold considered. For example, \textcite{Gmeiner:2008xq} found large numbers of three family models for $T^6/Z_6'$ orientifolds (the prime indicates a certain action of $Z_6$ on the torus $T^6$. On the other hand, 
 in an extensive study of
 free fermion orientifolds, no three family models were found (\cite{Kiritsis:2008mu}).

}
\else
See for example 
\textcite{Schellekens:1989wx,Fuchs:1989yv,GatoRivera:2010gv} versus 
\textcite{GatoRivera:2010xn, *GatoRivera:2010fi} for heterotic Gepner models; and \textcite{Gmeiner:2005vz}
versus \textcite{Rosenhaus:2009cs} for ${\mathbb Z}_2 \times {\mathbb Z}_2$ orientifold models; see   \textcite{Douglas:2006xy} for an 
analytical study of this case. 

In a systematic scan of a class of free fermion heterotic  models \cite{Faraggi:2006bc}
 three families occurred in about $15\%$ of all cases.  However,  in a study of Gepner orientifolds
 with Standard Model gauge groups \cite{Dijkstra:2004cc} the number of three family spectra was about two orders of magnitude
 less than those with two families. There are many other constructions giving three families, but usually no scanning is done for other values.
\fi 

Taking all these results together one may conclude that getting three families may be slightly more difficult than getting one or two, but
it is at worst a landscape naturalness problem at the level of a few percent, and even this  suppression may be due to the examples being too special.
Therefore it is legitimate at this point to view the number of families simply as a number that came out of a distribution, which requires no
further explanation.

 \subsubsection{Grand Unification in String Theory}

 \paragraph{Fractional Charges.}

A remarkable feature of the quark and lepton families is  the absence of fractional electric charges for color singlets.
There is no evidence that free fractionally charged
particles exist in nature, with a limit of less than $10^{-20}$ in matter \cite{Perl:2009zz}, under certain assumptions about their charges.
If indeed there are none, the global Standard Model gauge group is not $SU(3) \times SU(2)\times U(1)$, but
$S(U(3)\times U(2))$. The reason is that the former allows representations with any real values for the $U(1)$ charge, whereas in the
latter case the charges are restricted by the rule
\begin{equation}
\label{ChargeQuant}
\frac{t_3}3  + \frac{t_2}2 + \frac16 = 0\ {\rm mod}\ 1,
\end{equation}
where $t_3$ is the triality of the $SU(3)$ representation and $t_2$ the duality of $SU(2)$, twice the spin modulo integers.
This relation implies integral charges for color-singlet states. But this is just
an empirical rule. Nothing we know at present imposes such a relation. Anomaly cancellation restricts the allowed charges, but
arbitrary charges, even irrational ones, can be added in non-chiral pairs or as scalar fields. In fundamental theories one may expect
charges to come out quantized (due to Dirac quantization for magnetic monopoles), but that still does not imply that they are quantized in the correct way. 

Already for almost four decades we know an excellent explanation for the empirical fact (\ref{ChargeQuant}): Grand Unification, which embeds
the Standard Model in a single, simple gauge group $SU(5)$ \cite{Georgi:1974sy}. So far this idea remains just a theory. In
its simplest form it made a falsifiable prediction, the decay of the proton, and this was indeed falsified. 
\ifExtendedVersion {\color{darkred} It also predicts magnetic monopoles, but
these are too heavy to produce in accelerators, and any primordial ones would have been diluted by inflation.  Despite the falsification,
the basic idea is still alive, because it is not hard to complicate the theory (for example by making it supersymmetric) and avoid the falsification.  Low energy supersymmetry lifts the expected proton life time from $10^{30}$ years to $10^{36}$ years, just outside the
range of current experiments (there are other potential sources of proton decay in supersymmetric theories, which must be carefully
avoided). The idea of group-theoretic unification is too compelling to give up on, but to strengthen the argument we must convince 
ourselves that the absence of fractional charges  in our universe could not have an anthropic origin.

It is obvious that the existence of  fractionally charged particles with suitable chosen masses and abundances can be potentially
catastrophic. 
For example, the electron might decay into two or more lighter half-integral charges. But then one would still have to rule
out atoms using some of these particles instead of electrons. Obviously the lightest fractionally charged particle would be stable,
and this would seriously frustrate chemistry, stellar dynamics and big bang nucleosynthesis. But  it is hard to see how one can turn
this into an anthropic argument against any such particle, regardless of mass, abundance, and charge. 

}\fi

\ifExtendedVersion {\color{darkgreen}
Hence we must conclude that the structure of the Standard Model gauge group strongly suggests an embedding in $SU(5)$ or a larger group, at
a more fundamental level. For a while it seemed reasonable that one day a new, more symmetric theory would be found
with a built-in GUT structure. 

Indeed, if Grand Unification is a fundamental law of physics, one might hope to find a theory that unequivocally predicts it.
But string theory is not that theory. It seemed like that for a while in 1984, when GUTs came out
``naturally" from Calabi-Yau compactifications of the $E_8\times E_8$ heterotic string, but within a few years it became
clear that GUTs are by no means the only possible outcome, and that furthermore the GUTs obtained from Calabi-Yau
related compactifications do not generically break in the correct way to the Standard Model gauge group. Let us first understand why GUT gauge groups come out so easily.}\fi

 \ifExtendedVersion {\color{red}
}\else {
If Grand Unification is a fundamental law of physics, one might hope to find a theory that unequivocally predicts it.
String theory is not that theory. It seemed like that for a while in 1984, when GUTs came out
``naturally" from Calabi-Yau compactifications of the $E_8\times E_8$ heterotic string, but within a few years it became
clear that GUTs are by no means the only possible outcome, and that furthermore the GUTs obtained from Calabi-Yau
related compactifications do not generically break in the correct way to the Standard Model gauge group. 

}\fi

\paragraph{Heterotic Strings.}
There are two equivalent ways of understanding why Grand Unification emerges so easily in $E_8\times E_8$ heterotic strings.
 In Calabi-Yau compactification this comes from the embedding of the $SU(3)$ holonomy
group of the manifold in one of the $E_8$ factors, breaking it to $E_6$, an acceptable but not ideal GUT group.  In world-sheet
constructions this is a consequence of the ``bosonic string map" \cite{Lerche:1986cx} used to map the fermionic (right-moving) sector
of the theory into a bosonic one, in order to be able to combine it in a modular invariant way with the left-moving sector. 
\ifExtendedVersion {\color{darkred} The bosonic string map takes the fermionic sector of a heterotic or type-II string, and maps it to a bosonic sector.
The world-sheet fermions $\psi^{\mu}$ transform under the $D$-dimensional Lorentz group $SO(D\!-\!1,1)$. 
The bosonic string map replaces this by an $SO(D+6)\times E_8$ affine Lie algebra, which manifests itself as a
gauge group in space-time.  In \cite{Lerche:1986cx} this trick was used to map the problem of finding modular invariants to
the already solved problem of characterizing even self-dual lattices.} \fi
This
automatically gives rise to a four-dimensional theory with an $SO(10)\times E_8$ gauge group and chiral fermions in the spinor
representation of the first factor.  
\ifExtendedVersion {\color{darkred}

Finding modular invariant partition functions for interacting CFTs is a much harder problem. But one can start
with a canonical solution, a symmetric pairing of the left- and the right-moving degrees of freedom, which is automatically modular
invariant. Unfortunately that is not very suitable for heterotic strings, where the right sector is fermionic and the left sector
bosonic. This is where the bosonic string map comes in. To get a modular invariant heterotic string one can start with a 
symmetric and modular invariant type-II string, and map the fermionic sector to a bosonic sector. In the process, one obtains
an $SO(D+6) \times E_8$ gauge symmetry for free. This was indeed the method used in \cite{Gepner:1987qi}. For $D=4$ this
yields $SO(10)$ (we ignore the $E_8$), and furthermore it is automatic that space-time fermions belong to the spinor
representation of that group. This is the perfect GUT theory. The 16-dimensional spinor of $SO(10)$ contains precisely
one family of the Standard Model, with a right-handed neutrino. }\fi

This $SO(10)$ group is seen by many as the ideal GUT group. The somewhat less  ideal $E_6$ appearing in typical
Calabi-Yau compactifications is an artifact of these constructions. 
\ifExtendedVersion{\color{darkred}
This has to do with
space-time supersymmetry. If one requires the spectrum to be supersymmetric in space-time, an additional is needed.
This goes by many names, such as GSO-projection, $\beta$-projection, spectral flow, or adding a spinor root to the chiral algebra. 
But in any cases it implies a modification of the right, fermionic sector of the theory. This violates modular invariance
unless we also change the left, bosonic sector. The canonical way of doing that is by adding a spinor root to
$SO(10)$, turning it into $E_6$. But there are other ways than the canonical one. In many cases, one can find
another operator that transforms under modular transformations exactly as the spinor root, but has a different
conformal weight (the eigenvalue of $L_0$, which is observed as mass). Consequently it is not visible in the
massless particle spectrum, and in particular $SO(10)$ is not extended to $E_6$. 

Therefore the appearance of $E_6$ in early string constructions is probably best viewed as a ``lamp-post"
effect. It is what happens in the most easily accessible fully symmetric constructions, but it is not generic.
The generic gauge group is $SO(10)$ with matter in the spinor representation.}

These $SO(10)$ GUTs are the best hope for believers in the uniqueness paradigm. There is indeed something unique about it:
the $(16)$ of $SO(10)$ is the smallest anomaly free complex irreducible representation for any Lie-algebra. But this
is spoiled a little because it occurs three times. Still, with only slight exaggeration one can state that this ideal GUT group
emerges uniquely from the heterotic string. All we had to do is specify the space-time dimension, $D=4$, and apply 
the bosonic string map, and we get $SO(10)$ for free. \fi

But this is as good as it gets. Nothing in the structure of the Standard Model comes out more convincingly than this. A
mechanism to break $SO(10)$ to
$SU(3)\times SU(2)\times U(1)$ can be found, but it does not come out automatically. Furthermore, it works less nicely
than in field theory GUTs. The heterotic string spectrum does not contain the Higgs representation used in field theory.
The breaking can instead be achieved by adding background fields (Wilson lines).

But in that case the full spectrum of these heterotic strings  will never satisfy (\ref{ChargeQuant}),
and it is precisely the deep underlying structure of string
theory that is the culprit. In a string spectrum every state is relevant, as is fairly obvious from the modular invariance
condition. Removing one state destroys modular invariance. In this case, what one would like to remove are the
extra gauge bosons in $SU(5) \subset SO(10)$ in comparison to $SU(3) \times SU(2)\times U(1)$. To do this one has to add something
else to the spectrum, and it turns out that the only possibility is to add something that violates (\ref{ChargeQuant}) and
hence is fractionally charged \cite{Schellekens:1989qb}. The possible presence of fractional charges in string spectra
was first pointed out by \textcite{Wen:1985qj}  and the implications were discussed further in \textcite{Athanasiu:1988uj}.

\ifExtendedVersion {\color{darkred}

 \paragraph{Fractional charges in Heterotic spectra.}

The occurrence of fractional charges in heterotic string spectra has been studied systematically for 
free fermion constructions 
and for heterotic Gepner models.  All these models realize the gauge group
in the canonical heterotic way, as a subgroup of $SO(10)$ (which may be further extended to $E_6$). There is a total
of four distinct subgroups that one may encounter within $SO(10)$. 
These subgroups are further subdivided into several classes, distinguished by the minimal electric charge quantum that occurs in their spectra. These
charge quanta are {\it not} determined by group theory in quantum field theory, but by affine Lie algebras in string theory. This gives a total of 
eight possibilities, with charge quanta given in curly brackets:
\begin{eqnarray*}
SU(3)\times SU(2) \times U(1) \times U(1)   &\ \ \ \{\frac{1}{6}, \frac{1}{3}, \frac12\} \\
SU(3)\times SU(2)_L \times SU(2)_R\times U(1) &\ \  \ \{\frac16,\frac13\} \\
SU(4)\times SU(2)_L \times SU(2)_R &\ \ \  \{\frac12\}
 \end{eqnarray*}
plus $SU(5)\times U(1)$ and $SO(10)$, which automatically yield integer charges.
This classification applies to all constructions in the literature where the Standard Model
is realized with level 1 affine Lie algebras, with a standard $Y$ charge normalization,
embedded via an $SO(10)$ group. The minimal electric charge {\it must} be realized in the
spectrum, but it is in principle possible that fractionally charged particles are vector-like (so
that they might become massive under deformations of the theory), have Planck-scale masses or
are coupled to an additional interaction that confines them into integer charges, just as QCD does
with quarks. 
}

But how often does this happen?
In \cite{Assel:2010wj}  a large class of free fermionic  theories with Pati-Salam spectra. These authors did find
examples with three families where all fractionally charged particles are at the Planck mass, but only in about $10^{-5}$ 
of the chiral spectra. In \cite{GatoRivera:2010gv,GatoRivera:2010xn,GatoRivera:2010fi,Maio:2011qn} a similar small fraction
was seen, but examples were only found for even numbers of families. These authors also compared the total number of spectra
with chiral and vector-like fractional charges, and found that about in $5\%$ to $20\%$ of the chiral,  non-GUT spectra the
fractional charges are massless, but vector-like. They also found some examples of confined fractional charges. 

If one assumes that in genuine string vacua vector-like particles will always be very massive, this is a mild landscape naturalness problem.
But avoiding fractional charges by chance is an unattractive solution. There may be a better way out. 
In orbifold models $SO(10)$ is broken using background gauge 
fields on Wilson lines. In this process fractional charges must appear, and therefore they must be in the twisted sector of the orbifold model.
If the Wilson lines correspond to freely acting discrete symmetries of the manifold (see \cite{Witten:1985xc}), the twisted sector
fields are massive, and hence all fractionally charge particles are heavy.  This method is commonly used in Calabi-Yau based constructions,
{\it e.g.} \cite{Anderson:2009mh}, but is chosen for phenomenological reasons, and hence this does not answer the question why nature would have chosen
this option. Also in the heterotic mini-landscape an example was found \cite{Blaszczyk:2009in}, but only after numerous examples with massless, vector-like
fractional charges. But these authors suggested another  rationale for using freely acting symmetries, namely that otherwise the Standard Model $Y$ charge
breaks if the orbifold singularities are ``blown up". It is not clear how that would impact models at the exact orbifold point without blow-up, but at least
it suggests a solution.

\else {

A possible  way out is that the fractional charges may all have Planck masses. They may also be vector-like, which means
that they may become massive under perturbations of the spectrum.
But how often does this happen?
 \textcite{Assel:2010wj}  have made a  survey of a large class of free fermionic  theories with Pati-Salam spectra. These authors did find
examples with three families where all fractionally charged particles are at the Planck mass, but only in a fraction of $10^{-5}$ 
of the chiral spectra. In \textcite{GatoRivera:2010gv,GatoRivera:2010xn,*GatoRivera:2010fi,Maio:2011qn} a similar small fraction
was seen, but examples were  found only for even numbers of families. These authors also compared the total number of spectra
with chiral and vector-like fractional charges, and found that in about  $5\%$ to $20\%$ of the chiral,  non-GUT spectra the
fractional charges are massless, but vector-like. 
They also found some examples of  fractional charges confined by an additional
gauge group ({\it i.e.} not QCD). 

If one assumes that in genuine string vacua vector-like particles will always be very massive, this is a mild landscape naturalness problem.
But avoiding fractional charges by chance is an unattractive solution. There may be a better way out. 
In orbifold models $SO(10)$ is broken using background gauge 
fields on Wilson lines. In this process fractional charges must appear, and therefore they must be in the twisted sector of the orbifold model.
If the Wilson lines correspond to freely acting discrete symmetries of the manifold (see \textcite{Witten:1985xc}), the twisted sector
fields are massive, and hence all fractionally charged particles are heavy.  This method is commonly used in Calabi-Yau based constructions,
{\it e.g.} \textcite{Anderson:2009mh}, but is chosen for phenomenological reasons, and hence this does not answer the question why nature would have chosen
this option. Also in the heterotic mini-landscape an example was found \cite{Blaszczyk:2009in}, but only after numerous examples with massless, vector-like
fractional charges. But these authors suggested another  rationale for using freely acting symmetries, namely that otherwise the Standard Model $Y$ charge
breaks if the orbifold singularities are ``blown up". It is not clear how that would impact models at the exact orbifold point without blow-up, but at least
it may point towards a solution.

}\fi
\ifExtendedVersion {\color{darkred}

Another disappointment from the perspective of the uniqueness paradigm  is that the natural appearance of $SO(10)$ is
really just a lamppost effect as well. Generic heterotic strings constructed using free fermion or bosonic methods can have
many other gauge groups. If $SO(10)$ comes out, that is just by choice.

}\fi


\ifExtendedVersion\else
In heterotic strings, the problem of fractional charges 
can also be avoided by considering  realizations of the gauge groups in terms of higher level affine Lie algebras
\cite{Lewellen:1989qe}.  One can even get GUT gauge groups \cite{Kakushadze:1996iw} with adjoint Higgses.
But this comes out only by choice, and the same is true for the fermion representations. Generically, these will have massless
higher rank tensor matter representations, which cannot occur for level 1 affine algebras.
\fi

\ifExtendedVersion {\color{darkred}

}\fi 

\ifExtendedVersion {\color{darkgreen}
 \paragraph{Higher Level Heterotic String GUTs.}

Within the context of heterotic strings there is another way of dealing with the unification problem. It is possible to
construct heterotic string theories with affine Lie algebras of levels higher than 1. The first example was the 
aforementioned $E_8$ level 2 in 10 dimensions, which is non-supersymmetric and tachyonic. In four dimension 
one can construct such theories as well
(\cite{Lewellen:1989qe}) and even get GUT gauge groups (\cite{Kakushadze:1996iw}). This removes one problem 
of the level 1 GUTs, namely that the gauge group can now be broken in the standard way used in field theory
GUTs, by means of a Higgs mechanism. By emulating  field theory GUTs one can reproduce their success  
The canonical case is the breaking of $SU(5)$ to $SU(3)\times SU(2)\times U(1)$.
This requires a Higgs boson in the adjoint representation of $SU(5)$, and matter in that representation be massless if
$SU(5)$ is realized as a level 1 affine algebra. Adjoints can only appear in the gauge boson sector, either as gauge bosons
or as gauginos, but not in the matter sector. Allowing higher levels solves that problem, but at a price. Adjoint representations
are now allowed, but so are other tensor representations. The beauty of the canonical heterotic GUTs is that only 
fundamental representation of $SU(3)$ and $SU(2)$ are allowed as massless states. This is a huge improvement over
quantum field theory, where there is no restriction on the representations. But this is partly lost if one considers higher
levels. The Standard Model can be ``accommodated", but there is no construction where it really comes out as naturally as
one might have hoped.
}\fi

 \paragraph{GUTs and Intersecting Brane Models.}

\ifExtendedVersion {\color{darkred} 
Yet another possibility to get the Standard Model is by means of stacks of intersecting branes or similar constructions, as
discussed in section \ref{OrientifoldsAndIntersectingBranes}. The three main classes discussed there allow various GUT groups, such as
the Pati-Salam group, trinification or $SU(5)$. Fractional charges are automatically avoided for open strings with both
ends on a Standard Model stack, in all classes. }\else {
In all three classes  of intersecting branes  depicted in Fig. \ref{GUTTrinification}, fractional charges are automatically avoided for open strings with both
ends on a Standard Model stack. }\fi
But this is partly by design: these brane configurations are constructed to give
at least all the particles in a Standard Model family, and then it turns out that there is no room anymore for additional matter.
\ifExtendedVersion {\color{darkred}
If additional branes are added that do not contribute to the Standard Model gauge group (as ``hidden" or ``dark matter" sectors), the
intersection of these branes with the Standard Model does give rise to particles with fractional electric charge, except in the $SU(5)$ class, where
charges are integer as in group-theoretical $SU(5)$ models. The fractional charges are 
half-integer
for the class of Fig. \ref{GUTTrinification}(a)  and $\pm x$ modulo integers for the  class of Fig. \ref{GUTTrinification}(c))}\else {
But if additional branes are added that do not contribute to the Standard Model gauge group (as ``hidden" or ``dark matter" sectors), they
carry a fractional charge $\pm x {\rm~mod}~1$ (with $x$ defined in Eq. (\ref{OrientifoldOptions})), so that only in the $SU(5)$ class all charges 
are integer.
}\fi

But even in this case, one cannot speak of true unification: intersecting brane
models in this class include cases (presumably the vast majority) where the $U(5)$ stack is pulled apart into a $U(3)$ and a $U(2)$
stack. This works equally well for getting the Standard Model representations, but without any  $SU(5)$ GUT group. This is essentially a realization of the $S(U(3)\times U(2))$ group that is
sufficient  to explain electric charge integrality for color singlets.
This substantially weakens any claim that understanding the structure of a 
Standard Model family requires a full GUT group. \ifExtendedVersion {\color{red}}\else { Furthermore intersecting brane GUTs allow massless symmetric rank-2
tensors \cite{Cvetic:2002pj}, which can only be avoided by carefully hand-picking spectra that do not contain them \cite{Anastasopoulos:2006da}.  }\fi
\ifExtendedVersion {\color{red}}\else {

In F-theory, GUT spectra were found only about twelve years after the invention of F-theory, and
it is therefore hard to argue that GUTs appear naturally. F-theory GUTs can be thought of as non-perturbative generalizations of the
intersection brane GUTs mentioned above, and similar remarks apply. In particular, they are an option, and not a prediction of string theory.
However, after making this choice and putting in some information about quark masses and mixings, a truly remarkable group-theoretic structure
emerges, which we will discuss in section \ref{LanSym}.
}\fi

\ifExtendedVersion {\color{darkred}
Intersecting brane $SU(5)$ 
also have a {\it disadvantage} with respect to
heterotic models. In heterotic GUTs a basic property of affine Lie algebra representations 
guarantees that only vectors and asymmetric tensors can appear -- precisely what is needed. But  in brane models
symmetric tensors are also allowed in principle, and indeed, in the first examples of brane $SU(5)$ models  \cite{Cvetic:2002pj} all spectra
had chiral symmetric tensors. In later work  \cite{Anastasopoulos:2006da} this problem was solved, but only by selecting spectra where the chiral symmetric tensors are absent. Since there is no obvious anthropic argument against symmetric tensors, the conclusion is once again
that the Standard Model group only comes out as a phenomenological constraint.
}\fi

\ifExtendedVersion {\color{darkred}
\paragraph{F-theory GUTs}

In F-theory, GUT spectra were found only about twelve years after the invention of F-theory, and
it is therefore hard to argue that they appear naturally. Since all research has focused on getting the Standard Model out -- with some
beautiful and fascinating results -- little can be said about alternative possibilities. However the situation is presumably comparable
to the that of intersecting brane $SU(5)$ GUTs, which is a limiting case: GUTs are input, not output.

}\fi

\ifExtendedVersion
\subsubsection{Coupling Constant Unification}\label{ConstUnif}\paragraph{Convergence.}
\else
\paragraph{Coupling Constant Unification}\label{ConstUnif}\fi
It has been known for decades that the three running gauge coupling constants converge to roughly the same value
at an energy scale a few orders of magnitude below the Planck scale. 
\ifExtendedVersion {\color{darkgreen}
This works provided one rescales the Standard Model 
$U(1)$ by  factor $\frac{3}{5}$  computable from the embedding in $SU(5)$.
After the precision measurements of LEP this
statement required some adjustment: the supersymmetric partners of all quarks and leptons have to be taken into account
above a scale just above 1 TeV. Even though no
evidence for supersymmetry has been found, the sensitivity to their mass scale is logarithmic, and hence coupling unification
may still be correct. Currently, this empirical fact still holds at the level of a few percent. Since any new physics at the unification
scale will introduce threshold corrections at the GUT scale, one can never do better than this from just low energy data. 
}\else {
This requires a GUT-motivated normalization of the $U(1)$ coupling and the assumption of low-energy supersymmetry.
}\fi

\ifExtendedVersion {\color{darkred}

\paragraph{The GUT scale}
An important consequence of GUT unification is proton decay. The lifetime of the proton depends on $M_{\rm GUT}^4$, where
$M_{\rm GUT}$ is the unification scale. In the early 80's the proton lifetime was predicted to be about $10^{30}$ years in the
simplest GUT models, tantalizingly just a little bit above the bounds known at the time.
But proton decay was not found, and the current limit is at about $10^{32}$ years. 
Supersymmetric unification moves $M_{\rm GUT}$ up towards the Planck scale, and enhances the proton lifetime to
about $10^{36}$ years, above the current bounds (but it also introduces new mechanisms for proton instability, which
must be avoided). The fact that the scale moved towards the Planck scale is fascinating in itself, and might  be taken as a hint
for a relation between Planck scale and GUT physics. Nevertheless, a gap of about three orders of magnitude remains.
}\fi

\begin{center}
\begin{figure}
\includegraphics[width=3.3in]{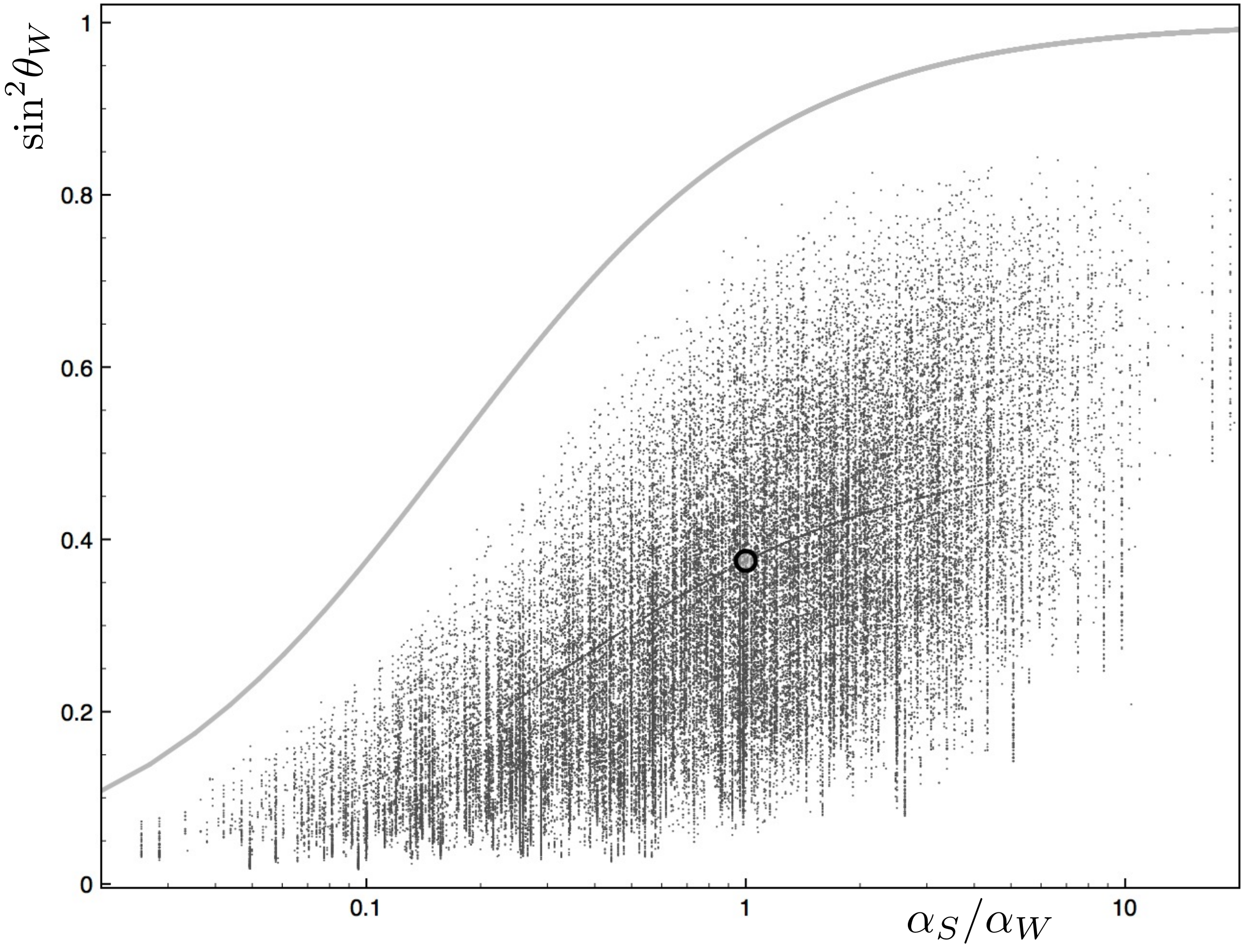}
\caption{Distribution of Standard Model Couplings in a class of intersecting brane models.}
\label{Couplings}
\end{figure}
\end{center}

\ifExtendedVersion \paragraph{Coupling Unification in String Theory.} \fi
Just as group theoretic unification, gauge coupling unification is not an automatic consequence of string theory, but
a phenomenological input. This is illustrated \ifExtendedVersion {\color{darkblue} for  a class of orientifold models}\fi  in Fig. \ref{Couplings}. Here a distribution of $\alpha_s/\alpha_w$ is plotted
versus ${\rm sin}^2 \theta_w$ for about 200.000 intersecting brane models obtained in \textcite{Dijkstra:2004cc}. These spectra
are of the Madrid model type depicted in  Fig. \ref{GUTTrinification}(a). Since the gauge couplings are not related, 
one would not expect them to respect gauge coupling unification, and indeed they do not. One gets a broad cloud of 
points around the GUT point, indicated by the black circle. In this corner of the landscape, coupling unification is a mere
coincidence.

In corners of the landscape with {\it group-theoretic} GUT unification, {\it coupling} unification is often problematic. This can perhaps
be attributed to the fact that string theory is simply more constraining than field theory, but it is still an indication that the perfect
string-GUT has not yet been found.

Heterotic  GUTs predict a  value for the unification scale that is substantially too large.
\ifExtendedVersion {\color{darkred}
In F-theory GUTs  the breaking of the $SU(5)$ GUT  group is usually achieved neither by Higgses in the $({\bf 24})$ (as in field theory)
nor by Wilson lines (as in heterotic strings) but  by $U(1)$ flux in the hypercharge direction  (see however \cite{Marsano:2012yc} for an
F-theory example with Wilson line breaking).
}\else {
In F-theory the breaking of the $SU(5)$ GUT  group is usually achieved neither by Higgses in the $({\bf 24})$ (as in field theory)
nor by Wilson lines (as in heterotic strings) but  by $U(1)$ flux in the hypercharge direction  (see however \textcite{Marsano:2012yc} for an
F-theory example with Wilson line breaking). }\fi
This may help solving
the notorious doublet-triplet splitting problem, but also spoils coupling unification (see \textcite{Blumenhagen:2008aw} and also
\textcite{Donagi:2008kj} for a discussion of various contributions to thresholds).
Since there
are often exotics that can contribute to the running it may still be possible 
to match
the observed low energy couplings, but this turns the apparent convergence into a strange accident. 

\ifExtendedVersion\paragraph{Unification versus anthropic arguments.}\fi
Coupling constant unification could lead to a clash between anthropic tuning and fundamental symmetries.
\ifExtendedVersion {\color{darkred}
 If the low energy values of the
three couplings $g_1, g_2$ and $g_3$ are all tightly anthropically constrained, it would be strange if they could still meet each other at a higher scale.
Put differently, to optimize the Standard Model for life, it would be better not to be constrained by a coupling constant relation, unless this is
an inevitable feature of a fundamental theory. In the string landscape, it is not.}\else {
To optimize the Standard Model for life, it would be better not to be constrained by a coupling constant relation, unless this is
an inevitable feature of a fundamental theory. In the string landscape, it is not.
}\fi

Of the three constants, $g_3$ is indeed anthropically constrained. It determines $\Lambda_{\rm QCD}$ and the proton mass. We will discuss this in section 
\ref{SMScales}. The weak coupling $g_2$ is much less constrained: thresholds of weak
decays are much more important than the decay rates themselves. The constraints on $g_1$, or almost equivalently on $\alpha$, are discussed
below. It does not
appear to be tightly constrained, except perhaps in fine-tunings of certain nuclear levels. Unless these are much more severe than we currently
know, coupling unification would not get in the way of anthropic constraints. It has two free parameters, a mass scale and the value of the unified
coupling at that scale, which allow sufficient freedom to tune both $\Lambda_{\rm QCD}$ and $\alpha$. Alternatively, one could argue that the value of $\Lambda_{\rm QCD}$ is tuned to its anthropic value by means of tuning of $\alpha$, {\it assuming} Grand Unification
\cite{Carr:1979sg,Hogan:1999wh}. 

\paragraph{Just a Coincidence?}

Standard model families have an undeniable GUT structure. One might have hoped
that a bit more of that structure would emerge from a fundamental theory in a ``natural" way,
even taking into account the fact that part of this structure has  anthropic relevance. 
GUTs can be found in several areas of string theory;
see \textcite{Raby:2011jt} for a review. But a compelling top-down argument in favor of GUTs is missing. Both group-theoretical and coupling unification
are options in string theory, not predictions.
Nevertheless, one could still speculate that Grand Unification is chosen in the string landscape either because
GUTs are statistically favored -- despite suggestions that symmetry is {\it not} favored \cite{Douglas:2012bu} -- or that it offers anthropic advantages.
For example, it might turn out to play a r\^ole in  inflation or baryogenesis after all, although the originally proposed 
GUT-based baryogenesis mechanism does not work.

But  is it just a coincidence that the three running coupling constants seem to converge to a single point, close to, but just below 
 the Planck scale?  It would not be the only one. The little-known mass formula for leptons pointed out by \textcite{Koide:1983qe},
 $m_e+m_{\tau}+m_{\mu}=\frac23 (\sqrt{m_e}+\sqrt{m_{\mu}}+\sqrt{m_{\tau}})^2$,
 is seen by most people as a coincidence,
 because it relates pole masses at different mass scales. But it predicts the $\tau$ mass correctly with $0.01\%$ accuracy, a whole lot better than
 the few percent accuracy of GUT coupling unification.
 Another  potential
coincidence, allowed  by the current data within two standard deviations,   is that the 
self-coupling of the Higgs boson might run towards zero  with vanishing $\beta$-function, exactly at the Planck mass \cite{Bezrukov:2012sa},
a behavior predicted in the context of asymptotically safe gravity (see however  \textcite{Hebecker:2012qp} for an alternative idea in string theory).
Note that this coincidence is incompatible with GUT coupling unification: the latter requires
low-energy supersymmetry, but the former requires a pure Standard Model. So at least one of these two coincidences must be just that.

 \subsubsection{\label{Alpha}The Fine-structure Constant}

The fine-structure constant enters in nearly all anthropically relevant formulas, but it is often not very sharply constrained. 
Rather than tight constraints, one gets a large number of hierarchies of scales, such as sizes
of nuclei, atoms, living beings, planets, solar systems and galaxies, as well as time scales and 
typical energies of relevant processes. See \textcite{Press:1983rs,Carr:1979sg,BarrowTipler,Bousso:2009ks} for attempts to express these scales in terms of fundamental parameters, usually including $\alpha$.

An example of a hierarchical condition is the requirement that the Bohr radius should be
substantially larger than  nuclear radii, {\it i.e.} $\alpha (m_e/m_{p}) \ll 1$, presumably anthropically
required, but not a very strong restriction on $\alpha$. A  stronger condition follows from
the upper and lower limits of stellar masses \cite{BarrowTipler}
\begin{equation}\label{StellarMassWindow}
 \left( \frac{\alpha^2 m_p}{m_e}\right)^{3/4} N m_p \lesssim M_{\star} \lesssim 50\ N m_{p} \ ,
\end{equation}
where $N$ is the typical number of baryons in a star, $N=(M_{\rm Planck}/m_p)^3$. Requiring that the upper limit
be larger than the lower one yields $\alpha^2 \lesssim 200 (m_e/m_p)$, or $\alpha \lesssim 0.3$.
See \textcite{Barnes:2011zh} and chapter IV of \textcite{Tegmark:1997qn} for fascinating plots of many other limits.

The value of $\alpha$ is constrained from above by the competition between strong and electromagnetic interactions.
The electromagnetic contribution to the neutron-proton mass difference is about 0.5 MeV and proportional to $\alpha$. Changing
$\alpha$ by a factor of three destabilizes the proton, but this is far from determining $\alpha$. In nuclei, total strong interaction binding energies
scale with the number of nucleons $N$, electromagnetic repulsion energy scales as $\alpha N^2/R$, and $R$ scales as $N^{1/3}$.  Hence
the maximum number of nucleons in a nucleus scales as $\alpha^{-3/2}$ \cite{Hogan:1999wh}. Increasing $\alpha$ by a factor of three 
implies drastic changes, but also here a tight bound is hard to obtain. The precise location of nuclear levels is much more sensitive to $\alpha$, 
and might give tight lower and upper bounds, for example via the Beryllium bottleneck. But to draw any conclusions one would have to recompute all potentially
relevant nuclear levels and all types of nucleosynthesis. As a function of $\alpha$, levels may not just move {\it out of} convenient locations, but also
{\it into} convenient locations.

A lower bound on $\alpha$ can be derived from limits on the CMB fluctuations $Q$  \cite{Tegmark:1997in}. In our universe, $Q\approx 10^{-5}$.
If $Q$ is too large, galaxies would be too dense and planetary orbits would be disrupted too frequently; if $Q$
is too small the galaxies could be unable to form stars or retain heavy elements after a supernova explosion. 
Clearly these are not strict limits, but taking them at face value one finds that
 the anthropic upper limit on $Q$ is
$\approx 10^{-4}$, and scales with $\alpha^{16/7}$, whereas the lower limit is $Q \approx 10^{-6}$, scaling with 
$\alpha^{-1}[{\rm ln}(-\alpha)]^{-16/9}$. For smaller $\alpha$ the upper limit decreases and the lower limit
increases. The window closes if $\alpha$ is about a factor five smaller than $1/137.04$.
This assumes everything
else is kept fixed. 
Although the origin of the $\alpha$-dependence is a complicated matter, the fact that a lower bound is
obtained is ultimately traceable  to the need for electromagnetic cooling of matter in galaxy formation, and
the r\^ole of electromagnetic radiation in the functioning of the sun. Obviously, switching off electromagnetism is bad
for our health.


\ifExtendedVersion {\color{darkred}
Although the fine-structure constant is obviously 
relevant in chemistry, the lack of competition with another force makes this an unlikely area for fine-tunings. 
In \textcite{PhysRevA.81.042523} the dependence
of the chemistry of life-supporting molecules on $\alpha$ as well as the electron/proton mass ratio is studied. This includes
the bond angle and bond length of the water molecule, and many reaction energies. Substantial changes are
found only if $\alpha$ is increased by an order of magnitude or more. Ultimately the special properties of water
(expansion upon freezing) may be destroyed, but there is no useful limit here. It cannot be excluded that
somewhere in the chemical reaction chain leading to complex biochemical molecules   there are bottlenecks
that strongly depend on $\alpha$, but we may never know.}\fi

\ifExtendedVersion {\color{darkred}
\textcite{Lieb:1988hn} proved the existence of an upper bound on $\alpha$  based on the stability of many-body
systems ({\it i.e.} many nuclei and many electrons), but the value of that bound is too uncertain to be of any
relevance.
}
\fi

The competition between gravity and electromagnetism in stars is another place to look for 
anthropic relations. An interesting one concerns the surface 
temperature of typical stars compared to the ionization temperature of molecules, $T_{\rm ion} \approx \alpha^2 m_e$.
 These two temperatures are remarkably close. 
Since the former temperature depends on the relative strength of gravity and the latter does not, the coincidence implies
a relation between the strength of the two interactions.
Equating these temperatures gives the fascinating relation
\begin{equation}\label{CarterCoincidence}
\alpha^{6}\left(\frac{m_e}{m_{p}}\right)^2 \approx \left(\frac{m_{p}}{M_{\rm Planck}}\right).
\end{equation}
Numerically, both sides of this relation are $4.5 \times 10^{-20}$ and $7.7\times 10^{-20}$. Although this
is
 close, the actual temperatures are proportional to the fourth root of these numbers so that the sensitivity is less
than the formula suggests (often the square of this relation is presented, making it look even more spectacular).
But does the closeness of those two temperatures have any anthropic significance?
Carter has conjectured that it might.  Due to the temperature coincidence,  typical stars are on the dividing
line between radiative and convective, and he argued that this might be linked to their ability to form  planetary systems
(see \textcite{Carr:1979sg,BarrowTipler} for a discussion). Perhaps a more credible relation was suggested by \textcite{Press:1983rs}, who argued
that   solar radiation would either be too damaging or not useful for photosynthesis if
these temperatures were very different. 
\ifExtendedVersion {\color{darkred}

In a substantial part of the anthropic literature, starting with \textcite{Carr:1979sg}, GUT relations are used to link the value
of $\alpha$ to the mass of the proton via logarithmic running. But it in the spirit of the discussion in sec. \ref{WhatVaries} it is
better not to do that. In the Standard Model there is a clear decoupling between known physics and speculative physics
(GUTs or strings), and one should therefore consider unconstrained variations of $\alpha$. Such variations are physically
meaningful even if we find evidence for GUTs. Since the only landscape we can talk about, the string theory landscape, 
does not impose GUT unification, there is even less reason to impose it as a constraint on valid low energy physics.
In the string landscape, there is no justification for the statement that ratio $m_{p}/M_{\rm Planck}$ is 
forced to small values by tuning $\alpha$ (see {\it e.g.} \cite{Hogan:1999wh}.}\fi

 \subsection{Masses and Mixings}\label{QuarkMasses}
\def\MEV{{\rm MeV}}

 \subsubsection{Anthropic Limits on Light Quark Masses}

In the Standard Model quark masses are eigenvalues of Yukawa coupling matrices $\lambda$ multiplied by the Higgs vev $v$. 
Therefore anthropic constraints on these masses take the form of long elongated regions in the Standard Model $(\lambda,v)$ parameter space, with rescalings
in $\lambda$ compensating those of $v$. 
All constraints come from the effect of changes in the quark masses on QCD, and do not depend on the origin of these masses. 
\ifExtendedVersion {\color{darkred}

Several slices through this parameter space have been considered in the literature. One can vary Yukawa couplings while keeping $v$ fixed. This allows
masses of up to 200 GeV in order to avoid Landau poles in the couplings. One may also vary $v$ while keeping the Yukawa couplings fixed. This allows 
arbitrarily large quark masses, but the electron mass also varies, so that the sign of quantities like $m_d-m_u-m_e$ cannot be flipped. In all cases a decision
has to be made what to do with $\Lambda_{\rm QCD}$. Some authors ({\it e.g.} \textcite{Agrawal:1997gf}) assume a fixed coupling at a high scale (inspired by GUTs) and take into account
renormalization group corrections to $\Lambda_{\rm QCD}$ caused by the other parameter changes. Others vary $\Lambda_{\rm QCD}$ so that 
average nucleon masses remain unchanged  \textcite{Jaffe:2008gd}. One may also simply leave $\Lambda_{\rm QCD}$ unchanged, and vary quark masses freely.
Any choice of masses can be realized in the Standard Model, so one could in principle explore the full space of possibilities for all six quarks.
If one also allows the charged lepton and neutrino masses to cover the full range, one gets a  complicated patchwork of regions with different degrees of habitability.
But most research in this area has focused on our own neighborhood, assuming two or three quarks are light. }\fi
An early discussion of the environmental impact of fermion masses \ifExtendedVersion {\color{darkred} -- carefully avoiding mentioning anthropic implications -- }\fi  can be found in \textcite{Cahn:1996ag}. 

The only admissible variations in hadronic and nuclear physics are those that can be derived from variations
in the relevant Standard Model parameters: the QCD scale $\Lambda_{\rm QCD}$, and the dimensionless ratios
\begin{equation}
\label{QCDpars}
\frac{m_u}{\Lambda_{\rm QCD}},\ \frac{m_d}{\Lambda_{\rm QCD}},\ \frac{m_s}{\Lambda_{\rm QCD}} \ ,
\end{equation}
although we will often just write $m_u, m_d$ and $m_s$. 
The strange quark is light enough to make a sizable contribution to nucleon masses  by virtual processes (see \textcite{Kaplan:1989fc}) and some authors 
take its variation into account \cite{Jaffe:2008gd}, even allowing it to become as light as the $u$ and $d$ quarks.
In the limit $m_u=m_d=0$, the chiral limit, the theory
has an exact $SU(2)_L\times SU(2)_R$ symmetry, which is spontaneously broken. In this limit the pion, the Goldstone boson of the
broken symmetry, is exactly massless. In the real world it has a mass proportional to $\sqrt{\Lambda_{\rm QCD} (m_u+m_d)}$, and the pions are
the only hadrons whose mass vanishes in the chiral limit. All other hadron masses are proportional to $\Lambda_{\rm QCD}$. 

\ifExtendedVersion\else
{
In the parameter plane (\ref{QCDpars}) one would like to know the location of several interesting anthropic boundary lines: the
stability line of ${\rm (}^1{\rm H)}$, the combined stability line of {\it all} forms of hydrogen, including deuterium and tritium, the stability lines of di-nucleons, and the
stability lines of all elements thought to be anthropically essential, as well as contour plots of all abundances. We are still very far
from all that, and one can also argue about anthropic necessities. For example, deuterium and tritium can take over the r\^ole  of ${\rm (}^1{\rm H)}$
in biochemistry. If deuterium and all other di-nucleons are unstable, synthesis of all elements from nucleons would  have to start with three-body
processes, but hydrogen stars could simply get hotter and denser until this happens. Keeping all these caveats in mind, let us see where
some of these lines are.
}
\fi


\ifExtendedVersion {\color{darkgreen}
Ideally, one would like to have a contour plot of the various anthropic constraints in the parameter plane (\ref{QCDpars}).
Many older papers studying these effects discuss them in terms of strong interaction parameters that cannot be varied independently in QCD. 
The superior method for QCD computations is lattice gauge theory, because it can in principle fully simulate the full non-perturbative theory. In practice, however, it
is limited to relatively simple operators, and has difficulties reaching the chiral limit $m_u, m_d\rightarrow 0$ because the quark Compton wave length exceeds the
lattice size. The next best technique in this limit is chiral perturbation theory, which treats the quark masses as small perturbations of the chiral limit theory. Other 
techniques that are used include the MIT bag model, the Skyrme model and meson-exchange nucleon-nucleon potentials.

The quark mass dependent anthropic bounds are related to the existence and stability of matter relevant
for complex chemistry and biology, the abundances of this matter due to 
big bang and stellar nucleosynthesis, stellar lifetimes and energy
production in stars, roughly in order of decreasing anthropocentricity. Some important potentially catastrophic boundary lines one may cross when changing parameters are:

\begin{itemize}
\item{Instability or absence of hydrogen ${\rm (}^1{\rm H)}$.  At this boundary line our kind of life ceases to exist, but there is no good reason why deuterium (or even tritium) could not take
over its r\^ole in biochemistry.  Some life forms on earth tolerate heavy water quite well, and did not even evolve in a pure deuterium environment.
A bigger worry is stellar burning, which in our universe relies heavily on hydrogen fusion, and would be drastically different.
Even without hydrogen there are still plenty of alternatives, but it is not
clear whether such stars would have the right lifetimes and other properties to allow biochemical evolution. Finally, beyond the hydrogen stability line the neutron becomes stable, and
on free neutron background can have a variety of nasty consequences \cite{Hogan:2006xa,Cahn:1996ag}. Note that Hydrogen instability by electron capture occurs before free proton instability, because
the latter costs $2m_ec^2$ more energy.
}
\item{Instability of all di-nucleons. Beyond this line any kind of synthesis of heavier elements from nucleons would have to start with three-body processes. In our universe stable deuterium serves as a stepping stone. 
Its importance can be observed in the deuterium bottleneck in nucleosynthesis.
Although deuterium is stable, it is disintegrated by photons. Therefore $^4$He synthesis only starts after the photon density has dropped sufficiently.
Even without stable deuterium, nucleosynthesis in stars may still be possible in extremely dense environments. Furthermore, even an
unstable, but long-lived deuterium would still enhance the rates. But in any case, beyond the di-nucleon stability line we are in
terra incognita.
}
\item{Instability or all hydrogen isotopes.  Beyond this line there is no stable hydrogen, deuterium or tritium (nor higher isotopes).
One would have to argue that life can be based purely on complex molecules without hydrogen of any kind. } 
\item{Instability of all heavy elements.  Beyond this line any  elements beyond Helium become unstable.
It does not matter much which one we choose, because 
their instability boundaries are close to each other. So one can use ${12\atop 6}{\rm C}$ (one of the most strongly bound ones) as a benchmark.
The instability of these elements would deprive us of any scenario for the existence of complexity. Low abundance of these elements is a less serious issue, because it is hard to decide what the minimal
required abundance is.}
\end{itemize}

Now let us see how these lines are crossed if we vary Standard Model parameters. It is convenient to consider
variations along the  $m_u-m_d$ and $m_u+m_d$  axes (isospin violating and isospin preserving), because the effects of these variations are rather different.

}\fi
\paragraph{The proton-neutron mass difference.}

\ifExtendedVersion {\color{darkgreen}
The most obvious feature of the quark masses is
 the extremely small up quark mass. This is especially noteworthy since in the two heavy families the charge $\frac23$ quarks are considerably heavier than the $-\frac{1}3$ 
quarks. If the up and down quark masses were equal the proton is expected to be heavier than the neutron because of the larger electromagnetic repulsion. 
The relevant parameter for weak decay of the proton or neutron is $\Delta=m_n-m_p-m_e=.782$, in our universe.  We will keep neutrino masses negligibly small in this section. 
To relate $\Delta$ to quark masses we have to overcome the problem that quarks are confined, so that their masses 
can only be measured indirectly.
Furthermore, like all Standard Model parameters, the quark masses depend on the energy scale at which they are measured. The Particle data group gives the
following masses for the three light quarks (in the $\overline{\rm MS}$ scheme at a scale of $\approx 2$ GeV)
\begin{eqnarray*}
&m_u &= 2.3^{+0.7}_{-0.5}
  \MEV\ \   \\
&m_d &= 4.8^{+0.7}_{-0.3}\  \MEV   \\
&m_s &= 95 \pm 5 \MEV 
\end{eqnarray*}

The scale at which these are defined is not the right one for computing nucleon masses. For example, the proton neutron mass difference has a contribution
equal to $m_u-m_d$, provided one uses quark masses at the correct scale. But the exact scale is hard to determine, and running the quark masses to low
energy scales is complicated because perturbation theory breaks down. 
An empirical way of dealing with this is to scale the $m_u-m_d$ mass difference so that it equals the proton neutron mass difference
minus the estimated electromagnetic contribution.  The latter is $\epsilon_{\rm EM}\approx .5\ \MEV$, and is to first approximation proportional to $\alpha\Lambda_{\rm QCD}$ (see \cite{Quigg:2009xr} for more details).
Hence the mass difference, in terms of the quarks masses given above, is
\begin{equation}\label{Zscaling}
m_{n}-m_{p}=Z(m_d-m_u)-\epsilon_{\rm EM}
\end{equation}
}
Here $Z$ is an empirical scale factor, relating quark masses defined at some high scale to the observed mass difference. This parametrizes
renormalization group running, which cannot be reliably calculated at low energy. The electromagnetic mass
difference  $\epsilon_{\rm EM}\approx 0.5\ \MEV$ is to first approximation proportional to $\alpha\Lambda_{\rm QCD}$ (see \textcite{Quigg:2009xr} for more details). For the quark masses at 2 GeV quoted by the Particle Data Group  \cite{PDG} 
one gets $Z=0.7$. 

{\color{darkgreen}
The aforementioned hydrogen stability line is crossed when this quantity changes sign. What happens after that
is {\it qualitatively} clear. As we move towards more negative values all nuclei become unstable, because the proton-neutron mass difference overcomes
the binding energy and protons inside nuclei can decay. Analogously, if we increase $m_n-m_p-m_e$ the neutron becomes less stable and can decay within nuclei. 
Since nuclei with only protons do not exist, this implies also that all nuclei decay. 
}
\else
 {
The most obvious feature of the quark masses is
 the extremely small up quark mass. This is important, because the Coulomb interaction tends to make the neutron lighter than the proton, and
 the $m_d-m_u$ quark mass difference overcomes that. The proton-neutron mass difference can be parametrized as follows  \cite{Damour:2007uv}.
 \begin{equation}\label{Zscaling}
m_{n}-m_{p}=Z(m_d-m_u)-\epsilon_{\rm EM}.
\end{equation}
Here $Z$ is an empirical scale factor, relating quark masses defined at some high scale to the observed mass difference. This parametrizes
renormalization group running, which cannot be reliably calculated at low energy. The electromagnetic mass
difference  $\epsilon_{\rm EM}\approx 0.5\ \MEV$ is to first approximation proportional to $\alpha\Lambda_{\rm QCD}$ (see \textcite{Quigg:2009xr} for more details). For the quark masses at 2 GeV quoted by the Particle Data Group  \cite{PDG} 
one gets $Z=0.7$. 
}\fi

If $m_d-m_u$ is {\it increased}, the neutron becomes less stable, so that it starts decaying within nuclei. Since neutrons
are required for nuclear stability, this eventually implies instability of all nuclei. 
 If $m_d-m_u$ is {\it decreased}, the proton 
becomes unstable. First the hydrogen atom becomes unstable against electron capture, for  a slightly higher value the free proton can
decay, and eventually all nuclei become unstable. It is convenient to express all limits in terms of the available energy, $\Delta=m_n-m_p-m_e$
in neutron decay. We will assume that neutrino masses remain negligible.
From electron capture and $\beta$ decay of nuclei one gets respectively the following limits
\begin{equation*}
M(A,Z)-M(A,Z\!-\!1)  < \delta(\Delta) < M(A,Z\!+1\!)-M(A,Z).
\end{equation*}
The masses $M(A,Z)$ used here are {\it atomic} masses, and hence include electron masses. 
\ifExtendedVersion {\color{darkred}
Changes in atomic binding energies are ignored. Furthermore it is assumed here that the proton or neutron
are not ejected from the nucleus. This would lead to only marginally different bounds, {\it c.f.} \textcite{Jaffe:2008gd}.
Near the nuclear stability line the first bound is usually negative and the second usually positive.
If one of these limits is exceeded, the atom $(A,Z)$ is not stable.

To obtain exact {\it quantitative results} for these thresholds one has to take into account changes in binding energies as a function of the quark masses. 
This effect is probably small (and zero if only $m_e$ is varied) because the result depends only on {\it differences} in binding energy of nuclei with the same numbers of nucleons
and the strong force is isospin invariant. 
One may hope that the changes are minimized if $m_u+m_d$ is kept fixed, because then the mass of the pions and several other relevant mesons does not change, and hence
binding energies may not change much either. But it is not always possible to keep $m_u+m_d$ fixed, and vary $m_u-m_d$ while keeping the masses positive.

Keeping all these caveats in mind, we get the following limits for some nuclei of interest:
\begin{eqnarray*}
^1{\rm H}   ~~~~~~~  &   ~~~~~0 ~~~ <  ~~~~~~~ \Delta&   \\
^2{\rm H}   ~~~~~~~  &   -2.2\ \MEV < \Delta&  <  2.2\ \MEV \\
^3{\rm H}    ~~~~~~~  &  -8.5\ \MEV  <   \Delta& < .762\ \MEV \\
^4{\rm He}    ~~~~~~~  &  -22.7\ \MEV  <   \Delta& < 23.6\ \MEV \\
^{12}_{~6}{\rm C}   ~~~~~~~  &  -12.6\ \MEV <   \Delta&\ < 18.12\  \MEV \\
^{14}_{~7}{\rm N}   ~~~~~~~  &  \phantom{-}.62\ \MEV <   \Delta&\ < 5.92\  \MEV \\
^{16}_{~8}{\rm O}   ~~~~~~~  &  -9.6\ \MEV <   \Delta&\ < 16.2\  \MEV
\end{eqnarray*}
}\fi
\ifExtendedVersion {\color{darkred} To obtain absolute bounds one should also consider proton-rich or neutron-rich
nuclides that are not stable in our universe, but may be stable for different values
of $\Delta$, such as $^{9}_{6}{\rm C}$.  In \textcite{Agrawal:1997gf} arguments are given against the stability of proton rich ($Z \gg N$) nuclei. Even taking all this into account,} the maximum variation in $\Delta$ is about $\pm 25 \MEV$ (which translates
to $\pm 35$ for the quark masses given earlier). Beyond that no stable nuclei exist. This is a very conservative bound. Long 
before reaching this bound catastrophic changes occur, and there is no guarantee that the few stable nuclei can actually be synthesized.

\else 
{
The maximum variation in $\Delta$ is about $\pm 25\ \MEV$ (which translates
to $\pm 35$ MeV for the quark mass differences). Beyond that point no stable nuclei exist. This is a very conservative bound, which 
does not depend much on details of nuclear binding. Long 
before reaching this bound catastrophic changes occur, and there is no guarantee that the few stable nuclei can actually be synthesized.}\fi

\ifExtendedVersion {\color{darkred}
Note that only the linear combination $m_n-m_p-m_e$ is bounded by these arguments. \cite{Hogan:1999wh} has observed that the reaction
\begin{equation}
p+p \rightarrow D + e^+ + \nu
\end{equation}
is sensitive to $m_n -m_p +m_e$, and shuts down if this quantity is increased by more than $.42\  \MEV$. 
This reaction is a step in hydrogen burning in the sun and hence this is definitely a boundary of {\it our} domain in parameter space.
However, \cite{Weinberg:2005fh} has pointed out that one replace the outgoing $e^+$ by
an ingoing $e^-$. The resulting three-body interaction also does the job (although less efficiently), but only restricts $m_n -m_p -m_e$. 
}\fi

\paragraph{Nuclear binding.} 
While it is intuitively obvious that increasing or decreasing $m_u-m_d$ by a few tens of $\MEV$ in both directions 
will lead to instability of all nuclei, this is far less obvious for variations in $m_u+m_d$.  An intuitive argument is suggested
by the lightness of the pion. The pion mass increases with $\sqrt{m_u+m_d}$, which decrease the range of the one-pion
exchange potential, and this could make nuclei less stable. 
But one-pion exchange is not a correct description of nuclear physics. In the literature, estimates have been given of the
effect of quark mass changes on binding of heavy nuclei  based on effective field theory and models for nuclear matter. 
In \textcite{Damour:2007uv} the binding energy per nucleon for heavy nuclei \ifExtendedVersion {\color{darkred} ($^{16}{\rm O}$ and $^{208}{\rm Pb}$ were used as
benchmarks) }\fi
was studied  as a function of scalar and vector contact interactions. \ifExtendedVersion {\color{darkred}
These give contributions to the binding energies that are large, and opposite in sign.  
Scalar interactions give a negative contribution that is an order of magnitude larger than the actual binding energy, and
receive important contributions from two-pion exchange. The latter decreases with increasing pion mass. Because of the
large cancellations, only a moderate increase in pion mass would make the positive contributions overwhelm the negative one
and destabilize nuclei. These arguments may not convince QCD purists, but do they provide a rationale for substantial 
dependence of nuclear binding on $m_u+m_d$.  }\fi According to these authors, a conservative estimate for the
maximum allowed increase in $m_u+m_d$ is about $64\%$. 

\paragraph{Bounds on the Higgs vev.}\label{HiggsVev}

The limits discussed above are often expressed in terms of allowed variations of the Higgs vacuum expectation value, under the
assumption that the Yukawa couplings are kept fixed. 
The upper bound of $\Delta$ of 25\ $\MEV$ translates into an upper bound on $v/v_0$ (where $v_0$ is the observed value) of about 20. The negative lower bound has no effect, because $v$ cannot be negative. But if one just requires stability of hydrogen $^1{\rm H}$ under
electron capture, the bound is $\Delta > 0$, which  implies 
(but note that the error in $m_d-m_u$ is huge) 
\begin{equation}
\frac{v}{v_0} > \frac{\epsilon_{\rm EM}}{Z(m_d-m_u)-m_e}  \approx 0.4\ .
\end{equation}
Here we used the method of \textcite{Damour:2007uv}; in \textcite{Hogan:2006xa}  the lower bound was estimated as $0.6 \pm 0.2$ using
lattice results on isospin violation \cite{Beane:2006fk}.
If we also use the more model-dependent nuclear binding bounds, the window for $v/v_0$ is quite small, $0.4 < v/v_0 < 1.64$.

Limits on $v/v_0$ were first presented by \textcite{Agrawal:1997gf}, who estimated an upper limit $v/v_0 < 5$, from a combination of the two 
arguments  on stability of nuclei discussed above. 
In this work the Higgs mass parameter $\mu^2$ is varied over its entire range, from $-M_{\rm Planck}^2$  to $+M_{\rm Planck}^2$, while
keeping all other parameters in the Lagrangian fixed. Then if  $\mu^2$ is negative, $v=\sqrt{-\mu^2/\lambda}$, and 
$v/v_0$ can lie anywhere between $0$ and $10^{17}$ GeV.  The anthropic range is 
 in any case extremely small in comparison to the full allowed range. 
 \ifExtendedVersion {\color{red}
 }
 \else 
 {
 Note that for $v/v_0 > 10^3$ a qualitative change occurs, because the stable particle will be the 
$\Delta^{++}$ instead of the proton; however this is not expected to improve the odds for complex life.

The interesting and important case $\mu^2 > 0$ -- no Higgs mechanism, but quarks and leptons getting a mass from the pion vev -- is also
discussed in these papers; see also \textcite{Quigg:2009xr}. The arguments against this case rest on the electron mass becoming too small, so that
all matter increases in size and decreases in average density and typical biochemical temperatures are reduced.
}\fi

 \ifExtendedVersion {\color{darkgreen}
 Because of the possible implications for the gauge hierarchy problem, it is important to exclude the entire range, not just the small region around us. If $v$ is increased to much larger values, there is still an important qualitative change.
As the $m_d-m_u$ mass differences increases, at some point the lightest three up-quark state will be lighter than the proton. This
is the $\Delta^{++}$, about 300 MeV heaver than the proton because of QCD effects.  There is a small range where both the
proton and the $\Delta^{++}$ are stable.  
If there are no nuclear bound states (as seems plausible) of these
objects this universe is comparable to a pure helium or helium plus hydrogen universe, and probably lifeless \cite{Agrawal:1997gf}. }\fi
\ifExtendedVersion {\color{red} 
}
\else 
{
An updated discussion of  bounds on quark masses can be found in \textcite{Barr:2007rd}. They also consider the possibility
of having separate up and down quark Higgs bosons, each  with variable scales, while the Yukawa couplings are kept fixed. }\fi
\ifExtendedVersion {\color{red}}\else {   }\fi

\ifExtendedVersion {\color{darkgreen}
One may apply analogous arguments  in the two-Higgs case. In many models, there are separate Higgs bosons for up and down quarks.
Then an interesting variation is to keep all Yukawa couplings fixed and allow the two Higgs scales to vary independently.  Then larger
hierarchy in the up-quark sector is taken as a given. \cite{Barr:2007rd} argue that rough equality of $m_u$ and $m_d$ is anthropically
required, so that the large ratio of the top quark and bottom quark mass are predicted.].

Although the weak scale is linked to the Higgs mechanism, which operates for $\mu^2 < 0$, it is also interesting to see if 
the mechanism itself is anthropically needed, in other words if life plausibly exists for $\mu^2 > 0$.
For positive $\mu^2$ the Higgs potential does not have the familiar ``Mexican hat" shape, but now the weak interaction symmetry is broken by
quark condensates, $\langle \bar q q\rangle \propto f_{\pi}^3$, where $f_{\pi}$ is the pion decay constant. (see \textcite{Quigg:2009xr} for a interesting discussion of
closely related cases.) 
This is the usual chiral symmetry breaking mechanism of QCD, except that all quark flavors participate democratically, because
they are all massless prior to chiral symmetry breaking.
These vacuum expectation values feed into the Higgs system via the Yukawa couplings, and the dominant contribution
comes from the top quark. This generates a vev for the Standard Model Higgs boson, which in its turn generates all quark and lepton masses
via the Yukawa couplings. The Higgs vacuum expectation value is given by $v=\lambda_t (f_{\pi}^3/\mu^2$) for $\mu^2 >> f_{\pi}^2$ (for simplicity we will 
focus on this case here). The pion decay constant is proportional
to $\Lambda_{\rm QCD}$, which, in keeping with the philosophy behind this work, is kept fixed. Hence all quark and lepton masses are computable numbers,
and they are smaller by a factor $\approx 10^{-9}$ than the values observed in our universe. In particular the electron mass is reduced by a factor $10^9$, and hence
the characteristic time scale of chemistry becomes larger by the same factor, while biochemical energies and temperatures are reduced.
\textcite{Agrawal:1997gf}  argued that in that case it would take
too long for  the universe to cool off sufficiently to allow biochemical life; stars would have burned out already. They also point out that most
protons would have decayed, but this can be circumvented easily by not having Grand Unification. }

{\color{darkred}
An important loophole in this argument was 
pointed out by \textcite{Barr:2007rd}. The original discussion ignored accelerated expansion, which had not been discovered yet. In a universe
with accelerated expansion cooling occurs much more rapidly, which invalidates the argument. However \textcite{Barr:2007rd} point another limit: if the electrons mass decreases by a factor $y$ planets increase in size by a factor $y^3$, and their average density can drop below the cosmological constant density. In that case the expansion would presumably rip them apart. Note that the density of matter in Planck units is 
$\approx (\alpha m_e)^3 m_p = 2.2 \times 10^{-93}$. 
}\fi

\paragraph{Big Bang Nucleosynthesis.}

\ifExtendedVersion {\color{darkred}
Understanding the synthesis of heavy elements in our universe has been one of the great successes of last century
physics. It is a complicated and delicate process, starting with production of some light elements during the hot phase
of the Big Bang, followed by synthesis of heavier elements in stars and supernovae. Several of these processes are rather
sensitive to the values of Standard Model parameters, such as the light quark masses, $\Lambda_{\rm QCD}$, $\alpha$, and the scale and strength of the weak interactions. It is important to distinguish observational and anthropic constraints. 
}\fi

In our kind of universe Big Bang Nucleosynthesis  (BBN) leads  mainly to production
of $^4{\rm He}$, $^1{\rm H}$, and small amounts of deuterium, tritium and lithium. 
The main potential impact of BBN
is therefore a destructive one: there might be too little hydrogen left.  A hydrogen-less universe is anthropically challenged, but there are no obvious arguments against the other extreme, a helium-less universe \cite{Carr:1979sg}.
Helium is needed as a stepping stone to heavier elements, but can also be made in stars.

In which extreme we end up is to a large extent determined by the electroweak freeze-out temperature (the temperature where the rate of 
electroweak $n \leftrightarrow p$ conversions drops below the expansion rate)
\begin{equation}
T_{\rm f}  \approx  \left(\frac{G_N}{G_F^4}\right)^{\frac16} = (v/M_{\rm Planck})^{\frac13}v \approx 0.66\ \MEV\ ,
\end{equation}
where $v$ is the Higgs vev.
At temperatures above $T_{\rm f}$ protons and neutrons are in thermodynamic equilibrium, and their ratio is given by a Boltzmann factor,
$n/p={\rm exp}[-(m_n-m_p)/T_{\rm f}]$. 
At $T_{\rm f}$  the ratio $n/p$  is ``frozen", and only decreases slightly because of neutron decay.  After freeze-out, the outcome of BBN is
determined only by strong interactions, which conserve flavor. They burn essentially all remaining baryons into helium, removing equal amounts of $p$ and $n$.
Hence one ends 
up with a fraction of hydrogen  equal to $(p-n)/(p+n)$ at freeze-out. This fraction approaches the  danger zone (no $^1$H) if
\begin{equation}
\left(\frac{m_n-m_p}{v}\right)\left(\frac{M_{\rm Planck}}{v}\right)^{\frac13} \rightarrow 0.
\end{equation}
This remarkable quantity involves all four interactions, since $m_n-m_p$ receives contributions from quark mass differences  \ifExtendedVersion {\color{darkred} (proportional to $v$) }\fi
and electromagnetic effects. The latter are proportional to $ \Lambda_{QCD}$, and in this way BBN 
is sensitive to changes in  that scale
\cite{Kneller:2003xf}.

There are two remarkable order of magnitude coincidences here: $T_{\rm f} \approx m_n-m_p$, and the neutron lifetime $\tau_n$ is of order the duration of nucleosynthesis. It is not clear if these have  any anthropic relevance. Increasing $m_n-m_p$ and decreasing $\tau_n$ to more ``natural" values
leads to a {\it larger} fraction of $^1{\rm H}$.  It is almost as if these quantities are {\it anti}-anthropically tuned! The hydrogen fraction is only
moderately sensitive to increases of $v$, since for large $v$ the dependence cancels out in the first factor, and  the neutron
lifetime decreases. Even if we ignore the latter, an increase of $v$ 
by a factor 1000
decreases the mass fraction of hydrogen  from $75\%$ to $6\%$.
It is hard to argue that this would not be enough. 
\ifExtendedVersion {\color{darkred} 
It is fascinating that $T_{\rm f}$ and $m_n-m_p$ have the same order of magnitude, even though they have a completely different origin.
However, the anthropic impact of this coincidence is not so clear, as long as no thresholds 
are crossed (especially nucleon and di-nucleon stability, see below)}. Note that to get into the danger zone one would have to make $m_n-m_p$ even smaller than 
it already is:  this would be  {\it anti}-anthropic tuning! The other natural  way to get into the danger zone is to enlarge $v$. 
But for large $v$ the dependence on $v$ cancels out in the first factor.
Consequently an increase of $v$ by a factor $10^3$ still gives a universe with about $2\%$ of hydrogen, and it is hard to argue convincingly that this is not enough. 
Another interesting fact is that in our universe the free neutron decays with a life-time of about 880 seconds, remarkably close to the duration of Big Bang Nucleosynthesis, given the wide range of weak interaction lifetimes. If the neutron decayed much faster,
we would end up with a universe consisting  mostly out of hydrogen. 

No strict limit will follow from such an argument, but it is quite possible that the number of observers per baryon drops off sharply in the limit of vanishing hydrogen abundance.
In our universe we have ended up with about 75$\%$ hydrogen and 25$\%$ Helium. It is possible that the optimum is somewhere near this point.\fi

\paragraph{Few-Nucleon systems.}

The stability properties of two and three nucleon systems certainly {\it look} fine-tuned in our universe: 
Deuterium is just bound by $1.1$ MeV per nucleon, di-protons and di-neutrons are just not bound by about 60-70 keV. Tritium is much more strongly bound than deuterium
but $\beta$-decays to $^3$He. But a decrease of the neutron-proton mass difference 
by a mere 20 keV(!) would make it stable.  Once $\beta$-decay is forbidden, tritium may be stable even after the deuterium stability line has been crossed, because of its higher binding energy.

Possible consequences of tritium stability 
on stars, apart from its potential r\^ole in chemistry, were discussed by \textcite{Gould:2012qp}. This author speculates that 
changes in fusion processes in stars could affect the formation of planets. 

\ifExtendedVersion {\color{darkgreen}
In much of the literature on anthropic tuning one finds the claim that stability of the di-proton would have a huge effect on BBN. The
idea was that all protons would be fused to di-protons
(which then decays to Deuterium, which is then burned into Helium), so that no $^1{\rm H}$ Hydrogen 
would be left (a stable di-neutron
may also have {\it observable} effects \cite{Kneller:2003ka}, but they are less likely to have {\it anthropic} implications.)

This claim is based on the mistaken assumption that the di-proton production
cross-section should be comparable to that of deuterium. However, there are important differences: Coulomb repulsion for $pp$
and identical particle effects \cite{Bradford,MacDonald:2009vk}. In particular, deuterium and di-proton/di-neutron have different
spin wave functions, because the $S=1$ state of the deuteron (the lowest state for dynamical reasons) is forbidden for two
identical nucleons. 
}
{\color{darkred}
Although these authors consider illegitimate variations of the ``old-fashioned effective strong coupling constant" ({\it not} $\alpha_s$)
or nuclear potentials, the main point appears to be a valid one: even if di-nuclei are stable, this will not strongly affect big bang
nucleosynthesis. But these arguments do not demonstrate that there are any valid parameter values for which the di-nucleons are
stable.
}
\else
{

Claims about the important impact of di-proton stability on BBN, in much of the literature on anthropic tuning,  are probably
exaggerated, as they incorrectly assume that the  di-proton production
cross-section would be comparable to that of deuterium \cite{Bradford,MacDonald:2009vk}.}\fi

Stability of di-nuclei {\it does} have a huge impact on stars. If the di-proton were stable, the deuteron production rate could be ten orders
of magnitude larger than in our universe, with unknown consequences \cite{Bradford}. So the di-proton stability
line -- if it exists at all -- marks the end of our region and the beginning of terra incognita.

The tritium stability line can undoubtedly be crossed by changing the quark masses, but for the other  stability lines
this cannot be  decided without a more detailed look at nuclear binding. The dependence of binding on 
quark masses is still uncertain. For instance, it is not clear if the  deuteron is bound in the chiral limit; see
\textcite{Beane:2002xf,Beane:2002vs,Epelbaum:2002gb}. For recent results and references on the impact
of variations of quark masses on nuclear forces and BBN see  \textcite{Berengut:2013nh}\footnote{Many papers studying the impact of variations
on BBN 
or the triple-alpha process  consider
 {\it observational} constraints, for the purpose of detecting variations in constants of nature.  
This should not be confused with {\it anthropic} constraints. Another source of confusion is that some authors
convert variations in the strong force to variations in $\alpha$ via an assumed GUT relation, as explained 
in \textcite{Calmet:2001nu,Langacker:2001td}. This greatly enhances the sensitivity to variations in $\alpha$, see
{\it e.g.} \textcite{Ekstrom:2009ef}.}.

\ifExtendedVersion {\color{darkred}
There are many studies of the effect
of quark masses on nuclear binding in order to 
constrain possible variations of parameters since the time of BBN and/or solve the primordial Lithium problem.
According to  \textcite{Flambaum:2007mj} an upward change 
of the pion mass of 
60\% would make the deuteron unbound, whereas a downward change by 15\% would  make the di-neutron bound;
see also \textcite{CarrilloSerrano:2012ja}}\fi

\ifExtendedVersion {\color{darkred}
Lattice computations are still in their infancy, and
cannot get close enough to small quark masses \cite{Ishii:2006ec,Chen:2010yt,Beane:2011iw,Yamazaki:2011nd}. A result that might have implications in this context
was presented in \cite{Braaten:2003eu}, who
pointed out that
QCD is remarkable close to a critical infrared limit cycle in the three nucleon system, which can possibly be reached by small
quark mass variations.  
}\fi

Properties of few-nucleon systems are potentially anthropically relevant, and appear to be fine-tuned, but
too little is known about either to draw firm conclusions.

\paragraph{The triple alpha process.}\label{TripleAlpha}

BBN ends with a universe consisting mainly of protons, electrons and $\alpha$-particles. Fusion to heavier
elements is inhibited because there are no stable nuclei with $A=5$ or $A=8$. Hence there are no paths with only two-particle reactions leading to heavier nuclei. 
The most obvious path to $^{12}{\rm C}$ is $\alpha+\alpha\rightarrow\ \!\!{^8{\rm Be}}$, followed by  $\ \!\!{^{~8}{\rm Be}}+\alpha  \rightarrow\ \!\!{^{12}{\rm C}}$.
But $^8{\rm Be}$ is unstable with a  lifetime of about $10^{-16}$ seconds, so this does not look promising. 

There are at least three remarkable facts that  improve the situation. First of all, the $^8{\rm Be}$ ground state 
is a very narrow resonance in the $\alpha\alpha$-channel, enhancing the first process. The narrowness of this resonance is due
to a remarkable tuning of strong versus electromagnetic interactions \cite{Higa:2008dn}. 
Secondly, there is a resonance of $^{12}{\rm C}$  (the second excitation level) that
enhances the second process. Finally, a logical third step in this
chain, $ {^{12}{\rm C}}+\alpha\rightarrow {^{16}{\rm O}}$, is {\it not}  enhanced by a resonance. If that were the case
 all $^{12}{\rm C}$
would be burned to $^{16}{\rm O}$. Indeed,  there is a  resonance in $^{16}{\rm O}$ (at 7.10 MeV)  that lies close to, but just {\it below} the $^{12}{\rm C}+\alpha$ threshold at 7.16 MeV.

The reaction rate of the triple-$\alpha$ process is proportional to \cite{Burbidge:1957vc}
\begin{equation}
r_{3\alpha} \propto \Gamma_{\gamma} \left(\frac{N_{\alpha}}{k_B T}\right)^3 e^{-\epsilon/k_B T}, 
\end{equation}
where $\epsilon \approx 397$ keV is the energy of the $^{12}{\rm C}$ resonance above the $3\alpha$ threshold,  $\Gamma_{\gamma}$ is de width of its radiative
decay into $^{12}C$ and $N_{\alpha}$ is the $\alpha$-particle number density. \ifExtendedVersion {\color{darkgreen} Since the participating $\alpha$-particles are from the tail of a thermal distribution, raising the resonance energy 
decreases the $^{12}{\rm C}$ production rate rapidly, lowering it increases $^{12}{\rm C}$ production. }\fi This formula 
enters into the calculation of element abundances, which can be compared with observations. Assuming $^{12}{\rm C}$ synthesis
takes place in the late stage of red giants at temperatures of order $10^8 K$ one can then fit $\epsilon$ to the observed abundances, 
by moving the resonance along the exponential tail. This was done by \textcite{Hoyle} and led to a prediction for $\epsilon$, which in its turn led to a  prediction of an excited level of  $^{12}{\rm C}$ 
at $7.65$ MeV above the ground state. This resonance (now known as the ``Hoyle state") was indeed found.
For an excellent account of the physics and the history see \textcite{HoyleHistory}.

\ifExtendedVersion

Since the abundance of Carbon is at stake, it is tempting to draw anthropic conclusions. {\color{darkgreen}
But Hoyle's result is merely an {\it observational} statement, not an {\it anthropic} statement. Determining the true anthropic relevance of the Hoyle state is a difficult matter. 
Changing the resonance changes the stars themselves, because the triple-alpha process is their source
of energy. One must take into account not only $^{12}{\rm C}$ production, but also the burning of $^{12}{\rm C}$ to $^{16}{\rm O}$. These processes are all strongly temperature
dependent, and may occur in a variety of types of stars of very different masses. \fi 
\ifExtendedVersion   Even if most $^{12}{\rm C}$ {\it we} observe is made in red giants, one must allow for the possibility that in different
universes other processes dominate. Furthermore, an energy level of $^{12}{\rm C}$ is not a free parameter; one would really have to study the full,
combined effect of changing Standard Model parameters on all relevant processes. One has to make assumptions about the chances for life in an
environment with different $^{12}{\rm C}$ to $^{16}{\rm O}$ ratios. Even if all $^{12}{\rm C}$ is burned to $^{16}{\rm O}$ and heavier elements, one has to rule out the possibility of Carbon-less life
based on heavier elements. It is  therefore not surprising that different authors have come to rather different conclusions. }\fi

\ifExtendedVersion\else
Since the abundance of Carbon is at stake, it is tempting to draw anthropic conclusions. But there are several caveats.
Carbon production is obviously not maximized for the observed value of $\epsilon$: for smaller $\epsilon$ the rate is even larger. 
One cannot assume that if $\epsilon$ is changed, $T$ remains fixed. Since the triple-$\alpha$ process must provide energy to counterbalance gravitational pressure, it is inevitable that the star compresses
to higher densities and temperatures if $\epsilon$ is increased.  Furthermore one should also
take oxygen production into account. At higher temperatures $^{16}{\rm O}$ production starts becoming more important.
The net effect is that if $\epsilon$ is increased, a larger fraction of Helium is burned to $^{16}{\rm O}$ and a smaller fraction to $^{12}{\rm C}$. To compute an
optimum, one would have to know the optimal Carbon/Oxygen ratio for life, and without a theory, and only our own kind of life as data, this
is impossible. An additional complication is that for smaller $\epsilon$ red giant type stars would produce very little $^{16}{\rm O}$, but more massive, hotter
stars can take over.  Even if no $^{12}{\rm C}$ is formed or all of it is destroyed, there would still be heavier elements, and perhaps there can be
complexity and life without Carbon. \fi

Without the Hoyle state the third excited state of $^{12}{\rm C}$ at $9.64$ could take over its r\^ole, but then stars would burn at such high temperatures
that even primordial $^{12}{\rm C}$ would be destroyed \cite{LivioHoyle}. Hence the existence of  the Hoyle state is indeed important for our kind of
life. However, 
\ifExtendedVersion {\color{darkgreen}
according to \textcite{Weinberg:2005fh} the existence of the Hoyle state
in $^{12}{\rm C}$ can be understood on the basis of collective dynamics of $\alpha$-particles. The
same $\alpha\alpha$-potential that produces the $^8{\rm Be}$ ground state would give rise to $n\alpha$ oscillator-like excitations in 
$A=4n, Z=2n$ nuclei, appearing above the $n\alpha$ threshold. 
This demystifies the {\it existence} of the Hoyle state for $n=3$, but does not predict its location
relative to $(A-4,Z-2)+\alpha$ threshold. Both the sign and the size of that difference are crucial. 
}\else
according to \textcite{Weinberg:2005fh} the {\it existence} of the Hoyle state
in $^{12}{\rm C}$ can be understood on the basis of collective dynamics of $\alpha$-particles, and hence is not a major surprise.\fi

\ifExtendedVersion
{\color{darkgreen}
The quantitative effect of changes of the resonance en- ergy away from its value of 288 keV was studied by \textcite{LivioHoyle}. These authors varied the excitation level in large steps, and found that for an upward change of 277 keV or more very little 12C is produced, but for an in- crease of 60 KeV they did not find a significant change. A downward variation by 60 keV led to a four-fold increase in the 12C mass fraction, suggesting that we are not liv- ing at the anthropic optimum (``things could be worse, but they could easily be much better", in the words of the authors); but this statement does not take into ac- count the reduction of Oxygen. }
\else
{
The quantitative effect of changes of the resonance energy was studied by \textcite{LivioHoyle}. These authors  varied the excitation
level in large steps in numerical stellar nucleosynthesis models, and found
that for an upward change of 277 keV or more
very little $^{12}{\rm C}$ is produced. For an increase of 60 KeV there was no significant change, whereas a decrease of 60 keV led to a four-fold
increase in $^{12}{\rm C}$. \fi
\textcite{Schlattl:2003dy}, using more advanced stellar evolution codes that follow the entire evolution of massive stars,
found that in a   band of $\pm 100$ keV around the resonance energy the changes in abundances are small.
\ifExtendedVersion {\color{darkred}

Compared to $\epsilon =397$ keV this suggest a tuning of no more than about $20\%$, but the really relevant question is 
how $^{12}{\rm C}$ and $^{16}{\rm O}$ production vary as function of Standard Model parameters. This is an extremely complicated issue, since
a proper treatment requires keeping track of all changes in nuclear levels, the rates of all processes and the effect on models for stellar evolution.
Processes that are irrelevant in our universe may become dominant in others. Most work in this area has focused just on finding out how
the Hoyle state moves. 
}\fi

To decide how fine-tuned this is one would like to see the effect of Standard Model parameter changes. A first step in that direction
was made by \textcite{Oberhummer:2000zj}, who studied the effect on the resonance energy of rescalings of the nucleon-nucleon and Coulomb
potentials. They concluded that changes of $0.5\%$ and $4\%$ respectively led to changes in C or O abundances 
\ifExtendedVersion{\color{darkgreen}
by several orders of magnitude (factors 30 to a 1000). These changes correspond to shift in the Hoyle state energy of about 130 keV \cite{Schlattl:2003dy}, 
and hence this conclusion is in rough agreement with \textcite{LivioHoyle}. However, in \textcite{Schlattl:2003dy} the same group considered
a more sophisticated model of stellar evolution (including the possibility of C-production in helium flashes), and concluded
that their conclusions on fine-tuning were ``considerably weakened".  

Although rescaling of the nuclear force is
more meaningful then an {\it ad hoc} rescaling of a nuclear energy level, it is still not a valid parameter change. 
A step towards that goal was made by \textcite{Epelbaum:2012iu}. These authors investigate the quark mass dependence of 
the $^{12}{\rm C}$ and $^{16}{\rm O}$ productions rates using ``nuclear lattice simulations." This is a combination of chiral effective field
theory with   simulations of non-perturbative nucleon-pion theory on a lattice (not to be confused with lattice QCD). These authors
conclude that $^{12}{\rm C}$ and $^{16}{\rm O}$ production would survive a $2\%$ change in the light quark masses or the fine structure
constant. Beyond this band (which corresponds  to a change of around 100 keV in the Hoyle state energy) substantial changes 
can be expected.
}\else
by more than an
order of magnitude. However, in  \textcite{Schlattl:2003dy} these conclusions were weakened. Using  nuclear lattice simulations \textcite{Epelbaum:2012iu} 
conclude that $^{12}{\rm C}$ and $^{16}{\rm O}$ production would survive a $2\%$ change in the light quark masses or the fine structure
constant. This band corresponds  to a change of around 100 keV in the Hoyle state energy. Exactly how far one can venture outside that band
is an extremely complicated issue, since
a proper treatment requires keeping track of all changes in nuclear levels, the rates of all processes and the effect on models for stellar evolution.
Processes that are irrelevant in our universe may become dominant in others. 
}\fi

One can try to convert these survivability bands in terms of variations of the Higgs vev, the common scale of the quark masses. 
The naive expectation
is that enlarging the Higgs vev increases the pion mass, which weakens the nuclear potential, which, according to \textcite{Oberhummer:2000zj}, increases the resonance energy
and hence lowers the C/O ratio. If one focuses only on $^{12}{\rm C}$ (assuming Oxygen can be made elsewhere), this would put an upper 
limit on the Higgs vev $v$. 
\ifExtendedVersion {\color{darkgreen}

 Indeed, \textcite{Hogan:2006xa} concludes that there is an upper limit for $v$. Using the aforementioned simple potential model for collective oscillations of $\alpha$-particles  suggested
 by   \textcite{Weinberg:2005fh}, the author estimates that the energy gap between the Hoyle state and the 3$\alpha$
 threshold scales with $(v/v_0)^{\frac12}$. But since the triple alpha process is main energy source of the star, it must
 scale its temperature by the same factor to maintain equilibrium. Then the triple alpha rate, and hence the $^{12}{\rm C}$ production rate
 stays the same, but
  the change in temperature affects all 
 other reaction rates. For reaction rates that are not thought to be very sensitive to changes in $v$, like 
 $\alpha+ {^{12}{\rm C}}\rightarrow {^{16}{\rm O}}$ this implies a change in reaction rates by a factor
 ${\rm exp}[(E/T)((v/v_0)^{\frac12}-1)]$, where $E/T$ is estimated to be about 30. For an {\it increase} of $v/v_0$ by $5\%$ the
 $^{16}{\rm O}$ burning rate doubles, and  most $^{12}{\rm C}$ is expected to burn to $^{16}{\rm O}$. For a {\it decrease} of $v/v_0$ $^{16}{\rm O}$ production
 decreases, but perhaps $^{16}{\rm O}$ could still be produced in hotter stars. Therefore \textcite{Hogan:2006xa} only puts an {\it upper} limit
 on the allowed increase of $v$ of about $5\%$. This is still an order of magnitude tighter than the upper bound on $v$ from nuclear
 binding (see sect. (\ref{HiggsVev}), but there is a disturbing discrepancy in all these estimates. Sensitivities differ by an order of magnitude, and there is not even agreement on the existence of upper and lower bounds.}
\else
Indeed, \textcite{Hogan:2006xa}, using Weinberg's model of collective $\alpha$ particle excitations to determine
the $v$-dependence, found an {\it upper} bound on $v$ about $5\%$ above its observed value. 
\fi
But \textcite{Jeltema:1999na}, using the results
of \textcite{Oberhummer:2000zj} mentioned above, find a {\it lower} limit on $v$ about $1\%$ below its observed value. 
\ifExtendedVersion {\color{darkblue} Note that this lower bound is derived assuming the same physical requirement,
namely sufficient Carbon production. In other words, \textcite{Jeltema:1999na} and  \textcite{Hogan:2006xa} find {\it a different sign} for the dependence
of the Carbon production rate on $v$.}\fi
\ifExtendedVersion {\color{darkred}
Although the discrepancy may be due to 
the different treatment of nuclear forces, there is another difference: in the first work the strong interaction scale is kept fixed, whereas in the
second the strong coupling is kept fixed at the GUT scale. Then $\Lambda_{\rm QCD}$ changes due to changes in quark mass thresholds affecting
the running of $\alpha_s$. This appears to be the dominant effect. 
}
\else 
{
The discrepancy may be due to  a different treatment of nuclear forces or a different slice through the parameter space: in the first work $\Lambda_{\rm QCD}$ is kept fixed, whereas in the
second the strong coupling is kept fixed at the GUT scale. Then changes in $v$ affect $\Lambda_{\rm QCD}$ because of changes in quark mass thresholds.
}\fi

Expressed in terms of changes if $v$, the results of \textcite{Epelbaum:2012iu} indicate that  the Hoyle state energy goes up when $v$ is increased, but
there are contributing terms with different signs and large errors. Therefore the opposite dependence is not entirely ruled out. 

\ifExtendedVersion {\color{darkred}
The Hoyle resonance has also received attention because it may give bounds on temporal variations
of constants of nature.   If some constants were different during the early stages of the universe this would affect $^{12}{\rm C}$ or
$^{16}{\rm O}$ production, and this may lead to disagreement with observed abundances.
To investigate this 
\cite{Ekstrom:2009ef} have studied massive (15 and 60 $M_{\odot}$) population III stars believed to be the earliest ones formed in our universe.  To parametrize 
variations they allow a rescaling of both the strong and the Coulomb interaction potential. They find a sensitivity at the per mille
level to changes in the strong force. 
Note however that these are observational bounds and not anthropic bounds, and hence
it is not surprising that they are tighter than those mentioned above.
 These authors also give bounds on allowed fractional variations of the fine structure constant of order $10^{-5}$, but these are
 obtained from the strong interaction bounds using the assumption of Grand Unification. The direct bounds on $\alpha$ are
 far weaker than those of strong force variations. 
}\fi


\ifExtendedVersion {\color{darkred}
In \cite{Higa:2008dn} the $\alpha\alpha$ resonance is examined using
effective field theory, 
and these authors observe the scattering length is enhanced by a factor a 1000 in comparison 
to its natural value. This is due to fine-tuned cancellations between several strong
interaction contributions, as well as cancellations between strong and Coulomb contributions. Since the cross section
is proportional to the square of the scattering length, this is a huge effect.
They also observed that an increase
of the strong interaction range by 20\% would make $^8{\rm Be}$ ground state stable. This would remove the Beryllium bottleneck
altogether, and lead to such drastic changes that the consequences are hard to estimate. Perhaps $^{12}{\rm C}$ could already
be produced in Big Bang Nucleosynthesis.
If stability of $^8{\rm Be}$ can be achieved within the 
Standard Model parameter space, this would turn the anthropic question up-side down: why do we live in a universe with $^{12}{\rm C}$ production
bottleneck, given the absence of that bottleneck in others?
}\fi

Even the most conservative interpretation of all this work still implies that a minute change of $v$  with respect
to $\Lambda_{\rm QCD}$ in either direction has drastic consequences. Note that the full scale of $v/v_0$ goes up to $10^{17}$, and the  variations discussed above are by 
just a few percent.
 \subsubsection{\label{Top}The Top Quark Mass}

The top quark may not seem an obvious target for anthropic arguments, but it may well be important
 because of it large coupling to the Higgs boson, which
plays a dominant r\^ole in the renormalization group running of parameters. In supersymmetric theories, this large coupling
may drive the Higgs $\mu^2$ parameter to negative values, triggering electroweak symmetry breaking (see 
\textcite{Ibanez:1982fr}; since this work preceded the top quark discovery, the authors  could only speculate about its mass).

The large top quark mass may also play an important r\^ole in the Standard Model, although the mechanism is less clear-cut, see
\textcite{Feldstein:2006ce}. These authors argue that in a landscape the top quark mass is pushed to large values to
enhance vacuum stability. This issue was re-analyzed recently by \textcite{Giudice:2012vv} using the recent data on
the Higgs mass and under somewhat different assumptions. 
They conclude that the quark masses may be understood in terms of a broad distribution
centered around one GeV, with the light quark masses and the top quark mass as outliers, pushed
to the limits by anthropic (atomic or stability) pressures.

 \subsubsection{Charged Lepton Masses}

The electron mass is bounded from above by the limits from nuclear stability already discussed in
section \ref{QuarkMasses}. If the electron is a factor $2.5$ heavier, hydrogen $^1$H is unstable against
electron capture; if one can live with tritium the bound goes up to about $10$ MeV. Beyond that bound most
heavy nuclei are unstable as well. See \textcite{Jenkins:2009ux}  for other, less restrictive bounds, for example the fact 
that a much heavier electron  (by a factor $\gtrsim 100$) would give rise to electron-catalyzed fusion in matter.

There are several arguments for smallness of the electron mass in comparison to the proton mass.
The bound $(m_e/m_p)^{1/4} \ll 1$ is important for having matter with 
localized nuclei \cite{BarrowTipler}, but there is no clear limit. Limits on hierarchies of scales ({\it e.g.} Bohr radius
versus nuclear radius, see section \ref{Alpha}) are not very tight because the electron mass is multiplied with 
powers of $\alpha$.

There are also lower bounds on the electron mass, but mostly qualitative ones. Lowering the electron mass
enhances the Thomson scattering cross section that determines the opacity of stars. It affects the temperature
of recombination and all chemical and biological temperatures.  The  stellar mass window (\ref{StellarMassWindow})
gives a bound on $m_e$ because the lower limit must be smaller than the upper one: $m_e > 0.005\ \alpha^2 m_p \approx 250$ eV. 

If muon radiation plays an important r\^ole in DNA mutations, then the location of the muon mass just
below the pion mass would be important (see footnote 17 in \textcite{Banks:2003es}). 
But  the danger of anthro\-pocentrism is enormous here.

 \subsubsection{Masses and Mixings in the Landscape}\label{Mixing}

In theoretical ideas about quark masses one can clearly distinguish  two antipodes: anarchy versus symmetry.
In the former case one assumes that masses and mixings result from Yukawa couplings that are randomly selected from some distribution, 
whereas in the latter case one tries to identify flavor  symmetries or other structures that give the desired result.


The quark mass hierarchies are very unlikely to come out of a flat distribution of Yukawa couplings. However, one can get roughly  the right answer  from scale-invariant distributions \cite{Donoghue:1997rn}
\begin{equation}
f(\lambda)=\rho(\lambda)d\lambda \ ,  \  \  \  \ \rho(\lambda) \propto \frac{1}{\lambda},
\end{equation}
where $f(\lambda)$ is the fraction of values between $\lambda$ and $\lambda+d\lambda$. A flat distribution is obtained for
$\rho={\rm const}$. Scale invariant distributions are generated by exponentials of random numbers. In string theory, this
can come out very easily if the exponent is an action. A canonical example is a ``world-sheet instanton", where the
action is the area of the surface spanned between three curves in a compact space. In intersecting brane models of 
the Madrid type shown in Fig. \ref{GUTTrinification}(a) this is indeed how Yukawa couplings are generated from the branes whose intersections
produce the left-handed quarks, the right-handed quarks and the Higgs boson.
Note that both types of 
distributions require small and large $\lambda$ cut-offs in order to be normalizable. In the intersecting brane picture this comes
out automatically since on a compact surface there is a minimal and a maximal surface area.

\begin{center}
\begin{figure}
\includegraphics[width=3.4in]{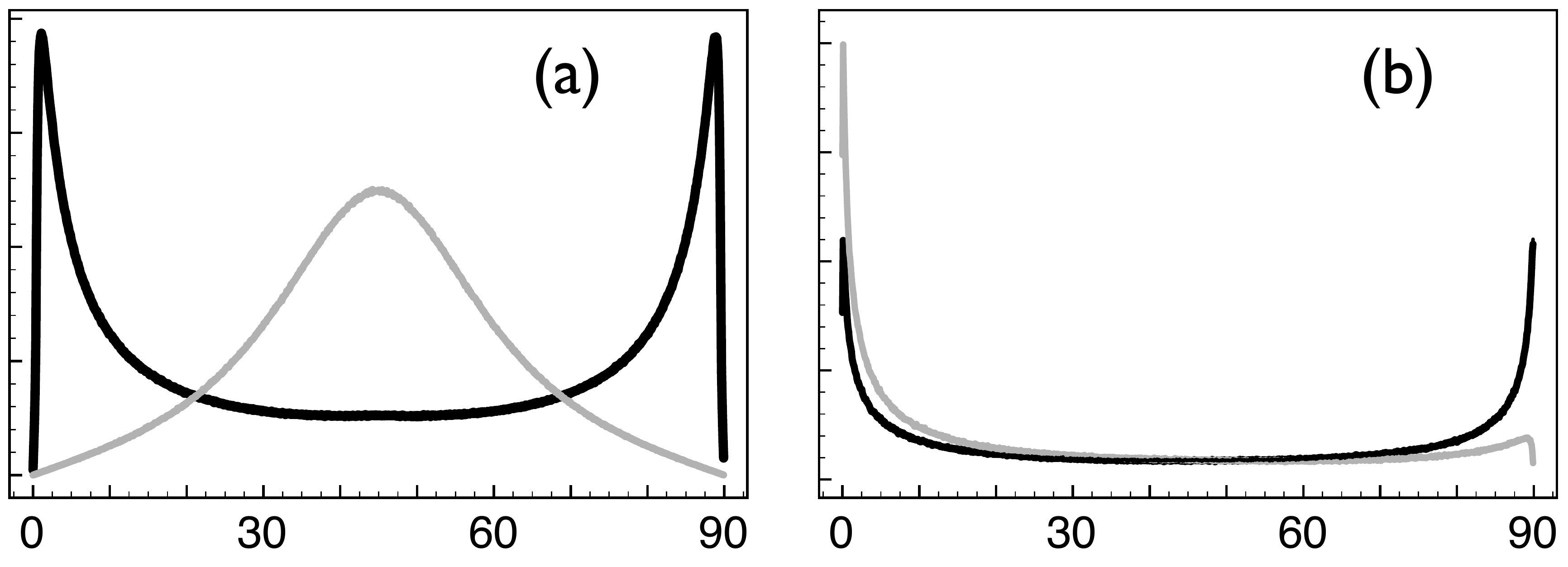}
\caption{Distribution of CKM angles at small and large angles for a scale invariant distribution. The black line is for $\theta_{12}$ and $\theta_{23}$, the gray line
is for $\theta_{13}$.}
\label{AngularDist}
\end{figure}
\end{center}

The smallness of the CKM angles makes a very convincing case against flat distributions. 
This is illustrated in Fig. \ref{AngularDist}(a).
Here $2 \times 2$ random complex matrices $M$ are considered, with entries chosen from two different distributions. What is
plotted is the distribution of the values of the rotation angle required to diagonalize the matrix (this requires separate left- and right
matrices, and the angle is extracted from one of them). The gray line is for a flat distribution of matrix elements, $M_{ij}=r_1+ir_2$, 
where $r_1$ and $r_2$ are random numbers  in the interval $[-1,1]$. The black line is
for a scale invariant distribution, $M_{ij}=e^{-s r_1}e^{2\pi i r_2}$, where $r_1$ and $r_2$ are random numbers between 0 and 1, and
$s$ is a real parameter. In the figure $s=5$ was used. As $s$ is increased, the angle distribution starts developing a peak at
small angles, but also near $90^{\circ}$. Clearly, small angles are unlikely for flat distributions, but not for
scale invariant ones. 

This is easy to understand. If a random matrix is generated with a scale invariant distribution, typically one matrix
element will be much larger than all others, and will select the required rotation. If it is on the diagonal, no rotation is
needed, and if it is off-diagonal one of the two matrices will have to make a $90^{\circ}$ rotation. 

This becomes a bit more murky for $3\times 3$ matrices, but the 
main trait persists in the full CKM matrix.  In Fig. \ref{AngularDist}(b) we show the distribution for the three angles in the CKM matrix,
with $M_u$ and $M_d$ distributed as above, but 
with $s=12$ \ifExtendedVersion {\color{darkred} (these dimensionless numbers are multiplied with a Higgs vev to
define the mass scale; we only look at angles and mass ratios here)}\fi. 
Only one phenomenological constraint was put in,  namely that the top quark mass must be at
least ten times the bottom quark mass; all other combinations of $M_u$ and $M_d$ are rejected. The largest mass was scaled to $m_t$ by means
of a common factor (the Higgs vev).
The distributions
for $\theta_{12}$ and $\theta_{23}$ are indistinguishable and symmetric on the interval $[0^{\circ}, 90^{\circ}]$ and are peaked
at both ends, while the distribution for $\theta_{13}$ is more strongly peaked and only near $\theta_{13}=0$. There is a large plateau
in the middle, and for $\theta_{12}$ and $\theta_{23}$ the peak is 40  times above the value at $45^{\circ}$. For larger values
of $s$ the peaks become more pronounced, and move towards the asymptotes at $0^{\circ}$ and $90^{\circ}$.

\begin{center}
\begin{figure}
\includegraphics[width=3.4in]{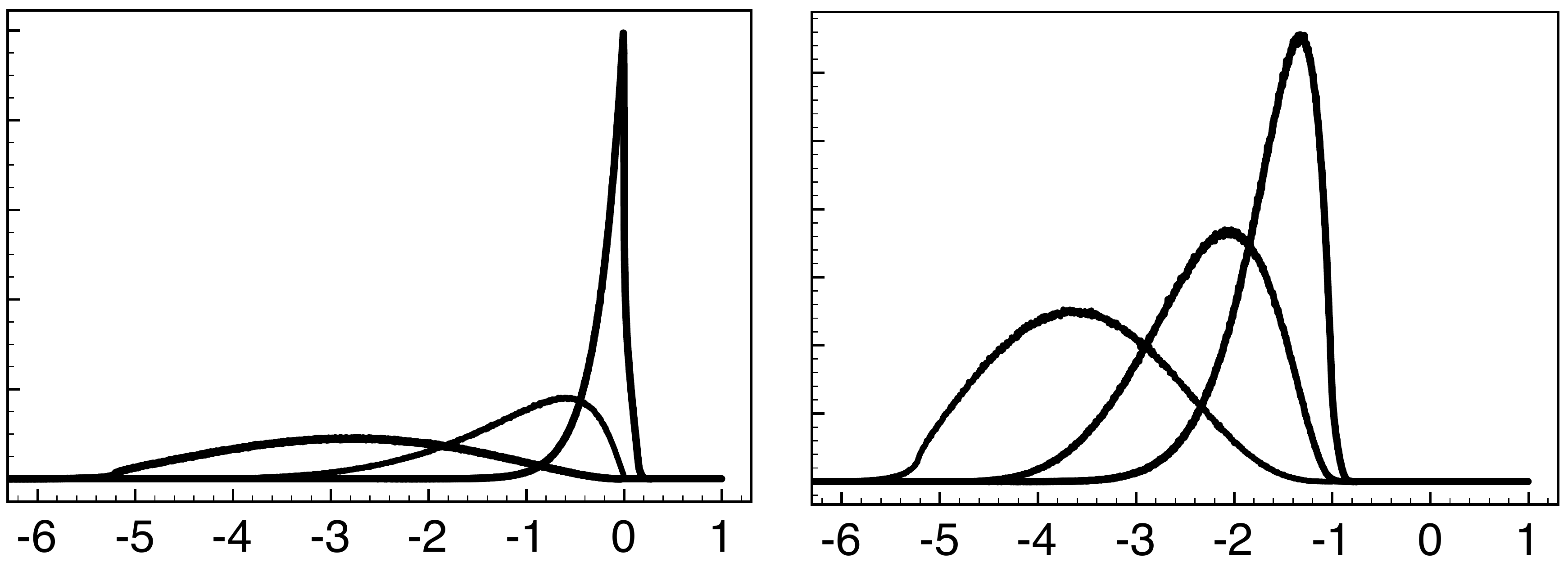}
\caption{Distribution of up-type (u,c,t) and down-type (d,s,b) masses. On the horizontal axis powers of ten are indicated.}
\label{QuarkMassesDist}
\end{figure}
\end{center}

The eigenvalue distribution is even more interesting and is shown in Fig. \ref{QuarkMassesDist}. No special effort was made to fit the single parameter $s$
to the observed quark masses and mixings; the value $s=12$ was chosen just to get roughly in the right ballpark, for illustrative
purposes only. Note that the difference between the two plots is entirely due to the requirement $m_t > 10\ m_b$. Renormalization group running
was not taken into account.  This might favor large top quark masses because of the infrared fixed point of the Yukawa couplings \cite{Donoghue:1997rn}.

The angular distributions easily accommodate the observed values $\theta_{12}=13^{\circ}$, $\theta_{23}=2.38^{\circ}$ and $\theta_{13}=0.2^{\circ}$, and the mass distributions have no difficulties with the observed mass hierarchies. Furthermore, the lowest eigenvalues
have very broad distributions, so that they can easily accommodate the anthropic requirements for $m_u$, $m_d$ and the
electron mass. Note that the angular distributions predict that two of the three angles are just as likely  to be large ($\approx 90^{\circ}$) as small. 
Hence the observation that all three are small comes out in about one quarter of all cases. Furthermore there are large central
plateaus. 

A much more complete analysis, including renormalization group running, was done by \textcite{Donoghue:2005cf}. 
These authors consider more general distributions, $\rho(\lambda)=\lambda^{-\delta}$,  determine the optimal distribution
from the quark masses, and compute the median values of the CKM matrix elements. They do indeed obtain the correct hierarchies in
the angles. They also work out the distribution of the Jarlskog invariant and find that it peaks at roughly the right value. 
The latter invariant was also considered by \textcite{Gibbons:2008jz}, who introduced a natural measure on the 
4-dimensional coset space that
is defined by the CKM matrix, $U(1)^2\backslash SU(3)/U(1)^2$. Taking the observed quark masses into account, they obtained a likely
value for $J$ close to the observed one.


An analysis that is  similar in spirit was done by \textcite{Hall:2007zh,Hall:2007zj}. Instead of scale invariant distributions, these authors 
assume that Yukawa couplings derive from overlap integrals of Gaussian wave functions in extra dimensions,
using a mechanism due to \textcite{ArkaniHamed:1999dc} to generate hierarchies and small mixing from strongly localized wave functions in extra dimensions. 
An advantage of this  mechanism is  that wrong pairings (large mixing angles between up-type and down-type quarks of different families) are strongly suppressed.
This method also
accommodates all observed features of quark masses and mixings rather easily. 

\subsubsection{Landscape vs. Symmetries}\label{LanSym}

The landscape ideas discussed above suggest that elaborate symmetries are not needed
to understand the observed masses and mixings. 
\ifExtendedVersion {\color{darkred}
There are indications that in a landscape, symmetries are strongly
disfavored \cite{Douglas:2012bu}, and attempts to understand quark masses and mixing using family symmetries have
had only very limited success.
}\fi

But there might be structure in the Yukawa matrices.
An interesting suggestion is gauge-top unification, which is found to occur
in a subset of mini-landscape models. This singles out the top quark and relates its Yukawa couplings directly to the
gauge couplings at the unification scale. In addition there is a $D_4$ discrete symmetry relating the  first two families. See
\textcite{Pena:2012ki} for further discussion and references. 

In the simplest possible orientifold models, for examples the ones 
 depicted in Fig. \ref{GUTTrinification}, all families are on equal footing. But this is not always the case, and there are many examples
 where different families have their endpoints on different branes. This gives rise to Yukawa coupling matrices where some entries
 are perturbatively forbidden, but can be generated by D-brane instantons, giving rise to a hierarchy of scales. Several possibilities were
 investigated by \textcite{Anastasopoulos:2009mr}.

Almost the exact opposite of landscape anarchy has emerged in the context of F-theory. The most striking phenomenon
is a stepwise enhancement of symmetries towards $E_8$. 
Gauge fields live on D7 branes, which have an eight-dimensional
world volume. Four of these dimensions coincide with Minkowski space, and the other four wrap a four-dimensional volume 
in the eight-dimensional Calabi-Yau fourfold that defines F-theory. Two-dimensional intersection curves of the
four-dimensional curves correspond to matter, and point-like triple intersections  of matter curves correspond to Yukawa couplings.
This leads to fascinating enrichment of old GUT ideas into higher dimensions: gravity sees all dimensions, gauge groups live on
eight-dimensional surfaces, matter on six-dimensional surfaces, and three-point couplings are localized in four dimensions, or just a point
in the compactified space.

The properties of gauge groups and matter are determined by ADE-type singularities defined by the embedding of these surfaces
in the  elliptically fibered Calabi-Yau fourfold. To get the required GUT group one starts with seven-branes with an $SU(5)$ singularity.
The matter curves have an enhanced singularity; to get a $(\bar {\bf 5})$ of $SU(5)$ the singularity must enhance $SU(5)$ to $SU(6)$, and to get
a $(\overline{\bf 10})$ it must enhance it to $SO(10)$. Further enhancements occur for the point-like singularities that correspond to
Yukawa couplings: to get the ${\bf 10}.{\bf \bar 5}.{\bf \bar 5}$ down-quark couplings one needs an $SO(12)$ singularity, and to get the 
${\bf 10}.{\bf 10}.{\bf 5}$ up-quark couplings one needs $E_6$.

The Yukawa couplings are, to first approximation, rank-1 matrices, which implies that each has one non-vanishing eigenvalue ($t$, $b$ and $\tau$) and
two zero eigenvalues. But two arbitrary rank-1 matrices will have their
eigenvectors pointing in unrelated directions, and since the CKM matrix is defined by the relative orientation, it will in general not be close to 1,
as it should be. This can be solved by assuming that the top and down Yukawa points  lie very close to each other. If they coincide
the singularity is enhanced to $E_7$ (which contains both $E_6$ and $SO(12)$). Finally there are arguments based on
neutrino physics that suggest that the singularity must be further enhanced to $E_8$ \cite{Heckman:2009mn}. Although this fascinating group-theoretic
structure gained attention in recent F-theory  GUT constructions \cite{Heckman:2008qa},  it was described prior to that by \textcite{Tatar:2006dc} in a more general setting, applied to
heterotic strings, M-theory and F-theory. These authors derived the $E_7$ structure requiring the absence of baryon number violation dimension-4 operators.

To get non-zero values for the other masses, a mechanism like the one of \textcite{Froggatt:1978nt} was proposed. This works by
postulating one or more additional $U(1)$'s and assigning different charges to the different families. \ifExtendedVersion {\color{darkred}
The $U(1)$'s are spontaneously
broken by charged scalar fields.  In the unbroken theory some Yukawa couplings are forbidden by charge conservation, but there
will in general be allowed non-renormalizable  couplings (defining a scale $M_{\rm FN}$) involving powers of scalars. If the scalars get a vev, the forbidden couplings
are generated with coefficients proportional to  powers of that vev, normalized by powers $M_{\rm FN}$. This gives rise to reasonable hierarchies if the scalar
vevs are $\epsilon M_{\rm FN}$, with $\epsilon$ of order $.1$.

}\fi
\textcite{Heckman:2008qa} showed that similar $U(1)$ symmetries automatically exist in certain F-theory compactifications,
and that they could lead to the required hierarchies  and small mixing angles. These are parametrized in terms of a small
parameter $\epsilon \approx \sqrt{\alpha_{\rm GUT}} \approx 0.2$.
But to actually obtain deviations from rank-1 matrices has
been a fairly long struggle, since some expected contributions turned out to respect the exact rank-1 structure.
For  recent work  and further references see \textcite{Font:2012wq}.

But important questions remain. Why would we find ourselves at or close to an $E_8$ point in the landscape? A CKM matrix close to 1 is
{\it phenomenologically}, but not {\it anthropically} required. It is not clear how the exact values  are distributed.
One should also ask the question if, in any of the methods discussed, the quark mass
hierarchies and mixings would have been even roughly predicted, if we had not known them already.

 \subsubsection{\label{NeutrinoMasses}Neutrinos}

There is a lot to say about neutrino masses in string theory and other theories, but here we will focus on landscape and anthropic issues. 
For  for a summary  of what is known about neutrinos see chapter \ref{StandardModelSection}, and for a recent review of various new ideas see
\textcite{Langacker:2011bi}.

\ifExtendedVersion {\color{darkred}
Neutrinos differ from quarks and the other leptons in several important ways: they are much lighter, some of
their mixing angles are large, they do not have a similar hierarchical mass spectrum and they are not 
charged, so that they can have Majorana masses. From the landscape perspective, the objective is not only
to find concrete examples where all these features are realized, but to understand also why we find ourselves
in such a universe.
Here there is an interesting confrontation between ``new physics" and anthropic arguments.

\paragraph{The seesaw mechanism.}}\else {

\paragraph{The seesaw mechanism.}
Neutrinos offer an interesting confrontation between ``new physics" and anthropic arguments. }\fi
On the one hand,  small neutrino masses are explained convincingly by the seesaw mechanism, which requires nothing
more than a number of singlet fermions, Yukawa couplings between these singlets and  the lepton doublets and Majorana masses for the singlets. 
In the string landscape the singlets are
generically present because most Standard Model realizations are $SO(10)$-related and because
singlets are abundant in nearly all string compactifications. Unlike $SO(10)$-related singlets, generic singlets usually do not have Yukawa
couplings with charged leptons, but those couplings may  be generated by scalar vevs; see 
\textcite{Buchmuller:2007zd} for an explicit heterotic string example.

Majorana masses tend to be a bigger obstacle.  
 It is not obvious that string theory satisfies the QFT lore that ``anything that is allowed is obligatory", which would imply that all allowed masses
are non-zero, and in particular that all singlets must have Majorana masses. In an extensive study of the superpotential 
of a class of heterotic strings, 
 \textcite{Giedt:2005vx} found no examples of such mass terms. Even if such examples were found in other cases ({\it e.g.} \textcite{Buchmuller:2007zd,Lebedev:2007hv}), this
still casts doubts on the {\it generic} presence of Majorana masses. 
But perhaps the examples are
too special, and perhaps all singlet fermions have large masses in generic, non-supersymmetric, fully stabilized vacua.
If not, string theory is facing the serious problem of predicting, generically, a plethora of massless or light singlet fermions.
Even if they do not have Dirac couplings and hence do not participate in a neutrino see-saw, this is a  problem in its own right.

Just as Yukawa couplings, Majorana masses can  be generated by scalar vevs, but one can also obtain
 Majorana masses in exact string theory. In the context of orientifold models of
the Madrid type this can in principle be achieved as follows. In these models there is always a $B\!-\!L$
symmetry. Usually this symmetry is exact and leads to a massless gauge boson \cite{Dijkstra:2004cc}.
This is in disagreement with experiment, and since massless $B\!-\!L$ gauge bosons are ubiquitous in string theory,
it is reasonable to ask why we do not see one in our universe. The answer may be anthropic: $B\!-\!L$ gauge bosons
lead to a repulsive force between protons and neutrons and may destabilize nuclei. There would also be
drastic changes in atoms and chemistry. 
\ifExtendedVersion {\color{darkgreen}
But let us take this for granted and consider the small set of cases where
the $B\!-\!L$ symmetry is broken because of mixing with axions (this is analogous to the Green-Schwarz mechanism
for anomalous $U(1)$'s, but in some cases this mechanism breaks anomaly-free $U(1)$'s as well). }\else {
But let us take this for granted and consider the small set of cases where
the $B\!-\!L$ symmetry is broken. }\fi
In those cases a Majorana mass may be
generated by non-perturbative effects due to D-brane instantons \cite{Florea:2006si,Ibanez:2006da,Blumenhagen:2006xt,Cvetic:2007ku,Argurio:2007vqa}. This does indeed work, but  in practice the relevant instanton contributions are nearly 
always
killed by a surplus of zero-modes  \cite{Ibanez:2007rs}. Even if one assumes that this is an artifact of special models, there is still another problem: instanton
generated terms have logarithmically distributed scales. Since D-brane instantons have mass-scales that are unrelated to those of the
Standard Model gauge group, their scale is not linked to the Standard Model scale. But there is also no particular reason why it would
be the large scale needed for small neutrino masses. 

If a large number of singlet neutrinos is involved in the see-saw 
mechanism, as string theory suggests, this may have important benefits.
It raises the upper limit for leptogenesis \cite{Eisele:2007ws}  and also raises the 
 seesaw scale \cite{Ellis:2007wz}.  

\paragraph{Anthropic arguments.}
Neutrinos are not constituents of matter, so that
they do not have to obey ``atomic" anthropic bounds.
Nevertheless, they have a number of potential anthropic implications. In our universe, neutrinos play a r\^ole
in big bang nucleosynthesis, structure formation, supernova explosions, stellar processes, the decay of the neutron,
pions 
and other particles, 
the mass density
of the universe and possibly leptogenesis.

Many of these processes would change drastically if neutrino masses were in the typical range of charged leptons,
but one should not jump to anthropic arguments too quickly. The fact that universes may exist where weak interactions -- including
neutrinos -- are not even necessary \cite{Harnik:2006vj}  underscores that point. But there are a few interesting limits nonetheless.

If the  sum of all neutrino  masses exceeds 40 eV they would overclose the universe. But there is no need to argue if this is an
observational or an anthropic constraint, because
for 
much larger masses (larger than the pion mass) they would all be
unstable, invalidating any  such argument. An interesting limit  follows from
leptogenesis \cite{Fukugita:1986hr}, which sets an upper bound to neutrino masses of $0.1$ eV 
\cite{Buchmuller:2003gz}. If this is the only available mechanism for generating a net baryon density this would
imply an anthropic upper bound on neutrino masses.

\textcite{Tegmark:2003ug} gave a rationale for small neutrino masses based on galaxy formation. They argued
that fewer galaxies are formed in universes with larger neutrino masses. If the distribution of neutrino masses
does not favor very small values, this leads to an optimum at a finite value, which is about 1 eV (for $\sum m_\nu$).
This is barely consistent with the aforementioned leptogenesis limit. Note that this mechanism favors Dirac masses. 
The seesaw mechanism with GUT-scale Majorana masses gives distributions that are too strongly peaked at zero. 

\paragraph{Landscape distributions.}

In the neutrino sector one can still make predictions. Until recently, this included the angle $\theta_{13}$, which until 2012 was
consistent with zero, an implausible  value from the landscape perspective.

The other opportunities for prediction are the masses, or at least their hierarchy. Generically, any model that gives the required large 
quark and lepton mass hierarchies will tend to produce hierarchies in the neutrino sector as well. Therefore it is not surprising that
all work listed below prefers a normal hierarchy (the inverted hierarchy requires two relatively large,
nearly degenerate masses).

 The two large neutrino mixing angles are an obvious challenge for distributions that produce small quark mixing angles.
 But there are several  ways in which neutrino masses could be different from quark and charged lepton masses. First of all,
  right-handed neutrinos might not belong to families the way quarks and leptons do. Secondly, there may be hundreds of them, not just three, and
  thirdly the origin of their Majorana mass matrix is not likely to be related to that of the Higgs coupling.  
  
  \textcite{Donoghue:2005cf} studied neutrino mixing angle distributions using Dirac couplings distributed like those of quarks, and with three right-handed
  neutrinos. These were assumed to have
a Majorana matrix with random matrix elements, with various distributions. These authors find that with these minimally biased
  assumptions the likelihood of getting the observed mixing angles is only about 5\% to 18\%, with the latter value occurring for a small Majorana scale
  of about $10^7$ GeV. They strongly predict a normal hierarchy, a wide distribution of $\theta_{13}$ disfavoring the value zero, and a Majorana neutrino mass
  (as would be observed in neutrinoless double-beta decay) of  order $0.001$ eV.

 The approach studied by \textcite{Hall:2007zj,Hall:2008km}, mentioned above for quarks, can accommodate neutrino mixing by assuming that
 wave functions of lepton doublets are less localized than those of quarks. The Majorana mass matrices are generated using overlap integrals of
 randomized gaussian wave functions. This works, but is  more biased towards the observed result.

Neutrino masses and mixings have also been studied in F-theory \cite{Bouchard:2009bu}. 
An interesting prediction is that the hierarchy is
not just normal, but more concretely $m_1:m_2:m_3 \approx \alpha_{\rm GUT}  : \sqrt{\alpha_{\rm GUT}} : 1$ with $\alpha_{\rm GUT} \approx 0.04$. 
Using the two mass splittings this
gives neutrino masses of approximately $2, 9$ and $50$  meV. The predicted value for $\theta_{13}$ is equal to  $\sqrt{\alpha_{\rm GUT}}$, and
is compatible with the recently observed vale.

 \subsection{\label{SMScales}The Scales of the Standard Model}

The classic Standard Model has two scales, the strong and the weak scale. To first approximation
the strong scale, $\Lambda_{QCD}$, determines the proton mass, and the weak scale determines the masses of the quarks and leptons.
The proton mass owes less than 1\% of its mass  to the up and down quarks. Indeed, the proton mass
is non-vanishing in the limit of vanishing quark masses, and would be only a little bit smaller in that limit. 

The weak scale and the strong scale have a 
rather different origin in the Standard Model. The former is directly related to the only  dimensionful parameter in the Lagrangian, the
parameter $\mu^2$, whereas the latter 
comes out as a pole in the  running of the QCD coupling constant
towards the IR region. This produces a dimensionful parameter, $\Lambda_{\rm QCD}$, from a dimensionless one, 
$\alpha_s=g_s^2/4\pi$. This is known as ``dimensional transmutation".
At one loop order, the logarithmic running of  $\alpha_s$ determines $\Lambda_{\rm QCD}$ in the following way
\begin{equation}
\alpha_s(Q^2) = \frac{1}{\beta_0\ {\rm ln}(Q^2/\Lambda_{\rm QCD}^2)},
\end{equation} 
with $\beta_0=(33-2 N_f)/12\pi$, where $N_f$ is the number of quark flavors, $N_f=6$.
Here $Q$ is the relevant energy scale. If we measure the function at one scale,
it is determined at any other scale. One can invert this relation to obtain
\begin{equation}
\label{QCDscale}
\Lambda_{QCD}=Q\ e^{-1/({2\beta_0 \alpha(Q^2)})},
\end{equation} 
Note that $\Lambda_{QCD}$ is a free parameter, which can be traded for 
$\alpha_s(Q^2) $ at some fixed scale, if desired.

Two things are remarkable about the weak and  strong scales. Both are very much smaller than
the Planck scale
\begin{equation}    M_{Planck}= \sqrt{\hbar c^5 \over G_N}  = 1.2209 \times 10^{19} ~\rm{GeV} ,
\end{equation}   
and they are within about two or three orders of magnitude from each other.
The smallness of {\it both} scales  is responsible for the extreme weakness of gravity in comparison to the other forces.
This fact has important anthropic implications. 

There are many ways of varying these scales while keeping other parameters fixed. 
Many  papers on anthropic arguments in astrophysics, such as \textcite{Carr:1979sg}, study the effect of varying
$m_p/M_{\rm Planck}$.  However, $m_p$ is not  a Standard Model parameter. It is mainly determined by
$\Lambda_{QCD}$, but it is ultimately also affected by the weak scale. If we move up that scale by  a few orders of magnitude while
keeping the Yukawa couplings fixed, the quark masses rather than $\Lambda_{\rm QCD}$ dominate the proton mass.  Many
other things change as well, making it hard to arrive at a clean conclusion. If we enlarge the proton mass by enlarging
$\Lambda_{QCD}$, it is not just the proton mass that changes, but also the strength of the strong coupling.  

 \subsubsection{Changing the Overall Scale}

The cleanest way of studying the effect of varying
the QCD scale is to vary all Standard Model scales by the same factor
$L$ with respect to  $M_{\rm Planck}$.
This  keeps all of nuclear physics and chemistry unchanged,
except for the overall scale. No thresholds are crossed, and every allowed process remains allowed in rescaled
universes.  Hence the chemistry of life is unaffected. 

\ifExtendedVersion {\color{darkred}
This change keeps  us within realm of quantum field theory, except for 
a few caveats. To keep the argument clean we should rescale {\it everything} by $L$, including new scales
that have not been observed yet, such as a GUT scale or a Majorana mass scale for neutrinos. But we should
avoid pushing any scale beyond the Planck scale. This would imply a maximum value for $L$. This caveat
applies only if such a new scale is relevant for the existence of life, and that does not seem very likely. 
The other caveat is the fact that the Planck scale itself enters
into the argument, because we are comparing the Standard Model scale to it.  Strictly speaking, this
implies that we are making implicit assumptions about the fundamental theory of gravity, and not just quantum field theory. 
The only way this could be relevant is if for some values of $L$
no valid quantum field exists. This is the case in theories with large extra dimensions, where the weakness of gravity is
explained by making the gravitational field expand into a higher dimensional space that opens up 
as soon as we probe some short distance. This distance could be as large as $.1 {\rm mm}$ \cite{ArkaniHamed:1998rs}. If nature is {\it fundamentally}
like this, the Planck scale is just an illusion, and there would be a maximum to the allowed rescaling of the Standard Model; there would simply
be no theory that holds for values larger than that maximum. If large extra dimensional models are just realized in a corner in a landscape, but not due to some presently unknown fundamental restriction, this caveat is irrelevant.
}\fi 

\ifExtendedVersion {\color{darkred}

If we increase all mass scales by a factor $L$, all time scales will be reduced
by a factor $L$, and hence we can expect evolution to go $L$ times faster in Planck time units.
But the lifetime of stars like our sun changes by a factor $L^{-3}$ \cite{Carr:1979sg} in the same units.  For a modest
rescaling of $L \approx 10$, this means that our sun would last a factor 100 less compared to evolution, and dies before evolution has produced 
anything of interest. To make use of this argument, we
will have to assume that the time evolution needed on earth is about typical, and at least not atypically slow. Furthermore
one can replace the sun by a star with less brightness and a longer lifetime. For increasing $L$ one will then slowly
move out to the tail of the distribution of stars. In the words of \textcite{Carr:1979sg}: ``{\em there is no basis for being at all
quantitative}". But it seems clear that somewhere in the range between $10^{-19}$ and $1$ there will be an anthropic boundary
on the $m_p/M_{\rm Planck}$ axis.

Indeed, if we go to extremes and make $L$ is so large that $m_p$ approaches $M_{\rm Planck}$
we encounter very adverse conditions.  
}
\else
 {
It is not hard to establish the existence of an anthropic bound. }\fi
Basic kinematics  implies a maximum for the number of nucleons in objects with gravitation balanced by internal pressure.
This maximum is  $\approx (M_{\rm Planck}/m_{p})^3$, and determines the maximum number of nucleons in stars to within a factor of order 10 \cite{Carr:1979sg}. If we increase $m_p$ {\color{darkblue} (by increasing $L$)} we will reach a point where the maximum is smaller than the number of nucleons in a 
human brain, which means that brain-sized objects collapse into black holes. 
If we set the necessary number of nucleons in a brain conservatively at about $10^{24}$,
we find a limit of $m_{p} \ll 10^{-8} M_{\rm Planck}$. 
\ifExtendedVersion {\color{darkgreen}
Here we
are almost certainly beyond the boundary.}\fi

These objects are just clusters of nucleons, not necessarily hot enough to have nuclear fusion. 
It is probably not too anthropocentric to assume that stars should ignite, not just to have stars as sources of energy but even more importantly as 
processing plants of elements heavier than Lithium. \ifExtendedVersion {\color{darkgreen}
The conditions for this to happen have been analyzed
by \textcite{Adams:2008ad}. In this paper, the boundaries for existence of stars are determined as a function of three parameters,
the gravitational fine structure constant $\alpha_G=(m_p/M_{\rm Planck})^2$, the electromagnetic fine structure $\alpha=e^2/4\pi$,
and a composite parameter $\cal C$  that determines nuclear reaction rates. The boundaries are shown in Fig. 5 of that paper, as a function
of ${\rm log}(G/G_0)$ and ${\rm log}(\alpha/\alpha_0)$, where $G$ is Newton's constant and the subscript $0$ indicates the
value observed in our universe. 
Three different boundary lines are shown for three values of ${\cal C}$.  The maximum value of all
curves occurs for a value of $G/G_0$ of about $10^7$. Since a variation of $G$ while keeping $m_p$ fixed 
is equivalent to a variation of
$m_p^2$  keeping $G$ fixed, this would imply that $m_p$ could be three to four orders of magnitude larger than observed
without a catastrophic change in the existence of stars. Some authors \cite{StengerBook} interpret this as evidence {\it against} anthropic fine-tuning, However, as discussed in \cite{Barnes:2011zh}, one can also look at it in the opposite way, namely that
this argument offers an anthropic understanding for about 80\% of the smallness of  $(m_p/M_{\rm Planck})$,
on a logarithmic scale. It all depends on
the definition of {\it fine}-tuning and which measure one chooses.   Furthermore, if we consider the scale variation discussed
above, the limit get considerably tighter. The quantity ${\cal C}$ has the dimension $[{\rm mass}]^{-3}$. It depends in a complicated
way on the strong and the weak scale, but if we enlarge both by the same factor while keeping everything else fixed, it is
reduced by a factor $L^3$. From Fig. 5 of  \cite{Adams:2008ad} we can then estimate the consequences. 
}
\else
 {
Conditions for existence of stars in other universes where investigated by \textcite{Adams:2008ad}. 
}\fi
The result is that the combined Standard Model scale cannot be enlarged by more than about a factor 10 without 
losing nuclear fusion in stars\footnote{\ifExtendedVersion{\color{darkgreen} Thanks to Fred Adams for clarifying this point.}\else Note that \textcite{Adams:2008ad} allows variations of nuclear reaction rates beyond QCD, and hence finds
a larger allowed variation. Tracing the
scale dependence in the computation leads to a much smaller effect.\fi}. 
\ifExtendedVersion {\color{darkgreen}
This argument depends more strongly on keeping everything but the scale fixed than the first one, but the 
broader variations considered
in \textcite{Adams:2008ad}  -- including some that may not be realizable in any underlying quantum field theory -- cover many possibilities already.

It is  more difficult to find a {\it lower} bound on the overall scale. For smaller values, stars continue to exist, but their average
size increases with respect to those of beings built out of the same number of protons as we are. An interesting
argument that get closer to fixing the scale is the ``Carter coincidence" (\ref{CarterCoincidence}), especially 
in the interpretation of \textcite{Press:1983rs}. For fixed values of $\alpha$, this
implies that if the overall Standard Model scale is lowered, typical stars would become too hot for
biochemistry. The trouble with arguments of this kind is however that parameters of stars have a distribution,
and one could move towards the tail of that distribution. Then probabilistic arguments are needed, with all
the inherent caveats.}\fi

Variation of all Standard Model mass scales with respect to the Planck mass was studied by
\textcite{Graesser:2006ft}. These authors consider the effect of changing the Planck mass on several
cosmological processes, such as inflation, baryogenesis, big bang nucleosynthesis, structure formation and
stellar dynamics, and find that the  anthropic window on the scale is narrow (less than an order of magnitude in either direction), if other cosmological parameters
are kept fixed.

\ifExtendedVersion {\color{darkgreen}
The most important point is the following. In texts about particle physics one often finds the statement: ``it is a big mystery
why the scale of the Standard Model is so much smaller than the Planck mass". Some authors make that statement only about
the weak scale, because they consider the smallness of the strong scale understood in terms of (\ref{QCDscale}). It is indeed true
that (\ref{QCDscale}) makes it much easier to get a small scale, if one tunes the scale by tuning $\alpha_s(M_{\rm Planck})$. 
But that does not change the fact that the smallness of {\it both} scales is anthropically required. The fact that the strong scale
is distributed logarithmically is not in  dissonance with  anthropic reasoning, which only requires logarithmic tuning. 
Note that we are not assuming GUT coupling constant unification here, unlike most of the anthropic literature, {\it e.g.}
\textcite{Carr:1979sg} and \textcite{Hogan:1999wh}. We will return to that subject later.  
}
\else 
{
Therefore the smallness of the ratio $m_p/M_{\rm Planck}$ -- in the sense of a  variation of the overall scale of the Standard Model -- is undoubtedly
needed anthropically. The true distribution of the
scale depends ultimately on the landscape distributions at the string scale. 
The fact that the strong scale
seems distributed logarithmically because of ``dimensional transmutation"  ({\it i.e.} Eq. (\ref{QCDscale})) is not in dissonance with  anthropic reasoning, which only requires logarithmic tuning to the right order of magnitude. It is harder to establish a {\it lower} bound on the overall scale, but
big changes do occur if it is lowered, since astrophysical sizes, times and temperatures scale differently than biological ones. See for example
the discussion of the Carter conjecture in section \ref{Alpha}.}\fi

 \subsubsection{The Weak Scale}\label{GaugeHierarchy}

The smallness of the weak scale, also known as the gauge hierarchy problem, is not just a matter of very small ratios, 
but now there is also a fine-tuning problem. 
The small parameter $\mu^2$ gets contributions from quantum corrections or  re-arrangements 
of scalar potentials
that are proportional to $M^2$, where $M$ is the relevant  large scale.  Hence it looks like  these terms must be tuned to
thirty significant digits so that they add up to the very small $\mu^2$ we observe. \ifExtendedVersion {\color{darkred}
This is clearly a ``why" problem. Even if one decomposes  $\mu^2$ into several terms, it remains a parameter in the Lagrangian
which can take any real value without violating any consistency condition. Some formulations of the hierarchy problem
even violate basic quantum mechanics.  It makes no sense to say that $\mu^2$ gets large quantum corrections, because
that statement would imply that individual quantum corrections are observables. It is also dangerous to use language like
``stabilizing the hierarchy", because there is no instability. Unfortunately this is often confused with the stability bound on the Higgs potential,
see chapter (\ref{StandardModelSection}). If the Higgs boson were 10 GeV heavier, the latter problem would not exist, but
the gauge hierarchy fine-tuning is unaltered.
On the other hand, although the large gauge hierarchy is
clearly anthropic, as argued above, one should not jump to the conclusion that therefore the hierarchy 
problem does not require a solution in terms of fundamental physics. }\fi

 \paragraph{Anthropic Bounds on the Weak Scale.}
 
 \ifExtendedVersion {\color{darkgreen}
An obvious effect of changing the weak scale is a changing the mass of the $W$ and $Z$ boson, and hence the
strength of the weak interactions. But the anthropic relevance of the weak interactions is questionable, as discussed in section 
(\ref{OtherHabitable}).
A much more important anthropic implication of changing the weak scale is that this changes all quark and charged lepton masses by a common factor. The idea that the weak scale might be anthropically determined was suggested for the first time (at least in public) by \textcite{Agrawal:1998xa}.
All bounds follow from the anthropic bounds on quark and lepton masses
discussed in section (\ref{QuarkMasses}), if we keep the Yukawa couplings fixed and vary $\mu^2$. This is certainly a legitimate change in the Standard Model effective field theory.
It has been argued that under certain 
conditions, this kind of variation occurs generically in string theory landscapes (see section (\ref{PredLandsubsection})
But what happens if we allow the Yukawa couplings
to vary as well? 
}
\else 
{
The idea that the weak scale might be anthropically determined was suggested for the first time (at least in public) by \textcite{Agrawal:1998xa}.
They considered
anthropic bounds on the weak scale following from   changes in quark masses, keeping   the Yukawa couplings   fixed, as discussed
in section \ref{QuarkMasses}.
But what happens if we allow the Yukawa couplings
to vary as well?
}\fi

\ifExtendedVersion {\color{darkgreen}
One needs additional assumptions on the distribution of Yukawa couplings in the landscape to arrive at a conclusion. If very small Yukawa couplings
are extremely unlikely, then the conclusion that the weak scale must be small for anthropic reasons remains valid. Otherwise it is merely an interesting observation about the value of $\mu^2$ given
everything else. }\fi
\textcite{Donoghue:2009me} compute a likelihood function for the Higgs vev using a scale invariant distribution function of the Yukawa couplings, determined from the observed distribution of quark masses. Using this distribution, and a flat distribution in $v$, both the Higgs vev and the Yukawa couplings are allowed to vary, under the assumption
that the Yukawa distribution  does not depend on $v$. The conclusion is that values close to the observed vev are favored.

However, \textcite{Gedalia:2010iy} make different assumptions. These authors also consider, among others,  scale invariant distributions.
But scale invariant distributions require a cutoff to be normalizable.  If one assumes that values as small as $\lambda_y=10^{-21}$ have a similar likelihood as values of order 1, then it is statistically
easier to get three small masses (for the $u$ and $d$ quarks and for the electron) using small Yukawa couplings and a large Higgs vev than the way it is done in our universe. If  furthermore one assumes a weakless universe as discussed in \textcite{Harnik:2006vj}, the conclusion would be that in the multiverse there are far more universes {\it without} than {\it with} weak interactions, given atomic and nuclear physics as observed. See however \textcite{Giudice:2012vv} for a
way of avoiding the runaway to small Yukawas and large Higgs vevs.

If indeed in the string landscape extremely small values of Yukawa couplings are not strongly suppressed,
and if weakless universes are as habitable as ours (which is not as obvious as  \textcite{Gedalia:2010iy} claim),
this provides one of the most convincing arguments in favor of a solution to the hierarchy problem: a
mechanism that tilts the distribution of $\mu^2$ towards smaller values. \ifExtendedVersion{\color{darkblue} Note that we are implicitly assuming here
that $\mu^2$ comes out of some ``landscape" distribution. We will argue below that this is the only sensible interpretation of essentially
all proposed solutions to the hierarchy problem.}\fi

\ifExtendedVersion \paragraph{Natural Solutions to the Hierarchy Problem.}\else  \paragraph{Low Energy Supersymmetry.}\fi
The fact that a logarithmic behavior works for the strong scale has led to speculation that  a similar phenomenon should be
expected for the weak scale. At first sight the most straightforward solution is to postulate an additional interaction that mimics QCD and
generates a scale by dimensional transmutation. The earliest idea along these lines is known as ``technicolor".
 Another possibility is that
there exist large extra dimensions, lowering the higher-dimensional Planck scale to the TeV region. 
But the most popular idea is low energy supersymmetry (susy). The spectacular
results from the LHC experiments have put all  these ideas under severe stress, but low energy susy
remains a viable possibility. For this reason this is the only option that we will consider more closely here.

\ifExtendedVersion \paragraph{Low Energy Supersymmetry.}\fi
Low energy susy does not directly explain the smallness of the Higgs parameter $\mu^2$, but rather the ``technical naturalness" problem.
In the Standard Model, the quantum corrections to $\mu^2$  are quadratically sensitive to high scales. In the supersymmetric Standard
Model, every loop contribution is canceled by a loop of a hypothetical particle with the same gauge quantum numbers, but
with spin differing by half a unit, and hence opposite statistics: squarks, sleptons and gauginos.  
None of these additional
particles has been seen so far. Supersymmetry is at best an exact symmetry at high energies.
\ifExtendedVersion {\color{darkred}
 Supersymmetry more than doubles the particle content of the Standard Model;
in addition to ``superpartners" for each known particle, an additional Higgs multiplet is needed. The up-type  quarks get their
mass from one of them, $H_u$, and down-type quarks the charged leptons get their mass from $H_d$. After weak symmetry breaking
five massive scalar fields remain, instead of the single Higgs boson of the Standard Model.  }\fi

Rather than a single dimensionful parameter $\mu^2$ the supersymmetrized Standard Model has at least two, a parameter
which, somewhat confusingly, is traditionally called $\mu$, and a scale $M_S$ corresponding to
susy breaking. The latter scale may be generated by dimensional transmutation, and
this is the basis for susy as a solution to the hierarchy problem. But the additional scale $\mu$, which can be thought 
of as a supersymmetric Higgs mass prior to weak symmetry breaking, requires a bit more discussion.
To prevent confusion
we will equip the supersymmetric $\mu$-parameter with a hat. 

\ifExtendedVersion {\color{darkred}
 All supermultiplets are split by an amount proportional to $M_S$, and there is a good reason why
only the superpartners are lifted, and not the Standard Model particles themselves: they can all get masses without
breaking $SU(2)\times U(1)$. The fact that none of the superpartners has been observed so far implies that $M_S$ must be
in the TeV range, although precise statements are hard to make without a much more detailed discussion.  
The parameter $\hat\mu$ determines the mass of the fermionic partners of the Higgs bosons, and it
also determines the masses of the Higgs bosons themselves, through mass terms of the form $\| \hat\mu \|^2 (h_u^{\dagger}h_u+
h_d^{\dagger}h_d)$, where $h_u$ and $h_d$ are the scalar components of the Higgs multiplets. Unlike the Standard Model
Higgs mass terms these terms are positive definite. The familiar ``Mexican hat" potential can only be generated after supersymmetry
breaking, which contributes additional terms that can be negative. }\fi

Since \ifExtendedVersion {\color{darkred} we argued above that }\fi  $\mu^2$, just as $\hat\mu$, is merely a parameter that can take any value, it may seem that
nothing has been gained. The difference lies in the quantum corrections these parameters get. For the $\mu^2$ parameter these quantum corrections take the 
(simplified) form
\begin{equation}\label{MuOne}
\mu^2_{\rm phys}=\mu^2_{\rm bare} + \sum \alpha_i \Lambda^2 + {\rm logarithms},
\end{equation}
whereas for $\hat\mu$ one finds
\begin{equation}\label{MuTwo}
\hat \mu_{\rm phys}=\hat\mu_{\rm bare} \left(1+   \sum \beta_i {\rm log} (\Lambda/Q) + \ldots\right).
\end{equation}
Here ``bare" denotes the parameter appearing in the Lagrangian and ``phys" the observable, physical parameter, defined and
measured at some energy scale $Q$; $\Lambda$ denotes some large scale at which the momentum integrals are cut off. 
\ifExtendedVersion {\color{darkred}
Note that $\hat \mu_{\rm phys}$ vanishes if  $\hat\mu_{\rm bare}$ vanishes. This is because
for vanishing $\hat\mu$ the theory gains an additional symmetry, a chiral symmetry for the fermions. All amplitudes violating that symmetry must vanish in the limit where the symmetry breaking parameter goes to zero. This is 't Hooft's criterion for technical
naturalness \cite{tHooft:1979bh}. No symmetry is gained if the Standard Model parameter $\mu^2$ is set to zero
(scale invariance is not a quantum symmetry), and hence the quantum corrections
to $\mu^2$ do not vanish in that limit. }\fi

\ifExtendedVersion {\color{darkred}
If one views this as a standard renormalization procedure where $\mu^2_{\rm phys}$ and $\hat \mu_{\rm phys}$ are renormalized
parameters there is strictly mathematically still no difference between these two situations, because the left-hand sides are all
that can be measured, and $\Lambda$ is irrelevant. But intuitively there are two important differences, the quadratic $\Lambda$-dependence of (\ref{MuOne}) and the proportionality of (\ref{MuTwo}) with $\hat \mu$. Usually, discussions of naturalness 
do not go beyond this intuitive level. However, this can be made precise if we assume that all quantities are defined in a 
landscape where all quantities are well-defined and $\Lambda$ is some large but finite scale, as would be the case in the
string theory landscape. Then (\ref{MuOne}) tells us that  $\mu^2_{\rm phys}$ is an infinite sum of terms proportional to $\Lambda^2$ with coefficients $\alpha_i$ that are numbers of order 1 depending on the location in the landscape. If we now consider the
complete distribution of values of $\mu^2_{\rm phys}$ over the landscape, we find that  $\mu^2_{\rm phys}$ is some number
of order 1 times $\Lambda^2$. If the distribution is flat and $\Lambda \approx M_{Planck}$, 
we need a total number of about $10^{35}$ points to have a chance to
encounter one point in the landscape where $\mu^2_{\rm phys}$ is close to the observed value. On the other hand, the same 
argument applied to $\hat \mu$ gives us no information about how it is distributed. The logarithms are of order one, and do not
play a r\^ole in the discussion, 
and we need additional information about the distribution of  $\hat\mu$ in the fundamental theory in order to be able to say
anything more. That distribution could be logarithmic, or even a delta-function at zero, or it could even be flat in $\hat\mu$ (which
is still better than flat in $\hat\mu^2$).  The question why $\hat\mu$ is of order the weak scale and not much larger
is known as the ``$\hat\mu$-problem". The advantage of low energy supersymmetry is that at least there is a chance of finding a mechanism to make $\hat\mu$ small, whereas without low energy supersymmetry we have no other option than assuming 
that  our universe was picked out of a large enough distribution. 

If there were nothing anthropic about $\mu^2$ it would be preposterous to claim that we observe a very unlikely
small value just by chance. This might appear to lead to the conclusion  \cite{Weinberg:2005fh} ``{\em If the electroweak symmetry breaking scale is anthropically fixed, then we can give up the decades long search for a natural solution of the hierarchy problem}". 
But the anthropic and the natural solution are not mutually exclusive. 
}
\else { }\fi
\ifExtendedVersion {\color{darkred}
Before jumping to conclusions, we should ask ourselves why we want a solution to the hierarchy problem. 

The answer depends on whether one adopts the uniqueness paradigm or the landscape paradigm. Advocates of the former
might be perfectly happy if $\mu^2/M_{\rm Planck}^2$ came out as a combination of 
powers of $\pi$ and other fundamental numbers, but they would not expect that. Hence they believe that there
must exist a missing ingredient that determines that scale. Low energy supersymmetry, with all of its parameters fixed by
some fundamental theory is then an excellent candidate. If the $\hat\mu$-parameter as well as the scale of supersymmetry
breaking are determined by something analogous to dimensional transmutation, it is entirely possible that the small number
comes out as an exponential of numbers of order 1, which conceivably could be computed exactly. However, using just
low energy physics it is not possible to make this scenario precise. The scale of supersymmetry breaking (which determines
the mass of the superpartners, the squarks, sleptons and gauginos) might be several orders of magnitude above the weak
scale, with the gap being bridged by computable small numbers or powers of coupling constants. In this case at least {\it three} numbers 
must come out just right to get the anthropically required small mass scale: the strong scale, the value of $\mu$ and the
supersymmetry breaking scale.  

But most discussions of naturalness, especially in the technical sense, are implicitly assuming a landscape. 
}
\else 
{

The difference between these two kinds of quantum corrections is most easily understood if one thinks of them in terms of distributions, {\it i.e.} a landscape.
Indeed, 
the concept of naturalness, especially in the technical sense,  implicitly assumes a landscape, a point also emphasized by \textcite{Hall:2007ja}. }\fi
\ifExtendedVersion {\color{darkred}
As stated in \cite{Hall:2007ja}: ``{\em It is very important to notice that in talking about naturalness, we are dealing, either explicitly or implicitly, with an ensemble in which parameters of the theory are varied according to some definite distribution}. }\fi
If one adopts the landscape paradigm, the rationale
for a natural solution of the hierarchy problem would be that the unnatural solution comes  at a high statistical price,
$\mu^2/M_{\rm Planck}^2 \approx 10^{-35}$. 
\ifExtendedVersion\else
This holds  for the Standard Model with a flat distribution of values of
$\mu^2$ between $0$ and $M_{\rm Planck}^2$, as suggested by the renormalization of $\mu^2$.
On the other hand, the renormalization of $\hat \mu$, proportional to $\hat\mu$ itself, gives no information about its distribution.\fi

\paragraph{The Supersymmetry Breaking Scale.}
Low energy susy lowers the statistical price by replacing $M_{\rm Planck}$ by $M_{\rm susy}$,  the susy
breaking scale.  Here we define it as the typical scale of super multiplet mass splittings\footnote{At least two distinct definition of the 
susy breaking scale
are used in the literature. 
Furthermore there exist several mechanisms for ``mediation" of susy breaking, such as gauge and gravity 
mediation.  The discussion here is only qualitative, and does not depend on this.
See \textcite{Douglas:2006es} for further details.}. This suggests that 
the statistical price for  a small weak scale can be minimized by setting $M_{\rm susy} \approx \mu$. This is the
basis for two decades of predictions of light squarks, sleptons and gauginos, which, despite being much more sophisticated than this, have led to two decades of wrong expectations. But in a landscape, the likelihood $P({\mu})$ for a weak scale $\mu$ is
something like  
\begin{equation}\label{SusySuppression}
P(\mu)=P_{\rm nat}(\mu,M_{\rm susy}) P_{\rm landscape}(M_{\rm susy}).
\end{equation}
The first factor is the naive naturalness contribution, $P{\rm nat}(\mu,M_{\rm susy})\propto \mu^2/M_{\rm susy}^2$, and
the second one is the fraction of  vacua with a susy breaking scale $M_{\rm susy}$.

During the last decade there have been several attempts to determine $P_{\rm landscape}(M_{\rm susy})$. One such argument, suggested 
by \textcite{Susskind:2004uv,Douglas:2004qg} suggested that it increases with a power given by the number of
susy breaking parameters ($F$ and $D$ terms). If true, that would rather easily overcome the  $(M_{\rm susy})^{-2}$ dependence
of the first factor. However, this assumes that all these sources of susy breaking are independent, which is not necessarily
correct \cite{Denef:2004cf}. Other arguments depend on the way susy is broken (called ``branches" of the landscape in
\textcite{Dine:2005yq}). The arguments are presented in detail in section V.C of \textcite{Douglas:2006es}. 
An important contributing factor that was underestimated in earlier work is the fact that vacua with broken susy
are less likely to be stable. This can lead to a huge suppression \cite{Marsh:2011aa,Chen:2011ac}. There are large factors going
in both directions, but the net result is uncertain at present.

One might expect intuitively that there should be another suppression factor $\Lambda^4/M_{\rm susy}^4$ in Eq. (\ref{SusySuppression}) due
to the fact that unbroken susy can help fine-tuning the cosmological constant $\Lambda$ just as it can help fine-tuning $\mu$ 
\cite{Susskind:2004uv,Banks:2003es}. But this is wrong, basically because it is not true that $\Lambda=0$ in supergravity. In general
one gets $\Lambda \leq 0$, which must be canceled to 120 digit precision just as in the non-supersymmetric theories. There is a branch
with $\Lambda = 0$ before susy breaking, but this requires a large (R-)symmetry, which is statistically unlikely
\cite{Dine:2005gz}.

Despite the inconclusive outcome
there is an important lesson in all this. Conventional bottom-up naturalness arguments that
make no mention of  a landscape are blind to all these subtleties. If these arguments fail in the only 
landscape we are able to discuss, they should be viewed with suspicion. Even if in the final analysis all uncertain
factors conspire to favor low energy susy in the string theory landscape, the naive naturalness arguments would have been correct only by pure luck.

\paragraph{Moduli.} 
There is another potentially crucial feature of string theory that conventional low energy susy arguments are missing:
moduli (including axions). This point was made especially forcefully by \textcite{Acharya:2012tw} and
earlier work cited therein.

It has been known for a long time that moduli can lead to cosmological problems 
\cite{Coughlan:1983ci,deCarlos:1993jw,Banks:1993en}. If they are  stable or long-lived they can overclose the universe; if they decay
during or after BBN they will  produce additional baryonic matter and destroy the successful BBN
predictions. For fermionic components of moduli multiplets these problems may sometimes be solved by 
dilution due to inflation. But bosonic moduli have potentials, and will in general
be displaced from their minima. Their time evolution is governed by the equation
\begin{equation}\label{HubbleFriction}
\ddot \phi + 3 H \dot\phi +\frac{\partial V}{\partial\phi}=0,
\end{equation}
where $H$ is the Hubble constant. If $V=\frac12 m^2 \phi^2 + \hbox{higher order}$  and $H \gg m$ 
then the second term dominates over the third, and $\phi$ gets frozen at some constant value (``Hubble friction").
This lasts until $H$ drops below $m$. Then the field starts oscillating in its potential, and releases its energy.
The requirement that this does not alter BBN predictions leads to a lower bound on the scalar moduli mass
of a few tens of TeV (30 TeV, for definiteness). 

Furthermore one can argue \cite{Acharya:2010af} that the mass of the lightest modulus is of the same order of magnitude as the
gravitino mass, $m_{3/2}$. The latter mass is generically of the same order as the soft susy breaking
scalar masses: the squarks and sleptons searched for at the LHC. This chain of arguments leads  to the
prediction that the sparticle masses will be a few tens of TeV, out of reach for the LHC,  probably even after
its upgrade. But there was also  a successful (though fairly late and rather broad) prediction of the Higgs 
mass\footnote{The Higgs mass, $\approx 126$ GeV was also correctly predicted in finite unified theories, see  \textcite{Heinemeyer:2007tz}
and on the basis of asymptotically safe gravity, see \textcite{Shaposhnikov:2009pv}.   
Bottom-up supersymmetric models, ignoring moduli, suggested an upper limit of at most 120 GeV.}
 \cite{Kane:2011kj}. 

However, there are loopholes in each step of the chain. Light mo\-duli can be diluted by ``thermal inflation" \cite{Lyth:1995ka}, and
the mass relation between gravitinos and sparticles can be evaded in certain string theories. The actual result of  \textcite{Acharya:2010af} is
that the lightest modulus has a mass smaller than $m_{3/2}$ times a factor of order 1, which can be large in certain cases.
Hence this  scenario may be {\it generic}, but is certainly not {\it general}.


The relation between $m_{3/2}$ and fermionic super particles (Higgsinos and gauginos) is less strict and more
model-dependent. They might be lighter than $m_{3/2}$ by one to two orders of magnitude and accessible at the 
LHC.   Gaugino mass suppression in fluxless M-theory compactifications is discussed by \textcite{Acharya:2007rc}. This was also seen in
type-IIB compactifications, with
typical suppression factors of order ${\rm log} (M_{\rm Planck}/m_{3/2})$ \cite{Choi:2004sx,Choi:2007ka,Conlon:2006us}.

A susy scale of $30$ TeV introduces an unnatural fine-tuning of five orders of magnitude\footnote{In comparison with
a weak scale of $\approx 100$ GeV and expressed in terms of the square of the scale, in accordance with
the scale dependence of quantum corrections.}, the ``little hierarchy". This tuning requires an explanation beyond the mere 
phenomenological necessity.  The explanation could be anthropic, which would be much better than
observational. A universe that seems fine-tuned for our existence makes a lot more sense than a universe
that seems fine-tuned just to misguide us.

Could this explain the 30 TeV scale? Statements 
like ``the results of BBN are altered"  or ``the universe is overclosed" if moduli are lighter do indeed sound potentially anthropic. But it is
not that simple. Constraints from BBN are mostly just observational, unless one can argue that
all hydrogen would burn to helium. Otherwise, what BBN can do, stars can do better. Overclosure just
means disagreement with current cosmological data. Observers in universes just like ours in all other respects
might observe 
that they live in a closed universe with $\Omega \gg 1$, implying recollapse in the future. But the future
is not anthropically constrained. The correct way to compare universes with light moduli anthropically to ours is  to
adjust the Hubble scale so that after inflation $\Omega \approx 1$. This would give a universe with different ratios
of matter densities, but it is not at all obvious that those ratios would be catastrophic for life. Without such an argument, the claim
that moduli require a 30 TeV susy scale is much less convincing. See also  \textcite{Giudice:2006sn} for a different view on a possible
anthropic origin of the little hierarchy.

\paragraph{The Cost of Susy.}
Another anthropically relevant implication of low-energy susy is stability of baryons. Supersymmetry
allows ``dimension-4" operators that violate baryon number and lepton number that do not exist in the Standard Model: they are group-theoretically allowed, but contain an odd number of fermions. If all these operators are
present with ${\cal O}(1)$ coefficients they give rise to anthropically disastrous proton decay. 
 This can be solved by postulating 
 a discrete symmetry that forbids the dangerous couplings
(most commonly R-parity, but there are other options, see \textcite{BerasaluceGonzalez:2011wy} 
for a systematic summary). In the landscape
global symmetries are disfavored, but R-parity may be an exception \cite{Dine:2005gz}. Landscape studies
of intersection brane models indicate that they occur rarely \cite{Ibanez:2012wg,Anastasopoulos:2012zu}, 
but since they are anthropically required  
one can tolerate a large statistical price. 

But apart from {\it anthropically} required tunings, susy is also {\it observationally} fine tuned. 
There are dimension five operators that can give rise to observable but not catastrophic proton
decay. A generic supersymmetric extension of the Standard Model gives rise to large violations
of flavor symmetry: for general soft mass term, the diagonalization of squark matrices requires unitary rotations that are
not related to those  of the quarks. There are also substantial contributions to CP-violating processes. All of these
problems can be solved, but at a statistical price that is hard to estimate, and hard to justify. Moving the
susy breaking scale to 30 TeV ameliorates some of these problems, but does not remove them.

Since susy has failed to fully solve the hierarchy problem, we must critically examine the other arguments 
supporting it. The so-called ``WIMP-miracle", the claim that stable superpartners precisely give the required amount
of dark matter, has been substantially watered down in recent years. On closer inspection, it is off by a few orders of magnitude
 \cite{ArkaniHamed:2006mb},
and a ``non-thermal" WIMP miracle has been suggested \cite{Acharya:2009zt} in its place. Although this is based on WIMPs produced in out of equilibrium
decays of moduli, and fits nicely with string theory, two miracles is one too many.
Axions are a credible dark matter candidate, and several
authors have suggested scenarios where both kinds of dark matter are present \cite{Tegmark:2005dy,Acharya:2012tw}. But then we could also do
without WIMPs altogether.
Furthermore dark matter is constrained anthropically.
Although crude arguments based on structure formation of \textcite{Hellerman:2005yi} still allow a rather large window of five orders of magnitude,
this is not much larger than the uncertainty of the WIMP miracle. Furthermore it is far from obvious that life would flourish equally well in dense dark matter
environments so that the true anthropic bound might be much tighter.
The other main argument, gauge coupling unification, has already been discussed in section \ref{ConstUnif}. It is more seriously
affected by problems at the string scale than by the upward motion of the susy scale, on which it only depends logarithmically.

Ideas like {\it split} supersymmetry  (a higher mass scale just for the superpartners of fermions) and {\it high scale} supersymmetry 
(a larger susy scale) are becoming more and more
{\it salonf\"ahig} in recent years.
Perhaps counter-intuitively, their scales are constrained from {\it above} by the Higgs mass measurement
\cite{Giudice:2011cg}: in supersymmetric theories the Higgs self-coupling cannot become negative, as it appears to be doing.
 It is hard to avoid the idea that the most natural scenario is {\it no} supersymmetry. But that would also imply that
everything we think we know about the landscape is built on quicksand. This is a huge dilemma that we will hear a lot more about
in the future.

%
%
%
%
%
%
%
%
%

 \subsection{\label{Axions}Axions}

\ifExtendedVersion {\color{darkred}
Arguably the most serious problem of the Standard Model -- ignoring cosmology -- is the absence of any explanation why the  CP-violating
angle $\bar\theta$ of the strong interactions is as small as it is, see section \ref{StandardModelSection}. }\fi
Unlike the large gauge hierarchy, the extreme smallness of the strong CP-violating
angle $\bar\theta$ has few anthropic implications. 
Apart from producing as yet unobserved nuclear dipole moments, 
$\bar\theta$ {\it can}  have substantial effects on nuclear physics, including anthropically
 relevant features like deuteron binding energies and the triple-alpha process. In
 \textcite{Ubaldi:2008nf} the reaction rate of the triple-alpha process was found to be ten times larger if $\bar\theta=0.035$. But at best this would  explain two to three of the observed
 ten orders of magnitude of fine tuning.
 
\ifExtendedVersion {\color{darkred}
In a huge landscape we could attribute the small value of $\bar\theta$ to pure chance, but accepting that possibility undermines 
any discussion of other fine-tuning problems. 
Another option is that one of the quark masses (which could only be $m_u$) vanishes or is 
extremely small, but that option is in $4\sigma$ disagreement with the data. Furthermore this just shifts the fine-tuning problem from $\bar\theta$ to one
of the quark masses. There are other options, but one solution stands out because of its simplicity, the mechanism discovered by }\fi
There are several possible solutions, but one  stands out because of its simplicity: the mechanism discovered by
\textcite{Peccei:1977hh}. 
 It requires nothing more than adding a scalar $a$ and a non-renormalizable coupling:
\begin{equation}\label{AxionAction}
\Delta{\cal L} = \frac12 \partial_{\mu} a \partial^{\mu} a  + \frac{a}{32\pi^2 f_a} \sum_a F^a_{\mu\nu} F^a_{\rho\sigma}\epsilon^{\mu\nu\rho\sigma},
\end{equation}
 where $f_a$ is the ``axion decay constant".   Since $F\tilde F$  (where $\tilde F_{\mu\nu}=\frac12 \epsilon_{\mu\nu\rho\sigma}F^{\rho\sigma}$)
  is a total derivative, after integration by parts   the second term is
 proportional to $\partial_{\mu} a$. Hence there is a shift symmetry $a \rightarrow a+\epsilon$. This allows us
to shift $a$ by a constant $-\bar \theta f_a$ so that the $F\tilde F$ term (\ref{StrongCP}) is removed from the action.  However, the shift symmetry is anomalous 
with respect to QCD because the $F\tilde F$ term 
is a derivative of a gauge non-invariant operator. Through non-perturbative effects the anomaly generates a potential with a minimum at $a=0$ of the form
\begin{equation}\label{AxionPot}
V(a) \propto \Lambda_{\rm QCD}^4 \left( 1 - {\rm cos}(a/f_a)\right).
\end{equation} 
Note that $\bar\theta$ is periodic with period $2\pi$, so that the shift symmetry is globally a $U(1)$ symmetry.
It was pointed out by \textcite{Weinberg:1977ma,Wilczek:1977pj} that this breaking of the $U(1)$ symmetry
leads to a pseudo-scalar pseudo-Goldstone boson, which was called ``axion". 
The mass of this particle is roughly $\Lambda_{\rm QCD}^2/f_a$, but if we take into account the proportionality factors in (\ref{AxionPot})
the correct answer is 
\begin{equation}\label{AxionMass}
m_a=\frac{m_{\pi}  f_{\pi}}{f_a} F(m_q),
\end{equation}
 where $f_{\pi}$ is the pion decay constant and $F(m_q)$ a function of the (light) quark masses that is proportional to their product.  
The scale $f_a$ was originally assumed to be that of the weak interactions, leading to a mass prediction of order 100 KeV,
that is now ruled out.  But soon it was realized 
 that  $f_a$ could be chosen freely, and in particular much higher, making the axion ``harmless" or  ``invisible" (see \textcite{Kim:1986ax} and references therein).
This works if
the coupling $f_a$ is within a narrow window. For small $f_a$ the constraint is due to the fact that 
supernovae or
white dwarfs would cool too fast by axion emission. This gives a lower limit $f_a > 10^9$ GeV. 

The upper limit is cosmological. In the early universe
the axion field would be in a random point $\theta_0$ in the range $[0,2\pi]$ (``vacuum misalignment").
The potential (\ref{AxionPot}) is irrelevant at these energy scales. During the expansion and cooling
of the universe, the field remains at that value until the Hubble scale drops below the axion mass. Then the field starts oscillating in its potential, releasing the stored
energy, and contributing to dark matter densities. 
The oscillating axion field can be described as a Bose-Einstein condensate of axions. Despite the small axion mass, this is cold dark matter:  the axions were not thermally produced. Axions may in fact be the ideal dark matter
candidate \cite{Sikivie:2012gi}.

The axion contribution to dark matter density is proportional to 
\begin{equation}\label{Omegaa}
\Omega_a \propto (f_a)^{1.18} {\rm sin}^2(\frac12 \theta_0),
\end{equation}
 (see \textcite{Bae:2008ue} for a recent update and earlier references). 
The requirement that this does not exceed the observed dark matter density leads to a limit $f_a < 10^{12}$ GeV, unless $\theta_0 \approx 0$.    
 This results in a small allowed
window for the axion mass: $6\ \mu{\rm eV} < m_a < 6\ {\rm meV}$. Observing such a particle is hard, but one may use the fact that axions couple 
(in a model-dependent way) to two photons. Several attempts are underway, but so far without positive results.
The location of the axion window is fascinating. It is well below the GUT and  Planck scales, but roughly in the
range of heavy Majorana masses in see-saw models for neutrinos. It is also close to the point where the extrapolated Higgs
self-coupling changes sign, although there are large uncertainties.

There are many string-theoretic, landscape and anthropic issues related to axions. Candidate axions occur abundantly in string theory (see \textcite{Svrcek:2006yi} for details and earlier references). \ifExtendedVersion {\color{darkred}
Indeed, they were already discussed in the very first paper on 
heterotic string phenomenology \textcite{Witten:1984dg}.
There is a ``universal" or ``model-independent" axion related to the $B_{\mu\nu}$ field that
occurs in any string theory. In compactified strings one has, in addition to these, zero-modes of the compactified ten-dimensional
$B_{\mu\nu}$ field with indices $\mu$ and $\nu$ in the compactified directions. Then the number of zero-modes depends on
the topology of the compactification manifold. These are called ``model-dependent" axions. 
In type-II strings Kaluza-Klein modes of compactified RR anti-symmetric tensor fields yield many many more axions. 
Furthermore the required couplings are present in string theory, and can be derived from the tree-level or one-loop
terms that cancel anomalies in the Green-Schwarz mechanism. }\fi

But exact global symmetries, like
axion shift symmetries, are not supposed to exist in theories of quantum gravity, and hence they are not expected 
to exist in string theory. Therefore one expects all the candidate axions to acquire a mass.  The Peccei-Quinn (PQ) mechanism
can only work if a light axion survives with couplings to QCD, and with a mass contribution from other sources that
is much smaller than the QCD-generated mass.

\ifExtendedVersion {\color{darkred}
There is a second worry. In supersymmetric theories (which includes all string theories we can discuss)
 axions can be thought of as imaginary parts of complex
scalar fields  in chiral multiplets. The real part is sometimes called the saxion, and is a modulus in string theory. Moduli must be stabilized. Because axions have derivative couplings, they are far less constrained than moduli.
In particular, they are not constrained by fifth force limits, nor do they decay to affect BBN abundances. For
moduli, those constraints give a lower mass limit of order 10 TeV (although there ways out). Mechanisms that
give such masses to moduli may give the same masses to the axionic components, which is fatal for their
r\^ole as PQ axions. Axion-components of flux-stabilized moduli get stabilized as well, and hence acquire a
large mass. The same is true for stabilization due to instanton induced terms in the superpotential, of the form ${\rm exp}(a\rho)$; these
terms stabilize both the real as the imaginary component of the complex field $\rho$.  Furthermore some axions are projected
out by orientifold projections, while others can be ``eaten" by vector bosons in a Stueckelberg mechanism that give mass
to anomalous $U(1)$'s  (and often even to non-anomalous $U(1)$'s. 
}\fi

\ifExtendedVersion {\color{red}}\else {
Axions are imaginary parts of moduli, which must be stabilized,  and they must somehow escape getting a mass from the stabilization.
They must also survive orientifold projections and not be eaten by vector bosons in a Stueckelberg mechanism. }\fi
However, in most
string theories there exist candidate axions that are exactly massless to all orders in perturbation theory, and
which must therefore get their masses from non-perturbative effects. These effects can be expected to
give rise to  axion masses proportional to $e^{-S}$, where $S$ is an instanton action. 

It is not likely that a light axion exists just for QCD. From the string theory perspective, it would seem strange
that out of the large number of candidate axions just one survives. From the gauge theory perspective, many
different gauge groups with many different non-abelian factors are possible. 
Either they generically come with axions, or QCD is a special case for no apparent reason.
\ifExtendedVersion {\color{darkred}
 Even the Standard Model
itself has a second non-abelian factor. 
Although $SU(2)$ has no observable
$\theta$-angle, it would seem absurd that a PQ-mechanism exists just to make the observable $\theta_{\rm QCD}$ 
parameter vanish. 
}\fi

This has led to the notion of an ``axiverse" \cite{Arvanitaki:2009fg}, a plethora of axions,
with masses spread logarithmically over all scales;  only the mass of the QCD axion
is determined by (\ref{AxionMass}).
Realizations of an axiverse have been discussed in fluxless M-theory compactifications \cite{Acharya:2010zx}
and in type-IIB models in the LARGE Volume Scenario \cite{Cicoli:2012sz}. Both papers consider compactifications
with many K\"ahler moduli that are stabilized by a single non-perturbative contribution rather than 
a separate contribution for each modulus. Then all  K\"ahler moduli can be stabilized, but just
one ``common phase" axion acquires a large mass. All remaining ones get tiny masses from other instantons.
For supersymmetric moduli stabilization (such as the KKLT scenario, but unlike LVS) a no-go theorem was proved by \textcite{Conlon:2006tq}, 
pointing out that for each massless axion there would be a tachyonic saxion after up-lifting.  But in
\textcite{Choi:2006za} a generalization of the KKLT scenario was considered where this problem is avoided.
Axions in the heterotic mini-landscape were discussed by \textcite{Choi:2009jt}. They consider
discrete symmetries that restrict the superpotential, so that the lowest order terms have accidental $U(1)$ symmetries 
that may include a PQ symmetry. 

The upper limit $f_a < 10^{12}$ GeV is problematic for axions in string theory, which  
generically prefers a higher scale \cite{Svrcek:2006yi}. A way out of this dilemma is to assume that the misalignment angle in
Eq. (\ref{Omegaa})
is small. This is an option if the PQ phase transition occurred before inflation, so that we just observe a single
domain of a multi-domain configuration with a distribution of values of $\theta_0$. If the phase transition occurred after
inflation, we would instead observe an average of ${\rm sin}^2 \theta_0$, equal to $\frac12$. 
To allow an increase of $f_a$ to the GUT or string scale of about $10^{16}$ GeV a value of $\theta_0 \approx 10^{-3}$ would be
sufficient. One could even assume that this value came out ``by accident", which is still a much smaller accident than 
required for the strong CP problem.  However, the fact that the upper limit on $f_a$ is due to  the axion's contribution
to dark matter has led to the suggestion that we live in an inflated domain with small $\theta_0$ not by accident,
but for anthropic reasons \cite{Linde:1991km}. Furthermore, the fact that this parameter is an angle and that axions are 
not strongly coupled to the rest of the landscape makes it an ideal arena for anthropic reasoning \cite{Wilczek:2004cr}. 
This was explored in detail by \textcite{Tegmark:2005dy} and \textcite{Freivogel:2008qc}. The upper bound on the axion decay constant can be raised
if there is a non-thermal cosmological history, for example caused by decay of $\approx 30$ TeV moduli \cite{Acharya:2012tw}.

Whatever solution is proposed for the strong CP problem, it should not
introduce a fine-tuning problem that is worse. Therefore models specifically constructed and tuned to have a QCD axion in the allowed window, 
but which are rare within their general class, are
suspect.  This appears to be the case in all models suggested so far. The ``rigid ample divisors" needed in the
M-theory and type-II constructions mentioned above are not generic, and the discrete symmetries invoked
in heterotic constructions may be a consequence of the underlying mathematical simplicity of the orbifold
construction. But it is difficult to estimate the amount of fine tuning that really goes into these models.

The anthropic tuning required to avoid the upper bound on $f_a$ was discussed by \textcite{Mack:2009hv}. This author
concludes that avoiding constraints from isocurvature fluctuations in the CMB, which are observational and
not anthropic, requires
 tuning of both $\theta_0$ and the inflationary Hubble scale to small values. The amount of tuning is more than the
 ten orders of magnitude needed to solve the strong CP problem.
This problem increases  exponentially if there are many
axions \cite{Mack:2009hs}.

There are numerous possibilities for experiments and observations that may shed light on
the r\^ole of axions in our universe, and thereby provide information on the string theory landscape. The observation
of tensor modes in the CMB might falsify the axiverse \cite{Fox:2004kb,Acharya:2010zx}. See  
\textcite{Arvanitaki:2009fg,Marsh:2011bf,Ringwald:2012hr} for a variety of possible signatures, ongoing experiments
and references.

 \subsection{\label{VarCons}Variations in Constants of Nature}

If we assume that constants of nature can take different values in {\it different} universes, it is natural to ask if they might
also take different values within {\it our own} universe. In the Standard Model the parameters are fixed   (with a computable 
energy scale
dependence) and cannot take different values at different locations or times without violating the postulate of translation invariance.

There is a lot of theoretical and observational interest in variations of constants of nature, and for good reasons. 
The observation of such a variation would have a huge impact on current ideas in particle physics and cosmology. See
\textcite{Langacker:2001td} for a concise review and \textcite{Uzan:2002vq} for a  more extensive one, and \textcite{Chiba:2011bz} for
an update on recent bounds and observations. 
\ifExtendedVersion {\color{darkred}
The observations include high precision experiments using atomic clocks, 
the Oklo nuclear site (a natural occurrence of nuclear fission, about 2 billion years ago), quasar absorption  spectra, the CMB and Big Bang
Nucleosynthesis. Note that these phenomena occur on hugely different timescales, so that they cannot  be easily
compared. 
Some  observations explicitly look for time dependence, but for distant objects like quasars one cannot disentangle
time and space dependence so easily. }\fi \ifExtendedVersion {\color{darkred}The observed phenomena depend on various combinations of constants, and the results are most often presented
in terms of variations in $\alpha$ or the electron/proton mass ratio $\mu=m_e/m_p$. In both cases
the relevant limits or reported signals are of order $10^{-15}$ per year. 
}\else {
The results are most often presented
in terms of variations in $\alpha$ or the electron/proton mass ratio $\mu=m_e/m_p$. }\fi
The best current limits on $\Delta\alpha/\alpha$ are about
$10^{-17}$ per year, from atomic clocks and from the Oklo natural nuclear reactor. Recently a limit  $\Delta\mu/\mu  < 10^{-7}$ was found
by comparing transitions in methanol in the early universe (about 7 billion years ago) with those on earth at present \cite{Ubachs}.
\ifExtendedVersion {\color{darkred}
This is for a look-back time of 7 billion years, 
so this correspond to a limit on the  average variation of $1.4 \times 10^{-17}$ per year. 
There are also constraints on
the variation of Newton's constant from a variety of high precision observations, such as lunar laser ranging, binary systems, the Viking mars lander, Big Bang Nucleosynthesis and the CMB. These give limits of order $10^{-12}$ per year. 
}\fi

But in addition to limits there have also been positive observations. Using the Keck observatory in Hawaii and the Very Large Telescope (VLT)
in Chili, \textcite{Webb:2010hc} reported a {\it spatial} variation of $\alpha$. Earlier observations at Keck of a smaller value of $\alpha$, 
 at that time interpreted as a temporal variation \cite{Webb:2000mn}, 
combined with more recent VLT observations of a larger value,  fit a dipole distribution in the sky. These results have a statistical 
significance of  4-5$\sigma$. 
\ifExtendedVersion {\color{darkred}
There are also reports of variations in $\mu$ and the cosmological constant along the same
dipole direction, but with not more than 2$\sigma$ significance, see \textcite{Berengut:2010xp,Damour:2011fa}. Note hat the size of
the $\Delta{\alpha}/\alpha$ variation in a single direction is about $10^{-15}$ per year, and hence would disagree with the
atomic clock and Oklo bounds, assuming a linear time dependence. 
But there may be no such discrepancy if it is interpreted as a {\it spatial} variation, even taking into account the Earth's motion 
through the dipole  \cite{Flambaum:2010zz}.
It is not clear why $\alpha$ should vary more than other parameters, especially $\mu$. The latter is sensitive to $\Lambda_{\rm QCD}$ and
the Higgs vev. The former is expected to vary much more strongly than $\alpha$ if one assumes GUTs \cite{Langacker:2001td,Calmet:2001nu}; the former is expected, in certain 
landscape toy models, to vary much strongly more than dimensionless parameters.
}\else {
Because these results would imply a {\it spatial} and not a {\it temporal} variation,  
a clash with other, negative, results is avoided.}\fi
 
There are no good theoretical ideas for the expected size of a variation, if any. 
In string theory, and quite generally in theories with extra dimensions, the couplings are functions
of scalar fields, and are determined by the vacuum expectation value of those
fields, subject to equations of motion of the form (\ref{HubbleFriction}).
This makes it possible to maintain full
Poincar\'e invariance and relate the variations to changes in the vacuum. 
For example, the action
for electrodynamics takes the form
\begin{equation} \label{AlphaVar}
{\cal L}=  -\frac{1}{4 e^2}     e^{-\phi/M_{\rm Planck}} F_{\mu\nu}F^{\mu\nu},
\end{equation} 
where $\phi$ is the dilaton field or one of the other moduli. Variations in $\phi$ lead to variations in $\alpha$
\begin{equation}
\Delta \alpha \propto \frac{\delta{\phi}}{M_{\rm Planck}}
\end{equation}
All other parameters of the Standard Model have a dependence on scalar fields as well.
Although this formalism allows variations in $\alpha$, it is clearly a challenge to explain why they would be as small as $10^{-15}$ per year. Note that this is
about $10^{-66}$ in Planck units, the natural
units of a fundamental theory like string theory.  

\ifExtendedVersion {\color{darkred}
This is reminiscent of the cosmological constant problem and the flatness problem, where it seemed reasonable to
assume that a ridiculously small number is actually zero. But we have strong indications that at least for the cosmological
constant this assumption is wrong.  It is clear that Planck size variations in couplings are anthropically not acceptable, nor are far smaller variations.
Parameters should stay within the absolute anthropic bounds, and even much smaller variations than that could be catastrophic. If
parameters like $\alpha$ and $\mu$ were to change substantially during evolution, there would not be enough time for
organisms to adapt to the changes in energy levels of their molecules. 
Although it is hard to arrive at a strict limit from such an argument, it seems clear that changes 
far larger than $10^{-15}$ per year would be acceptable, so that the near constancy of parameters cannot be
explained anthropically. It also seems preposterous to assume that a complete {\it function}, not just a value, is
fine-tuned for the existence of life.
Furthermore any such argument would still allow almost arbitrarily large spatial variations. 

While fast changes have a negative impact, slow changes might be beneficial. It is unlikely that
the Standard Model parameters are at their optimal anthropic value for BBN, Stellar Nucleosynthesis and biological evolution
simultaneously. If parameters could change slowly, we might expect to find ourselves in a universe where the parameters were
optimized for each cosmological epoch. Apparently that is not the case, because even during BBN the parameters differed at most
a few percent from their current values. 

Hence the observation of a variation in any Standard Model parameter would imply a huge fine-tuning problem, with little hope of an
anthropic explanation. Then the most attractive way out is that {\it within} our universe these parameters really are constants, although
they must vary in the multiverse.}\else {
The observation of a variation in any Standard Model parameter would imply a huge fine-tuning problem, with little hope of an
anthropic explanation: variations of fundamental parameters might have adverse effects on the evolution of life, but there is no reason
why the variation has to be as small as it is. }\fi
Then the most attractive way out is that {\it within} our universe these parameters really are constants, although
they must vary in the multiverse. The string theory landscape solves this problem in an elegant way, because each of its ``vacua" is at the bottom of a deep potential, 
completely suppressing any possible variations of the  moduli at sub-Planckian energies. 

This can be seen by considering the effect of changes in vevs of moduli fields on  vacuum energy. 
Here one encounters the problem that
contributions to vacuum energy in quantum field theory are quartically divergent. But this cannot be a valid reason to ignore them completely, as is often done
in the literature on variations of constants of nature. \textcite{Banks:2001qc} have pointed out that if a cut-off $\Lambda_{\rm cutoff}$
is introduced in quantum field theory, then the effect of a change in $\alpha$ on vacuum energy $V$ is
\begin{equation}\label{AlphaVac}
\delta V \propto \Delta{\alpha} (\Lambda_{\rm cutoff})^4.
\end{equation}
With $\Lambda_{\rm cutoff}=100$ MeV, the QCD scale, and assuming that vacuum energy should not dominate at the earliest stages of galaxy 
formation (corresponding to the time when quasar light was emitted), this gives a bound of $\Delta\alpha/\alpha < 10^{-37}$. If one assumes 
that $\delta V$ depends on $\Delta\alpha$ with a power higher than 1, this bound can be reduced, but a power of at least 8  is required to accommodate
the observed variation. This can only be achieved by a correspondingly extreme tuning of the scalar potential. 
Spatial variations are restricted by similar
arguments, although less severely.


\ifExtendedVersion {\color{darkred}
This is a general field theoretic argument, but it explicitly assumes a temporal, and not a spatial variation. It is assumed that quasars observed today 
give us information about the early stages of our own galaxy. In field theoretic models smooth spatial variations are harder to obtain than temporal variations \cite{Olive:2007aj}, but
it is possible the variation is due to the existence of different domains, with changes of $\alpha$ across a domain wall, which one can try to localize \cite{Olive:2012ck}.

In the string theory landscape $\Lambda_{\rm cutoff}$ is of order the Planck scale, and the cosmological constant is a sum of terms of order $M^4_{\rm Planck}$ which cancel
with 120 digit precision. Even the tiniest change in one of the contributions would completely ruin the cancellation, and the corresponding part of space-time would either
collapse or expand exponentially. 
It $\alpha$ changes by $10^{-5}$ in comparison to the value on earth is simply not possible that we observe a quasar there, and this is true for both temporal and spatial
variations (for temporal changes the limit  is $\Delta\alpha/\alpha < 10^{-104}$ \cite{Banks:2001qc}).

One may still entertain the thought that we are witnessing a entirely different vacuum of the string landscape, with a vacuum energy tuned to a very small value by
a different combination of fluxes (or other features)  than in our own domain. But there is no good reason why such a distinct vacuum would only differ from ours by a minor
variation of $\alpha$, and nothing more substantial. It is not even clear why it should couple to the same photon.  Another possibility is that one modulus has escaped moduli
stabilization and has remained extremely light, providing an almost flat direction along which several ``constants" vary, canceling each other's contribution to the cosmological
constant. But light moduli that have significant couplings tend to generate ``fifth forces" violating the equivalence principle. This is a general problem associated with variations in constants of nature, as observed
a long time ago by \cite{RevModPhys.29.355}. For a recent discussion see \cite{Damour:2011fa}.
}\else {

There are also constraints from ``fifth forces" violating the equivalence principle. This is a general problem associated with variations in constants of nature, as observed
a long time ago by \textcite{RevModPhys.29.355}. For a recent discussion see \textcite{Damour:2011fa}.
}\fi

Currently the observation of variations in constants of nature is still controversial, but there is a lot at stake. Evidence for variations would be good news for half of this review, and bad news for the other half.
If the parameters of the Standard Model already vary within our own
universe, the idea that they are constants can be put into the dustbin of history, where it would be joined almost certainly by the string theory landscape. String theory would be set back by about two
decades, to the time where it was clear that there were many ``solutions", without any interpretation as ``vacua" with a small cosmological constant.

\section{\label{EternalInflation}Eternal Inflation}

If string theory provides a huge ``landscape" with a large number of ``vacua",  how did we end up in one particular one? \ifExtendedVersion {\color{darkred} Some process where 
new universes are created from already existing is evidently required. This notion precedes the string theory landscape by many years, 
see for example \textcite{Linde:1986fe}, and especially Fig. 3. }\fi
The answer is eternal inflation, a nearly inevitable implication of most theories of inflation.
 \ifExtendedVersion {\color{darkred}  In theories of inflation it is nearly inevitable that 
inflation is eternal. In slow-roll inflation this happens if inflation stops only in certain regions of the universe, whereas
the rest continues expanding. It may then happen that several distinct vacua may be reached as the endpoint of inflation.
This is called {\em slow-roll eternal inflation} (SREI). Classically, the inflating universe is described by a dS space-time. 
But dS universes are not believed to be absolutely stable \cite{Susskind:2003kw}. Locally, decays can occur, when
bubbles of other vacua form and expand. This is called {\em false vacuum eternal inflation} (FVEI). Whether inflation really is eternal
depends on the decay rate: 
the total volume of false vacuum must grow faster than it decays or stops inflating.  If in the string theory landscape there
is at least one such vacuum,  as soon as  the multiverse ends up in such a configuration it will start inflating eternally. Even our own
universe could be that vacuum, provided it lives an order of magnitude longer than its current age.
Then ``typical" universes will have a long period of eternal inflation in their past. }\fi See \textcite{Guth:2000ka};  
\textcite{Linde:2002gj};  and \textcite{Freivogel:2011eg} 
for more discussion and references. If there is a possibility for transitions to  other universes, then this  would inevitably trigger
an eternal process of creation of new universes.
\ifExtendedVersion {\color{darkred}
Here ``eternal" means future eternal. Whether this also implies past-eternal has been a matter of debate, see 
\textcite{Mithani:2012ii, Susskind:2012yv,Susskind:2012xf}. }\fi

For different views on eternal inflation or on populating the landscape see respectively \textcite{MersiniHoughton:2012ez}
and \textcite{Hawking:2006ur}.

\subsection{Tunneling}

\ifExtendedVersion {\color{darkred} In the case of FVEI the decays can take place in various ways. }\else { Vacuum decay can take place in various ways. }\fi
The best known process were described by
\textcite{Coleman:1980aw} and 
by \textcite{Hawking:1981fz}. The former describes tunneling between false vacua, and the latter
 tunneling of a false vacuum to the top of the potential. \ifExtendedVersion {\color{darkred}
 The precise interpretation of these processes requires more discussion, see
 {\it e.g.} 
 \textcite{Affleck:1980ac,Weinberg:2006pc,Brown:2007sd}. }\fi  
 These processes
generate the nucleation of
bubbles of other vacua which expand, and then themselves spawn bubbles of still more vacua \cite{Lee:1987qc}.  
Tunneling between dS vacua may occur in both directions, up and down in vacuum energy, although up-tunneling is strongly suppressed 
with respect to down-tunneling (see {\it e.g.}  \textcite{SchwartzPerlov:2006hi})
\begin{equation}
\Gamma_{i \rightarrow j}=\Gamma_{j\rightarrow i}\ {\rm exp}\left(24\pi^2\left[\frac{1}{\Lambda_j}-\frac{1}{\Lambda_i}\right]\right).
\end{equation}
The endpoint of tunneling may be another dS vacuum, but it
may also be a Minkowski or AdS vacuum. Whether tunneling from Minkowski to
AdS is possible is disputed in \cite{Dvali:2011wk,Garriga:2011we}.
Minkowski vacua do not inflate,  and  AdS universes  collapse classically in a finite amount of time. 
Up-tunneling from these vacua to dS space is impossible, and therefore
they  are called terminal vacua.  They are ``sinks in the probability flow" \cite{Ceresole:2006iq,Linde:2006nw}.
According to \textcite{Bousso:2011aa,Susskind:2012pp} 
their existence in the landscape may be essential for understanding the arrow of time and for avoiding the
Boltzmann Brain problem (see below). Even though a  large portion of an eternally expanding universe ends up in a terminal vacuum, the rest continues expanding forever. A typical observer is expected to have a long period of eternal inflation in his/her/its past
\cite{Freivogel:2011eg}.

\subsection{\label{MeasureProblem}The Measure Problem.}

The word ``eternal" suggests an infinity, and this is indeed a serious point of concern. As stated in many papers:
{\em ``In an eternally inflating universe, anything that can happen will happen; in fact, it will happen an infinite number of times"}.
This, in  a nutshell, is the measure problem (see reviews by
\textcite{Vilenkin:2006qf}; \textcite{Guth:2007ng}; \textcite{Freivogel:2011eg}; and \textcite{Nomura:2012nt}).
If we want to compute the relative probability for events A and B, one may try
 to define it by counting the number of occurrences of A and those of B, and taking the ratio. But both numbers are
infinite. \ifExtendedVersion {\color{darkred} 

At this point one may even question the very concept of probability, in a situation where we will
never more than one item in a set. But cosmology is in exactly the same situation, and this is what gives rise to the 
``cosmic variance" problem. In fact, all arguments in particle physics that rely on naturalness make
implicit assumptions about probabilities.  The same is true for all predictions in particle physics.
Suppose a class of models fits all the Standard Model data, and a large fraction of that class has a certain property, for
example an additional $Z'$ boson. We look like then to predict the existence of this boson. 
But  without a notion of probability and a measure on the class of models, such statements are, strictly speaking, meaningless.
Usually the existence of a notion of probability is take for granted, and some naive measure is used.

Infinities are not unusual in physics, especially when we wander into unknown territory. 
They may sometimes be taken care of by
some kind of regularization, where the infinity is cut off. This then defines a ``measure".
One may hope that at the end one can take the cutoff to infinity. 
During the past two decades many such regularizations  (``measures") have been tried. A first test is to compute some ratios
of probabilities, and check if they even make sense. 
The earliest attempts led to absurd paradoxes.

%

Infinities may
also indicate that something fundamental is missing. For example, in quantum field theory several kinds of infinities occur.
Infrared divergences merely tell us that we should use appropriate initial states; renormalizable ultraviolet divergences tell us
that some parameters in the theory are ``external" and cannot be determined by the theory itself; Landau poles in the
running of couplings towards large energies point to new physics. In the present case, it could well be that the infinities signal that entirely new ingredients are needed that do not follow
from the underlying theory (string theory and/or quantum mechanics). The final verdict about this is not in yet. }\fi

It is not that hard to think of definitions that cut off the infinities, but many of them make disastrous  predictions. For example, they may
predict that observers -- even entire solar systems with biological evolution -- created by thermal or quantum fluctuations (``Boltzmann Brains") vastly
outnumber ones like ourselves, with a cosmological history that can be traced back in a sensible way. Or they may predict that universes just a 
second younger than ours are far more numerous (the ``Youngness paradox"). 
\ifExtendedVersion{\color{darkred} See section \ref{Paradoxes} for a more detailed explanation of these paradoxes.}\fi
If these predictions go wrong, they go wrong by double exponentials, and a formalism
that gives this kind of a prediction cannot be trusted for {\it any} prediction. 

\subsubsection{The Dominant Vacuum}
An ingredient 
that could very well be missing is a theory for the initial conditions of the multiverse. It would be unduly pessimistic to assume that
this is a separate ingredient that cannot be deduced from string theory (or whatever the theory of quantum gravity turns out to be). 
If it cannot be deduced by logical deduction, it might be impossible to get a handle on it. 

But eternal inflation may make this 
entire discussion unnecessary, provided all vacua are connected by physical processes. 
In that case, successive tunneling events may drive all of them  to the same ``attractor",  the longest lived dS vacuum whose occupation numbers dominate
the late time distribution. This is called the ``dominant vacuum" \cite{Garriga:1997ef,Garriga:2005av,SchwartzPerlov:2006hi}. Since tunneling rates are exponentially suppressed, this vacuum may dominate by a huge factor.
Then the overwhelming majority of vacua would have this attractor vacuum in its
history. This would erase all memory of the initial conditions. 
\ifExtendedVersion {\color{darkgreen} 

If the space of vacua is not physically connected, it falls apart into
disconnected components, each with a separate dominant vacuum. Then we are faced with the problem of determining in which component our universe is located. A disconnected landscape is a nightmare scenario. It is like having a multitude of ``theories of everything", and given the numbers one usually encounters the multitude might be a huge number.
But in \cite{Brown:2011ry} it was argued that despite some potential problems (vacua not connected by instantons, or only  connected through sinks \cite{Clifton:2007en} --  all dS vacua are reachable with non-zero transition rates. This
result holds for minima of the same potential, but arguments were given for parts of the landscape with different topologies as well. See \textcite{Danielsson:2006xw}; 
\textcite{Chialva:2007sv}; and \textcite{Ahlqvist:2010ki} for a discussion of connections between Calabi-Yau flux vacua.
}
\else
 {
Furthermore \textcite{Brown:2011ry} have argued that despite some potential problems -- vacua not connected 
by instantons, or only  connected through sinks \cite{Clifton:2007en} --  all dS vacua are reachable with non-zero transition rates. This
result holds for minima of the same potential, but arguments were given for parts of the landscape with different topologies as well. See \textcite{Danielsson:2006xw}; 
\textcite{Chialva:2007sv}; and \textcite{Ahlqvist:2010ki} for a discussion of connections between Calabi-Yau flux vacua.
 }\fi

The ``dominant vacuum" may sound a bit like the old dream of a selection principle. Could this be the mathematically unique vacuum
that many people have been hoping for? Since it can in principle be determined from first principles (by computing
all vacuum transition amplitudes) it is not very likely that it would land exactly in an anthropic point in field theory space, see Fig.
\ref{LandscapeDistributions}. 
\ifExtendedVersion {\color{darkred}
\textcite{Douglas:2012bu} argues that the dominant vacuum may be the dS vacuum with the lowest supersymmetry
breaking scale, since broken supersymmetry destabilizes the vacuum. That scale is likely to be many orders of magnitude below the
Standard Model scale, and is assumed to be not anthropic. Otherwise it would lead to the clearly wrong prediction that the vast majority of observers in the string theory landscape see an extremely small supersymmetry breaking scale. }\fi If the dominant vacuum
is not itself anthropic, the anthropic vacuum 
reached from it by the largest tunneling amplitude is now a strong candidate for describing our universe. With extreme optimism one
may view this as an opportunity to compute this vacuum from first principles \cite{Douglas:2012bu}.
Unfortunately, 
apart from the technical obstacles, there is a more fundamental problem: the dominant vacuum itself depends on the way the measure is defined.

\subsubsection{Local and Global Measures}

The earliest attempts at defining a measure tried to do so globally for all of space-time
by defining a time variable and imposing a cut-off. Several measures of this kind have been proposed, which we will
not review here; see the papers cited above and references therein.

But a comparison with black hole physics provides an important
insight why this may not be the right thing to do.  There is a well-known discrepancy between
information disappearing into a black hole from the point of view of an infalling observer or a distant observer. In the former case
information falls into the black hole with the observer, who does not notice anything peculiar when passing the horizon, whereas
in the latter case the distant observer will never see anything crossing the horizon. A solution to this paradox is to note that the
two observers can never compare each others observations. Hence there is no contradiction, as long as one does not try to
insist on a global description where both pictures are simultaneously valid. This is called {\em black hole complementarity}  (and
has come under some fire recently; see \textcite{Braunstein:2009my,Almheiri:2012rt} and later papers for further discussion).

The same situation exists in eternal inflation. The expanding dS space, just like a black hole, also has a horizon.
In many respects, the physics is in fact analogous \cite{Gibbons:1977mu}.
If it is inconsistent to describe black hole physics simultaneously from the distant and infalling observer perspective, the same
should be true here. 
This suggests that one should only count observations  within the horizon.  
This idea has been implemented by several
authors in somewhat different ways. The causal patch measure \cite{Bousso:2006ev} only takes into account
observations in the causal past of the future endpoint of a word line. Several variations on this idea exist which we
will not attempt to distinguish here. Remarkably, in some cases these local measures are equivalent to global ones
(local/global duality), see \textcite{Bousso:2008hz,Bousso:2009mw}. 

Using only quantum mechanical considerations, 
\textcite{Nomura:2011dt} has  developed a picture that only includes observations by a single observer. In the end, probabilities
are then defined as in quantum mechanics, as squares of absolute values of coefficients of a quantum state.  In this approach, {\em ``the multiverse lives in probability space"}, and this is claimed to be tantamount to the many-world interpretation 
of quantum mechanics. 
Such  a relation has been pointed out by others as well \cite{Susskind:2003kw, Bousso:2011up,Aguirre:2010rw,Tegmark:2009pj}, but
it is too early to tell whether all these ideas are converging.

The current status can be summarized by two quotes from recent papers. \textcite{Nomura:2012nt} states emphatically 
{\em ``The measure problem in eternal inflation is solved"}, 
whereas  just a year earlier \textcite{Guth:2011ie} concluded 
{\em ``We do not claim to know the correct answer to the measure question, and so far as we know, nobody else does either."}

%
%
%
%
%
%
%
%
%
%
%
%
%

\ifExtendedVersion {\color{darkred}

\subsection{Paradoxes}\label{Paradoxes}

\subsubsection{The Youngness Problem}

Attempts at regulating the infinities in the definition of the measure can easily  lead to bizarre results. One approach, the proper time
cutoff  \cite{Linde:1993nz,Vilenkin:1994ua} is to choose a cutoff in some global time variable. This implies that there is a global
notion of time that applies not only to the eternally inflating space, but also to all bubble universes created in it. This obviously involves
an arbitrary choice of reference frame, and it has to be demonstrated that this does not matter. Now multiplicities of event can be made 
finite by only counting
events prior to the
cutoff. This leads to the problem that after our universe started, the eternally inflating space around it continues to
expand, producing other universes. While our universe goes through its evolution, an immense number of
others got started. Because of the global time definition, this number is well-defined and calculable.
If we analyze the resulting set of universes at the present time, then we would find that our own universe is vastly 
out-numbered by younger ones.  The use of a measure comes with a notion of typicality, sometimes called the ``principle
of mediocrity" \cite{Vilenkin:1994ua}. One would like predict features of our own universe by postulating that among all universes
the one we live in is typical, given the required conditions for life. Given a measure we can compute the relative rates
of various features, and check if we are typical or not. In the present case, we would find that most observers similar to ourselves
would find themselves in a much younger universe \cite{Guth:2007ng}. If such arguments go wrong, they usually do not go wrong
by a small percentage. In this particular case, 
our own universe is outnumbered by universes just a second(!) younger than ours by a factor 
${\rm exp}[10^{37}]$ (note the double exponential, which occurs frequently in this kind of argument).  
This means that either we are highly atypical (and with extremely high probability the only civilization in our own
universe, as \cite{Guth:2000ka} remarks jokingly), or that there is something terribly wrong with the measure. 

\subsubsection{Boltzmann Brains}

Another notorious problem is the Boltzmann Brain paradox \cite{Dyson:2002pf,Page:2006dt,Bousso:2006xc}.
If our universes continues an eternal expansion it will
eventually enter a stage where all stars have burned out, all life has disappeared,  all black holes have evaporated
and all protons have decayed (see \cite{Dyson:1979zz}\cite{Adams:1996xe} for a early accounts). 
Even in the Standard Model the proton eventually
decays due to weak instantons or gravitational effects, and therefore after a mere $10^{100}$ years there would be no proton left.

To appreciate the relevant time scales one can compare this to the Poincar\'e recurrence time \cite{Dyson:2002pf}.
This is defined by analogy with finite systems, which
cycle through all available states in a finite time. The number of states, and hence the entropy of dS space 
is indeed finite, and given by $S_{\rm dS}=A/4G_N$ \cite{Gibbons:1977mu}, where 
$A$ is the area of the even horizon. The area is $4\pi L^2= 4\pi/\Lambda$. Therefore $S_{\rm dS}$ is roughly equal to $10^{120}$
(the inverse of the cosmological constant in Planck units), and this gives a recurrence time of roughly
${\rm exp}[10^{120}]$ years (there is no need to be precise about times scales here, because expressed in Planck seconds or Hubble times the number is essentially the same). On this time scale, the first $10^{100}$ years during which the visible universe lives and dies are irrelevant.

On such 
enormous time scales even the most unlikely phenomena will eventually occur, and this includes the fluctuation of observers, or even part of the
universe out of the vacuum. This can happen due to quantum or thermal fluctuations. The latter (like a broken teacup that 
spontaneously re-assembles itself from its fragments)
are possible because 
an empty  dS  universe has a non-zero
temperature 
$T_{dS}=H_{\Lambda}/2\pi$, where $H_{\Lambda}=\sqrt{\Lambda/3}$ is the Hubble parameter associated with the expansion. 
For our universe, this gives a temperature of about $2\times 10^{-30} K$, so that $kT_{\rm}$ corresponds to $10^{-69}\ {\rm kg}$
in mass units. Hence the Boltzmann suppression for fluctuating a single hydrogen atom out of the vacuum is about 
${\rm exp}(-10^{42})$ (this ignores suppression factors due to baryon number violation, which give irrelevant contributions to the
exponent). If we define a minimal observer as a human brain, this gives a suppression factor of about ${\rm exp}(-10^{69})$.
This sort of thermal fluctuation occurs ${\rm exp}(10^{120})\times {\rm exp}(-10^{69}) \approx {\rm exp}(10^{120})$ times per
recursion time.
Quantum fluctuations are much less suppressed if one only requires the state to exist for a very short time; see \cite{Page:2006hr}
for a discussion of various options. However the status of quantum fluctuations as observers is more controversial \cite{Carlip:2007id}\cite{Davenport:2010jy}. 

Such a ``Boltzmann brain" could be in any on the quantum states a brain can have: most
of them completely non-sensical, but an extremely small fraction would have the same memories as observers living today. The
problem is that if our universe continuous expanding eternally, freak observers outnumber ordinary ones by
fantastically large numbers. 

To some this argument may seem absurd. Freak observers appear to be in no way comparable to us. One may, for example, call their 
consciousness, scientific curiosity, or their chances of survival into question. Perhaps the flaw in the argument is a misuse
of the notion of typicality? See \textcite{Hartle:2007zv,Garriga:2007wz,Page:2007bt,Bousso:2007nd} for different views on this. 
But one can make the argument a lot less bizarre by assuming that our entire solar system, or if necessary the entire galaxy,
pops out of the vacuum in the state it had 4 billion years ago (replacing a single brain  by the entire Milky Way galaxy changes the second exponent from 69 to 112, but there would still be about
${\rm exp}[10^{120}]$ freak galaxies per recurrence time). 
Then ordinary evolution can take place, and intelligent beings would evolve that develop science and that can in all respects be
compared to ourselves, except that they observe no other galaxies and no CMB. Such observers would still outnumber us by a double exponential. 
As much as one would like to reject such reasoning as nonsensical, it is difficult to find a convincing counter argument.

But there are several ways out. The measure problem is one possible culprit. In an infinite universe, both the number of ordinary
observers and the number of freak observers becomes infinite. This infinity must be regulated. In a landscape, one has to consider
not just our universe, but {\it all} universes that can produce freak observers (this includes universes that would not
be anthropic in the conventional sense, because freak observers can avoid most evolutionary bottlenecks; however
complexity is still a requirement). 
For discussion of the implications of the measure on the Boltzmann brain problem see \textcite{Vilenkin:2006qg,DeSimone:2008if,Bousso:2007nd}. In addition to the  choice of measure, the lifetime of the universe is evidently relevant.
For our universe, attempts to derive a limit have led to vastly different estimates
from as little as 20 billion years \cite{Page:2006dt} to ${\rm exp}(10^{40\pm 20})$ years (``the least precise prediction in the
history of science", as \textcite{Freivogel:2008wm} proudly remark). The huge difference is due to a different choice of measure: \textcite{Page:2006dt} includes all
the volume created by the exponential expansion. This includes regions that are not in causal contact, and that perhaps should not
be included (see below).  If the limit would be 20 billion years, it seems very unlikely that all relevant vacua in the string landscape
satisfy it, but for a doubly-exponential limit chances are much better. This is still a very strong constraint: every single vacuum in the 
landscape should have a decay width per volume that is larger than the production rate of freak observers. In KKLT-type vacua this
bound seems to be satisfied \cite{Westphal:2007xd,Dine:2007er,Freivogel:2008wm}.
There is yet another way out. If the fundamental parameters of the Standard Model are slowly
varying, we may gradually move out of the anthropic region of parameter space \cite{Carlip:2007id}, so that for most of the future
of our universe observers can simply not exist. However, slowly-moving parameters are not expected in the string landscape.

}\fi

\section{The Cosmological Constant in the String Landscape}\label{CosmoConstTWO}

The anthropic explanation for the smallness of $\Lambda$ requires a fundamental theory with a distribution 
of values of $\Lambda$, realizable in different universes. In string theory, this is provided by the 
Bousso-Polchinski discretuum (see section \ref{BoussoPolchinski}). This yields a dense set of 
$10^{\rm hundreds}$  discrete points over the full Planckian range\footnote{The smoothness of this distribution near zero is important, and requires further discussion, see \textcite{SchwartzPerlov:2006hi} and \textcite{Olum:2007yk}.} of $\rho_{\Lambda}$.
If this set does indeed exist,
it would be fair to say that string theory combined  with anthropic arguments explains the first  120 digits of $\rho_{\Lambda}$ on a particular
slice through parameter space.  \ifExtendedVersion {\color{darkred} Unfortunately, the point is weakened by the fact that all those digits are zero, and that there is no sound prescription 
for going beyond that. 
is not like computing fifth order corrections in QED to get the next digits of $g\!-\!2$.}\else {
But of course all those digits are zero. }\fi

To go  beyond  this  we need better control of inflation, to deal with variations in $Q$\ifExtendedVersion {\color{darkred}, better control of other aspects of cosmology to
take into account the effect of }\else { and }\fi other parameters. We also need  a solution to the measure problem and a better understanding  of the issues of typicality
and the definition of observers.  
At this moment the subject is still very much in a state of flux, without clear convergence to  a definitive answer. For example,
using different assumptions about the measure and different ways of parametrizing observers, \textcite{Bousso:2007kq}, \textcite{DeSimone:2008bq} and \cite{Larsen:2011mi} 
obtained cosmological constant distributions that peak  closer to the observed value than earlier work using the Weinberg bound. The first 
authors used the amount of entropy produced in a causal patch as a proxy for observers. The second used a global measure, and the last group  used the solution to the measure problem proposed by \textcite{Nomura:2011dt}; the latter two use conventional 
anthropic criteria.

An important test for solutions to the problem is whether they can explain coincidences (see {\it e.g}. \textcite{Garriga:2002tq}). 
The most famous of these is the ``why now" problem: why do we live fairly
close (within a few billion years) to the start of vacuum energy domination. By its very definition, this is an anthropic question. Another striking coincidence is
the order of magnitude of the absolute value of upper and lower bounds on $\Lambda$ ({\it c.f.} Eq. (\ref{LowerBound})). In other words, the life span of typical stars
is comparable to the age of the universe and the starting time of vacuum energy domination.  This depends on an apparent  coincidence between cosmological parameters
and Standard Model parameters,
$\rho_{\Lambda} \approx (m_p / M_{Planck})^6$. 

In essentially all work determining $\Lambda$ one of the coincidences is input, and determines the scale for the $\Lambda$ distribution.
For example in work based on galaxy formation, the quantity $Q^3 \rho_{\rm eq}$ determines that scale, but the ``why now" coincidence is
not solved. On the other hand, in \textcite{Bousso:2007kq} the time of existence of observers is the input scale, so that the ``why now" problem is
solved if $\rho_{\Lambda}$ peaks near 1 on that scale. This then turns the proximity of the maximum $\rho_{\Lambda}$ for galaxy formation, {\it i.e.} the
Weinberg bound, into a unexplained coincidence. If the cosmological constant can be computed as a pure number, as suggested
for example by \textcite{Padmanabhan:2012gv}, all these coincidences remain unexplained. The same is true if $\rho_{\Lambda}$ can be expressed in terms
of some Standard Model parameters, or if it is determined by the lowest possible value in the discretuum (see below). In all cases additional
arguments will be needed to explain these coincidences, or they will remain forever as unsolved naturalness problems.

Still more coincidences are listed  in \textcite{Bousso:2009ks}. These authors attempt to explain them by arguing
that landscape  distributions may drive us towards the intersection of multiple catastrophic boundaries, beyond which life is impossible.
The boundaries are computed using traditional anthropic arguments in universes with Standard-Model-like particle physics. They conjecture   that the gauge hierarchy, via
the aforementioned stellar lifetime coincidence, might be related to the cosmological constant hierarchy. The latter may then find an explanation in the discreteness of the landscape,
a possibility also suggested by \textcite{Bousso:2010zi}. This requires a total number of (anthropic) string vacua of about $10^{120}$.
A very different approach to coincidences is used by
\textcite{Bousso:2010im}, who argue that the coincidences can be understood entirely in terms of the geometry of cutoffs that define the measure in eternal
inflation.  They use a  minimal anthropic assumption, namely that observers  are made out of matter.


Several authors hope to avoid the anthropic argument, even though they accept the existence of a landscape, by suggesting that
the probability distribution of $\rho_{\Lambda}$ is peaked at zero. However,  strong peaking near zero for pure dS spaces is not likely to work. 
Only gravity can measure the cosmological  constant, and in the early universe,  when the ground state is selected,  its value is negligible in comparison to all other contributions.  
See  \textcite{Polchinski:2006gy} for a more extensive explanation of this point.

Despite this  objection, some authors speculate that somehow
the cosmological constant is driven to the lowest positive value $\Lambda_{\rm min}$.
The value of $\Lambda_{\rm min}$ is then roughly equal to the inverse of  $N$, the total number of vacua. 
For variations on this idea see \textcite{Kane:2005cd,Linde:2010nt}.
A different proposal was made
in   \textcite{Kobakhidze:2004gm}, who  suggest $\Lambda_{\rm min} = 1/N^2$.
In \textcite{HenryTye:2006tg,Sarangi:2007jb}, it is argued that due to ``resonance tunneling" all vacua have very short lifetimes, except some with very small $\Lambda$. 
Ideas of this kind would leave all apparent anthropic tunings unexplained. 

In the full set of string vacua, not just pure dS but including matter, there may well exist a unique vacuum, defined by having the
  smallest positive $\Lambda$.
 But this is not likely to be   our universe, since a unique vacuum will not satisfy the
other anthropic requirements. Even if for some reason it is strongly selected, this will generate run-away behavior in other variables, or leads to the kind of catastrophic predictions explained in section \ref{Catastrophic}.

Some authors use an analogy with solid state physics to argue that because of tunneling the true ground state 
wave function is a Bloch wave. But there is an important difference. In solid state physics observation times are much larger than
tunneling times, whereas in the landscape it is just the other way around. If observations are made at times much shorter than the 
tunneling time, this leads to collapse of the wave function and decoherence. Furthermore, in the landscape there must exist tunneling processes that change gauge groups,
representations and parameters. These can therefore not be treated as superselection sectors. The best one could hope to get is a linear combination
of amplitudes with different values of all Standard Model and cosmological parameters, which does not solve the problem of determining them.

Should we expect to understand why $\Lambda > 0$ in our universe, or is the sign just selected at random? On the one hand, from the perspective of 
vacuum energy in
quantum
field theory the point $\Lambda=0$ is not special. Nor is it special from the anthropic perspective:  life  with $\Lambda < 0$  seems perfectly possible.
On the other hand, classical physics and cosmology at late times are extremely sensitive  to the sign: the universe either collapses or expands. 
\ifExtendedVersion {\color{darkred} One might say that the
sign of $\Lambda$ matters, but the sign of $\rho_{\Lambda}$ does not. }\fi
The difference in sign implies important differences in
quantum physics.
The definition of the S-matrix in
quantum field theory (and string theory) is problematic in dS. \ifExtendedVersion {\color{darkred} There is an AdS/CFT correspondence but  no (known) dS/CFT correspondence. }\fi Tunneling
amplitudes between vacua are singular for $\Lambda \rightarrow 0$ (see section \ref{EternalInflation}).
In AdS spaces  any possibility of life finishes
at the crunch, and it matters how closely one can approach it; in dS spaces life is not limited by a crunch, but by the burning out of stars within the Hubble horizon
(see \textcite{Peacock:2007cw}  for an interesting discussion). Note that many authors consider only positive values for $\Lambda$, and some that do not
 ({\it e.g.} \textcite{Bousso:2010im}) actually predict negative $\Lambda$ more strongly than positive $\Lambda$.   The differences between AdS and dS are too large to assume blindly  that we ended up in a dS universe purely
by chance.



Many other aspects of the cosmological constant problem and possible solutions are reviewed by \textcite{Weinberg:1988cp}; \textcite{Polchinski:2006gy}; and \textcite{Bousso:2007gp}.

\section{Conclusions}

Barring surprises, we are facing a choice between two roads. One of them,
the traditional symmetry-based road of particle physics, may ultimately lead nowhere. A uniquely determined theory
of the universe and all of its physics leaves us with profound conundrums regarding the existence of life. The other
road, leading towards a huge landscape, is much more satisfactory in this respect, but is intrinsically much harder
to    confirm. Low energy supersymmetry might have helped, but is a luxury we may not have. The Susy-GUT idea, the lamppost of the
symmetry road, is losing its shine. 
GUTs do not fit as comfortably in the string landscape as most people believe, and susy does not fit well with the data; 
the ways out
are increasingly becoming epicyclical.
Confusingly, the opposite is also true: GUTs still look as attractive as ever from a low energy perspective,
and the landscape, despite many arguments going both ways, may  prefer  low energy susy after all.

Will we ever know?
Here are some possible future developments that  would cast serious doubts on the string theory landscape
\begin{itemize}
\item{The evidence for a well-distributed and connected dS landscape in string theory crumbles.}  
\item{Low-energy supersymmetry is strongly  predicted, but not seen at LHC (or vice-versa).}
\item{Solid evidence for variations of constants of nature emerges.}
\end{itemize}
There is movement on all of these fronts, and in twenty years we will probably have a different view on all of them. 
There are plenty of other  possibilities  for game-changing developments.

In the string theory landscape, the key concept  linking all these issues is: {\em Moduli}. This is where all lines meet: 
supersymmetry breaking and its scale, variations of constants, axions and the strong CP problem, (eternal) inflation,  dark matter, the
cosmological constant and/or quintessence, and ultimately the existence and features of the string landscape itself. 

But suppose there is no convincing experimental falsification on any of these  issues, {\it then} will we ever know?
Ultimately the convincing evidence may have to come from theory alone. Of all the  open theoretical issues, the
measure problem of eternal inflation is probably the biggest headache. But not everything hinges on that.
In the context of string theory, the following problems can be addressed without it.
\begin{itemize}
\item{Derive string theory from a  principle of nature.}
\item{Establish its consistency.}
\item{Prove that it has a landscape.}
\item{Prove that the Standard Model is in that landscape.}
\item{Show that all quantities are sufficiently densely distributed to explain all anthropic fine-tunings.}
\item{Confirm that these vacua are connected by some physical process, so that they can all be sampled.}
\end{itemize}

Perhaps this is as far as we will ever be able to go. We may never be able to derive our laws of physics, but 
we may  just feel comfortable with our place in the landscape. This requires understanding our environment, not
just the point where we live, but also the region around it.
This can fail dramatically and cast severe doubts on certain landscape assumptions.
Therefore a large part of this review has been devoted to all the impressive work that has been done in this area during the last decade. There is great physics in anthropic reasoning!

\begin{acknowledgments}
The author would like to thank Fred Adams, Pascal Anastasopoulos, Luke Barnes, Reinier de Adelhart Toorop, Shanta de Alwis, Raphael Bousso,  Michele Cicoli, Kiwoon Choi, John Donoghue, Michael Douglas,  Ben Freivogel, Gian Giudice, Mariana Gra\~na, Stefan Groot-Nibbelink, Arthur Hebecker, Lennaert Huiszoon, Renato Higa, Gordy Kane, Andre Lukas, 
Ulf Meissner, Sander Mooij, Piet Mulders, Hans-Peter Nilles, Kalliopi Petraki, Fernando Quevedo, Robert Richter, Marc Sher, Gary Shiu, Rob Timmermans, Patrick Vaudrevange, Jordy de Vries, Timo Weigand and Bernard de Wit for discussions, e-mail exchanges and for pointing
out misprints, and apologizes for 
all the details of those discussions that did not make in into this review.  Special thanks to Beatriz Gato-Rivera and Marieke Postma for carefully
reading the manuscript, critical comments and discussions.
Also thanks to the Newton Institute, Cambridge, and  the CERN theory group for hospitality. This work has been partially 
supported by funding of the Spanish Ministerio de Econom\'\i a y Competitividad, Research Project
FIS2012-38816, and by the Project CONSOLIDER-INGENIO 2010, Programme CPAN
(CSD2007-00042). 
\end{acknowledgments}

\newpage

\bibliography{Landscape}{}

\end{document}